\documentclass[final,epsfig]{habi}
\usepackage{master}
\usepackage{graphicx}
\usepackage{subfigure}
\usepackage{amssymb}
\usepackage{amsmath}
\usepackage{epsfig}
\usepackage{latexsym}
\begin{document}
\frontmatter

\maketitle
\tableofcontents
%\listoffigures
%\listoftables

\mainmatter
\bibliographystyle{plain}
%%%%%%%%%%%%%%%%%%%%%%%%%%%%%%%%%%%%%%%%%%%%%%%%%%%%%%%%%%%%%%%%%%%%%%%%%

\chapter{Introduction}
One of the most outstanding problems of theoretical physics is to
unify our picture of electroweak and strong interactions with
gravitational interactions. We would like to view the attraction of
masses as appearing due to the exchange of particles (gravitons)
between the masses. In conventional perturbative quantum field theory
this is not possible because the theory of gravity is not
renormalizable. A promising candidate providing a unified picture is
string theory. In string theory, gravitons appear together with the
other particles as excitations of a string.

On the other hand, also from an observational point
of view gravitational interactions show some essential
differences to the other interactions . Masses always attract each
other, and the strength of the 
gravitational interaction is much weaker than the electroweak and
strong interactions. A way how this difference could enter a theory is
provided by the concept of ``branes''. The expression ``brane'' is
derived from 
membrane and stands for extended objects on which interactions are
localized. Assuming that gravity is the only interaction which is not
localized on a brane, the special features of gravity can be
attributed to properties of the extra dimensions where only gravity
can propagate. (This can be either the size of the extra dimension or
some curvature.) 

The brane picture is embedded in a natural way in string
theory. Therefore, string theory has the prospect to unify gravity
with the strong and electroweak interactions while, at the same time,
explaining the difference between gravitational and the other 
interactions. 

This set of notes is organized as follows. In chapter \ref{chap:pert},
we briefly introduce the concept of strings and show that 
quantized closed strings yield the graviton as a string excitation. 
We argue that the quantized string lives in a ten dimensional target
space. 
It is  shown that an effective field theory description of strings is
given by (higher dimensional supersymmetric extensions of) the Einstein
Hilbert theory.
The concept of compactifying extra dimensions is introduced and
special stringy features are emphasized. 
Thereafter, we introduce the orbifold fixed planes as higher
dimensional extended objects where closed string twisted sector
excitations are localized. The quantization of the open string will
lead us to the concept of D-branes, branes on which open string
excitations live. We compute the tensions and charges of D-branes and 
derive an effective field theory on the world volume of the D-brane.
Finally, perturbative string theory contains orientifold planes as
extended objects. These are branes on which excitations of unoriented
closed strings can live. Compactifications containing orientifold
planes and D-branes are candidates for phenomenologically interesting
models. We demonstrate the techniques of orientifold compactifications
at a simple example.  

In chapter \ref{chap:nonpert}, we identify some of the extended
objects of chapter \ref{chap:pert} as stable solutions of the
effective field theory descriptions of string theory. These will be
the fundamental string and the D-branes. In addition we will find
another extended object, the NS five brane, which cannot be
described in perturbative string theory.

Chapter \ref{chap:appl} discusses some applications of the properties
of branes derived in the previous chapters. One of the problems of
perturbative string theory is that the string concept does not lead to
a unique theory. However, it has been conjectured that all the
consistent string theories are perturbative descriptions of one
underlying theory called M-theory. We discuss how branes fit into this
picture. We also present branes as tools for illustrating duality
relations among field theories. Another application, we are
discussing is based on the twofold description of three dimensional
D-branes. The perturbative description leads to an effective conformal
field theory (CFT) whereas the corresponding stable solution to
supergravity contains an AdS space geometry. This observation results
in the AdS/CFT correspondence. We present in some detail, how the
AdS/CFT correspondence can be employed to compute Wilson loops in
strongly coupled gauge theories.
An application which is of phenomenological interest is the fact that 
D-branes allow to construct models in which the string scale is of
the order of a TeV. If such models are realized in nature, they should
be discovered experimentally in the near future.

Chapter \ref{chap:braneworlds} is somewhat disconnected from the rest
of these notes since it considers brane models which are not directly 
constructed from strings. Postulating the existence of branes on which
certain interactions are localized, we present the construction of 
models in which the space transverse to the brane is curved. We
discuss how an observer on a brane experiences gravitational
interactions. We also make contact to the AdS/CFT conjecture for a
certain model.
Also other questions of phenomenological relevance are addressed. These
are the hierarchy problem and the problem of the cosmological
constant. We show how these problems are modified in models containing
branes. 

Chapter \ref{furtherreading} gives hints for further reading and
provides the sources for the current text. 

Our intention is that this review should be self contained and be
readable by people who know some quantum field theory and general
relativity. We hope that some people will enjoy reading one or the
other section.

\newpage
\chapter{Perturbative description of branes}\label{chap:pert}
\section{The Fundamental String}\label{section1}
\subsection{Worldsheet Actions}
\setcounter{equation}{0}
\subsubsection{The closed bosonic string}\label{classbo}

Let us start with the simplest string -- the bosonic string. The
string moves along a surface through space and time. This surface is
called the worldsheet (in analogy to a worldline of a point
particle). For space and time in which the motion takes place we will
often use the term target space. Let $d$ be the number of target space
dimesnions. 
The coordinates of the target space
are $X^\mu$, and the worldsheet is a surface $X^\mu\left( \tau,
  \sigma\right)$, where $\tau$ and $\sigma$ are the time and space
like variables parameterizing the worldsheet. String theory is
defined by the requirement that the classical motion of the string
should be such that its worldsheet has minimal area. Hence, we choose
the action of the string proportional to the worldsheet. The
resulting action is called Nambu Goto action. It reads
\begin{equation}\label{nambugoto}
S = -\frac{1}{2\pi \alpha^\prime}\int d^2\sigma\, \sqrt{-g}.
\end{equation}   
The integral is taken over the parameter space of $\sigma$ and
$\tau$. (We will also use the notation $\tau = \sigma^0$, and $\sigma
= \sigma^1$.) The determinant of the induced metric is called $g$. The
induced metric depends on the shape of the worldsheet and the shape
of the target space,
\begin{equation}\label{induced}
g_{\alpha\beta} = G_{\mu\nu}\left(X\right)\partial_\alpha X^\mu
\partial_\beta X^\nu ,
\end{equation}
where $\mu , \nu$ label target space coordinates, whereas $\alpha ,
\beta$ label worldsheet parameters.
Finally, we have introduced a constant $\alpha^\prime$. It is the inverse
of the string tension and has the mass dimension $-2$. The choice of
this constant sets the string scale.  By construction, the action
(\ref{nambugoto}) is invariant 
under reparametrizations of the worldsheet.

Alternatively, we could have introduced an independent metric
$\gamma_{\alpha \beta}$ on the worldsheet. This enables us to write
the action (\ref{nambugoto}) in an equivalent form,
\begin{equation}\label{polyakov}
S = -\frac{1}{4\pi \alpha^\prime} \int d^2\sigma
\sqrt{-\gamma}\gamma^{\alpha\beta}
G_{\mu\nu}\partial_\alpha X^\mu\partial_\beta X^\nu .
\end{equation}
For the target space metric we
will mostly use the Minkowski metric $\eta_{\mu\nu}$ in the present
chapter.
Varying (\ref{polyakov}) with respect to $\gamma_{\alpha\beta}$ yields
the energy momentum tensor,
\begin{equation}\label{eptensor}
T_{\alpha\beta} =
-\frac{4\pi\alpha^\prime}{\sqrt{-\gamma}}\frac{\delta S}{\delta
  \gamma^{\alpha\beta}} = \partial_\alpha X^\mu \partial_\beta X_\mu
  -\frac{1}{2}
  \gamma_{\alpha\beta}\gamma^{\delta\gamma}\partial_{\delta}X^\mu
  \partial_{\gamma} X_\mu ,
\end{equation}
where the target space index $\mu$ is raised and lowered with
$G_{\mu\nu} = \eta_{\mu\nu}$.
Thus, the $\gamma_{\alpha\beta}$ equation of motion,
$T_{\alpha\beta}=0$, equates
$\gamma_{\alpha\beta}$ with the induced metric (\ref{induced}), and
the actions (\ref{nambugoto}) and (\ref{polyakov}) are at least
classically equivalent. If we had just used covariance as a guiding
principle we would have written down a more general expression for
(\ref{polyakov}). We will do so later. At the moment, (\ref{polyakov})
with $G_{\mu\nu}=\eta_{\mu\nu}$ describes a string propagating in the trivial
background. Upon quantization of this theory we will see that the
string produces a spectrum of target space fields. Switching on non
trivial vacua for those target space fields will modify
(\ref{polyakov}). But before quantizing the theory, we would like to
discuss the symmetries and introduce supersymmetric versions of
(\ref{polyakov}).  

First of all, (\ref{polyakov}) respects the target space symmetries
encoded in $G_{\mu\nu}$. In our case $G_{\mu\nu}=\eta_{\mu\nu}$ this
is nothing but $d$ dimensional Poincar\'e invariance. From the two
dimensional point of view, this symmetry corresponds to field
redefinitions in (\ref{polyakov}). The action is also invariant
under two dimensional coordinate changes
(reparametrizations). Further, it is Weyl invariant, i.e.\ it does not
change under
\begin{equation}\label{weyl}
\gamma_{\alpha\beta} \rightarrow e^{\varphi\left(\tau ,\sigma \right)}
  \gamma_{\alpha\beta} .
\end{equation}
It is this property which makes one dimensional objects special.
The two dimensional coordinate transformations together with the Weyl
transformations are sufficient to transform the worldsheet metric locally
to the Minkowski metric,
\begin{equation}\label{minkowski}
\gamma_{\alpha\beta} = \eta_{\alpha\beta} .
\end{equation}
It will prove useful to use instead of $\sigma^0, \sigma^1$ the light
cone coordinates,
\begin{equation}\label{redef}
\sigma^- = \tau -\sigma \,\,\, ,\mbox{and}\,\,\, \sigma^+ =\tau
+\sigma .
\end{equation}
So, the gauged fixed version\footnote{Gauge fixing means imposing
  (\ref{minkowski}).} of (\ref{polyakov}) is
\begin{equation}\label{fixed}
S=\frac{1}{2\pi\alpha^\prime} \int d\sigma^+ d\sigma^- \partial_-
X^\mu\partial_+ X_\mu .
\end{equation}
However, the reparametrization invariance is not completely
fixed. There is a residual invariance under the conformal coordinate
transformations, 
\begin{equation}\label{conf}
\sigma^+ \rightarrow
\tilde{\sigma}^+\left(\sigma^+\right)\,\,\,\,\,\, , \,\,\,\,\,\, 
\sigma^-\rightarrow \tilde{\sigma}^-\left(\sigma^-\right) .
\end{equation}
This invariance is connected to the fact that the trace of the energy
momentum tensor (\ref{eptensor}) vanishes identically, $T_{+
  -}=0$\footnote{The corresponding symmetry is called conformal
  symmetry. It means that the action is invariant under conformal
  coordinate transformations while keeping the worldsheet metric
  fixed. In two dimensions this is equivalent to Weyl invariance.} .
However, the other $\gamma_{\alpha\beta}$ equations are not identically
satisfied and provide constraints, supplementing (\ref{fixed}),
\begin{equation}\label{constraint}
T_{++}=T_{--}=0.
\end{equation}
The equations of motion corresponding to (\ref{fixed})
are\footnote{For the time being we will focus on closed strings. That
  means that we impose periodic boundary conditions and hence there
  are no boundary terms when varying the action. We will discuss open
  strings when turning to the perturbative description of D-branes in
  section \ref{D-branes}.}
\begin{equation}\label{eom}
\partial_+\partial_- X^\mu =0
\end{equation}
Employing conformal invariance (\ref{conf}) we can choose $\tau$ to be
an arbitrary solution to the equation $\partial_+\partial_-\tau =
0$. (The combination  of (\ref{conf}) and (\ref{redef}) gives 
\begin{equation}
\tau
\rightarrow \frac{1}{2}\left(\tilde{\sigma}^+\left(\sigma^+\right)
  +\tilde{\sigma}^-\left(\sigma^-\right)\right) ,
\end{equation}
which is the general solution to (\ref{eom})). Hence, without loss of
generality we can fix
\begin{equation} \label{lightconegauge}
X^+ = \frac{1}{\sqrt{2}}\left( X^0 + X^{1}\right) = x^+ + p^+\tau , 
\end{equation}
where $x^+$ and $p^+$ denote the center of mass position and momentum
of the string in the + direction, respectively. The constraint
equations (\ref{constraint}) can now be used to fix 
\begin{equation}
X^- = \frac{1}{\sqrt{2}}\left( X^0 - X^{1}\right)
\end{equation}
as a function of $X^i$ ($i=2,\ldots ,d-1$)
uniquely up to an integration constant corresponding to the center
of mass position in the minus direction.
Thus we are left with $d-2$ physical degrees of freedom $X^i$. Their
equations of motion are (\ref{eom}) without any further constraints.
By employing the symmetries of (\ref{polyakov}) we managed to reduce
the system to $d-2$ free fields (satisfying (\ref{eom})). Since these
symmetries may suffer from quantum anomalies we will have to be
careful when quantizing the theory in section \ref{quantization}. 

\subsubsection{Worldsheet supersymmetry}\label{worldshitsusy}

In this section we are going to modify the previously discussed
bosonic string by enhancing its two dimensional symmetries. We will
start from the gauge fixed action (\ref{fixed}) which had as residual
symmetries two dimensional Poincar\'e invariance and conformal
coordinate transformations (\ref{conf}).\footnote{Alternatively, we
  could start from the action (\ref{polyakov}). This we would modify
  such that it becomes locally supersymmetric. Finally, we would fix
  symmetries in the locally supersymmetric action.}
A natural extension of Poincar\'e invariance is
supersymmetry. Therefore, we will study theories which are
supersymmetric from the two dimensional point of view. In order to
construct a supersymmetric extension of (\ref{fixed}) one should first
specify the symmetry group and then use Noether's method to build an
invariant action. We will be brief and just present the result,
\begin{equation}\label{susy}
S=\frac{1}{2\pi\alpha^\prime}\int d\sigma^+ d\sigma^-\left( \partial_-
  X^\mu \partial_+ X_\mu +\frac{i}{2}\psi_+ ^\mu\partial_-\psi_{+\mu}
  +\frac{i}{2}\psi_- ^\mu\partial_+\psi_{-\mu}\right) ,
\end{equation}
where $\psi_\pm$ are two dimensional Majorana-Weyl spinors. To see
this, we first note that
\begin{equation}
i\psi_+\partial_-\psi_+ + i\psi_-\partial_+ \psi_- = -\frac{1}{2}
\left(\psi_+ , -\psi_-\right)\left( \rho^+\partial_+
  +\rho^-\partial_-\right) \left( \begin{array}{c} \psi_- \\
    \psi_+\end{array} \right) ,
\end{equation}
where 
\begin{equation}\label{twodgammas}
\rho^{\pm} = \rho^0 \pm \rho^1,
\end{equation}
with
\begin{equation}
\rho^0 = \left( \begin{array}{cc} 0 & -i \\ i & 0\end{array}\right)
\,\,\, \mbox{and}\,\,\, 
\rho^1 = \left( \begin{array}{cc} 0 & i \\ i & 0\end{array}\right) .
\end{equation}
It is easy to check that the above matrices form a two dimensional
Clifford algebra,
\begin{equation}
\left\{ \rho^\alpha , \rho^\beta\right\} = -2\eta^{\alpha\beta} .
\end{equation}
Also, note that $i\left(\psi_+ , -\psi_-\right)$ is the Dirac conjugate
of $\left(\begin{array}{c}\psi_- \\ \psi_+\end{array}\right)$ for real
$\psi_{\pm}$, i.e.\ of the Majorana spinor
$\left(\begin{array}{c}\psi_- \\ \psi_+\end{array}\right)$ . In
addition to two dimensional Poincar\'e invariance and invariance under
conformal coordinate transformations (\ref{conf})\footnote{Under the
  transformation (\ref{conf}) the spinor components transform as
$\psi_\pm\rightarrow
\left(\tilde{\sigma}^{\pm\prime}\right)^{-\frac{1}{2}}\psi_\pm$ .} the
action (\ref{susy}) is invariant under worldsheet supersymmetry,
\begin{equation}\label{suvarb}
\delta X^\mu =\bar{\epsilon}\psi^\mu = i\epsilon_+\psi_- ^\mu
-i\epsilon_-\psi_+ ^\mu ,
\end{equation}
\begin{equation}\label{compact}
\delta \psi^\mu = -i\rho^\alpha \partial_\alpha X^\mu \epsilon .
\end{equation}
In components (\ref{compact}) gives rise to the two equations
\begin{eqnarray}
\delta\psi^\mu _- & = & -2\epsilon_+ \partial_- X^\mu ,\label{suvarm}\\
\delta\psi^\mu _+ & = & 2\epsilon_-\partial_+ X^\mu . \label{suvarp}
\end{eqnarray}
When checking the invariance of (\ref{susy}) under (\ref{suvarb}),
(\ref{suvarm}), (\ref{suvarp}) one should take into account that
spinor components are anticommuting, e.g.\ $\epsilon_+\psi_- =
-\psi_-\epsilon_+$.
Since the supersymmetry parameters $\epsilon_\pm$ form a non chiral
Majorana spinor, the above symmetry is called $(1,1)$
supersymmetry. (In the end of this section we will also discuss the
chiral $(1,0)$ supersymmetry.)
To summarize, the action (\ref{susy}) has the following two dimensional
global symmetries: Poincar\'e invariance and supersymmetry. The
corresponding Noether currents are the energy momentum tensor,
\begin{eqnarray} 
T_{++}& = & \partial_+ X^\mu\partial_+ X_\mu + \frac{i}{2} \psi_+ ^\mu
\partial_+\psi_{+\mu} ,\label{enmopp}\\
T_{--}& = & \partial_- X^\mu\partial_- X_\mu + \frac{i}{2} \psi_- ^\mu
\partial_-\psi_{-\mu} ,\label{enmomm}
\end{eqnarray}
and the supercurrent
\begin{eqnarray}
J_+ = \psi_+ ^\mu\partial_+ X_\mu ,\label{sucup}\\
J_- = \psi_- ^\mu \partial_- X_\mu .\label{sucum}
\end{eqnarray}
The vanishing of the trace of the energy momentum tensor $T_{+-}\equiv
0$ is again a consequence of the invariance under the (local)
conformal coordinate transformations (\ref{conf}). The supercurrent is
a spin--$\frac{3}{2}$ object and naively one would expect to get four
independent components. That there are only two non-vanishing
components is a consequence of the fact that the supersymmetries
(\ref{suvarb}), (\ref{suvarm}), (\ref{suvarp}) leave the action invariant
also when we allow instead of constant $\epsilon_{\pm}$ for
\begin{equation}\label{losusy}
\epsilon_- = \epsilon_-\left( \sigma^+\right)\,\,\, \mbox{and} \,\,\,
\epsilon_+ = \epsilon_+\left(\sigma^-\right) ,
\end{equation}
i.e.\ they are ``partially'' local symmetries. Once again, the
vanishing of the energy momentum tensor is an additional constraint on
the system. We did not derive this explicitly here. But it can be
easily inferred as follows. In two dimensions the Einstein tensor
vanishes identically. Thus, if we were to couple to two dimensional
(Einstein) gravity, the constraint $T_{\alpha\beta}=0$ would
correspond to the Einstein equation. Similarly, the supercurrents
(\ref{sucup}), (\ref{sucum}) are constrained to vanish. (If the theory
was coupled to two dimensional supergravity, this would correspond to
the gravitino equations of motion.)

As in the bosonic case we can employ the symmetry (\ref{conf}) to fix
\begin{equation}\label{gaugefixbo}
X^+ = x^+ + p^+\tau .
\end{equation}
The local supersymmetry transformation (\ref{compact}) with
$\epsilon$ given by  (\ref{losusy}) can be used to gauge 
\begin{equation}\label{gaugefixfer}
\left(\begin{array}{c} \psi_- \\ \psi_+\end{array}\right)^{\mu = +} =
0.
\end{equation}
(We have written here the target space (light cone) index as $\mu = +$
in order to avoid confusion with the worldsheet spinor indices.)
Note, that the gauge fixing condition (\ref{gaugefixfer}) is
compatible with (\ref{gaugefixbo}) and the supersymmetry
transformations (\ref{suvarb}), (\ref{compact}), as
(\ref{gaugefixfer}) implies the supersymmetry transformation
\begin{equation}
\delta X^+ = 0 .
\end{equation}
The constraints (\ref{enmopp}), (\ref{enmomm}), (\ref{sucup}),
(\ref{sucum}) can be solved for $X^-$, and $\psi_\alpha ^{\mu = -}$
(here, $\alpha$ denotes the worldsheet spinor index). Therefore, after fixing
the local symmetries completely we are left with $d-2$ free bosons and
$d-2$ free fermions (from a two dimensional point of view).   

We should
note that in the closed string case (periodic boundary conditions in
bosonic directions) we have two choices for boundary conditions on the
worldsheet fermions. Boundary terms appearing in the variation of the
action vanish for either periodic or anti periodic boundary
conditions on worldsheet fermions. (Later, we will call the solutions
with antiperiodic fermions Neveu Schwarz (NS) sector and the ones with
periodic boundary conditions Ramond (R) sector.

Going back to (\ref{susy}), we note that alternatively we could have
written down a $(1,0)$ supersymmetric action by setting the left
handed fermions $\psi_+ ^\mu =0$. The supersymmetries are now given by
(\ref{suvarb}) and (\ref{suvarm}), only. The parameter $\epsilon_-$
does not occur anymore, and hence we have reduced the number of
supersymmetries by one half. More generally one can add left
handed fermions $\lambda^A _+$ which do not transform under
supersymmetries. Therefore, they do not need to be in the same
representation of the target space Lorentz group as the $X^\mu$
(therefore the index $A$ instead of $\mu$). Summarizing we obtain the
following $(1,0)$ supersymmetric action
\begin{equation} \label{hetero}
S=\frac{1}{2\pi\alpha^\prime}\int d\sigma^+d\sigma^-
\left(\partial_-X^\mu \partial_+X_\mu +\frac{i}{2} \psi^\mu _-\partial_{+}
  \psi_{-\mu} +\frac{i}{2}\sum_{A=1}^{N} \lambda_+ ^A
  \partial_-\lambda_{+A} \right).
\end{equation} 
this will turn out to be the worldsheet action of the heterotic
string. The energy momentum tensor is as given in (\ref{enmopp}),
(\ref{enmomm}) with $\lambda_+ ^A$ replacing $\psi_+ ^\mu$ in
(\ref{enmomm}). There is only one conserved supercurrent
(\ref{sucup}). 

Finally, we should remark that there are also extended versions of two
dimensional supersymmetry (see for example \cite{West:1990tg}). We will not be
dealing with those in this review.

\subsubsection{Space-time supersymmetric string}\label{greenschwarzstring}

In the above we have extended the bosonic string (\ref{polyakov}) to
a superstring from the two dimensional perspective. We called this
worldsheet supersymmetry. Another direction would be to extend
(\ref{polyakov}) such that the target space Poincar\'e invariance is
enhanced to target space supersymmetry. This concept leads to the
Green Schwarz string.
Space time supersymmetry means that the bosonic coordinates $X^\mu$
get fermionic partners $\theta^A$ (where $A$ labels the number of
supersymmetries $N$) such that the targetspace becomes a superspace.
In addition to Lorentz symmetry, the supersymmetric extension mixes
fermionic and bosonic coordinates,
\begin{eqnarray}
\delta \theta^A & = & \epsilon^A ,\label{tsusya}\\
\delta \bar{\theta} & = & \bar{\epsilon}^A ,\label{tsusyb}\\
\delta X^\mu & = & i\bar{\epsilon} \Gamma^\mu \theta^A ,\label{tsusyc}
\end{eqnarray}
where the global transformation parameter $\epsilon^A$ is a target
space spinor and $\Gamma^\mu$ denotes a target space Dirac matrix.
In order to construct a string action respecting the symmetries
(\ref{tsusya}) -- (\ref{tsusyc}) one tries to replace
$\partial_\alpha X^\mu$ by the supersymmetric combination
\begin{equation}\label{Pi}
\Pi^\mu _\alpha = \partial_\alpha X^\mu - i \bar{\theta}^A \Gamma^\mu
\partial_\alpha \theta^A .
\end{equation}
This leads to the following proposal for a space time supersymmetric
string action
\begin{equation}\label{proposal}
S_1 = -\frac{1}{4\pi\alpha^\prime} \int d^2\sigma \sqrt{-\gamma}
\gamma^{\alpha\beta} \Pi^\mu _\alpha \Pi_{\beta \mu }.
\end{equation}
Note that in contrast to the previously discussed worldsheet
supersymmetric string, (\ref{proposal}) consists only of bosons
when looked at from a two dimensional point of view.
The action (\ref{proposal}) is invariant under global target
space supersymmetry, i.e.\ Lorentz transformations plus the
supersymmetry transformations (\ref{tsusya}) -- (\ref{tsusyc}).
From the worldsheet perspective we have reparametrization invariance
and Weyl invariance (\ref{weyl}). This is again enough to fix the
worldsheet metric $\gamma_{\alpha\beta} = \eta_{\alpha\beta}$ (cf
(\ref{minkowski})). The resulting action will exhibit conformal
coordinate transformations (\ref{conf}) as residual symmetries. The
energy momentum tensor ((\ref{eptensor}) with $\partial_\alpha X^\mu$
replaced by $\Pi_\alpha ^\mu$ (\ref{Pi})) is again traceless. Like in
section \ref{classbo}, the vanishing of the energy momentum tensor gives two
constraints. We have seen that in the non-supersymmetric case fixing
conformal coordinate transformations and solving the constraints 
leaves effectively $d-2$ (transversal) bosonic
directions.\footnote{Since the field equations are different for
(\ref{proposal}) the details of the discussion in the bosonic case
will change. The above frame just gives a rough motivation for a
modification of (\ref{proposal}) carried out below.} 
In order for the target space supersymmetry not to be spoiled in
this process, we would like to reduce the number of fermionic
directions $\theta^A$ by a factor of
$$ \frac{2^{\frac{\left[ d-2\right]}{2}}}{ 2^{\frac{\left[
        d\right]}{2}}} = \frac{1}{2} $$
simultaneously.
So, we need an additional local symmetry whose gauge fixing will
remove half of the fermions $\theta^A$. The symmetry we are looking
for is known as $\kappa$ symmetry. It exists only in special
circumstances. First of all, the number of supersymmetries should not
exceed $N=2$ (i.e.\ $A=1,2$). Then, adding a further term 
\begin{eqnarray} \label{kappaterm}
S_2 & = &  \frac{1}{2\pi\alpha^\prime} \int d^2\sigma \left\{
  -i\epsilon^{\alpha\beta} \partial_\alpha X^\mu\left(
    \bar{\theta}^1\Gamma_\mu\partial_\beta \theta^1
    -\bar{\theta}^2\Gamma_\mu \partial_\beta
    \theta^2\right)\right. \nonumber \\
 & & \; \left. +\epsilon^{\alpha\beta}
  \bar{\theta}^1\Gamma^\mu\partial_{\alpha}\theta^1
  \bar{\theta}^2\Gamma_\mu\partial_\beta \theta^2\right\}  
\end{eqnarray}
to (\ref{proposal}) results in a $\kappa$ symmetric action. (We will give
the explicit transformations below.) In (\ref{kappaterm})
$\epsilon^{\alpha\beta}$ denotes the two dimensional Levi Civita
symbol. If one is interested in less than $N=2$ one can just put the
corresponding $\theta^A$ to zero. The requirement that adding $S_2$ to
the action does not spoil supersymmetry (\ref{tsusya}) --
(\ref{tsusyc}), leads to further constraints,\\
\begin{tabular}{l l}
$(i)$ & $d=3$ and $\theta$ is Majorana \\
$(ii)$ & $d=4$ and $\theta$ is Majorana or Weyl\\
$(iii)$ & $d=6$ and $\theta$ is Weyl \\
$(iv)$ &  $d=10$ and $\theta$ is Majorana-Weyl.
\end{tabular}

It remains to give the above mentioned $\kappa$ symmetry transformations
explicitly. By adding $S_1$ and $S_2$ one observes that the kinetic
terms for the $\theta$'s (terms with one derivative acting on a
fermion) contain the following projection operators
\begin{equation}
P_{\pm}^{\alpha\beta} = \frac{1}{2}\left( \gamma^{\alpha\beta} \pm
  \frac{\epsilon^{\alpha\beta}}{\sqrt{-\gamma}}\right) .
\end{equation}
The transformation parameter for the additional local symmetry is
called $\kappa^A _\alpha$. It is a spinor from the target space
perspective and in addition a worldsheet vector subject to the
following constraints
\begin{eqnarray}
\kappa^{1\alpha} & = & P_- ^{\alpha\beta}\kappa_\beta ^1 , \\
\kappa^{2 \alpha} & = & P_+ ^{\alpha\beta}\kappa_\beta ^2 , \\
\end{eqnarray}
where the worldsheet indices $\alpha , \beta$ are raised and lowered
with respect to the worldsheet metric $\gamma_{\alpha\beta}$. Now, we
are ready to write down the $\kappa$ transformations,
\begin{eqnarray}
\delta\theta^A &=& 2i \Gamma^\mu \Pi_{\alpha\mu}
\kappa^{A\alpha},\label{kappaa}\\ 
\delta X^\mu & = & i\bar{\theta}^A \Gamma^\mu \delta\theta^A ,\\
\delta\left(\sqrt{-\gamma}\gamma^{\alpha\beta}\right) &=& -16
\sqrt{-\gamma}\left( 
  P_- ^{\alpha\gamma}\bar{\kappa}^{1\beta}\partial_\gamma \theta^1 +
  P_+
  ^{\alpha\gamma}\bar{\kappa}^{2\beta}\partial_\gamma\theta^2\right). 
\label{kappae}      
\end{eqnarray}
For a proof that these transformations leave $S_1 + S_2$ indeed invariant we
refer to\cite{Green:1987sp} for example.

Once we have established that the number of local symmetries is
correct, we can now proceed to employ those symmetries and reduce the
number of degrees of freedom by gauge fixing. We will go to the light
cone gauge in the following. Here, we will discuss only the most
interesting case of $d=10$. As
usual we use reparametrization and Weyl 
invariance to fix $\gamma_{\alpha\beta} =\eta_{\alpha\beta}$. We can
fix $\kappa$ symmetry (\ref{kappaa})--(\ref{kappae}) by the choice
\begin{equation}\label{kappafix}
\Gamma^+\theta^1 = \Gamma^+\theta^2 = 0,
\end{equation}
where
\begin{equation}
\Gamma^{\pm} = \frac{1}{\sqrt{2}}\left( \Gamma^0 \pm \Gamma^9\right).
\end{equation}
This sets half of the components of $\theta$ to zero. 
With the $\kappa$ fixing condition (\ref{kappafix}) the equations of
motion for $X^+$ and $X^i$ ($i=2,\ldots, d-1$) turn out to be free
field equations (cf (\ref{eom})). 
The reason for this can be easily seen as follows. After imposing
(\ref{kappafix}), out of the fermionic terms
only those containing $\bar{\theta^A}\Gamma^-\theta^A$ remain in the
action $S_1 + S_2$. Especially, the terms fourth order in $\theta^A$
have gone. 
The above mentioned terms with $\Gamma^-$ couple to
$\partial_\alpha X^+$, and hence they will only have influence on the
$X^-$ equation (obtained by taking the variation of the action with
respect to $X^+$).  
Thus we can again fix the conformal
coordinate transformations by the choice (\ref{lightconegauge}). The
$X^-$ direction is then fixed (up to a constant) by imposing the
constraint of vanishing energy momentum tensor. Since the coupling of
bosons and fermions is reduced to a coupling to $\partial_\alpha X^+$,
there is just a constant $p^+$ in front of the free kinetic terms of
the fermions.

In the light-cone
gauge described above the target space symmetry has been fixed up to
the subgroup $SO(8)$, where the $X^i$ and the $\theta^A$
transform in eight dimensional representations.\footnote{A
  Majorana-Weyl spinor in ten dimensions has 16 real
  components. Imposing (\ref{kappafix}) leaves eight.} For $SO(8)$
there are 
three inequivalent eight dimensional representations, called
$\mbox{\bf 8}_{\mbox{\scriptsize \bf v}}$, $\mbox{\bf
    8}_{\mbox{\scriptsize \bf s}}$ , and  $\mbox{\bf
      8}_{\mbox{\scriptsize \bf c}}$. The group indices are chosen as
      $i,j,k$ for the  $\mbox{\bf 8}_{\mbox{\scriptsize \bf v}}$,
        $a,b,c$ for the $\mbox{\bf 8}_{\mbox{\scriptsize \bf s}}$, and
          $\dot{a},\dot{b},\dot{c}$ for the $\mbox{\bf
            8}_{\mbox{\scriptsize \bf c}}$. In particular, $X^i$
            transforms in the vector representation $\mbox{\bf
              8}_{\mbox{\scriptsize \bf v}}$.   
For the target space spinors we can choose either $\mbox{\bf
  8}_{\mbox{\scriptsize \bf s}}$ or $\mbox{\bf 8}_{\mbox{\scriptsize
    \bf c}}$.
Absorbing also the constant in front of the kinetic terms in a field
redefinition we specify this choice by the following notation
\begin{eqnarray}
\sqrt{p^+}\theta^1  &\rightarrow S^{1a}& \,\,\,\,\,\,
\mbox{or}\,\,\,\,\,\, S^{1\dot{a}} \\
\sqrt{p^+}\theta^2  &\rightarrow S^{2a}& \,\,\,\,\,\,
\mbox{or}\,\,\,\,\,\, S^{2\dot{a}} .
\end{eqnarray}
Essentially, we have here two different cases: we take the same
$SO(8)$ representation for both $\theta$'s or we take them mutually
different. The first option results in type IIB theory whereas the
second one leads to type IIA. 

So, the gauge fixing procedure simplifies the theory substantially. The
equations of motion for the remaining degrees of freedom are just free
field equations. For example for the type IIB theory they read,
\begin{eqnarray}
\partial_+\partial_- X^i &=& 0 , \\
\partial_+ S^{1a} & = & 0, \label{spinorone}\\
\partial_- S^{2a} & = & 0. \label{spinortwo}
\end{eqnarray}
They look almost equivalent to the equations of motion one obtains
from the worldsheet supersymmetric action (\ref{susy}) after
eliminating the $\pm$ directions by the light cone gauge. Especially,
(\ref{spinorone}) and (\ref{spinortwo}) have the form of two
dimensional Dirac equations where $S^1$ and $S^2$ appear as 2d
Majorana-Weyl spinors. An important difference is however, that in
(\ref{susy}) all worldsheet fields transform in the vector
representation of the target space subgroup $SO(d-2)$. 

In the rest of this chapter we will focus only on the worldsheet
supersymmetric formulation. There, target space fermions will appear
in the Hilbert space when quantizing the theory. We will come back to
the Green Schwarz string only when discussing type IIB strings living
in a non-trivial target space ($AdS_5\times S^5$) in section
\ref{adssection}. 

\subsection{Quantization of the fundamental string}\label{quantization}
\setcounter{equation}{0}

\subsubsection{The closed bosonic string}
Our starting point is equation (\ref{eom}). 
\begin{equation}
\partial_+ \partial_- X^i = 0.
\end{equation}
Imposing periodicity under shifts of $\sigma^1$ by $\pi$ we obtain the
following general solutions\footnote{Frequently, we will put
  $\alpha^\prime =\frac{1}{2}$. Since it is the only dimensionfull
  parameter (in the system with $\hbar = c =1$), it is easy to
  reinstall it when needed.}
\begin{equation}
X^\mu = X_R ^\mu\left( \sigma^-\right) + X_L ^\mu\left( \sigma^+\right) ,
\end{equation}
with 
\begin{eqnarray}
X_R ^\mu & = & \frac{1}{2} x^\mu + \frac{1}{2} p^\mu \sigma^- +
\frac{i}{2}\sum_{n\not= 0} \frac{1}{n} \alpha^\mu _n
e^{-2in\sigma^-},\label{leftsol} \\
X_L ^\mu & = & \frac{1}{2} x^\mu + \frac{1}{2} p^\mu \sigma^+ +
\frac{i}{2}\sum_{n\not= 0} \frac{1}{n} \tilde{\alpha}^\mu _n
e^{-2in\sigma^+}.\label{rightsol} 
\end{eqnarray}
Here, all $\sigma^\alpha$ dependence is written out explicitly, i.e.\
$x^\mu$, $p^\mu$, $\alpha_n ^\mu$, and $\tilde{\alpha}_n ^\mu$ are
$\sigma^\alpha$ independent operators. Classically, one can associate
$x^\mu$ with the center of mass position, $p^\mu$ with the center of mass
momentum and $\alpha_n ^\mu$ ($\tilde{\alpha}_n ^\mu$) with the amplitude
of the $n$'th right moving (left moving) vibration mode of the string
in $x^\mu$ direction. Reality of $X^\mu$ imposes the relations
\begin{equation}\label{reality}
\alpha_n ^{\mu\dagger} = \alpha_{-n}^\mu \,\,\, \mbox{and}\,\,\,
\tilde{\alpha}_n ^{\mu\dagger} = \tilde{\alpha}^\mu _{-n}.
\end{equation} 
We also define a zeroth vibration coefficient via
\begin{equation}
\alpha^\mu _0 = \tilde{\alpha}^\mu _0 = \frac{1}{2} p^\mu.
\end{equation}

Since the canonical momentum is obtained by
varying the action (\ref{fixed}) with respect to
$\dot{X}^\mu$ (where the dot means derivative with respect to $\tau$) we
obtain the following canonical quantization prescription. The equal
time commutators are given by
\begin{equation}\label{cancoma}
\left[X^\mu\left( \sigma\right) , X^\nu\left(\sigma^\prime\right)\right] =
\left[\dot{X}^\mu\left( \sigma\right) ,
  \dot{X}^\nu\left(\sigma^\prime\right)\right] = 
0,
\end{equation}
and
\begin{equation}\label{cancomb}
\left[ \dot{X}^\mu\left( \sigma\right) ,
  X^\nu\left(\sigma^\prime\right)\right] = 
-i\pi \delta\left( \sigma - \sigma^\prime\right)\eta^{\mu\nu}
\end{equation}
where the delta function is a distribution on periodic
functions. Formally it can be assigned a Fourier series
\begin{equation}
\delta\left(\sigma\right) = \frac{1}{\pi} \sum_{k=-\infty}^\infty
e^{2ik \sigma} . 
\end{equation}
With this we can translate the canonical commutators (\ref{cancoma}) and
(\ref{cancomb}) into commutators of the Fourier coefficients appearing
in (\ref{leftsol}) and (\ref{rightsol}),
\begin{eqnarray}
\left[ p^\mu ,x^\nu\right] & = & -i\eta^{\mu\nu} ,\label{oalga} \\
\left[ \alpha_n ^\mu , \alpha_k ^\nu\right] & = & n \delta_{n+k}
\eta^{\mu\nu} , \label{oalgb}\\
\left[\tilde{\alpha}_n ^\mu, \tilde{\alpha}_k ^\nu\right] & = & n \delta_{n+k}
\eta^{\mu\nu} \label{oalgc},
\end{eqnarray}
where $\delta_{n+k}$ is shorthand for $\delta_{n+k,0}$.
So far, we did not take into account the constraints of vanishing
energy momentum tensor (\ref{constraint}). To do so we go again to the
light cone gauge (\ref{lightconegauge}), i.e.\ set 
\begin{equation}
\alpha_n ^+ = \tilde{\alpha}_n ^+ = 0 \,\,\, \mbox{for}\,\,\, n\not=
0.
\end{equation}
Now the constraint (\ref{constraint}) can be used to eliminate $X^-$
(up to $x^-$), or alternatively the $\alpha^- _{n}$ and $\tilde{\alpha}_n
^-$,  
\begin{eqnarray}
p^+\alpha_n ^ - &=&  \sum_{m= -\infty} ^{\infty}
: \alpha_{n-m}^i\alpha_m ^i :\, - 2a\delta_n, \label{constsola}\\
p^+\tilde{\alpha}_n ^- &=&  \sum_{m= -\infty} ^{\infty}
: \tilde{\alpha}_{n-m}^i\tilde{\alpha}_m ^i :\, - 2a\delta_n \label{constsolb}
\end{eqnarray}
where a sum over repeated indices $i$ from $2$ to $d-1$ is understood.
The colon denotes normal ordering to be specified below. We have
parameterized the ordering ambiguity by a constant $a$. (In principle
one could have introduced two constants $a$, $\tilde{a}$. But this
would lead to inconsistencies which we will not discuss here.)
Equations (\ref{constsola}) and (\ref{constsolb}) are not to be read as
operator identities but rather as conditions on physical states which
we will construct now. 
We choose the vacuum as an eigenstate of the $p^\mu$
\begin{equation} \label{vacuumeigen}
p^\mu \left| k\right> = k^\mu\left| k\right> ,
\end{equation}
with $k^\mu$ being an ordinary number. Further, we impose that the
vacuum is annihilated by half of the vibration modes,
\begin{equation} \label{vacuumanil}
\alpha_n ^i\left| k\right> = \tilde{\alpha}_n ^i \left| k\right> = 0
\,\,\, \mbox{for}\,\,\, n > 0.
\end{equation}
The rest of the states can now be constructed by acting with a certain
number of $\alpha_{-n} ^i$ and $\tilde{\alpha}_{-n} ^i$ ($n>0$) on the
  vacuum. But we still need to impose the constraint
  (\ref{constraint}). Coming back to (\ref{constsola}) and
  (\ref{constsolb}) we can now specify what is meant by the normal
  ordering. The $\alpha^i _k$ ($\tilde{\alpha}^i _k$) with the greater Fourier
  index $k$ is written to the right\footnote{E.g.\ for $k>0$ this
    implies that $:\alpha_k ^i \alpha_{-k}^i: = \alpha_{-k}^i\alpha_k
    ^i$, i.e.\ the annihilation operator acts first on a state.}. For
  $n\not=0$ (\ref{constsola}) and 
  (\ref{constsolb}) just tell us how any $\alpha^- _n$ or
  $\tilde{\alpha}^- _n$ can be expressed in terms of the $\alpha_k ^i$
  and $\tilde{\alpha}_l ^i$.  The nontrivial information is contained
  in the $n=0$ 
  case. It is convenient to rewrite (\ref{constsola}) and
  (\ref{constsolb}) for $n=0$,
\begin{equation}\label{masshell}
2p^+p^- - p^i p^i = 8(N-a) = 8(\tilde{N}-a),
\end{equation} 
where (doing the normal ordering explicitly)
\begin{eqnarray}
N &=& \sum_{n=1}^{\infty} \alpha_{-n}^i\alpha_n^i , \label{number}\\
\tilde{N} & = & \sum_{n=1}^{\infty} \tilde{\alpha}_{-n}^i
\tilde{\alpha}_n ^i . \label{numbertilde}
\end{eqnarray}
The $N$ ($\tilde{N}$) are number operators in the sense that they count
the number of creation operators $\alpha_{-n} ^i$
($\tilde{\alpha}_{-n}^i$)
acting on the vacuum. To be precise, the $N$ ($\tilde{N}$) eigenvalue of
a state is this number multiplied by the index $n$ and summed over all
different kinds of creation operators acting on the vacuum (for left
and right movers separately).
Interpreting the $p^\mu$ eigenvalue $k^\mu$ as the momentum of a
particle (\ref{masshell}) looks like a mass shell condition with the
mass squared $M^2$ given by
\begin{equation}\label{massformula}
M^2 = 8( N -a ) = 8(\tilde{N} -a) .
\end{equation}
The second equality in the above equation relates the allowed right
moving creation operators acting on the vacuum to the left moving
ones. It is known as the level matching condition.

For example, the first excited state is
\begin{equation}\label{firstex}
\alpha^i _{-1} \tilde{\alpha}^j _{-1} \left| k\right> .
\end{equation}
By symmetrizing or antisymmetrizing with respect to $i,j$ and splitting
the symmetric expression into a trace part and a traceless part one
sees easily that the states (\ref{firstex}) form three irreducible
representations of $SO(d-2)$. Since we have given the states the
interpretation of being particles living in the targetspace, these
should correspond to irreducible representations of the little
group. Only when the above states are massless the little group is
$SO(d-2)$ (otherwise it is $SO(d-1)$). Therefore, for unbroken covariance
with respect to the targetspace Lorentz transformation, the states
(\ref{firstex}) must be massless. Comparing with (\ref{massformula})
we deduce that the normal ordering constant $a$ must be one,
\begin{equation}\label{normalo}
a \stackrel{!}{=} 1.
\end{equation}

In the following we are going to compute the normal ordering constant
$a$. Requiring agreement with (\ref{normalo}) will give a condition on
the dimension of the targetspace to be 26. The following calculation
may look at some points a bit dodgy when it comes to computing the
exact value of $a$. So, before starting we should note that the
compelling result will be that $a$ depends on the targetspace
dimension. The exact numerics can be verified by other methods which
we will not elaborate on here for the sake of briefness. We will
consider only $N$ since the calculation with $\tilde{N}$ is a very
straightforward modification (just put tildes everywhere). The initial
assumption is that naturally the ordering in quantum expressions would
be symmetric, i.e.\
\begin{equation}
N- a = \frac{1}{2}\sum_{n=-\infty,n\not=0} ^{\infty}\alpha_{-n} ^i\alpha_n
^i .
\end{equation}
By comparison with the definition of $N$ (\ref{number}) and using the
commutation relations (\ref{oalgb}) we find
\begin{equation}\label{mumpitz}
a= - \frac{d-2}{2} \sum_{n=1}^\infty n.
\end{equation}
This expression needs to be regularized. A familiar method of
assigning a finite number to the rhs of (\ref{mumpitz}) is known as
`zeta function regularization'. One possible representation of the
zeta function is
\begin{equation}\label{zeta}
\zeta\left(s\right) = \sum_{n=1}^\infty n^{-s}.
\end{equation}
The above representation is valid for the real part of $s$ being
greater than one. The zeta function, however, can be defined also for
complex $s$ with negative real part. This is done by analytic
continuation. The way to make sense out of (\ref{mumpitz}) is now to
replace the infinite sum by the zeta function
\begin{equation}
a = -\frac{d-2}{2}\zeta\left( -1\right) = \frac{d-2}{24}.
\end{equation}
Comparing with (\ref{normalo}) we see that we need to take
\begin{equation}
d=26
\end{equation}
in order to preserve Lorentz invariance. This result can also be
verified in a more rigid way. Within the present approach one can
check that $a=1$ and $d=26$ are needed for the target space Lorentz
algebra to close. In other approaches, one sees that the Weyl symmetry
becomes anomalous for $d\not= 26$. 

Since $N$ and $\tilde{N}$ are natural numbers we deduce from
(\ref{massformula}) that the
mass spectrum is an infinite tower starting from $M^2 = -8 =
-4/\alpha^\prime$ and going up in steps of $8 = 4/\alpha^\prime$. The
presence of a tachyon (a state with negative mass square) is a
problem. It shows that we have looked at the theory in an unstable
vacuum. One possibility that this is not complete nonsense could be
that apart from the massterm the tachyon potential receives higher
order corrections (like e.g.\ a power of four term) with the opposite
sign. Then it would look rather like a Higgs field than a tachyon, and
one would expect some phase transition (tachyon condensation) to occur
such that the final theory is stable. For the moment, however, let us
ignore this problem (it will not occur in the supersymmetric theories
to be studied next). 

\begin{figure}
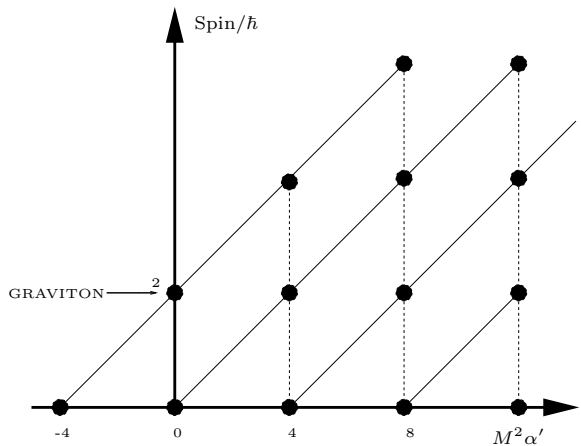
  
\begin{center}
\input MASSSPECTRUM.pstex_t
\end{center}
\caption{Mass spectrum of the closed bosonic string}
\label{massspektrum}
\end{figure}

The massless particles are described by (\ref{firstex}). The part
symmetric in $i,j$ and traceless corresponds to a targetspace
graviton. This is one of the most important results in
string theory. There is a graviton in the spectrum and hence string
theory can give meaning to the concept of quantum gravity. (Since
Einstein gravity cannot be quantized in a straightforward fashion
there is a graviton only classically. This corresponds to the
gravitational wave solution of the Einstein equations. The particle
aspect of the graviton is missing without string theory.) 
The trace-part of (\ref{firstex}) is called dilaton whereas the piece
antisymmetric in $i,j$ is simply the antisymmetric tensor field
(commonly denoted with $B$). A schematic summary of the particle
spectrum of the closed bosonic string is drawn in figure \ref{massspektrum}.

%%%%%%%%%%%%%%%%%%%%%%%%%%%%%%%%%%%%
%%%%%%%%%%%%%%%%%%%%%%%%%%%%%%%%%%%%

As a consistency check one may observe that the massive excitations
fit in $SO(25)$ representations, i.e.\ they form massive representations of
the little group of the Lorentz group in 26 dimensions.

As we have already mentioned, this theory contains a graviton, which
is good since it gives the prospect of quantizing gravity. On the
other hand, there is the tachyon, at best telling us that we are in
the wrong vacuum. (There could be no stable vacuum at all -- for
example if the tachyon had a run away potential.) Further, there are
no target space fermions in the spectrum. So, we would like to keep
the graviton but to get rid of the tachyon and add fermions. We will
see that this goal can be achieved by quantizing the supersymmetric
theories. 

\subsubsection{Type II strings}\label{typeII}

In this section we are going to quantize the (1,1) worldsheet
supersymmetric string. We will follow the lines of the previous
section but need to add some new ingredients. We start with the action
(\ref{susy}). The equations of motion for the bosons $X^\mu$ are
identical to the bosonic string. So, the mode expansion of the $X^\mu$
is not altered and given by (\ref{leftsol}) and (\ref{rightsol}). The
equations of motion for the fermions are,
\begin{eqnarray}
\partial_- \psi_+ ^\mu = 0 , \label{eomfermip}\\
\partial_+ \psi_- ^\mu =0. \label{eomfermim}
\end{eqnarray}
Further, we need to discuss boundary conditions for the worldsheet
fermions. Modulo the equations of motion (\ref{eomfermip}) and
(\ref{eomfermim}) the variation of the action (\ref{susy}) with respect
to the worldsheet fermions turns out to be\footnote{Again we put
  $\alpha^\prime = \frac{1}{2}$.}
\begin{equation}\label{boundaryterm}
\frac{i}{2\pi}\left.\left( - \psi_{+\mu}\delta \psi_+ ^\mu + \psi_{-\mu}
  \delta\psi_- ^\mu\right)\right|^{\pi}_{\sigma = 0}.
\end{equation}
For the closed string we need to take the variation of $\psi_+ ^\mu$
independent from the one of $\psi_- ^\mu$ at the boundary (because we do
not want the boundary condition to break part of the supersymmetry
(\ref{suvarm}) and (\ref{suvarp})). Hence, the spinor components can
be either periodic or anti-periodic under shifts of
$\sigma$ by $\pi$. The first option gives the Ramond (R) sector.
In the R sector the general solution to (\ref{eomfermip})
and (\ref{eomfermim}) can be written in terms of the following mode
expansion 
\begin{eqnarray}
\psi_- ^\mu =\sum_{n \in {\mathbb Z}}d_n ^\mu
e^{-2in\left(\tau -\sigma\right)} ,\\
\psi_+ ^\mu = \sum_{n \in {\mathbb Z}}\tilde{d}_n ^\mu
e^{-2in\left(\tau +\sigma\right)} .
\end{eqnarray}
The other option to solve the boundary condition is to take
anti-periodic boundary conditions. This is called the Neveu Schwarz
(NS) sector. 
In the NS sector the general solution to the equations of motion
(\ref{eomfermip}) and (\ref{eomfermim}) reads\footnote{The reality
  (Majorana) condition on the worldsheet spinor components provides
  relations analogous to (\ref{reality}).}
\begin{eqnarray}
\psi_- ^\mu = \sum_{r \in {\mathbb Z}+\frac{1}{2}}b_r ^\mu
e^{-2ir\left(\tau -\sigma\right)} ,\\
\psi_+ ^\mu = \sum_{r \in {\mathbb
    Z}+\frac{1}{2}}\tilde{b}_r ^\mu 
e^{-2ir\left(\tau +\sigma\right)} ,
\end{eqnarray}
where now the sum is over half integer numbers ($\ldots ,
-\frac{1}{2} , \frac{1}{2}, \frac{3}{2}, \ldots$).

For the bosons the canonical commutators are as given in
(\ref{cancoma}), (\ref{cancomb}). Hence, the oscillator modes satisfy again
the algebra (\ref{oalga}) -- (\ref{oalgc}). Worldsheet fermions commute
with worldsheet bosons. The canonical (equal time) anti-commutators for
the fermions are
\begin{eqnarray}
\left\{ \psi_+ ^\mu\left(\sigma\right) , \psi_+ ^\nu
  \left(\sigma^\prime\right)\right\} =
\left\{ \psi_- ^\mu\left(\sigma\right) , \psi_- ^\nu
  \left(\sigma^\prime\right)\right\} & = & \pi \eta^{\mu\nu} \delta\left(
  \sigma - \sigma^\prime\right) , \\
\left\{ \psi_+ ^\mu \left(\sigma\right) , \psi_-
  ^\nu\left(\sigma^\prime\right)\right\} & = & 0. 
\end{eqnarray}
For the Fourier modes this implies
\begin{equation}\label{anticom}
\left\{ b_r ^\mu ,b_s ^\nu\right\} = \left\{ \tilde{b}_r ^\mu
  ,\tilde{b}_s ^\nu\right\} = \eta^{\mu\nu}\delta_{r+s}
\end{equation}
in the NS sectors\footnote{We say NS sectors and not NS sector because
  there are two of them: a left and a right moving one.}, and
\begin{equation}\label{comramond}
\left\{ d_m ^\mu ,d_n ^\nu\right\} = \left\{ \tilde{d}_m ^\mu
  ,\tilde{d}_n ^\nu\right\} = \eta^{\mu\nu}\delta_{m+n}
\end{equation}
in the R sectors. Like the bosonic Fourier modes these can be split
into creation operators with negative Fourier index, and annihilation
operators with positive Fourier index. What about zero Fourier index?
For the NS sector fermions this does not occur. The vacuum is always
taken to be an eigenstate of the bosonic zero modes where the
eigenvalues are the target space momentum of the state. (This is
exactly like in the bosonic string discussed in the previous section.) 
The Ramond sector zero modes form a target space Clifford algebra (cf
(\ref{comramond})). This means that the Ramond sector states form a
representation of the $d$ dimensional Clifford algebra, i.e.\ they are
target space spinors. We will come back to this later. Pairing left
and right movers, there are
altogether four different sectors to be discussed: NSNS, NSR, RNS,
RR. In the NSNS sector for example the left and right moving
worldsheet fermions have both anti-periodic boundary conditions. The
vacuum in the NSNS sector is defined via (\ref{vacuumeigen}),
(\ref{vacuumanil}) and 
\begin{equation}\label{groundnsns}
b_r ^\mu \left| k\right> = \tilde{b}_r ^\mu \left| k\right> = 0 \,\,\,
\mbox{for} \,\,\, r>0.
\end{equation}
We can build states out of this by acting with bosonic left and right
moving creation operators on it. Further, left and right moving
fermionic creators from the NS sectors can act on
(\ref{groundnsns}). We should also impose the constraints
(\ref{enmopp}) -- (\ref{sucum}) on those states. As before, we do so
by going to the light cone gauge
\begin{equation}
\alpha^+ _n = \tilde{\alpha}^+ _n = b_r ^+ =\tilde{b}_r ^+ = 0.
\end{equation}
Then the constraints can be solved to eliminate the minus directions. The
important information is again in the zero mode of the minus
direction. This reads (\ref{masshell})
\begin{equation}\label{massshelf}
2p^+p^- - p^i p^i = 8(N_{NS}-a_{NS}) = 8(\tilde{N}_{NS} -a_{NS}).
\end{equation}
The expressions for the number operators are modified due to the
presence of (NS sector) worldsheet fermions
\begin{equation}\label{snumber}
N_{NS} = \sum_{n=1} ^\infty \alpha_{-n} ^i \alpha_n ^i +
\sum_{r=\frac{1}{2}}^\infty r b_{-r} ^i b_r ^i ,
\end{equation}
and the analogous expression for $\tilde{N}_{NS}$.
Its action on states is like in the bosonic case (see discussion below
(\ref{numbertilde})) taking into account the appearance of fermionic
creation operators. Again, we have put a so far undetermined normal
ordering constant in (\ref{massshelf}) and taken normal ordered
expressions for the number operators. Now, the first excited state is
\begin{equation}\label{excy}
b_{-\frac{1}{2}} ^i \tilde{b}_{-\frac{1}{2}}^j \left| k\right> .
\end{equation}
Its target space tensor structure is identical to the one of
(\ref{firstex}). In particular it forms massless representations of
the target space Lorentz symmetry. Thus, Lorentz covariance implies
that
\begin{equation}
a_{NS} = \frac{1}{2}
\end{equation}
should hold.

We compute now $a_{NS}$ by first naturally assuming that a symmetrized
expression appears on the rhs of (\ref{massshelf}). This gives (see also
(\ref{mumpitz})) 
\begin{equation} \label{mumpitzf}
a_{NS} = - \frac{d-2}{2}\sum_{n=1}^\infty n
+\frac{d-2}{2}\sum_{r=\frac{1}{2}}^\infty r .
\end{equation}
We use again the zeta function regularization to make sense out of
(\ref{mumpitzf}). For the second sum the following formula proves
useful
\begin{equation}\label{moddingformula}
\sum_{n=0}^\infty \left( n+c\right) = \zeta\left( -1,c\right) =
-\frac{1}{12}\left( 6c^2 -6c +1\right) .
\end{equation}
(Note, that splitting the lhs of (\ref{moddingformula}) into
$\zeta\left( -1\right) + c + c\zeta\left( 0\right)$ gives a different
(wrong) result. This is because we understand the infinite sum as an
analytic continuation of a finite one: $ \sum \left( n+a\right)^{-s}$
with real part of $s$ greater than one. For generic $s$ the above
splitting is not possible.) Anyway, with the regularization
prescription (\ref{moddingformula}) we get for (\ref{mumpitzf}) 
\begin{equation}
a_{NS} = \frac{d-2}{16}.
\end{equation}
We conclude that the critical dimension for the $(1,1)$ worldsheet
supersymmetric string is
\begin{equation}
d=10.
\end{equation}
Like in the bosonic string there are more rigid calculations giving
the same result. 

\begin{figure}
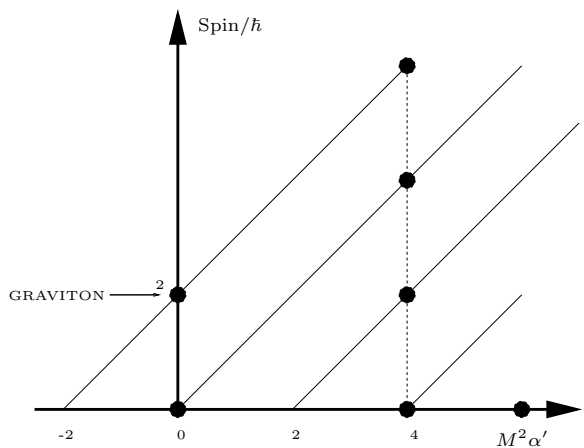
  
\begin{center}
\input NSfermion.pstex_t
\end{center}
\caption{NS-NS spectrum of the type II string. In comparison to figure
  \ref{massspektrum} the horizontal axis has been stretched by a factor
  of two.}
\label{NSspectrum}
\end{figure}

The massless spectrum from the NSNS sector is identical to the
massless spectrum of the closed bosonic string. Again, we have a
tachyon: the NSNS groundstate. Here, however this can be consistently
projected out. This is done by imposing the GSO (Gliozzi-Scherk-Olive)
projection. To specify what this
projection does in the NS sector we introduce fermion number operators
$F$ ($\tilde{F}$) counting the number of worldsheet fermionic NS
right (left) handed creation operators acting on the vacuum. In
addition, we assign to the right (left) handed NS vacuum an $F$
($\tilde{F}$) eigenvalue of one\footnote{This means that we can
  write $F= 1 +\sum_{r>0} b^i_{-r}b_r ^i$, and an analogous expression
  for $\tilde{F}$.} . Now, the GSO projection is carried
out by multiplying states with the GSO projection operator
\begin{equation} \label{gsoprojection}
P_{GSO} = \frac{1 +\left( -1\right)^F}{2} \frac{1 + \left(
    -1\right)^{\tilde{F}}}{2} .
\end{equation}
Obviously this does not change the first excited state (\ref{excy})
but removes the tachyonic NSNS ground state. There are several
reasons why this projection is consistent. At tree level\footnote{The
  worldsheet has the topology of a cylinder, or a sphere when Wick
  rotated to the Euclidean $2d$ signature.} for example one may check
that the particles which have been projected out do not reappear as
poles in scattering amplitudes. Imposing the GSO projection becomes
even more natural when looking at the one loop level. In the Euclidean
version this means that the worldsheet is a torus. Summing over all
possible spin structures (the periodicities of worldsheet fermions
when going around the two cycles of the torus) leads naturally to the
appearance of (\ref{gsoprojection}) in the string partition function
\cite{Seiberg:1986by} (see also \cite{Lust:1989tj}). The NSNS spectrum
subject to the GSO projection looks as follows. The number operator  
(\ref{snumber}) is quantized in half-integer steps. The GSO projection
removes half of the states, the groundstate, the first massive states,
the third massive states and so on. The NSNS spectrum of the type II
strings is summarized in figure \ref{NSspectrum}. 

%%%%%%%%%%%%%%%%%%%%%%%%%%%%%%%%%%%%
%%%%%%%%%%%%%%%%%%%%%%%%%%%%%%%%%%%%

We have achieved our goal of removing the tachyon from the spectrum
while keeping the graviton. We also want to have target space
spinors. We will see that those come by including the R sector into
the discussion. The most important issue to be addressed here is the
action of the zero modes on the R groundstate. By going to the
light-cone gauge, we can again eliminate the plus and minus (or the 0
and 1) directions leaving us with eight\footnote{We use here the
  previous result that we need to have $d=10$ in order to preserve
  target space Lorentz invariance.}  zero modes for the left and
right moving sectors each. We rearrange these modes into four complex
modes 
\begin{eqnarray}
D _1 & = & d_0 ^2 + i d_0 ^3 ,\label{creat1}\\
D _2 & = & d_0 ^4 + i d_0 ^5 ,\\
D _3 & = & d_0 ^6 + i d_0 ^7 ,\\
D _4 & = & d_0 ^8 + i d_0 ^9 .\label{creat4}
\end{eqnarray}
The only non-vanishing anti-commutators for these new operators are
($I=1,\ldots, 4$; no sum over $I$)
\begin{equation}\label{compcom}
\left\{ D_I , D^\dagger _I\right\} = 2.
\end{equation}
In particular, the $D_I$ and $D^\dagger _I$ are nilpotent. We can now
construct the right moving R vacuum by starting with a state which is
annihilated by all the $D_I$,\footnote{In this notation we suppress the
  eigenvalue $k^\mu $ of the bosonic zero modes.}
\begin{equation}\label{rvacuum}
D_I \left| -,-,-,-\right> = 0 \,\,\, \mbox{for all $I$}.
\end{equation}
Acting with a $D_I ^\dagger$ on the vacuum changes the $I$th minus
into a plus, e.g.\
\begin{equation}\label{oenschboens}
D_3 ^\dagger\left| -,-,-,-\right> = \left| -,-,+,-\right> .
\end{equation}
Acting once more with $D_3 ^\dagger$ will give zero. Acting with $D_3$ on 
(\ref{oenschboens}) will give back (\ref{rvacuum}) because of
(\ref{compcom}). Thus, we have a $2^4 =16$-fold degenerate
vacuum. This gives an on shell Majorana spinor in ten dimensions. For
the left movers the construction is analogous. (The above method to
construct the state is actually an option to construct (massless) spinor
representations when the $d_0^i$ are identified with the target space
Gamma matrices.) Without further motivation (which is given in the
books and reviews listed in section \ref{furtherreading}) we state how
the GSO projection is performed 
in the R sector. First, we define
\begin{equation}\label{fermionpar}
\left( -1\right)^F = 2^4\, d_0 ^2 d_0^3 d_0 ^4 d_0 ^5 d_0 ^6 d_0 ^7 d_0 ^8
d_0 ^9 \left( -1\right)^{\sum_{n>0}d^i _{-n}d^i_n} ,
\end{equation}
where the factor of $2^4$ has been introduced such that $\left(
  -\right)^{2F}=1$, ensuring that (\ref{rgso}) defines projection
operators. Note also that $\Gamma^\mu = \sqrt{2}d_0 ^\mu$ satisfies the
canonically normalized Clifford algebra $\left\{ \Gamma^\mu
  ,\Gamma^\nu\right\} = 2\eta^{\mu\nu}$.
For the groundstate this is just the chirality operator (the product
of all Gamma matrices) in the transverse eight dimensional space.
Now, we multiply the R states by one of the following projection
operators
\begin{equation}\label{rgso}
P_{GSO}^\pm = \frac{1\pm\left(-1\right)^F}{2}
\end{equation}
We perform the analogous construction in the left moving R
sector. There are essentially two inequivalent options: we take the same
sign in (\ref{rgso}) for left and right movers, or different
signs. Taking different signs leads to type IIA strings whereas the
option with the same signs is called type IIB. Multiplying the R
groundstate with one of the operators (\ref{rgso}) reduces the 16
dimensional Majorana spinor to an eight dimensional Weyl
spinor\footnote{The two different choices in (\ref{rgso}) give either
  the $\mbox{\bf
      8}_{\mbox{\scriptsize \bf s}}$ or the $\mbox{\bf
      8}_{\mbox{\scriptsize \bf c}}$ representation of $SO(8)$
    mentioned in section \ref{greenschwarzstring}}.

To complete the discussion of the R sector we have to combine left and
right movers, i.e.\ to construct the NSR, RNS, and RR sector of the
theory. Let us start with the NSR sector.  The mass shell condition 
(\ref{massshelf}) reads now
\begin{equation}
2p^+p^- - p^i p^i = 8\left(N_{NS} - \frac{1}{2}\right) = 8 \tilde{N}_R
,
\end{equation}
where the number operator in the R sector is given as
\begin{equation} \label{rnumber}
N_R = \sum_{n=1}^\infty \alpha_{-n}^i \alpha_n ^i + \sum_{n=1}^\infty
nd_{-n}^i d_n ^i ,
\end{equation}
and the analogous expression for the left movers. We have put the
normal ordering constant in the Ramond sector to zero. This can
easily be justified by replacing the half integer modded sum over $r$ by
an integer modded one in (\ref{mumpitzf}). Level matching implies that
the lowest allowed state in the NSR sector is massless and given by
\begin{equation}\label{NSR}
b^i _{-\frac{1}{2}}\left| k\right> u_a 
\end{equation}
where $u_a$ denotes the eight component Majorana-Weyl spinor comming
from the R ground states surviving the GSO projection. 
The 64 states contained in (\ref{NSR}) decompose into an eight
dimensional and a 56 dimensional representation of the target space
little group $SO(8)$. The 56 dimensional representation gives a 
gravitino of fixed chirality, whereas the eight dimensional one gives
a dilatino of fixed chirality.

The discussion of the RNS sector goes along the same line giving again
a gravitino and a dilatino either of opposite (to the NSR sector)
chiralities corresponding to type IIA theory, or of the same
chiralities when the type IIB GSO projection is imposed.

Finally, in the RR sector the lowest state is obtained by combining
the left with the right moving vacuum. This state is massless due to
the normal ordering constant $a_R =0$. It has 64 components. The
irreducible decompositions of the RR state depend on whether we have
imposed GSO conditions corresponding to type IIA or type IIB. In the
type IIA case the 64 states decompose into an eight dimensional vector
representation and a 56 dimensional representation. Thus in the type
IIA theory, the RR sector gives a massless $U(1)$ one-form gauge
potential $A_\mu$ and a three-form gauge potential $C_{\mu\nu\rho}$.
In the type IIB theory the 64 splits into a singlet, a 28 and a 35
dimensional representation of $SO(8)$. This corresponds to a
``zero-form'' $\Phi^\prime$, a two-form $B^\prime _{\mu\nu}$, 
and a four-form gauge potential with selfdual field strength $C^{*}
_{\mu\nu\rho\sigma}$. 
The particle content of the type II theories can be arranged in to
$N=2$ supermultiplets of chiral (type IIB) or non-chiral (type IIA)
ten dimensional supergravity.
The (target space bosons of the) massless spectrum of the type II
strings is summarized in table 
\ref{tab:tab2}. 

\subsubsection{The heterotic string}

Since the heterotic string is a bit out of the focus of the present
review we will briefly state the results. The starting point is the
action (\ref{hetero}). Without the $\lambda_+ ^A$ this looks like the
type II theories with the left handed worldsheet fermions
removed. 

Indeed, this part of the theory leads to the spectrum of the
type II theories with only the NS and R sector. The massless spectrum
corresponds to $N=1$ chiral supergravity in ten dimensions. 
It corresponds to the states (the $\tilde{\alpha}^i_n$ are the Fourier
coefficients for the left moving bosons, and the $b_r ^i$ for the
right moving fermions in the NS sector)
\begin{equation}\label{het1}
\tilde{\alpha}_{-1}^i b^j _{-\frac{1}{2}}\left|k\right>,
\end{equation}
in the NS sector, and
\begin{equation}\label{het2}
\tilde{\alpha}_{-1}^i \left|k\right> u_\alpha
\end{equation}
in the R sector, where we denoted again the GSO projected R vacuum
with $u_\alpha$. The above states must be massless since they form
irreducible representations of $SO(d-2)$. Focusing on the right moving
sector we can deduce that the right moving normal ordering constant
must be $\frac{1}{2}$ like in the type II case. Hence, the number of
dimensions (range of $\mu$) is ten.
As it stands the above spectrum leads to an anomalous theory. 
But there is still the option of switching on
the $\lambda_+ ^A$. Let us first deduce the number of additional
directions (labeled by $A$) needed. In the sector where the vacuum is
non degenerate due to the presence of the $\lambda_+ ^A$, we know that
we need the left moving normal ordering constant to be
one. (Otherwise the states  (\ref{het1}) would not be massless, but
still form $SO(d-2)$ representations.) The vacuum does not receive
further degeneracy in the sector where all of the $\lambda^+_A$ have
anti-periodic boundary conditions. In this sector the normal ordering
constant is (see also (\ref{mumpitzf})), the label $A$ stands for
anti-periodic
\begin{equation}
\tilde{a}_A = \frac{d-2}{24} + \frac{D}{48},
\end{equation} 
where we have called the number of additional directions $D$
($A=1,\ldots ,D$). The consistency condition $\tilde{a}_A =1$ tells us
that there must be 32 additional directions,
\begin{equation}
D=32 .
\end{equation}

Let us first discuss the simplest option, namely that all of the
$\lambda_+^A$ have always identical boundary conditions, either
periodic or antiperiodic. In the periodic sector one easily computes
that the normal ordering constant $\tilde{a}_P$ is negative
($-\frac{1}{3}$). Hence, there are no massless states in this
sector. In the NS sector we find in addition to (\ref{het1}) the
massless states (denoting with $\tilde{b}^A_{r}$ the Fourier
coefficients of $\lambda_+ ^A$ in the anti-periodic sector)
\begin{equation}\label{gaugefield}
\tilde{b}^A_{-\frac{1}{2}}\tilde{b}^B_{-\frac{1}{2}}b_{-\frac{1}{2}}^i
\left| k\right> .
\end{equation}
Since the $\tilde{b}^A$ anti-commute this is an anti-symmetric $32 \times 32$
matrix. In addition it is a target space vector (because of the index
$i$). Therefore, the state (\ref{gaugefield}) is an $SO(32)$ gauge field.
The corresponding R sector provides (after imposing the GSO
projection) fermions filling up an $N=1$
supermultiplet in ten dimensions. Together, with this $SO(32)$
Yang-Mills part the ten dimensional field theory with the same
massless content is anomaly free. The GSO projection in the periodic
sector is such that only states with an even number of left moving
fermionic creators survive. In the P sector it removes half of the
groundstates (leaving only spinors of definite chirality with respect
to the internal space spanned by the $A$ directions). 

Another option is to group the $\lambda^A_+$ into two groups of 16
directions. 
Then we would naturally split the state (\ref{gaugefield}) into three
groups: ${\bf (120,1)}$, ${\bf (1,120)}$, and ${\bf (16,16)}$,
depending on whether $A$ 
and $B$ in (\ref{gaugefield}) are both in the first half ($1,\ldots,
16$), both in the second half ($17,\ldots ,32$), or one of them out of
the first half and the other one out of the second half.
So far, this gave only a rearrangement of those states. 
But now we impose the GSO projection such that only states survive
where an even number of fermionic left moving creators act in each
half separately. This removes the ${\bf (16,16)}$ combination.
Further, when we
split the range of indices into two groups of 16 each, there will be
additional massless states. It is simple to check that in the sector
where half of the boundary conditions are periodic and the other half
is anti-periodic (the AP or PA sector), the left moving normal
ordering constant vanishes. Hence, the corresponding ground states
give rise to massless fields, provided right moving creation operators
act such that level matching is satisfied. This gives (removing half
of those states by GSO projection) ${\bf (128,1)}$ additional massless
vectors from the PA sector, and another ${\bf (1,128)}$ from the AP
sector.  
Together with the vectors from the AA sector this gives an $E_8\times
E_8$ Yang-Mills field. The R sector state fills in the fermions
needed for $N=1$ supersymmetry in ten dimensions. This corresponds to the other
known $N=1$ anomaly free field theory.

The bosonic parts of the massless spectra of the consistent closed
string theories in ten dimensions is summarized in table
\ref{tab:tab2}. We have added the number of supercharges $Q$ from a target
space perspective, and also the number of worldsheet supersymmetries
$\psi_\mu$, in 
the NSR formulation.

\begin{table}
\begin{center}\begin{tabular}[h]{|l|l|l|l|l|}
\hline 
   & \# of $Q$'s& \# of $\psi_\mu$'s & \multicolumn{2}{c|}{massless
   bosonic spectrum}\\  
\hline
IIA & 32 & 2& NSNS & $G_{\mu\nu}$, $B_{\mu\nu}$, $\Phi$ \\ \cline{4-5}
 & & %$\psi_{\mu L}$, $\psi_{\mu R}$ 
& RR & $A_{\mu}$, $C_{\mu\nu\rho}$
\\ \hline\hline
IIB & 32 & 2& NSNS &  $G_{\mu\nu}$, $B_{\mu\nu}$, $\Phi$ \\ \cline{4-5}
 & & %$\psi_{\mu L}$, $\psi_{\mu L}^{\prime}$ 
& RR & 
$C^{*}_{\mu\nu\rho\sigma}$, $B_{\mu\nu}^{\prime}$, $\Phi^{\prime}$ \\
\hline\hline
heterotic & 16 & 1 &\multicolumn{2}{c|}{$G_{\mu\nu}$, $B_{\mu\nu}$, $\Phi$}\\
$E_8\times E_8$ & & &\multicolumn{2}{c|}{$A_{\mu}^a$ in
 adjoint of $E_8\times E_8$ }
\\ \hline\hline
heterotic & 16 & 1 &\multicolumn{2}{c|}{$G_{\mu\nu}$, $B_{\mu\nu}$, $\Phi$}\\
$SO(32)$ & & &\multicolumn{2}{c|}{$A_{\mu}^a$ in
 adjoint of $SO(32)$ }
%\\ \hline\hline
%type I & 16 & 1 &NSNS  & $g_{\mu\nu}$,  $\phi$\\ \cline{4-5}
%$SO(32)$ & & &open string &$A_{\mu}^a$ in
%adjoint of $SO(32)$ \\ \cline{4-5}
% & & &RR & $b_{\mu\nu}$
\\ \hline\end{tabular}\end{center} 
\caption{Consistent closed string theories in ten dimensions.} \label{tab:tab2}
\end{table}

\subsection{Strings in non-trivial backgrounds}\label{betafunctions}
\setcounter{equation}{0}

In the previous sections we have seen that all closed strings contain
a graviton, a dilaton, and an antisymmetric
tensor field in the massless sector. This is called the universal
sector. So far, we have 
studied the situation where the 
target space metric is the Minkowski metric, the antisymmetric tensor
has zero field strength and the dilaton is constant. In order to
investigate what happens when we change the background, we need to
modify the action (\ref{polyakov}) as follows (this action is called
the string sigma model)
\begin{eqnarray}
S & =& -\frac{1}{4\pi\alpha^\prime}\int d^2 \sigma\left(
  \sqrt{\gamma}\gamma^{\alpha\beta} G_{\mu\nu}\left( X\right)
  \partial_\alpha X^\mu\partial_\beta X^\nu +
  i\epsilon^{\alpha\beta}B_{\mu\nu} \left( X\right)\partial_\alpha
  X^\mu\partial_\beta X^\nu\right)\nonumber \\
 & & \,\,\,  -\frac{1}{4\pi}\int d^2 \sigma\sqrt{\gamma}
  \Phi\left( X\right) R^{(2)},
\label{nonlinear}
\end{eqnarray}
where $R^{(2)}$ is the scalar curvature computed from
$\gamma_{\alpha\beta}$. Throughout this section we will consider a
Euclidean worlsheet signature. 
Note, that the dilaton term does not contain
$\alpha^\prime$. In general, the theory (\ref{nonlinear}) cannot be
quantized in an easy way. The best one can do is to take a
semiclassical approach. Since $\alpha^\prime$ enters like $\hbar$ in
ordinary field theories this will result in a perturbative expansion
in $\alpha^\prime$. The term with the dilaton can be viewed as a first
order contribution in this expansion. Without this term,
(\ref{nonlinear}) has again three local symmetries: diffeomorphisms
and Weyl invariance. The dilaton term breaks Weyl
invariance in general. We will be interested in the question under
which 
circumstances Weyl invariance remains unbroken in the semiclassically
quantized theory. To answer this, first note that $G_{\mu\nu}$,
  $B_{\mu\nu}$, and $\Phi$ can be viewed as couplings from a two
  dimensional perspective. Weyl invariance in particular implies
  global scale invariance. But scale invariance is related to
  vanishing beta functions in field theory. Thus, we will compute the
  beta functions of $G_{\mu\nu}$, $B_{\mu\nu}$ and $\Phi$ as a power
  series in $\alpha^\prime$. 
However, there is a subtlety here. Under field redefinitions
(infinitesimal shifts of $X$ by $\chi\left[ X\right]$) the couplings
change according to 
\begin{eqnarray}
\delta G_{\mu\nu} & = & 2D_{(\mu}\chi_{\nu)} \label{metricchange},\\
\delta B_{\mu\nu} & = & \chi^\rho H_{\rho\mu\nu}
+\partial_\mu L_\nu -\partial_\nu L_\mu \label{bchange},\\
\delta \Phi & = & \chi^\rho \partial_\rho \phi \label{phichange},
\end{eqnarray}
where we have defined
\begin{equation}\label{hfield}
H_{\rho\lambda\kappa} = \partial_\rho B_{\lambda\kappa} +
\partial_\lambda B_{\kappa\rho} +\partial_\kappa B_{\rho\lambda}
\end{equation}
and 
\begin{equation}
L_\kappa = \chi^\rho B_{\kappa\rho}.
\end{equation}
Expression (\ref{hfield}) defines a field strength corresponding to
the $B$ field. It is invariant under a $U(1)$ transformation 
\begin{equation}\label{uinv}
\delta B_{\mu\nu} = \partial_{[\mu}V_{\nu]} ,
\end{equation}
with $V_\mu$ being an arbitrary target space vector. It is easy to
check that also (\ref{nonlinear}) possesses the invariance
(\ref{uinv}). The symmetry
(\ref{uinv}) can be taken care of by allowing for arbitrary $L_\mu$
in (\ref{bchange}).
Thus the couplings and hence the beta functions are not unique.
But actually we will be
not just interested in vanishing beta functions. This would ensure
only global
scale invariance. The requirement of Weyl invariance is more strict
and will fix the arbitrariness. 
 
In order to compute the beta functions, we need to fix the worldsheet
diffeomorphisms.  
We leave the explicit form of the fixed metric $\gamma_{\alpha\beta}$
unspecified. The gauge fixing procedure introduces ghosts, the
diffeomorphism invariance is replaced by BRST invariance. The ghost
action depends only on the 2d geometry. Therefore, we expect that the
ghosts contribute only to the dilaton beta function. We will not treat
them explicitly but guess their contribution in the end of this
section. The 
semiclassical approach means that we start from some background string
$\bar{X}^\mu $ satisfying the equations of motion. We study the theory
of the fluctuations around this background string. Instead of using
the fluctuation in the coordinate field $X^\mu$ we will take the
tangent vector to the geodesic connecting the background value
$\bar{X}^\mu$ with the actual value $X^\mu$. This difference is
supposed to be small in this approximation. In order to
compute the tangent vectors we connect the background value and the
actual position of the string by a geodesic. The line parameter $t$
is chosen such that at $t=0$ we are at the background position and at
$t=1$ at the actual position. The geodesic equation is (the dot
denotes the derivative with respect to $t$),
\begin{equation}\label{geodesic}
\ddot{\lambda}^\mu + \Gamma_{\nu\rho}^\mu \dot{\lambda}^\nu
\dot{\lambda}^\rho = 0 
\end{equation}      
and the boundary conditions are 
\begin{equation}
\lambda^\mu\left( 0\right) = \bar{X}^\mu \,\,\, ,\,\,\,
\lambda^\mu\left( 1\right) = X^\mu .
\end{equation}
Note that the target space Christoffel connection
$\Gamma_{\nu\rho}^\mu$ depends on $X^\mu$. The first non-trivial
effects should come from terms second order in the fluctuations in
the action. (First order terms vanish when the background satisfies
the equations of motion.) We call the tangent vector to the geodesic
(at $\bar{X}^\mu$)
\begin{equation}
\xi^\mu = \dot{\lambda}^\mu\left( 0 \right) .
\end{equation}
One can solve (\ref{geodesic}) iteratively leading to a power series in $t$,
\begin{equation}\label{iter}
\lambda^\mu \left( t\right) = \bar{X}^\mu +\xi^\mu t -\frac{1}{2}
\Gamma^{\mu}_{\nu\rho} \xi^\nu\xi^\rho t^2 -\frac{1}{3!}\Gamma^\mu
_{\nu\rho\kappa}\xi^\nu\xi^\rho\xi^\kappa t^3 +\ldots ,
\end{equation}
where 
\begin{equation}
\Gamma^\mu_{\nu\rho\kappa} = \nabla_\nu \Gamma^\mu_{\rho\kappa} =
\partial_\nu \Gamma^\mu _{\rho\kappa}
-\Gamma^{\lambda}_{\nu\rho}\Gamma^\mu _{\lambda\kappa} -
\Gamma^{\lambda}_{\nu\kappa}\Gamma^\mu_{\rho\lambda}.
\end{equation}
Further, we may choose local coordinates such that only the constant
and the term  linear
in $t$ appears in (\ref{iter}) and all higher order terms vanish in a
neighborhood of $\bar{X}^\mu$. (This is done by spanning the local
coordinate system by tangent vectors to geodesics.) The
corresponding coordinates are called Riemann normal coordinates. In
these coordinates the Taylor expansion of the various terms in
(\ref{nonlinear}) around $\bar{X}^\mu$ takes the following form (up to
second order in the fluctuations),
\begin{eqnarray}
\partial_\alpha X^\mu & = & \partial_\alpha \bar{X}^\mu + D_\alpha
\xi^\mu +\frac{1}{3} {R^\mu} _{\lambda \kappa \nu}\left( \bar{X}\right)
 \xi^\lambda\xi^\kappa
\partial_\alpha \bar{X}^\nu , \\
G_{\mu\nu}\left( X\right) & = & G_{\mu\nu}\left( \bar{X}\right)
-\frac{1}{3} R_{\mu\rho\nu\kappa}\left( \bar{X}\right)\xi^\rho
\xi^\kappa ,\\
B_{\mu\nu}\left( X\right) & = & B_{\mu\nu}\left( \bar{X}\right) + D_\rho
B_{\mu\nu} \left( \bar{X}\right) \xi^\rho +\frac{1}{2}
D_{\lambda}D_{\rho} B_{\mu\nu}\left( \bar{X}\right)
\xi^\lambda\xi^\rho \nonumber \\ & & \,\,\,\,\,\,\, -\frac{1}{6}
{R^\lambda}_{\rho\mu\kappa}B_{\lambda\nu}\left(\bar{X}\right)\xi^\rho\xi^\kappa
+\frac{1}{6}{R^\lambda}_{\rho\nu\kappa}B_{\lambda\mu}\left(\bar{X}\right)
\xi^\rho\xi^\kappa ,\\
\Phi\left( X\right)  & = & \Phi\left( \bar{X}\right) +
D_\mu\Phi\left(\bar{X}\right)\xi^\mu + \frac{1}{2}D_\mu D_\nu
\Phi\left(\bar{X}\right) \xi^\mu\xi^\nu ,
\end{eqnarray}
where $D_\rho$ denotes the usual covariant derivative in target
space, and ${R^\mu}_{\nu\rho\sigma}$ is the target space Riemann
tensor
\begin{equation}
{R^\mu}_{\nu\rho\lambda} = \partial_\rho \Gamma^\mu_{\nu\lambda} -
\partial_\lambda \Gamma^\mu_{\nu\rho} + \Gamma^\omega
_{\nu\lambda}\Gamma^\mu _{\omega\rho} -\Gamma^\omega
_{\nu\rho}\Gamma^\mu _{\omega\lambda}.
\end{equation}
Note that in the Riemann normal coordinates the contributions
quadratic in the Christoffels vanish.
Further, we have defined
\begin{equation}\label{covder}
D_\alpha \xi^\mu = \partial_\alpha \xi^\mu +
\Gamma^\mu _{\lambda\nu}\xi^\lambda \partial_\alpha \bar{X}^\nu .
\end{equation}
Collecting everything, one can expand the action (\ref{nonlinear}) in
a classical contribution $S_0$ and a contribution due to
fluctuations. There will be no part linear in $\xi^\mu$ as long as
$\bar{X}^\mu$ satisfies the equations of motion. The first non-trivial
part is quadratic in the $\xi^\mu$. We denote it by
\begin{equation}
S^{(2)} = S^{(2)}_G + S^{(2)}_B +S^{(2)}_\Phi ,
\end{equation}
with (the background fields $G$, $B$ and $\Phi$ are taken at $\bar{X}^\mu$)
\begin{eqnarray}
S^{(2)}_G & = &
-\frac{1}{4\pi\alpha^\prime}\int d^2 \sigma
\sqrt{\gamma}\gamma^{\alpha\beta} \left( 
  G_{\mu\nu}D_\alpha \xi^\mu D_\beta \xi^\nu \right.\nonumber\\ & &
\,\,\,\,\,\,\,\,\,\,\,\,\,\,\,\,\,\,\,\,\,\, \left.
+ R_{\rho\mu\kappa\nu}
  \partial_\alpha \bar{X}^\mu \partial_\beta\bar{X}^\nu
  \xi^\rho\xi^\kappa\right) ,\label{2g}\\
S^{(2)}_B & = &  -\frac{1}{4\pi\alpha^\prime} \int d^2 \sigma
i\epsilon^{\alpha\beta} \left( \partial_\alpha \bar{X}^\rho H_{\rho\mu\nu}
  \xi^\nu D_\beta \xi^\mu\right. \nonumber\\
& & \,\,\,\,\,\,\,\,\,\,\,\,\,\,\,\,\,\,\,\,\,\,\,\,
\left. +\frac{1}{2} D_\lambda H_{\rho\mu\nu}
  \xi^\lambda \xi^\rho \partial_\alpha \bar{X}^\mu\partial_\beta
  \bar{X}^\nu\right)\label{2b}\\ 
S^{(2)}_\Phi & = & -\frac{1}{4\pi}\int d^2 \sigma
\sqrt{\gamma}R^{(2)}\frac{1}{2} D_\mu D_\nu \Phi \xi^\mu \xi^\nu
\label{2phi} . 
\end{eqnarray}
The next step is to redefine the fields $\xi^\mu$ in terms of a
vielbein,
\begin{equation}\label{lorentzi}
\xi^\mu = E^\mu _A \xi^A,
\end{equation}
with
\begin{eqnarray}
G_{\mu\nu} & = & E^A _\mu E^B _\nu \eta_{AB}, \\
E^\mu _A E_{\mu B} & = & \eta_{AB}.
\end{eqnarray}
In what follows, capital latin indices will be raised and lowered with
the Minkowski metric. The normal coordinate expansion is
useful not only to get the expressions (\ref{2g}), (\ref{2b}), (\ref{2phi})
in a covariant looking form. An important advantage of this method is
that the functional measure (in a path integral approach) for the
$\xi^A$ is the usual translation invariant measure. This will simplify
the computation of the partition function. In order to be able to do
the field redefinition (\ref{lorentzi}) in a meaningfull way we have to
ensure that the fluctuations are parameterized by target space
vectors. The tangent vectors to geodesics connecting the background
with the actual value are a natural choice. Before writing down the
action in terms of the $\xi^A$, we will absorb the first term in
(\ref{2b}) in an additional  connection in the kinetic term (the first
term in (\ref{2g})). That can be done by adding and subtracting a term
looking like 
$$\partial_\alpha \bar{X}^\rho \partial^\alpha \bar{X}^\kappa
{{H_\rho}^\lambda} _\mu H_{\kappa\lambda\nu} \xi^\mu\xi^\nu .$$    
We define the covariant derivative on $\xi^A$ by plugging
(\ref{lorentzi}) into ({\ref{covder}) and introducing an additional
  connection
\begin{equation}
{\cal D}_\alpha \xi^A = D_\alpha \xi^A
+\frac{i}{2}\frac{{\epsilon_\alpha}^\beta}{\sqrt{-\gamma}}\partial_\beta
\bar{X} ^\rho E_\mu ^A {{H_\rho}^\mu}_\nu E^\nu _B \xi^B ,
\end{equation}
where $D_\alpha \xi^A$ corresponds to the contribution from (\ref{covder}).
The part of the action quadratic in fluctuations finally takes the
form
\begin{equation}\label{quadra}
S^{(2)} = -\frac{1}{4\pi \alpha^\prime}\int d^2\sigma \sqrt{\gamma}\left(
\gamma^{\alpha\beta}{\cal D}_\alpha \xi^A {\cal D}_\beta \xi_A +
M_{AB}\xi^A\xi^B \right) ,
\end{equation}
where the potential is
\begin{equation}
M_{AB} = \gamma^{\alpha\beta}\partial_\alpha \bar{X}^\mu\partial_\beta
\bar{X}^\nu {\cal G}_{\mu\nu AB} + i
\frac{\epsilon^{\alpha\beta}}{\sqrt{\gamma}}\partial_\alpha
\bar{X}^\mu\partial_\beta 
\bar{X}^\nu  {\cal B}_{\mu\nu AB} + \alpha^\prime
R^{(2)}{\cal F}_{AB} .
\end{equation}
The matrices ${\cal G}$ , ${\cal B}$ and ${\cal F}$ do not have an
explicit dependence on the worldsheet coordinates and are given by
\begin{eqnarray}
{\cal G}_{\mu\nu AB} & = & E_A ^\rho E_B ^\kappa
\left( R_{\rho\mu\kappa\nu}-\frac{1}{4} {{H_\mu}^\lambda}_\rho
  H_{\nu\lambda\kappa }\right) \label{gope},\\
{\cal B}_{\mu\nu AB} & = & \frac{1}{2} D_\lambda
H_{\rho\mu\nu}E^\lambda _A E^\rho 
_B ,\label{bope}\\
{\cal F}_{AB} & = & \frac{1}{2}E^\mu _A E^\nu _B D_\mu D_\nu \Phi
.\label{pope} 
\end{eqnarray}
Since the action (\ref{quadra}) is quadratic in the fluctuations,
integrating over the fluctuations will result in the determinant of an
operator. For the general form of the operator in (\ref{quadra}) it is
very covenient to use known formul\ae\ from the heat kernel
technique. In the heat kernel approach the partition function 
$$ Z= \int {{\cal D} \xi^A}\, e^{iS^{(2)}} $$
can be expressed as a formal sum\cite{gilkey-book,Branson:1998ze}
\begin{equation}
\log{Z} =\frac{1}{2}\int \frac{dt}{t}e^{-{\cal O}t}=\frac{1}{2} 
\int_{\epsilon\mu^{-2}}^\infty
\frac{dt}{t}\sum_{n=-2}^\infty a_n 
t^{\frac{n}{2} -1}  ,
\end{equation}
where $\epsilon$ is a dimensionless UV cutoff and $\mu$ is a mass
scale introduced for dimensional reasons. The symbol ${\cal O}$ stands
for the operator whose determinant is of interest. We rescale $t$ by
$\alpha^\prime$ such that ${\cal O}$ has mass dimension 2.\footnote{The
  appearance of a power series in $\alpha^\prime$ is more obvious in a
  Feynman diagramatic treatment. There, the propagator goes like
  $\alpha^\prime$ whereas vertices go like $1/\alpha^\prime$. This
  relates directly the order of $\alpha^\prime$ in logarithmically
  divergent diagrams to the number of loops. The disadvantage of this
  approach is that the discussion for a general worldsheet metric $\gamma$
  is more involved. Fixing $\gamma$ to be the Minkowski metric results
  in problems when computing the dilaton beta function since $R^{(2)}$
  vanishes for this choice.} 
In order to compute the beta
functions, we are interested in the logarithmically divergent piece,
i.e.\ in $a_2$. This can be found in the
literature\cite{gilkey-book,Branson:1998ze} 
\begin{equation}\label{heatdiver}
a_2 = \frac{1}{4\pi}\int d^2\sigma \sqrt{\gamma}\left(- M_A ^A +
  \frac{d}{6}R^{(2)}\right) .
\end{equation}
The divergence can be cancelled by adding appropriate counterterms to
the action. This amounts to a replacement of the bare (infinite)
couplings $G_{\mu\nu}$, $B_{\mu\nu}$, $\Phi$ in the following way,
\begin{eqnarray}
G_{\mu\nu} & = & G_{\mu\nu} ^{ren} - \frac{\alpha^\prime}{2}\log
\left(\mu^2/\epsilon\right) 
{{\cal G}_{\mu\nu A}}^A ,\\
B_{\mu\nu} & = & B_{\mu\nu}^{ren} -\frac{\alpha^\prime}{2}\log\left(
  \mu^2/\epsilon\right) {{\cal B}_{\mu\nu A}}^A ,\\
\Phi & = & \Phi^{ren} -\frac{1}{2} \log\left(\mu^2/\epsilon\right) 
\left(-\frac{d+c_g}{6} + \alpha^\prime{{\cal F}_A}^A\right) \label{phiren},
\end{eqnarray}
where ${\cal G}$, ${\cal B}$ and ${\cal F}$ are as defined in
(\ref{gope}), (\ref{bope}) and (\ref{pope}) but now in terms of the
renormalized couplings. Further, we have included a possible
contribution of the diffeomorphism fixing ghosts. Their action depends
only on the intrinsic two dimensional geometry and neither on the
embedding in the target space nor on the form of the background fields
$G_{\mu\nu}$, $B_{\mu\nu}$ and $\Phi$. Therefore, the ghosts can
contribute only a 
constant renormalization of the dilaton $\Phi$ which we have
parameterized by $c_g$ in (\ref{phiren}). The beta functions can be
computed by taking the derivative of the renormalized couplings with
respect to $\log 
\mu$ using the $\mu$ independence of the bare couplings. Up to order
$\alpha^\prime$ this leads to (they are all expressed in terms of
renormalized quantities and we supress the corresponding superscript in
the following)
\begin{eqnarray}
\beta^{(G)}_{\mu\nu} & =& \alpha^\prime\left( R_{\mu\nu}
  -\frac{1}{4}{H_\mu}^{\lambda\rho} H_{\nu\lambda\rho}\right), \\
\beta^{(B)}_{\mu\nu} & = & \frac{\alpha^\prime}{2} D^\lambda
  H_{\lambda\mu\nu} ,\\
\beta^{(\Phi)} & = & -\frac{d+c_g}{6} + \frac{\alpha^\prime}{2} D^2 \Phi
  .
\end{eqnarray}  
Because of the ambiguities related to the field redifinitions
(\ref{metricchange}) -- (\ref{phichange}) we cannot just set the beta
functions to zero but only deduce that the model is Weyl invariant
(in first approximation) if
\begin{eqnarray}
\bar{\beta}_{\mu\nu}^{(G)} & = &\beta_{\mu\nu}^{(G)} + D_{(\mu} M_{\nu )}
= 0 , \label{weylmetric}\\
\bar{\beta}_{\mu\nu}^{(B)} & = & \beta_{\mu\nu}^{(B)} +\frac{1}{2}
{H_{\mu\nu}}^\lambda M_\lambda +\partial_{[\mu} L_{\nu ]} =0,\label{weylb} \\
\bar{\beta}^{(\Phi)} & = & \beta^{(\Phi)} +\frac{1}{2}\partial_\mu
\Phi \, M^\mu = 0 \label{weyldilaton}
\end{eqnarray}
The vectors $M_\mu$ and $L_\mu$ are not fixed by just checking for
global scale invariance. In order to compute them we would need to
impose (local) Weyl invariance. This could be done by computing the
expectation value of the trace of the energy momentum tensor. Instead
of doing so, we will choose a rather indirect way of fixing the
ambiguity. Implicitly, we will be using a theorem stating that
$\bar{\beta}^{(\Phi)}$ is constant if (\ref{weylmetric}) and
(\ref{weylb}) are satisfied\cite{Curci:1987hi}. In other words this
means that up to a constant contribution (\ref{weyldilaton}) should be
an integrability condition for the other two equations. Before
deriving such a condition we need to study of which form the vectors
$M_\mu$ and $L_\mu$ could be at the given order in $\alpha^\prime$. 
We want to express a vector in terms of our background fields
$G_{\mu\nu}$, $\Phi $ and $B_{\mu\nu}$. The field $B_{\mu\nu}$ should
enter only via the gauge invariant field strength $H_{\mu\nu\lambda}$
(since we have performed partial integegrations in $S^{(2)}$ such that
the beta functions come out in a gauge invariant form). With this
information it is easy to check that the only option we have is
($a$ is some constant)
\begin{equation}
M_\mu = a \partial_\mu \Phi \,\,\,\,\,\,\,\,\, , \,\,\,\,\,\,\,\,
L_\mu = 0 ,
\end{equation}
where we do not consider a gradient contribution in $L_\mu$ since this
would not be relevant. The next step is to take the divergence of
(\ref{weylmetric}). Using the Bianchi identity (i.e.\ the
vanishing divergence of the Einstein tensor), the identity
$$ H^{\rho\lambda\mu} D_\rho H_{\lambda\mu\nu} = \frac{1}{6}
D_\nu\left( H_{\mu\lambda\rho}H^{\mu\lambda\rho}\right), $$
and equations (\ref{weylmetric}) and (\ref{weylb}) one obtains
\begin{equation}\label{integrability}
D_\nu \left( \frac{a}{2}D^2 \Phi +\frac{\alpha^\prime}{12} H^2
  +\frac{a^2}{2\alpha^\prime} \left(\partial \Phi\right)^2\right) = 0 ,
\end{equation}
where we have defined
\begin{equation}
H^2 = H_{\rho\nu\lambda}H^{\rho\nu\lambda} .
\end{equation}
On the other hand, equation (\ref{weyldilaton}) implies
\begin{equation}\label{diladeriv}
D_\nu\left( \frac{\alpha^\prime}{2} D^2 \Phi +
  \frac{a}{2}\left(\partial \Phi\right)^2\right) = 0.
\end{equation}
Without the $H^2$ term, (\ref{integrability}) and (\ref{diladeriv})
would be the same. 
The $H^2$ term in the dilaton beta function is actually missing in our
computation since we took into account only one loop
contributions. Any counterterm in the action leading to an order
$\alpha^\prime$ contribution in the dilaton beta function should be
linear in $\alpha^\prime$. Since, at tree level the $B$ field enters
with a factor $\frac{1}{\alpha^\prime}$, the $H^2$ term in
$\beta^{(\Phi)}$ corresponds to a two loop contribution. In our
implicit approach we obtained this term (and in fact all order
$\alpha^\prime$ terms in $\beta^{(\Phi)}$) without doing a two loop
analysis.

We were not able to fix the value of the constant $a$, however. This is
because it could be absorbed in a rescaling of the field $\Phi$. But
this
would change the ratio of the constant contribution to the
dilaton beta function to the other contributions. Therefore, $a$ is
not arbitrary. The 
constant $a$ can be 
fixed for example by studying models with trivial metric and $B$ field
and a linear dilaton. These models are much easier to treat than the
generic one. The result is
\begin{equation}\label{dilfac}
a = 2\alpha^\prime .
\end{equation}
Let us discuss the case of trivial metric and vanishing $B$ field a
bit further. For a linear dilaton, the $\bar{\beta}^{(G)}$ and
$\bar{\beta}^{(B)}$ vanish identically. According to the previously
stated theorem (and to our result) the $\bar{\beta}^{(\Phi)}$ function
is constant in this case. Models with that feature are known as
conformal field theories. The constant dilaton $\bar{\beta}$ function
is related to an anomaly of the transformation of the energy momentum
tensor under conformal coordinate changes (while keeping the
worldsheet metric fixed). If we fix the worldsheet metric to be the
Minkowski metric the anomalous transformation of
the energy momentum tensor with respect to (\ref{conf})
reads
\begin{equation} \label{eptrafo}
\tilde{T}_{\tilde{\sigma}^+\tilde{\sigma}^+} = \left(
  \frac{d\sigma^+}{d\tilde{\sigma}^+}\right)^2
  T_{\sigma^+\sigma^+}\left(\sigma^+\right)
  +\frac{c}{12}S\left(\tilde{\sigma}^+ ,\sigma^+\right)
\end{equation}
where the second term denotes the Schwarz derivative
\begin{equation}
S\left(w,z\right) = \frac{z^\prime z^{\prime\prime\prime} -\frac{3}{2}\left(
  z^{\prime\prime}\right)^2}{\left(z^\prime\right)^2},
\end{equation}
where $z$ is a function of $w$ and the primes denote derivatives. The
Schwarz derivative has the following chain rule
\begin{equation}
S\left( w\left(v\left(z\right)\right), z\right) = \left(\frac{\partial
    v}{\partial w}\right)^2 S\left( v,z\right) + S\left(w,v\right).
\end{equation}
The transformation law (\ref{eptrafo}) is the most general possibility
such that associativity holds. An analogous consideration applies to $T_{--}$.
Now, from (\ref{eptrafo}) one can deduce part of the operator product
expansion (OPE) of two energy momentum tensors. To this end, one considers
infinitesimal transformations and uses the fact that they are generated
by $T_{++}$. One obtains the following OPE
\begin{eqnarray}
T_{\tilde{\sigma}^+\tilde{\sigma}^+}\left(\tilde{\sigma}^+\right)
 T_{\sigma^+\sigma^+}\left(\sigma^+\right) &=&
 \frac{c/2}{\left(\tilde{\sigma}^+ -\sigma^+\right)^4} +
 \frac{2}{\left(\tilde{\sigma}^+
 -\sigma^+\right)^2}T_{\sigma^+\sigma^+}\left(\sigma^+\right)\nonumber\\
& & \;  +
\frac{1}{\tilde{\sigma}^+
 -\sigma^+}\partial_{\sigma^+}T_{\sigma^+\sigma^+}\left(\sigma^+\right)
 +\ldots , 
\label{EPOPE}
\end{eqnarray}
where the dots stand for terms which are regular for $\tilde{\sigma}^+
= \sigma^+$.  For linear dilaton backgrounds this OPE can easily be
computed directly\footnote{One should first compute $T_{\alpha\beta}$
  by varying the action with respect to $\gamma_{\alpha\beta}$, and
  gauge fix $\gamma_{\alpha\beta} =\eta_{\alpha\beta}$ afterwards.},
leading to (\ref{dilfac}).  

It remains to fix the contribution coming from the gauge fixing
ghosts $c_g$. This can of course be calculated
directly\cite{Polyakov:1981rd,Polyakov:1981re}.
Here, we will guess it correctly, instead. From our discussion of the
quantized bosonic string in the light cone gauge in \ref{quantization}
we remember that the classical Lorentz covariance was preserved in
$d=26$. Comparing with (\ref{eptrafo}) we observe that our gauge
fixing procedure was justified only if $c=0$. Since, we did not have a
linear dilaton background there, this can happen only if
\begin{equation}\label{ghostcentral}
c_g = -26 .
\end{equation} 
Equation (\ref{ghostcentral}) can be confirmed by an explicit
computation (which can also be viewed as an alternative way of
deriving the critical dimension).

Given the fact, that a linear dilaton contributes to $c$, one may want
to go directly to $d=4$ by switching on a linear dilaton. 
One obvious problem with this is however  that target space
Lorentz covariance is broken explicitly -- there is a
distinguished direction in which the dilaton derivative points. The
more useful way of getting away from a 26 dimensional target space is
to replace 22 of the string coordinates by a conformal field theory
with central charge $d\rightarrow c=22$.  

To summarize, up to order $\alpha^\prime$ the action (\ref{nonlinear})
is Weyl invariant provided that the following set of equations holds,
\begin{eqnarray}
R_{\mu\nu} -\frac{1}{4} H_{\mu\rho\lambda}{H_\nu}^{\rho\lambda} +
2D_\mu\partial _\nu\Phi & = & 0,\label{metriceq}\\ 
-\frac{1}{2} D^\lambda H_{\lambda\mu\nu} +H_{\lambda\mu\nu}D^\lambda
\Phi  & = & 0,\label{beq}\\
\frac{1}{6}\left( d-26\right) -\frac{1}{2}\alpha^\prime D^2\Phi
+\alpha^\prime \left( \partial \Phi\right)^2 -\frac{\alpha^\prime}{24}
H^2 & = & 0.\label{dilatoneq} 
\end{eqnarray}
\subsection{Perturbative expansion and effective actions}
\label{effectiveactions} 
\setcounter{equation}{0}

In the previous section we have seen that imposing Weyl invariance
provides us with constraints on the background in which the string
propagates. These constraints can be viewed as equations of motion for
the background fields. Lifting those up to an action would then yield
an effective field theory description for the string theory. We have
discussed only the bosonic string, but an extension to the superstring
is possible. It may however be problematic. In the NSR formalism it is
for example not possible to include terms into the string sigma model
which would correspond to non-trivial RR backgrounds. Therefore, we
will sketch an alternative method of computing an effective action
here. We will not present any explicit calculations but just describe
the strategy. Starting from the spectrum and the amount of
supersymmetries belonging to a certain string theory one can write
down a general ansatz for an effective field theory action of the
string excitation modes. This ansatz can be further fixed by comparing
scattering amplitudes computed from the effective description to
amplitudes obtained from a string computation. The string amplitudes
can be described in a diagramatic fashion as depicted in figure \ref{feynman}.

%%%%%%%%%%%%%%%%%%%%%%%%%%%%%%%%%%%%
\begin{figure}
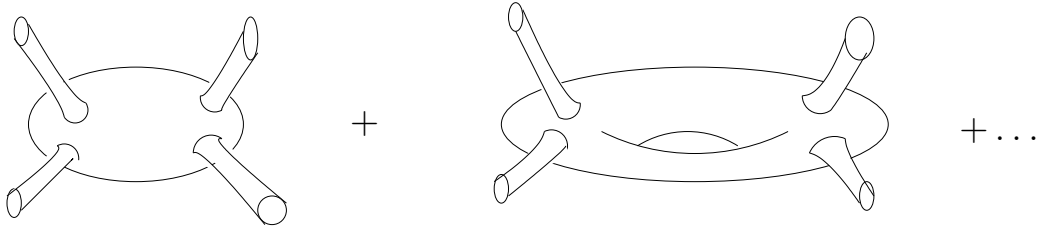
  
\begin{center}
\input feynman.pstex_t
\end{center}
\caption{Perturbative expansion of the four point function in a string
  computation}
\label{feynman}
\end{figure}
%%%%%%%%%%%%%%%%%%%%%%%%%%%%%%%%%%%%

The external four legs (hoses) correspond to the two incoming
particles scattering into two outgoing particles. The expansion is in
terms of the number of holes (the genus) of the worldsheet. The first
diagram in \ref{feynman} correponds to two incoming strings joining
into one string which in turn splits into two outgoing strings. In
that sense it contains two vertices. Analogously the second diagram
contains four vertices and so on. Assigning to each vertex one power of
the string coupling $g_s$, this gives a formal power series
\begin{equation}\label{pertexp}
{\cal A} = \sum_{n=0}^\infty g_s ^{2n+2} {\cal A}^{(n)}.
\end{equation}
This power series can be truncated after the first contributions as
long as $g_s \ll 1$. It remains to specify what $g_s$ is. To this end,
we first observe that the power of $g_s$ in the expansion terms in
(\ref{pertexp}) is nothing but minus the Euler number of a worldsheet
with $n$ handles and four boundaries. From (\ref{nonlinear}) we know
that the dilaton $\Phi$ couples to the Euler density of the
worldsheet. This follows immediately from the Gauss-Bonnet theorem
\begin{equation}
\frac{1}{4\pi}\int d^2\sigma \sqrt{\gamma} R^{(2)} = \chi \equiv
2\left( 1-n\right) -b ,
\end{equation}
where $n$ is the number of handles and $b$ is the number of boundaries
of the two dimensional worldsheet. (The calculation of the scattering
amplitudes is performed after the worldsheet signature has been Wick
rotated to the Euclidian one.) Thus, one can identify
\begin{equation}\label{stringcoupling}
g_s = e^{\langle \Phi\rangle},
\end{equation}
where $\langle \Phi\rangle$ denotes a constant vacuum expectation
value (VEV) of the dilaton.
(Remember from the previous section that a constant contribution to
$\Phi$ was not fixed by the conformal invariance conditions. This is
true for all string theories as can be easily seen by noticing that a
constant shift in $\Phi$ shifts the action (\ref{nonlinear}) by a
constant.) Therefore, the 
string coupling is an arbitrary parameter in the perturbative approach
to string theory. It is only restricted by the consistency requirement
that the perturbative expansion in figure \ref{feynman} should not break down,
i.e.\ $g_s \ll 1$.

There is also a second approximation in the computation of
the scattering amplitudes. Since the massive string states have
masses of the order of the Planck mass, they are ``integrated
out''. This means that we are interested in effects below the Planck
scale where those fields do not propagate. The effective field theory
actions contain only the massless modes. For consistency, one should
then restrict to processes where the momentum transfer is $p^2\ll
1/\alpha^\prime$. 

The (bosonic part of the) effective Lagrangians with at most two
derivatives and only the 
massless fields turn out to be
\begin{equation}
S = S_{univ} + S_{model}, 
\end{equation}
where $S_{univ}$ does not depend on which of the superstring theories
we are looking at, and $S_{model}$ is model dependent. The universal
sector has as bosonic fields the metric, the dilaton, and the $B$
field. The corresponding action is
\begin{equation}\label{universal}
S_{univ} = \frac{1}{2\kappa^2} \int d^{10} x \sqrt{-G} e^{-2\Phi}
\left( R +4\left(\partial \Phi\right) ^2 - \frac{1}{12} H^2\right) ,
\end{equation}
where $\kappa^2 \sim \left(\alpha^\prime\right)^4$ is the ten
dimensional gravitational constant.
Note that the set of equations of motion obtained from this action coincides
with the conformal invariance conditions (\ref{metriceq})--
(\ref{dilatoneq}), with the difference that for the superstring the
constant contribution in the dilaton equation vanishes for $d=10$. 

For type II strings there are additional contributions giving the
kinetic terms for the RR gauge forms and Chern-Simons interactions,
\begin{equation}\label{efftwo}
S_{model}^{II} = -\frac{1}{2\kappa^2}\int d^{10}x \sum_p
\frac{1}{2(p+2)!} F_{p+2}^2,
\end{equation}
where $F_{p+2}$ is the fieldstrength of a $p+1$ form RR gauge field
(plus --in some cases-- additional contributions which we will discuss
later), 
\begin{equation}\label{RRfieldstrength}
F_{p+2} = dA_{p+1}+ \ldots .
\end{equation}
The number $p$ is the spatial extension of an object which couples
electrically to the $p+2$ form gauge field. In the worldvolume action
of the corresponding $p$ dimensional object this coupling is
\begin{equation}
\int d^{p+1}\sigma \,\,\,
i\epsilon^{\alpha_1\cdots\alpha_{p+1}}\partial_{\alpha_1}X^{\mu_1}
\cdots \partial_{\alpha_{p+1}} X^{\mu_{p+1}} A_{\mu_1 \cdots \mu_{p+1}}.
\end{equation}
For a point particle ($p=0$) the above expression reads for example
$$ i\int d\tau \frac{dX^\mu}{d\tau} A_\mu .$$
From expression (\ref{nonlinear}) we observe that the fundamental
string is electrically charged under the NSNS $B$ field. We will meet
objects which are charged under the RR gauge forms when discussing
D-branes in section \ref{D-branes}. For the type IIA theory we have
$p=0,2,4$. Alternatively we could replace the field strength in
(\ref{efftwo}) by its Hodge dual\footnote{This can be done by adding
  the Bianchi identity $dF_{p+2} =d\left(\cdots\right)$ with a
  Lagrange multiplier to the 
  action and integrating out $A_{p+1}$. Because of covariance the
  Lagrange multiplier is a $7-p$ form and its field strength is an
  $8-p$ form.}
\begin{equation}
F_{8-p} = \star F_{p+2}.
\end{equation}
In the type IIA theory the definition (\ref{RRfieldstrength}) is
modified for the four form field strength
\begin{equation}
F_4 = dA_3 + A_1 \wedge H ,
\end{equation}
leading to a non standard Bianchi identity for the four form field
strength
\begin{equation}
dF_4 = F_2 \wedge H .
\end{equation}
Finally, the Chern-Simons interaction for type IIA is
\begin{equation}
S_{CS}^{IIA} = -\frac{1}{8\kappa^2}\int F_4 \wedge F_4 \wedge B .
\end{equation}

For type IIB theories one has $p=-1,1 ,3 $. For $p=-1$ the gauge form 
is a scalar, which is called axion. The
object which is electrically charged under this zero form is
localized in space and time. This is an instanton. The definition of
the field strength (\ref{RRfieldstrength}) receives further
contributions for $p=1$ and $p=3$
\begin{eqnarray}
F_3 & = & dA_2 - A_0 \wedge H , \\  
F_5 & = & dA_4 -\frac{1}{\sqrt{3}}A_2 \wedge H +
\frac{1}{\sqrt{3}}B\wedge F_3.
\end{eqnarray}
The Chern-Simons interaction for the type IIB theory is
\begin{equation}
S_{CS}^{IIB} = -\frac{9}{4\kappa^2} \int A_4\wedge H \wedge F_3 .
\end{equation}
 The five form field strength
has to be selfdual. This is not encoded in the action (\ref{efftwo})
but has to be added as an additional constraint,
\begin{equation}
F_5 =\star F_5  .
\end{equation}

In the heterotic string we have gauge fields transforming in the
adjoint of $SO(32)$ or $E_8\times E_8$. Their field strength is
defined as (we assign mass dimension one to the gauge fields $A$ --
this is related to a $\sqrt{\alpha^\prime}$ rescaling of $A$)
\begin{equation}
F = dA + A\wedge A.
\end{equation}
The definition of $H$ in
(\ref{universal}) needs to be modified\footnote{We present the
  effective action for the heterotic string just for completeness,
more details on differential geometry and anomaly cancelation in the
context of the effective heterotic theory can be found e.g.\ in
\cite{Green:1987mn}.} 
\begin{equation}\label{modh}
H = dB -\frac{\alpha^\prime}{4}\left(\omega_Y - \omega_L\right) .
\end{equation}
The Yang-Mills Chern-Simons form $\omega_Y$ is
\begin{equation}
\omega_Y = tr A\wedge dA + \frac{2}{3} 
tr A \wedge A \wedge A,
\end{equation}
where $A$ is the gauge connection of either $E_8 \times E_8$ or
$SO(32)$. The modification (\ref{modh}) implies that the $B$ field
transforms under gauge transformations and under local Lorentz
rotations in a non-trivial way such that
$H$ is gauge invariant. 
The Yang-Mills Chern-Simons form has the property that its
exterior derivative gives the instanton density (in a four dimensional
subspace with Euclidean signature), 
\begin{equation}
d\omega_Y = tr F\wedge F.
\end{equation}
The Lorentz Chern-Simons form is constructed from the spin connection
$\omega$, 
\begin{equation}
d\omega_L = tr \omega \wedge d\omega +\frac{2}{3} tr \omega
\wedge \omega \wedge \omega .
\end{equation}
If its exterior derivative takes values only on a four dimensional 
submanifold with Euclidean signature it corresponds to the Euler
density 
of that manifold,
\begin{equation}
d\omega_L = tr R\wedge R .
\end{equation}
If we take the ten dimensional geometry to consist of a direct product
of a six dimensional non compact and a four dimensional compact space
(with Euclidean 
signature) 
the modification (\ref{modh}) implies restrictions on the allowed
gauge bundles on the four dimensional compact space. The integration of
$dH$ over a compact space should vanish. It follows that the Euler
number of this space must be equal to the instanton number of the
gauge bundle.

In addition to the universal piece (\ref{universal}), the heterotic
action contains a gauge kinetic term and also the Green-Schwarz term
which ensures anomaly cancellation
\begin{equation}
S_{model}^{heterotic} = S_{gauge} + S_{GS},
\end{equation}
with
\begin{equation}\label{gkin}
S_{gauge} = -\frac{1}{2\kappa^2}\int d^{10} x e^{-2\Phi}
\frac{\alpha^\prime}{8} 
tr F^2, 
\end{equation} 
where again, the trace is taken over the gauge group ($E_8 \times E_8$
or $SO(32)$). The Green-Schwarz term is
\begin{equation}
S_{GS} = \frac{8\pi i}{\alpha^\prime}  \int B \wedge X_8,
\end{equation}
with (here, a power is meant with respect to the wedge product, e.g.\
$F^4 \equiv F\wedge F\wedge F\wedge F$)
\begin{equation}
X_8 =\frac{1}{2}\frac{1}{\left( 2\pi\right)^6} \frac{1}{48}\left(
    \frac{5}{4}tr 
    F^4 -\frac{1}{8}\left( tr F^2\right)^2 -\frac{1}{8} tr F^2 tr R^2
    +\frac{1}{8}tr R^4 +
\frac{1}{32} \left( tr R^2\right)^2\right)
\end{equation}
% 
%%%%%%%%%%%%%%%%%%%%%%%%%%%%%%%%%%%%%%%%%%%%%%%%%%%%%%%%%%%%%%%%%%%

To close this section on effective actions we identify the
different contributions with the worldsheet topologies they correspond
to. First, we observe that all the terms appearing in the effective
actions are of a structure such that they contain some power of $e^\Phi$
times a factor which is invariant under constant shifts in $\Phi$. 
In (\ref{stringcoupling}) we have identified the string coupling as a
constant part of $e^\Phi$. 
Thus, the leading term in the perturbative expansion in figure
\ref{feynman} enters the effective action accompanied with a factor of
$e^{-2\Phi}$. These are all terms in (\ref{universal}) and the gauge
kinetic term in the heterotic theory (\ref{gkin}). One may be tempted
to interpret the other terms (containing no $e^{-2\Phi}$ factor) as
one loop contributions. This is, however, misleading.
In order to simplify the Bianchi identities for the RR gauge forms we
have rescaled the RR gauge potentials by $e^\Phi$. Undoing this
rescaling means that for example the RR form $F_2$ receives a
further contribution
\begin{equation}
A_1  = e^{-\Phi} A_1 ^\prime\;\;\; \longrightarrow\;\;\; F_2 =
e^{-\Phi}\left(dA_1 ^\prime 
- d\Phi \wedge A_1 ^\prime\right) \equiv e^{-\Phi} F^\prime _2 ,
\end{equation}
and similar relations for the other RR field strengths. (If the terms
denoted by dots in (\ref{RRfieldstrength}) contain other RR field
strengths additional $\Phi$ derivatives will be picked up. But no
relative power of $e^\Phi$ will appear, since those terms always
contain one RR field strength or potential and an NSNS field strength
or potential. The NSNS fields are not rescaled.) After this rescaling
all terms in the type II thoeries are of the structure 
$$e^{-2\Phi} \left(\mbox{invariant under\ } \Phi \rightarrow \Phi +
  constant\right). $$
Since the rescaled (primed) fields correspond to the actual string
excitations, the effective type II actions given here contain only tree
level contributions.

As (implicitly) stated above, one loop contributions are multiplied by 
$g_s ^2$ and hence enter
the effective action with a factor of $e^{-2\Phi + 2\Phi} = 1$. 
In the type II examples, we have seen that due to field redefinitions
this correspondence may be changed. In the heterotic case, however,
there is no 
field redefinition such that all the terms in the effective action are
multiplied by the same power of $e^\Phi$. Indeed, the appearance of
the Green-Schwarz term corresponds to a torus amplitude from the
string theory perspective.
We excluded
also higher orders in $\alpha^\prime$ which would lead to higher
derivative terms and contributions with massive string
excitations.\footnote{An exception is the $\omega_L$ correction in
  (\ref{modh}) and the Green Schwarz term. They can be deduced by
  using supersymmetry and anomaly cancellation.}
As long as the string scale is much shorter (in length) than the
scale of the process we are interested in those terms can be
neglected.

\subsection{Toroidal Compactification and T-duality}\label{t-duality}
\setcounter{equation}{0}

In the previous sections we argued that perturbative superstring theories
are consistent provided that the target space is ten dimensional. 
As it stands, this cannot describe our observable (four dimensional)
world. 
At the end of section (\ref{betafunctions}), we  sketched as a possible
resolution to this problem the option to replace six of the target
space dimensions by a conformal field  theory with the desired 
central charge. One simple way to do so, is to replace a six
dimensional subspace of the ten dimensional Minkowski space by a
compact manifold. The coordinates of that compact manifold should
belong to a conformal field theory with a consistent central charge.
This restricts the set of possible compactifications. 
The easiest option is to compactify the additional directions on
circles (by periodic identification of the corresponding
coordinates). This clearly does not change the central charge
contribution of those directions, since the central charge depends only
on local features of the target space.

\subsubsection{Kaluza-Klein compactification of a scalar
  field}\label{KKSCALAR}

Before discussing some details of torus compactifications of string
theories we recall the Kaluza-Klein compactification of a free
massless scalar field. This will enable us to appreciate new
``stringy'' features which we will study afterwards.
Let us start with a free massless scalar living in a five dimensional
Minkowski space. We label the first four coordinates with a greek
index $\mu = 0,\ldots,3$ and call the fifth direction $y$. The five
dimensional field equation for the scalar $\varphi$ is
\begin{equation}\label{5dfree}
\left(\eta^{\mu\nu} \partial_\mu\partial_\nu + \partial_y ^2\right)
\varphi\left(x^\mu ,y\right) =0 .
\end{equation}
Now, we compactify the fifth direction on a circle of radius $R$
\begin{equation}\label{periodicity}
y\equiv y + 2\pi R .
\end{equation}
Solutions to (\ref{5dfree}) have to respect the periodicity
(\ref{periodicity}). Therefore, the $y$ dependent part of $\varphi$
can be expanded into a Fourier series of periodic functions. Focusing
on the $n$th Fourier mode, we find
\begin{equation}\label{fouriermode}
\varphi_n\left( x^\mu, y\right) = \varphi_n\left( x^\mu\right) e^{i
  \frac{n}{R} y},
\end{equation}
with integer $n$, i.e.\ the momentum in the fifth direction is
quantized.
Plugging (\ref{fouriermode}) back into (\ref{5dfree}) leads to
\begin{equation}
\left( \eta^{\mu\nu}\partial_\mu\partial_\nu - m^2
  _n\right)\varphi_n\left( x^\mu\right) = 0,
\end{equation}
with
\begin{equation}\label{KKmass}
m _n =  \frac{n}{R} ,
\end{equation}
i.e.\ the $n$th Fourier mode leads in the effective four dimensional
description to a Klein-Gordon field with mass (\ref{KKmass}). Since
the general solution of (\ref{5dfree}) is a superposition of all
Fourier modes the four dimensional description contains an infinite
Kaluza-Klein tower of massive four dimensional fields (depending only
on the $x^\mu$). There are two limits to be discussed. The
decompactification limit is $R\to \infty$. In this case all the
Kaluza-Klein masses (\ref{KKmass}) vanish. The four dimensional
description breaks down. The other limit is $R\to 0$ (or the
compactification radius becomes much shorter than the experimental
distance resolution). In this case, the KK masses (\ref{KKmass}) become
infinite except for $n=0$. Only the massless mode survives and no
trace from the fifth dimension is left. This picture is very different
in string theories as we will see now.

\subsubsection{The bosonic string on a circle}\label{boscirc}
\label{tdualsec} 

Even though the bosonic string is inconsistent because it contains a
tachyon, we will first study the compactification of the bosonic
string on a circle. The essential stringy properties will be visible
in this toy model. We compactify the 26th coordinate (the 25th spatial
direction),
\begin{equation}\label{period}
x^{25}  \equiv x^{25} + 2\pi R .
\end{equation} 
In the point particle limit string theory is just quantum mechanics of
a free relativistic particle. The plane wave solution contains the
factor $e^{ip_{25} x^{25}}$ where $p_{25}$ is the center of mass
momentum in the 25th direction. This wave function should be periodic
under (\ref{period}). This leads to a quantization condition for the
center of mass momentum in the compact direction
\begin{equation}
p_{25} = \frac{n}{R} ,
\end{equation}
with integer $n$ (the momentum number).
So far, everything is analogous to the free scalar field discussed
above. The new stringy property arises by observing that the
string can wind around the compact direction. Technically, this means
that the periodic boundary condition for the closed string is
modified
\begin{equation}
X^{25}\left( \tau ,\sigma + \pi\right) = X^{25}\left(\tau,
  \sigma\right) + 2\pi m R,
\end{equation}
where the integer $m$ denotes the winding number. With this
ingredients the mode expansions (\ref{leftsol}) and (\ref{rightsol}) are
\begin{eqnarray}
X^{25}_R & = & \frac{1}{2} x^{25} +\left(\frac{n}{2R} - mR\right) \sigma^-
+\frac{i}{2} \sum_{k\not= 0} \frac{1}{k} \alpha_k
^{25}e^{-2ik\sigma^-},\label{compmodl} \\
X^{25}_L & = & \frac{1}{2} x^{25} +\left( \frac{n}{2R} + mR\right)
\sigma^+
+\frac{i}{2}\sum_{k\not= 0} \frac{1}{k} \tilde{\alpha}^{25}_k
e^{-2ik\sigma^+}. \label{compmoder}
\end{eqnarray}
Taking into account the compact direction, the mass shell condition
has to be modified in a 
straightforward way,
\begin{equation}
\sum_{\mu =0}^{24} p_\mu p^\mu = -M^2 .
\end{equation}
Comparison with the constraints $T_{++} = T_{--} =0$
(\ref{constraint}) gives
\begin{equation}
M^2 = 4\left( \frac{n}{2R} - mR\right) ^2 + 8N -8 = 4\left(
  \frac{n}{2R} + mR\right)^2 + 8\tilde{N} -8\label{massenformel}
\end{equation}
where we have used the result of section \ref{quantization} for the
normal ordering. In particular, the level matching condition (the
second equality in (\ref{massenformel})) implies that
\begin{equation}\label{lefelmatsch}
N -\tilde{N} = nm .\end{equation}
Thus, for zero winding and momentum number the spectrum coincides with
the spectrum 
of the uncompactified string (see section \ref{quantization}). 
In the massless sector we have again a graviton, antisymmetric tensor
and 
dilaton which are obtained from the state
\begin{equation}
\alpha_{-1} ^i \tilde{\alpha}^j _{-1}\left| k\right> \,\,\,
,\,\,\, i,j \not= 25 .
\end{equation}
The target space interpretation of the remaining excitations
(containing creator(s) in 25th direction) is different. The two states
\begin{equation}\label{KKvectors}
\alpha^i _{-1} \tilde{\alpha}^{25}_{-1}\left| k\right> \,\,\, ,\,\,\,
\alpha^{25} _{-1} \tilde{\alpha}^{i}_{-1}\left| k\right> 
\end{equation}
are target space vectors.
They correspond to gauge fields of a
$U\left( 1\right) \times U\left( 1\right)$ gauge symmetry. Finally,
the state
\begin{equation}\label{KKscalars}
\alpha^{25} _{-1} \tilde{\alpha}^{25}_{-1}\left| k\right>
\end{equation}
describes a target space scalar.
The
spectrum is supplemented by a Kaluza-Klein and winding tower of
additional states as $n,m$ run through the integer numbers. An
interesting question is wether some of these additional states
are massless. For massless states the mass shell condition
(\ref{massenformel}) reads
\begin{equation}
2N -2 + \left( \frac{n}{2R} - Rm\right)^2 = 2\tilde{N} -2 +\left(
  \frac{n}{2R} + Rm\right)^2 = 0.
\end{equation}
These equations can be solved for nonvanishing $n$ or $m$ only at
special values of $R$. The most
interesting case is\footnote{Later, in section \ref{bosoorbi}, we will
  also discus the case $R^2 = 2$, where less states become massless.} 
\begin{equation}\label{selfdual}
R^2 = \frac{1}{2} = \alpha^\prime .
\end{equation}
One obtains the additional solutions listed in table \ref{tab:selfdspec}.
\begin{table}[h]
\begin{center}\begin{tabular}{|r|r|r|r|}
\hline $n$ & $m$ & $N$ & $\tilde{N}$ 
\\ \hline\hline
1& 1&1& 0\\
\hline
-1 & -1& 1 & 0\\
\hline
1 & -1 & 0 &1\\
\hline
-1 & 1 & 0 & 1\\
\hline
2 & 0 &0 &0\\
\hline
-2 & 0 &0 &0\\
\hline
0 & 2 &0 &0\\
\hline
0 & -2& 0 &0
 \\ \hline 
\end{tabular}\end{center}
\caption{Each line in this table gives a configuration of winding,
  momentum and occupation numbers leading to massless states at $R^2
  =\frac{1}{2}$.}\label{tab:selfdspec} 
\end{table}

Each of the first four states in table \ref{tab:selfdspec} contains
one creator. This gives four 
additional 
massless vectors (if the creator points into a non-compact direction)
and four massless scalars (if the creator points into the 25th
direction).
The latter four states in table \ref{tab:selfdspec} correspond to
massless scalars. 
Together with (\ref{KKvectors}) and (\ref{KKscalars}) we have
six vectors and nine scalars. The vectors combine into an $SU(2)\times
SU(2)$ gauge field whereas the scalars form a $\left(\mbox{\bf
    3},\mbox{\bf 3}\right)$ representation. For the special value
(\ref{selfdual}) the gauge group $U\left( 1\right)\times U\left(
  1\right)$ is enhanced to the non-abelian group $SU\left(
  2\right)\times SU\left( 2\right)$. The rank of the gauge group is
not changed.

An immediate question is: what is so special about (\ref{selfdual})?
To answer this, we rewrite (\ref{massenformel}) in a suggestive way 
\begin{equation} \label{windmommass}
M^2 = 4N +4\tilde{N} - 8 + \frac{n^2}{R^2} + 4m^2 R^2 ,
\end{equation}
where we already aplied (\ref{lefelmatsch}).
We observe that the spectrum is invariant under
\begin{equation}\label{tduality}
n \leftrightarrow m \,\,\, \mbox{and} \,\,\, R\leftrightarrow
\frac{\alpha^\prime}{R}.
\end{equation}
Recall that in the previous equations we have set $\alpha^\prime =
1/2$.  
The symmetry (\ref{tduality}) is called T-duality. Winding and
momentum numbers are interchanged and simultaneously the compactification
radius is inverted. If $R$ takes the value (\ref{selfdual}), the
spectrum is invariant under interchanging winding with momentum. This
radius is called the selfdual radius. Because of the symmetry
(\ref{tduality}) we can restrict the compactifications to radii equal
or larger than (\ref{selfdual}). This is an important difference to
the point particle discussed in the previous section. To make this
difference clearer let us take the compactification radius to
zero. All the Kaluza-Klein momenta diverge and only states with $n=0$
survive. This is similar to the point particle case. On the other
hand, 
all winding states degenerate. In order to make sense out of this
situation one can apply the T-duality tranformation
(\ref{tduality}). But then $R=0$ leads to the decompactification limit
and we are back at the 26 dimensional string. Therefore, in string
theory there are always traces of compact dimensions left.

Compactifying the string on a $D$ dimensional torus, the above
considerations lead to a ${\mathbb Z}_2 ^D$ symmetry in a
straightforward way. However, combining the T-duality along circles
with basis redefinitions of the torus lattice and integer shifts in
the internal $B$ field leads to an enhancement of the T-duality group
to $SO\left( D,D,{\mathbb Z}\right)$.

\subsubsection{T-duality in non trivial backgrounds}\label{tsigma}

In this section we will argue that the above described
T-dualiy is also a symmetry for non-trivial background
configurations. We closely follow\cite{Rocek:1992ps}. 
Our starting point is the non-linear sigma model
(\ref{nonlinear}). Compactification of one target space dimension is
possible if the sigma model is invariant under constant shifts in this
direction. For the first term in (\ref{nonlinear}) this implies that
the tangent to
the compactified direction is a Killing vector. The second term is
invariant provided that the Lie derivative of $B_{\mu\nu}$ in the
Killing direction is an exact two-form. For the last term to be
invaraint the Lie derivative of the dilaton $\Phi$ must vanish. We now
choose coordinates such that the isometry is represented by a translation
in the $d-1$ direction
\begin{equation}\label{globalsym}
X^{d-1} \rightarrow X^{d-1} + c .
\end{equation}
We call the other coordinates $X^i$.
The previously mentioned conditions on $B_{\mu\nu}$ and $\Phi$
imply that those fields are independent of $X^{d-1}$ (up to gauge
transformations). 

The next step is to gauge the symmetry (\ref{globalsym}) and to
``undo'' this by constraining the gauge fields to be of pure gauge. The
constraint is implemented with the help of a Lagrange multiplier
$\lambda$ which finally will replace $X^{d-1}$ in the T-dual model.
We introduce two dimensional gauge fields $A_\alpha$ changing under
(\ref{globalsym}) as
\begin{equation}
A_\alpha \rightarrow A_\alpha - \partial_\alpha c ,
\end{equation}  
and replace 
\begin{equation}
\partial_\alpha X^{d-1} \rightarrow D_\alpha X^{d-1} \equiv
\partial_\alpha X^{d-1} + A_\alpha .
\end{equation}
Together with the above mentioned constraint (implemented by a
Lagrange multiplier) this amounts to adding to (\ref{nonlinear})
(for simplicity we choose $\gamma_{\alpha\beta}
=\eta_{\alpha\beta}$)\footnote{For the Minkowskian worldsheet
  signature the $i$ in front of the $B_{\mu\nu}$ coupling in
  (\ref{nonlinear}) is replaced by one.}
a term 
\begin{eqnarray}
S_A & = & -\frac{1}{4\pi \alpha^\prime} \int d^2\sigma \left( G_{d-1,d-1}
  A_\alpha A^\alpha + 2 G_{d-1,\nu} A^\alpha\partial_\alpha
  X^\nu\right.
\nonumber \\ & & \;\;\;\;\;\;\;\;\;\;\;\;\;\; \left. + 2
  \epsilon^{\alpha\beta} B_{d-1,\nu} A_\alpha\partial_\beta X^\nu +
  2\lambda \epsilon^{\alpha\beta}\partial_\alpha A_\beta\right) .
\label{gaugeterm}
\end{eqnarray}
Integrating over $\lambda$ will result in the constraint of vanshing
field strength for the $A_\alpha$ which in turn imposes
\begin{equation}
A_\alpha = \partial_\alpha \varphi ,
\end{equation}   
with $\varphi$ being a worldsheet scalar. Shifting $X^{d-1}$ by $\varphi$
gives back the original sigma model (\ref{nonlinear}). Thus, adding
(\ref{gaugeterm}) does not change anything. However, there is a
subtlety here.
Compactifying the $d-1$ direction means that we identify $X^{d-1}$
with $X^{d-1} + 2\pi$ (this time we put the compactification radius
into the target space metric). In order to be able to absorb $\varphi$
into 
$X^\mu$, $\varphi$ should respect the same periodicity. This can be
ensured as 
follows. We continue the worldsheet to Euclidean signature and study
the sigma model for a torus worldsheet. Then we can assign two winding
numbers (corresponding to the two cycles of the torus) to the Lagrange
multiplier $\lambda$. Summing over these winding numbers (in a path
integral approach) will impose the required periodicity on the gauge
fields $A_\alpha$. Going through the details of this prescription
leads to the conclusion that the $\lambda$ ``direction'' is compact
$\lambda \equiv \lambda +2\pi$.

Instead of integrating out $\lambda$ (to check that we did not change
the model) we can integrate out $A_\alpha$. (Since $A_\alpha$ is not a
propagating field this can be done by solving the equations of
motion. As well, one can integrate out $A_\alpha$ in a path integral,
which is Gaussian.) This procedure leads us to a dual model
\begin{eqnarray}
S & =& -\frac{1}{4\pi\alpha^\prime}\int d^2 \sigma\left(
  \sqrt{-\gamma}\gamma^{\alpha\beta} \tilde{G}_{\mu\nu} 
  \partial_\alpha \tilde{X}^\mu\partial_\beta \tilde{X}^\nu +
  \epsilon^{\alpha\beta}\tilde{B}_{\mu\nu}\partial_\alpha
  \tilde{X}^\mu\partial_\beta \tilde{X}^\nu\right)\nonumber \\
 & & \,\,\,  -\frac{1}{4\pi}\int d^2 \sigma\sqrt{-\gamma}
  \tilde{\Phi} R^{(2)}.
\label{nonlineart}
\end{eqnarray}
The set of dual coordinates is $\left\{ \tilde{X}^\mu\right\} =\left\{
  \lambda, X^i\right\}$, and the dual background fields are,
\begin{equation}\label{tbg}
\tilde{G}_{d-1,d-1} = \frac{1}{G_{d-1,d-1}},\;\;\; \tilde{G}_{d-1,i} =
\frac{B_{d-1,i}}{G_{d-1,d-1}} ,
\end{equation}
\begin{equation}\label{tbg1}
\tilde{G}_{ij} = G_{ij}
  -\frac{G_{i,d-1}G_{d-1,j} +B_{i,d-1}B_{d-1,j}}{G_{d-1,d-1}},
\end{equation}
\begin{equation}\label{tbg2}
\tilde{B}_{d-1,i} =\frac{G_{d-1,i}}{G_{d-1,d-1}},\;\;\; \tilde{B}_{ij}
  = B_{ij} + \frac{G_{i,d-1}B_{d-1,j} +
  B_{i,d-1}G_{d-1,j}}{G_{d-1,d-1}} .
\end{equation}
To find the dual expression for the dilaton is a bit more
complicated. One can compute $\tilde{\Phi}$ in a perturbative way. To
this end, one requires that $\tilde{\Phi}$ is such that the conformal
invariance conditions (\ref{metriceq}), (\ref{beq}), (\ref{dilatoneq})
are satisfied whenever the original background satisfies them. This
leads to the following formula for the dual dilaton
\begin{equation}\label{tdil}
e^{-2\Phi} \sqrt{ -G} = e^{-2\tilde{\Phi}} \sqrt{-\tilde{G}} .
\end{equation}
From a path integral perspective the dilaton transformation can be
motivated as follows. The path integral measure for the $X^\mu$ is
covariant with respect to the metric $G_{\mu\nu}$. In the dual model
one would naturally use a measure which is covariant with respect to
the dual metric $\tilde{G}_{\mu\nu}$. The change of the measure
introduces a Jacobian which leads to (\ref{tdil}). To our knowledge
this is a rather qualitative statement which is difficult to prove
explicitly. 

To make contact with the simple case discussed in the previous section
we should take $G_{d-1,d-1} = R^2/\alpha^\prime$, $G_{ij} =
\eta_{ij}$, $B_{\mu\nu} = 0$ and $\Phi = const$. 
Then the T-duality formul\ae\ (especially (\ref{tbg})) imply that the
compactification radius is inverted. The dilaton receives a constant
shift and nothing else changes. This dilaton shift was not visible in
the discussion of the previous section. On the other hand, in the
present section we did neither see that T-duality interchanges winding
with momentum nor that there is an enhancement of gauge symmetry at
the selfdual radius,
because we did not study the spectrum of the general string theory. 

\subsubsection{T-duality for superstrings}\label{t-super}

In extending the discussion of section \ref{tdualsec} to the
superstring we will be sketchy and omit the technical details. Most of
the statements from section \ref{tdualsec} can be directly taken over
to the superstring. In the Ramond sector, some new ingredients
occur. First, we consider the type II superstring. Instead of the 26th
direction we compactify the tenth direction. Combining the
T-duality transformation (\ref{tduality}) with the mode expansions
(\ref{compmodl}) and (\ref{compmoder}) one realizes that
(\ref{tduality}) can be achieved by assigning $X^{9}= X^9 _L -X^9 _R$,
instead of $X^9 = X_L ^9 + X_R ^9$, or equivalently change
\begin{equation}\label{recepy}
\left( \mbox{right movers}\right) \longrightarrow -\left(\mbox{right
    movers}\right)
\end{equation}
while keeping the original prescription of combining left with right
movers. Carrying this prescription over to the fermionic sector we observe
that in the right moving Ramond sector (see (\ref{fermionpar}) for the
definition) 
\begin{equation}
\left( -\right)^{F} \longrightarrow 
-\left( -\right)^{F}.
\end{equation}
This in turn implies that the T-duality transformation takes us from the
type IIA to type IIB string and vice versa. Hence, T-duality is not a
symmetry in type II superstrings but relates the type IIA string with
type IIB. (This is true, whenever we perform T-duality in an odd
number of directions.) The type IIA string compactified on a circle
with radius $R$ is equivalent to the type IIB string compactified on a
circle with radius $\alpha^\prime /R$. This is consistent with the
observation that the massless spectra of circle compactified type IIA
and type IIB theories are identical as depicted in table
\ref{tab:spectra}, 
where $\mu,\nu = 0,\ldots,8$.
\begin{table}[h]
\begin{center}\begin{tabular}{|l|l|l|}
\hline & NS-NS & R-R 
\\ \hline\hline
IIA & $G_{\mu\nu}$, $B_{\mu\nu}$, $\Phi$, $G_{\mu 9}$, $B_{\mu 9}$ &
$A_\mu$, $A_9$, $C_{\mu\nu\rho}$, $C_{\mu\nu 9}$ \\ \hline
IIB & $G_{\mu\nu}$, $B_{\mu\nu}$, $\Phi$, $G_{\mu 9}$, $B_{\mu 9}$ &
$B_{\mu 9}^{\prime}$, $\Phi^{\prime}$, $C_{\mu\nu\rho 9}$, 
$B^{\prime}_{\mu\nu}$ \\ \hline 
\end{tabular}\end{center}
\caption{Massless type II fields in nine dimensions}\label{tab:spectra}
\end{table}

In order to discuss compactifications of the heterotic string, it is
useful to employ a formulation where the additional 32 left moving
fermions are bosonized into 16 left moving bosonic degrees of
freedom. We will not carry out this construction here. It can be found
in the books listed in section \ref{books:pert}.  The result which is
of interest in the current 
context is that those 16 left moving bosons are compactified on an even
selfdual lattice\footnote{There exist exactly two such lattices
  $\Gamma_8\times \Gamma_8$ and $\Gamma_{16}$ giving rise to the
  $E_8\times E_8$ and the $SO(32)$ string, respectively.}. 
That is, that even without further compactifications
from ten to less dimensions the heterotic string contains already a
left-right asymmetric compactification. The theory does not depend on
changing the basis of the `internal' lattice. Compactifying the tenth
dimension one observes another new feature which is present in the
heterotic string. In the previously discussed cases, there was one
modulus in circle compactfications, {\it viz.} the radius of the
circle. For the heterotic string we have 16 more moduli. These are
called Wilson lines. They arise from the possibility that the
non-abelian gauge fields can take constant vacuum expectation values
(vev)  in the Cartan subalgebra of the gauge group. The fact that (at
least) one of 
the ten directions has been compactified is important here. Otherwise,
a constant vev could be gauged away. To see this explicitly, let us
assume that the gauge field vev is (proportional to a generator in the
Cartan subalgebra)
\begin{equation}\label{wilsonline}
A_9 =   \frac{\Theta}{R} = e^{- \frac{x^9
    \Theta}{R}}\partial_9  e^{ \frac{x^9
    \Theta}{R}} ,
\end{equation}
where the second part of the equation shows that a constant vev is a
pure gauge configuration. However, in the compact case we have
identified $x^9$ with $x^9 + 2\pi R$ and hence only gauge
transformations which are periodic under this shift are allowed. 
This implies that the Wilson line (\ref{wilsonline}) can be gauged
away only if $\Theta$ is an integer. From this discussion it follows
that generically the gauge group is broken to $U(1)^{16}$ in the
compactification process. In addition there are the (abelian)
Kaluza-Klein gauge  fields $G_{9\mu}$ and $B_{9\mu}$ corresponding to
a $U(1)\times U(1)$ gauge symmetry. 
Thus, generically there is a $U\left( 1\right)^{18}$ gauge symmetry in
the circle compactified heterotic string. Depending on the moduli
(Wilson lines and compactification radius) there are special points of
stringy 
gauge group enhancement.
 
It can be proven that the $E_8\times E_8$ heterotic string and the
$SO(32)$ string are continuously connected in moduli space once they
have been compactified to nine dimensions. This can be shown by
observing that for a certain configuration of Wilson lines (where the
gauge group is broken to $SO(16)\times SO(16)$ in either theory)
T-duality maps the two compactifications on each other. (For details
see e.g.\ Polchinski's book\cite{polchinski-book}.) All other
compactifications can be reached by continuously changing the moduli.
Including the original ten dimensional theories as decompactification
limits we see that the two different heterotic strings belong to the
same set of theories sitting at different corners in moduli space.
For completeness we should mention that for compactifications of the
heterotic string on a $D$ dimensional torus one finds the T-duality
group $SO\left(16+D,D,{\mathbb Z}\right)$.

\section{Orbifold fixed planes}\label{sec:orbifold}
 
In the previous sections we have studied the theory of a one
dimensional extended object -- the string. One of the striking features
of this theory is that it automatically also describes objects 
which are extended along more than one space direction. As the
simplest example we will study now the orbifold fixed planes. Here, one
compactifies the string on a torus whose lattice has a discrete
symmetry, and gauges this symmetry\footnote{With gauging of a discrete
  symmetry we mean that only invariant states are kept.}. Thus, the
compact manifold is 
a $D$ dimensional torus divided by some discrete group. (We will
consider ${\mathbb Z}_2$ as such a group. It leaves an arbitrary
lattice invariant.) There are some points or ---when combined with the
other directions--- planes which are invariant under the discrete
group. These are the orbifold fixed planes. They present singularities
in the compact part of the space time. String theory gives a physical
meaning to orbifold fixed planes. We will see that certain string
excitations (particles or gauge fields from the target space
perspective) are confined to be located at the orbifold fixed
planes. By looking at an example where the orbifold can be
reached as a singular limit of a smooth manifold we will see that for string 
theory the singular nature of this limit is not ``visible''.  Instead
of discussing the general setups for orbifold compactifications we
will present two examples: the bosonic string on an orbicircle and the
type IIB string on 
$T^4/{\mathbb Z}_2$. We hope that this will demonstrate the general
idea with a minimal amount of technical complications. For more
details (and also orbifold compactifications of the heterotic string)
we recommend the review\cite{Nilles:1987uy}.

\subsection{The bosonic string on an orbicircle}\label{bosoorbi}
\setcounter{equation}{0}

Let us start by describing the target space geometry. We compactify
the 25th dimension on a circle like in section \ref{boscirc}. 
In addition, we identify opposite points on this circle. If we choose
the ``fundamental domain'' to be $-\pi R < x^{25} < \pi R$ this is done by
the ${\mathbb Z}_2$ identification: $x^{25} \equiv -x^{25}$. The
resulting target space is an interval in the 25th direction
as depicted in figure \ref{orbicircle}. Taking into account the
uncompactified dimensions, the end points of the interval (the fixed
points of the ${\mathbb Z}_2$) correspond to planes with 24 spatial
directions. Therefore, we call them orbifold-24-planes.

%%%%%%%%%%%%%%%%%%%%%%%%%%%%%%%%%%%%
\begin{figure}
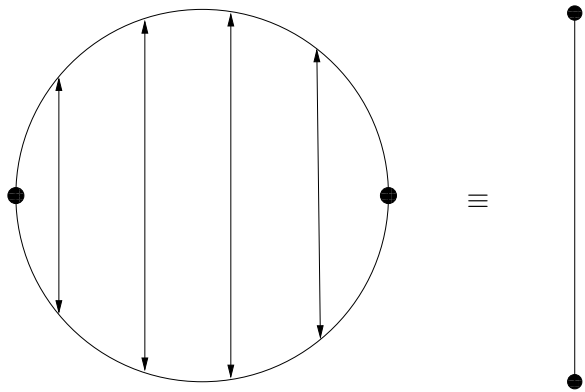
  
\begin{center}
\input orbicircle.pstex_t
\end{center}
\caption{The interval as an orbicircle. The fixed points (black dots)
  form the ends of the interval.}
\label{orbicircle}
\end{figure}
%%%%%%%%%%%%%%%%%%%%%%%%%%%%%%%%%%%%

We proceed by constructing the untwisted
spectrum. The term untwisted (in contrast to twisted) will become
clear later. It means that we construct the spectrum which is
invariant under the orbifold projection $x^{25} \rightarrow -x^{25}$. Since in
the bosonic string the right moving sector is identical to the left moving
one, we first write down the right moving states only. The result is
collected in table \ref{untwisted} ($i=2,\ldots,24$; the zeroth and
first direction are eliminated by the light-cone gauge). 

%%%%
\begin{table}
\begin{center}
\begin{tabular}[h]{|l|l|l|} 
\hline 
State & $ {\mathbb Z}_2$ & $24+1$ dim. rep. \\
\hline \hline
$\alpha_{-1}^i \left| 0\right>$ & $+$ & 1 vector \\
\hline
$\alpha^{25}_{-1}\left| 0\right>$ & $-$ & 1 scalar \\
\hline
\end{tabular}
\end{center}
\caption{Untwisted right moving states}\label{untwisted}
\end{table}
%%%%%%%%%%%%%%%%%%%%%%%%%%%

Now we need to combine left with right movers such that the resulting
state is invariant under the ${\mathbb Z}_2$. This is the case for 
the product of the vector with the vector and the scalar with the
scalar. 
Thus, we obtain the metric
$G_{ij}$, the antisymmetric tensor $B_{ij}$ and the dilaton
$\Phi$. The additional $U\left( 1\right)$ 
  vectors $G_{i25}$ and $B_{i25}$ are projected out in contrast to
  section \ref{boscirc}. The combination of the scalar from the left
  moving sector with the scalar from the right moving sector yields a
  target space scalar $G_{25\, 25}$. Since the groundstate is
  ${\mathbb Z}_2$ 
  invariant, the tachyon will survive the projection. If we are at the
  selfdual radius, there might be additional massless states (without
  imposing ${\mathbb Z}_2$ invariance these are listed in table 
  \ref{tab:selfdspec}). The action of the ${\mathbb Z}_2$ takes winding
  number to minus winding number and momentum number to minus momentum
  number as can be seen from the mode expansion (\ref{compmodl}),
    (\ref{compmoder}). This means that we can keep only invariant
    superpositions of states. From the first four entries in table
    \ref{tab:selfdspec} we obtain two additional massless
    vectors. These arise as follows. We add the first state to
    the second state of the listing and act with
    $\alpha_{-1}^i$, or we add the third to the fourth state
    and act with $\tilde{\alpha^i} _{-1}$. 
We can also subtract the second from the first state and act with
$\alpha_{-1}^{25}$, or we subtract the fourth from the third state and
act with $\tilde{\alpha}_{-1}^{25}$. This gives two massless scalars.
Adding the fifth to the sixth entry
    and the seventh to the eighth, we obtain two more scalars at the
    selfdual radius. This looks very unusual. Since we do not have any
    $U\left( 1\right)$ gauge fields away from the selfdual radius, now
    also the rank of the gauge group is enhanced at the selfdual
    radius. There are also additional tachyons at the selfdual
    radius. These are the two states which are obtained by adding the
    $n=0, m=1$ vacuum to the $n=0, m=-1$ vacuum. The second state is
    the same with $m$ and $n$ interchanged. These two additional tachyons
    have mass squared $M^2 = -6$, as can be easily computed from
    (\ref{windmommass}). 

Now, we come to a new feature which is unique to string theory. There
are additional twisted sector states. These states are periodic under
shifting $\sigma$ by $\pi$ only up to a (non-trivial) ${\mathbb Z}_2$
transformation. In our case this implies for the string
that its center of mass position has to be
located at a fixed plane and that the integer Fourier modes are
replaced by half-integer ones in the 25th direction. 
In the twisted sector we need to compute the groundstate
energy. This can be done by first modifying equation (\ref{mumpitzf})
in a straightforward way
\begin{equation}
a^{twisted} = - \frac{23}{2}\sum_{n=1}^\infty n -
\frac{1}{2}\sum_{r=\frac{1}{2}}^\infty r .
\end{equation}
Regularizing this expression according to the prescription
(\ref{moddingformula}) gives $a=\frac{15}{16}$. This implies that the
groundstate is tachyonic and also that there is no massless state
coming from this twisted ground state. There is one more tachyonic
state at the first level in the twisted sector. This is obtained by
acting with
$\alpha_{-\frac{1}{2}}^{25}\tilde{\alpha}_{-\frac{1}{2}}^{25}$. 
Collecting the results, we obtain one tachyon with $M^2 = -\frac{15}{2}$ and
one with $M^2 = -\frac{7}{2}$ at each fixed plane. Altogether, there
are four tachyons (and states with positive mass squared) located at
the fixed planes.  

The singular nature of the fixed points does not raise any problem in
string theory. It introduces twisted sector states which result in
additional particles which are located at the orbifold-24-planes in
target space.

It is interesting to note that the orbifold at the selfdual radius is
equivalent to the toroidally compactified bosonic string at twice the
selfdual radius.\footnote{Alternatively, we could use the T-dual
  version at half the selfdual radius.} For a detailed disussion of 
this equivalence we refer to Polchinski's book\cite{polchinski-book}. Here,
we will only check that the light (tachyonic and massless) spectra
coincide. Obviously, the gauge groups $U\left(1\right) \times
U\left( 1\right)$ are the same. For the bosonic string on the circle (with
$R^2 =2$) these are the off-diagonal metric and $B$-field components
$G_{i25}$, $B_{i25}$, whereas for the orbicircle compactification at
$R^2 = \frac{1}{2}$ these come from states with non-trivial winding
and momenta as discussed above. It remains to identify the four
additional massless 
scalars and the two tachyons (found in the non-trivial winding-momenta
sector) 
of the orbicircle compactification (at selfdual radius) and the four
additional tachyons from the twisted sector in the circle
compactification. Here, the special choice $R^2 = 2$ for the circle 
compactification comes into the game.   
With (\ref{windmommass}) evaluated at $R^2 = 2$ we find
exactly these missing states. At first, there are four massless scalars:
the vacua with $m=0$ and $n=\pm 4$, or $m=\pm 1$ and $n=0$.  
The two tachyons with $M^2 = -6$ are
obtained from the two vacua with $m=0$ and $n=\pm 2$. The two twisted
sector tachyons with $M^2 = -\frac{15}{2}$ correspond to the two vacua
with $m=0$ and $n=\pm 1$. The other two twisted sector tachyons with
$M^2 = -\frac{7}{2}$ can be identified in the circle compactification
as the vacua with $m=0$ and $n=\pm 3$.

The equivalence of the $S^1/ {\mathbb Z}_2$ compactification at the
selfdual radius and the $S^1$ compactification at twice the selfdual
radius shows that the moduli spaces of both compactifications are
connected at this point. This feature has a
stringy origin. From the 
target space perspective this is quite surprising. A field theory on
$24+1$ dimensional Minkowski space times an interval with certain fields
constrained to live at the endpoints of the interval is smoothly
connected to a field theory on $24+1$ dimensional Minkowski space
times a circle with all fields living in the whole space. However, due
to the tachyons both vacua are unstable. In the next section we will
see that similar things happen for the superstring which does not have
tachyons in its spectrum.

\subsection{Type IIB on $T^4/{\mathbb Z}_2$}\label{k3orbi}  
\setcounter{equation}{0}

Again, we start by describing the target space geometry. We compactify
the six, seven, eight and nine direction on a four dimensional torus. 
We view this four dimensional torus as the product of two
two-dimensional tori. The coordinates are labeled such that the sixth
and 
seventh direction form one $T^2$ and the eighth and ninth a second
$T^2$. Let us focus on this second $T^2$ with the understanding that
the same applies to the first $T^2$. 
In figure 
\ref{orbifoldbild} this is depicted by drawing a lattice in the
eight-nine plane. The fundamental cell is the parallelogram with edges
drawn with stronger lines. The lattice vectors are the lower and the
left 
edge of the fundamental cell. A two dimensional torus is obtained by
gluing together the
opposite edges of the fundamental cell. Shifts by lattice
vectors connect identified points.  

%%%%%%%%%%%%%%%%%%%%%%%%%%%%%%%%%%%%
\begin{figure}
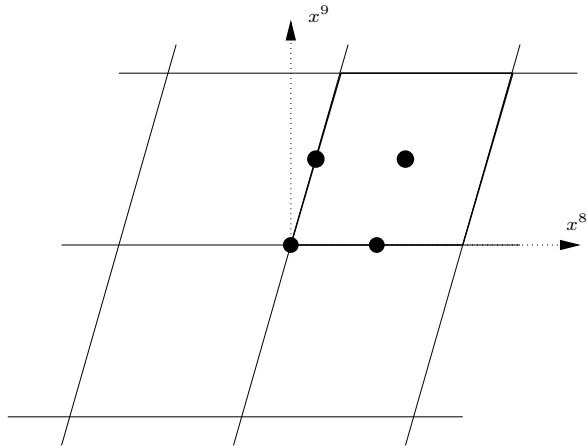
  
\begin{center}
\input orbifold.pstex_t
\end{center}
\caption{The orbifold $T^2/{\mathbb Z}_2$}
\label{orbifoldbild}
\end{figure}
%%%%%%%%%%%%%%%%%%%%%%%%%%%%%%%%%%%%

``Dividing'' the $T^4$ by ${\mathbb Z}_2$ means that in addition we
identify points via the prescription 
\begin{equation}\label{o-projection}
\left\{x^6, x^7 ,x^8 ,x^9\right\} \rightarrow \left\{-x^6,-x^7,-x^8
  ,-x^9\right\} . 
\end{equation}
This ${\mathbb Z}_2$ action leaves the four points indicated by black
dots in figure
\ref{orbifoldbild} times the four points in the first torus
invariant. Thus, we obtain sixteen
orbifold five-planes. At first, we construct the untwisted
spectrum.
Since in the type IIB case the left moving sector is identical to the
right 
moving one, we first write down the right moving states, only. The result is
collected in table \ref{untwisted2}. We choose the GSO projection such
that ---in the notation of (\ref{rvacuum})--- states with an odd number of
minus signs survive. The projection
(\ref{o-projection}) can be viewed 
as rotations by $180^{\circ}$ in the eight-nine
plane and simultaneously in the six-seven plane. This is 
useful for the identification of the behavior of the R-sector under
${\mathbb Z}_2$ transformations. We consider only states which lead to
massless particles when combined with the left movers.
%%%%%%%%%%%%%%%%%
\begin{table}
\begin{center}
\begin{tabular}[h]{|l|l|l|l|} 
\hline 

Sector & State & $ {\mathbb Z}_2$ & $5+1$ dim. rep. \\
\hline \hline
NS: & $\psi_{-\frac{1}{2}}^i \left| 0\right>$ & $+$ & 1 vector \\
    & $\quad i=2,\ldots, 5$ & & \\
 & $\psi^{6,7,8,9}_{-\frac{1}{2}}\left| 0\right>$ & $-$ & 4 scalars \\
\hline
R: &$\left| s_1 s_2 s_3 s_4\right>$ & $e^{\frac{i\pi}{2}\left(s_3
    +s_4\right)}$ &  \\
 & $\quad s_1=s_2$, $s_3 = -s_4$ &$+$ & 2 (anti-chiral) spinors \\
 & $\quad  s_1=- s_2$ , $s_3 =s_4$ & $-$& 2 (chiral) spinors\\
\hline
\end{tabular}
\end{center}
\caption{Untwisted right moving states}\label{untwisted2}
\end{table}
%%%%%%%%%%%%%%%%%%%%%%%%%%%

In the NSNS sector, we can combine the left moving vector with the right
moving one leading to the six dimensional graviton $G_{ij}$, the
antisymmetric tensor $B_{ij}$ and the dilaton $\Phi$.  Further, we can
combine scalars from the left moving sector with scalars from the
right moving one. This gives sixteen massless scalars corresponding to
$G_{ab}$ and $B_{ab}$ where the indices $a,b$ are internal, i.e.\ $a,b
= 6, \ldots ,9$. The target space vectors $G_{ia}$ and $B_{ia}$ are
projected out. In the RR sector, we can combine the chiral spinor with
the chiral one, and the anti-chiral with the anti-chiral one. This
leads to 32 massless (on-shell) degrees of freedom in the RR
sector. Tensoring a chiral spinor with a chiral spinor gives a
selfdual two-form potential (3 on shell components) and a scalar.
The tensor product of two antichiral spinors yields an anti-selfdual
two-form potential and a scalar.  We can perform four of those
combinations, each. 
With the 
notation of table \ref{tab:tab2} the RR states can be identified as follows:
\begin{itemize}
\item
$C^{*}_{ijkl}$ (or $C^{*}_{abcd}$) gives $\left( \begin{array}{c} 4 \\
    4\end{array}\right) =1$ degree of freedom (one scalar),
\item 
$C^{*}_{abij}$ gives 
3 anti-selfdual two-forms and 3 selfdual two-forms (18 degrees of freedom),
\item
$B^\prime _{ij}$ gives a two-form (6 degrees of freedom),
\item
$B^\prime _{ab}$ gives six scalars,
\item
$\Phi^\prime$ gives a scalar .
\end{itemize}
All other fields from the RR sector are projected out. Fermionic
degrees of freedom are obtained from the NSR and RNS sector. Combing
the vector with the anti-chiral spinors gives four
times\footnote{One factor of two arises because the NSR and RNS sector
  yield such a tensor product, each.  The second factor of two is due to
  the two anti-chiral spinors in table \ref{untwisted2}.}a $\left(\mbox{\bf 2},
  \mbox{\bf 2}\right) \otimes \left(
  \mbox{\bf 2}, \mbox{\bf 1}\right) = \left( \mbox{\bf 3},\mbox{\bf
    2}\right) \oplus \left( \mbox{\bf 1}, \mbox{\bf 2}\right)$ 
representation of the six dimensional little group $SO\left( 4\right)
= SU\left( 2\right) \times SU\left( 2\right)$.\footnote{A vector is in
the $\left(\mbox{\bf 2}, \mbox{\bf 2}\right)$, an anti-chiral spinor in the
$\left(\mbox{\bf 2},\mbox{\bf 1}\right)$, a chiral spinor in the 
$\left(\mbox{\bf 1},\mbox{\bf 2}\right)$, a selfdual twoform in the 
$\left(\mbox{\bf 3},\mbox{\bf 1}\right)$ and an anti-selfdual twoform in the 
$\left(\mbox{\bf 1},\mbox{\bf 3}\right)$.} Therefore, this
tensor product provides us with four chiral gravitini and four chiral
fermions.  

Combining the NS sector scalars with the chiral R sector spinors 
gives 16 chiral
spinors. From the existence of the four chiral gravitini we can guess
that the resulting low energy effective field theory has $N=4$ chiral
supersymmetry in six dimensions. (For a collection of supersymmetries
in various dimension see\cite{Salam:1989fm}.)

Before checking that also the rest of the massless states fit into
supersymmetric multiplets we should construct the twisted sector
states. The construction does not depend on the location of the fixed
plane. Therefore, we restrict the construction to one plane and
multiply the result by 16. In the twisted sector, the NS fermions with
an index corresponding to a compact dimension are integer modded
whereas the R sector fermions are half integer modded. Now, there are
NS sector zero modes forming a four dimensional Clifford algebra. The
twisted NS ground state is two-fold degenerate after imposing the GSO
projection. (We modify the notation of (\ref{rvacuum}) in a
straightforward way. Since we have only two creators and two
anihilators, the twisted NS groundstate has two entries. Performing the
GSO projection means that we keep those states with an odd number of
minus 
signs.) In the twisted R sector we do not have zero modes in the
compact direction. This lifts some of the vacuum degeneracy as
compared to the untwisted sector. The twisted R ground state is labeled
only by the first two entries. Again, we keep only states with an odd
number of minus signs.  
In order to deduce the masses of the states in the twisted NS sector,
we observe from (\ref{mumpitzf}) (and its regularisation) that
replacing four integer moded bosons by half integer moded ones changes
the normal ordering constant by $-\frac{4}{24}-\frac{4}{16}=
-\frac{5}{12}$. Changing the modding of four worldsheet fermions from
half-integer to integer gives another shift of
$-\frac{4}{16}+\frac{4}{24}=-\frac{1}{12}$. Thus, we arrive at
\begin{equation}
a_{NS}^{twisted} = a_{NS}^{untwisted} - \frac{1}{2} = 0.
\end{equation}
The twisted NS sector groundstate is massless. The R sector
groundstate is always massless, since fermions have the same modding
as bosons.  
The analogon of table \ref{untwisted2} for the
twisted sector is table \ref{twisted2}.
%%%%%%%%%%%%%%%%%
\begin{table}
\begin{center}
\begin{tabular}[h]{|l|l|l|l|} 
\hline 

Sector & State & $ {\mathbb Z}_2$ & $5+1$ dim. rep. \\
\hline \hline
NS: &$\left|s_3 s_4\right>$ & $e^{\frac{i\pi}{2}\left(s_3
    +s_4\right)}$ &  \\
 & $\quad s_3 = -s_4$ &$+$ & 2 scalars \\ \hline
R:  &$\left|s_1 s_2\right>$ &  &  \\
& $\quad s_1=- s_2$ ,  & $+$& 1 (chiral) spinors\\
\hline
\end{tabular}
\end{center}
\caption{Twisted right moving states}\label{twisted2}
\end{table}
%%%%%%%%%%%%%%%%%%%%%%%%%%%
Since all the twisted sector groundstates are invariant under the
${\mathbb Z}_2$ we can form all possible left-right tensor
products. Multiplying with 16 (the number of fixed planes) we obtain
64 scalars from the NSNS sector. The RR sector leads to 16
anti-selfdual two-forms and 16 scalars. The RNS and NSR sector give
rise to 64 chiral spinors. 

After we have obtained the full massless spectrum of the type IIB
string on $T^4/{\mathbb Z}_2$ we can fit it into super-multiplets of
$N=4$ chiral supergravity in six dimensions. The possible
supermultiplets are the gravity multiplet and the tensor
multiplet. The gravity multiplet contains the graviton and four
chiral gravitini from the untwisted sector. In addition, five selfdual
two-forms are in the gravity multiplet. A tensor multiplet is made out
of an anti-selfdual two-form, five scalars and four chiral
fermions.   
The five selfdual two-forms in the gravity multiplet we take from
$B_{ij}$, $B^\prime _{ij}$ and $C_{ijab} ^{*}$. After filling the
gravity multiplet, we are left with 21 anti-selfdual two-forms, 105
scalars and 84 chiral fermions. Thus, the remaining degrees of freedom
fit into 21 tensor multiplets.

To summarize, the massless spectrum of the type IIB string on
$T^4/{\mathbb Z}_2$ consists of one gravity multiplet and 21 tensor
multiplets of $D=6$ chiral $N=4$ supersymmetry. Some of the degrees of
freedom are confined to live on the orbifold-5-planes which fill
the $5+1$ dimensional non-compact space but are located in the
four dimensional compact space. In the next section we will argue, that
this setup is smoothly connected to compactifications without
orbifold-5-planes.  

\subsection{Comparison with type IIB on $K3$}
\setcounter{equation}{0}

In the previous section we compactified the type IIB string on
$T^4/{\mathbb Z}_2$. Among others, we obtained four chiral
gravitini. If we compactified on $T^4$ instead, the two ten
dimensional gravitini would give rise to four non-chiral
gravitini in six dimensions. Thus, our orbifolding removes half of the
supersymmetries. This is due to the fact, that the $T^4/{\mathbb Z}_2$
manifold belongs to a larger class of manifolds which are called
Calabi-Yau $n$-folds. Here, $n$ denotes the number of complex
dimensions, i.e.\ $n=2$ in our case. The Calabi-Yau twofolds are all
connected by smooth deformations and commonly denoted by $K3$. One
important feature of Calabi-Yau $n$-folds is that they possess
$SU\left(n\right)$ holonomy. This means that (for K3) going around
closed 
(non-contractable) curves induces an $SU\left( 2\right)$
transformation. In a toroidal 
compactification we split the ten dimensional spinor into a couple of
lower dimensional spinors. The possible values of the internal spinor
indices count the number of resulting lower dimensional spinors.
In a torus compactification, each value of the internal
indices gives rise to a massless spinor. This is because the internal
homogenous Dirac equation has always a solution -- any constant
spinor. If instead of a torus with trivial holonomy we compactify on
$K3$ with $SU\left( 2\right)$ holonomy, only spinors which do not
transform under the holonomy group give rise to massless six
dimensional spinors. This removes half of the internal components and
thus breaks half of the supersymmetry. Indeed, all $K3$
compactifications yield the same massless spectra. This is a
consequence of the fact that the number of zero-modes (of Laplace and
Dirac operators) does not change as we move from one $K3$ to another
one. The number of zero modes of the Laplace operators\footnote{There
  are several different Laplace operators whose form depends on the
  tensor structure of the object they act on.} are usualy
listed in Hodge diamonds. The Hodge 
diamond for $K3$ is 
%the Hodge diamond 
\begin{equation}\label{eq:hodge}
\begin{array}{c c c c c}
  &   & 1   &  & \\
  & 0\! &      &\! 0& \\
1\! &\!   & 20 & \! &\! 1\\
  & 0\! &      &\! 0&  \\
  &   &1    &  & 
\end{array}  .
\end{equation}
In the following we explain (roughly) how to read
(\ref{eq:hodge}). The $K3$ is a complex manifold. Therefore, we can choose
complex coordinates (and so we do). Then a tensor can have a couple of
holomorphic indices and a couple of anti-holomorphic indices. In
other words, there are $(p,q)$ forms on $K3$, where $p$ corresponds to
the number of holomorphic indices and $q$ to the number of
anti-holomorphic ones. Since the complex dimension of $K3$ is two, the
values of $p,q$
can be zero, one or two. We denote the number of zero modes
of a $(p,q)$ form with $h^{(p,q)}$. These (Hodge) numbers are arranged
into a Hodge diamond as follows\footnote{The symmetry of
  (\ref{eq:hodge}) is not accidental. The vertical symmetry is related
  to Hodge duality and the horizontal one to interchanging holomorphic
  with anti-holomorphic coordinates.}  
\begin{equation} 
\begin{array}{c c c c c}
            &              &  h^{0,0} &            &            \\
            & h^{1,0}\! &              &\! h^{0,1}\!&            \\
  h^{2,0}\! & \!             & h^{1,1} & \!            &\! h^{0,2}\\
            & h^{2,1} &              &h^{1,2}&             \\
             &              &h^{2,2}&              & 
\end{array}.
\end{equation}

From (\ref{eq:hodge}) we can deduce that an object which represents a
zero or a four form in the
internal space, has one zero mode. 
Such an object gives rise to one massless six-dimensional field.
A $p+q=2$ form possesses 22 zero modes, thus leading to 22 massless
fields in six dimensions. In order to write down the massless spectrum
of the $K3$ compactified type IIB string we need to know another
feature of the family of $K3$ manifolds. All Calabi-Yau manifolds (and in
particular $K3$) are Ricci flat. This means that the Ricci tensor
vanishes and hence we do not need any non-trivial background
configuration in order to satisfy the conformal invariance conditions
derived in section \ref{betafunctions}. This remains true under
certain deformations of the metric of $K3$. The space of such
non-trivial (not related to coordinate changes) metric deformations is
58 dimensional for the family of $K3$s.

Now, we are ready to derive the bosonic massless spectrum of the $K3$
compactified type IIB string. At first, we collect all zero forms of
$K3$. From the NSNS sector these are $G_{ij}$,$B_{ij}$,$\Phi$, and from
the RR sector $B^\prime _{ij}$, $\Phi^\prime$, $C^{*}_{ijkl}$. 
Since $h^{(0,0)} =1$ these appear once in the lower
dimensional spectrum. The $G_{ab}$ are not differential forms on $K3$
but metric deformations. They result in 58 massless scalars.
Since $h^{(p,q)}=0$ for $p+q$ odd, the Kaluza-Klein fields $G_{ia}$ and
$B_{ia}$ do not give rise to massless six dimensional fields.  

It remains to count the two-forms on $K3$. (The four form
$C^{*}_{abcd}$ we have already counted as $C^{*}_{ijkl}$ because of
selfduality.) The two-forms are $B_{ab}$, $B^\prime _{ab}$ and
$C^{*}_{ijab}$. $B_{ab}$ and $B^\prime _{ab}$ lead to 44 scalars in six
dimensions. The 22 zero-modes of $C^{*}_{ijab}$ can be decomposed into
three selfdual and 19 anti-selfdual twoforms in six
dimensions\cite{Aspinwall:1996mn}. Taking into account that the
$SU(2)$ holonomy breaks half of the supersymmetry (as compared to
$T^4$ compactifications) and that the fermionic zero modes are all of
the same chirality, we obtain the same massless spectrum as in the
$T^4/{\mathbb Z}_2$ case.

Indeed, $T^4/{\mathbb Z}_2$ corresponds to a limit in the space of
$K3$ manifolds where the $K3$ degenerates. As long as one considers
$K3$s very close to that point one obtains the same massless
spectrum. In string theory even the limit to the point where the $K3$
degenerates is well defined.

Let us see what happens when we repeat the $K3$ analysis for the
orbifold $T^4/{\mathbb Z}_2$. We will focus only on the bosonic
spectrum. First, we need to know the Hodge diamond for $T^4/{\mathbb
  Z}_2$. This can be easily ``computed'' without much knowledge of
algebraic geometry. On $T^4$ we obtain the Hodge numbers just by
counting independent components of the corresponding differential
forms, 
\begin{equation}
h^{p,q} = \left( \begin{array}{c} 2\\ p\end{array}\right) \left(
  \begin{array}{c} 2\\ q\end{array}\right) .
\end{equation}
The ${\mathbb Z}_2$ action is taken into account by removing forms
which are odd under the ${\mathbb Z }_2$. Thus, the Hodge diamond for
$T^4/{\mathbb Z}_2$ is
%the Hodge diamond 
\begin{equation}\label{eq:hodget4z2}
\begin{array}{c c c c c}
  &   & 1   &  & \\
  & 0\! &      &\! 0& \\
1\! &\!   & 4 & \! &\! 1\\
  & 0\! &      &\! 0&  \\
  &   &1    &  & 
\end{array}  .
\end{equation}
In six dimensions we obtain two twoforms $B_{ij}$ and $B^\prime _{ij}$
and three scalars $\Phi$, $\Phi^\prime$, $C^{*}_{ijkl}$. The internal
metric components $G_{ab}$ yield ten scalars. (Note, that constant
rescalings of the coordinates would change the range of those
coordinates, and hence are non-trivial deformations equivalent to a
constant change of the corresponding metric components.) From the
metric deformations we obtain 48 less scalars than in the $K3$
compactification. It remains to take into account the twoforms on
$T^4/{\mathbb Z}_2$: $B_{ab}$, $B^\prime _{ab}$ and $C^{*}_{ijab}$. We
obtain 12 massless scalars from $B_{ab}$ and $B^\prime _{ab}$,
together. On the $K3$ there were $32$ more massless scalars coming
from this sector. The $C^{*}_{ijab}$ combine into three selfdual and three
anti-selfdual twoforms. In the $K3$ compactification we obtained 16
more anti-selfdual two-forms.

The $T^4/{\mathbb Z}_2$ spectrum we computed here, would correspond to
the one which we had obtained in a field theory compactification. It
has 80 massless scalars and 16 anti-selfdual twoforms less than the
$K3$ compactified theory. In field theory, the spectrum jumps when we
take the singular orbifold limit in the family of $K3$s. From the
above construction it is obvious that we counted only untwisted states
from a string perspective. Indeed, the missing 80 scalars and 16
anti-selfdual twoforms are exactly what we obtained from the twisted
sector in the previous section. In string theory the spectrum of the
compactified theory does not feel the singular nature of the orbifold
limit. All that happens is that some part of the spectrum is localized
to 
the orbifold fixed planes. This localization appears in internal space
and is
not visible at experiments which cannot resolve the distances of the size
of the compact manifold. The energy scale of such experiments depends
on the type of interactions fields propagating into the compact
directions carry. For purely gravitational interactions it needs to be
much higher than e.g.\ for electro-magnetic interactions. We will come
back to this later.

To summarize we recall that string theory can be compactified on
singular manifolds. The moduli spaces of such compactifications can be
connected to compactifications on smooth manifolds. There are massless
states which are localized at singularities of the compact
manifold. These are the twisted sector states. They are of truly
stringy origin. 
%%%%%%%%%%%%%%%%%%%%%%%%%%%

\section{D-branes}\label{D-branes}

In this section we will present another kind of extended objects
resulting from string theory --- the D-branes. They are different from
the previously studied orbifold planes. D-branes can exist also
in uncompactified theories. They are dynamical objects, i.e.\ they
interact with each other and can move independent of the size of some
compact space. (The orbifold planes could move only if we changed the
size or shape of the compact manifold.) When we discussed the
fundamental string we did not consider the possibility of open
strings. We will catch up on that in the following. 

\subsection{Open strings}
\setcounter{equation}{0}

\subsubsection{Boundary conditions}\label{bces}

We recall the action of the superstring
\begin{equation}\label{superschtring}
S = \frac{1}{2\pi \alpha^\prime}\int d\sigma^+ d\sigma^- \, \left(
\partial_- X^\mu\partial_+X_\mu + \frac{i}{2} \psi_+
^\mu\partial_-\psi_{+\mu} +\frac{i}{2}\psi^\mu _-
\partial_+\psi_{-\mu}\right) .
\end{equation}
Now, we view the values $\sigma = 0,\pi$ as true boundaries of the
string worldsheet. Varying (\ref{superschtring}) with respect to
$X^\mu$ gives apart from the equations of motion (which are identical
to closed strings) boundary terms which should vanish separately,
\begin{equation}
\left. \delta X^\mu \partial_\sigma X_\mu\right| _{\sigma = 0} ^\pi =
0.
\end{equation}
For the closed string we have solved this equation by relating the
values at $\sigma =0$ with the ones at $\sigma = \pi$. This procedure
was local because we took the string to join to a closed string at
$\sigma =\pi$. Now, we proceed differently by not correlating the two
ends of the string, i.e.
\begin{equation}
\left. \delta X^\mu \partial_\sigma X_\mu\right| _{\sigma = 0} =
\left. \delta X^\mu \partial_\sigma X_\mu\right| _{\sigma = \pi} =0.
\end{equation}
Let us focus on the boundary at $\sigma =0$. We have two choices to
satisfy the boundary condition. If ---for $i= 0, \ldots ,
p$--- we allow for 
free varying ends 
($\delta X^i$ arbitrary at the boundary) we obtain Neumann boundary
conditions at those ends\footnote{Here, we use $i$ to label Neumann
  directions. After 
  fixing the light cone gauge we take $i=2,\ldots ,p$.}
\begin{equation}
\partial_\sigma X^i = 0 \;\;\; ,\;\;\; i =0,\ldots , p .
\end{equation}
For the remaining $d-p-1$ coordinates $X^a$ we choose to freeze the
end of the string --- the 
variation vanishes at the boundary. Hence, in those directions the
end of the string is confined to some constant position. The resulting
boundary conditions are Dirichlet conditions ($c^a$ is a constant vector),
\begin{equation}\label{dirichletcondition}
X^a = c^a \;\;\; ,\;\;\; a = p+1,\ldots , d.
\end{equation}
The end of the open string defines a surface which extends along
$p+1$ dimensions and is located in $d-p-1$ dimensions. This object is
called D-brane, where the ``D'' refers to the Dirichlet boundary
condition specifying its position. If we choose identical boundary
conditions for the other end of
the open string (at $\sigma =\pi$), we
describe an open string starting and ending on the same D-brane. For
different 
boundary conditions the open string stretches between two different
D-branes. The Neumann conditions imply that no momentum can flow out of
the ends of the open string. In the Dirichlet directions momentum can
leave the string through its end --- it is absorbed by the D-brane. The
target space Lorentz group is broken to $SO(p,1)$. 

Varying the action with respect to the worldsheet fermions results in
the same 
equations of motions as in the closed string case and in
the boundary conditions
\begin{equation}\label{fermionicrand}
\left. \left( -\psi_{+\mu}\delta \psi^\mu _+ +\psi_{-\mu}\delta
    \psi^\mu _-\right) \right|_{\sigma =0}^\pi = 0.
\end{equation}
In the closed string case we have solved this by assigning either
periodic or anti-periodic boundary conditions to the worldsheet
fermions. Since now the ends of the string are separated in the target
space this would imply some non-locality. Therefore, we impose the
boundary conditions (\ref{fermionicrand}) at each end separately
\begin{equation}\label{fermionicrander}
\left. \left( -\psi_{+\mu}\delta \psi^\mu _+ +\psi_{-\mu}\delta
    \psi^\mu _-\right) \right|_{\sigma =0} = 
\left. \left( -\psi_{+\mu}\delta \psi^\mu _+ +\psi_{-\mu}\delta
    \psi^\mu _-\right) \right|_{\sigma =\pi} =
0.
\end{equation}
Let us focus again on the boundary at $\sigma = 0$. We can solve
(\ref{fermionicrander}) by one of the options
\begin{equation}
\psi ^\mu _+ = \pm \psi^\mu _-\;\;\; \mbox{at} \;\;\; \sigma =0 .
\end{equation}
However, there is a correlation with the bosonic boundary conditions
via worldsheet supersymmetry. To be specific, we choose the plus sign
for Neumann conditions
\begin{equation}\label{fermneu}
\psi ^i _+ =  \psi^i _- \;\;\; \mbox{at} \;\;\; \sigma =0.
\end{equation}
The supersymmetry transformations (in particular (\ref{suvarm}) and
(\ref{suvarp})) should not change this boundary condition. Since
$\partial_\tau X^i$ is not specified by the boundary conditions this
yields  
\begin{equation}\label{unbrokenpart}
\epsilon_+ = -\epsilon_- \;\;\; \mbox{at}\;\;\; \sigma = 0,
\end{equation}
which implies that due to the boundary (at least) half of the
worldsheet supersymmetry is broken. 
(If we had started with $(1,0)$ worldsheet supersymmetry ---as we did
for the heterotic string--- the boundary would break worldsheet
supersymmetry completely.)
In order to ensure that not all of
the supersymmetry is broken we have to choose the opposite (compared to
(\ref{fermneu})) 
boundary conditions for worldsheet fermions in Dirichlet
directions
\begin{equation}\label{fermdi}
\psi ^a _+ = - \psi^a _- \;\;\; \mbox{at}\;\;\; \sigma =0.
\end{equation}
We could also interchange the fermionic boundary conditions in Dirichlet and
Neumann directions. Then another worldsheet supersymmetry would
survive. There is no physical difference between the two
choices. Nevertheless  it
is important, that we take the boundary conditions in the
Neumann directions to be ``opposite'' to the ones in Dirichlet
directions. One may also check that the open string action is
invariant under the worldsheet supersymmetries (\ref{suvarb}),
(\ref{suvarm}) and (\ref{suvarp}) provided that the worldsheet
fermions satisfy the boundary conditions (\ref{fermneu}),
(\ref{fermdi}) and (\ref{unbrokenpart}) is fulfilled. (Partial
integration introduces boundary integrals which vanish if these
additional constraints hold.) Recall also that the functional form of
the worldsheet supersymmetry parameter is restricted by the chirality
conditions (\ref{losusy}).

In the following we are going to discuss the boundary conditions at
the other end of the open string at $\sigma = \pi$. Going back to the
closed string we deduce from (\ref{suvarm}) that for anti-periodic
supersymmetry parameter $\epsilon_+$ the fermions are anti-periodic
for periodic bosons and vice versa. This means that anti-periodic
$\epsilon_+$ belongs to the NS sector and periodic ones to the R
sector. From the discussion of the
boundary conditions at $\sigma =0$ we infer that the supersymmetry
parameter has to satisfy one of the following conditions,
\begin{equation}\label{ghjk}
\epsilon_+ = \pm \epsilon_- \;\;\; \mbox{at}\;\;\; \sigma = \pi .
\end{equation}
In order to relate this to something like periodicity or
anti-periodicity we perform the so called doubling trick. This means
that we define a function $\varepsilon$ on the interval $0\leq
\sigma < 2\pi$. This is done in the following way (we indicate only
the $\sigma$ dependence),
\begin{equation}\label{doubled}
\varepsilon = \left\{ \begin{array}{ccc}
\epsilon_+\left( \sigma\right) & , & 0\leq \sigma < \pi \\
\pm \epsilon_-\left( 2\pi -\sigma\right) & ,& \pi \leq \sigma < 2\pi
\end{array} \right. . 
\end{equation}
The sign in the second line of (\ref{doubled}) is correlated to the
sign in (\ref{ghjk}) by the requirement of continuity at $\sigma
=\pi$.
Hence, $\varepsilon$ is (anti)-periodic for the lower (upper) sign in
(\ref{ghjk}) (taking into account the sign in
(\ref{unbrokenpart})). 

Now, let us perform this doubling trick also on the worldsheet bosons
and fermions. For the bosons it is useful to rewrite the boundary
conditions. Dirichlet boundary conditions mean that
\begin{equation}
\partial_+ X^a = -\partial_- X^a \;\;\; \mbox{at}\;\;\; \sigma^+
-\sigma^- = 0. 
\end{equation}
Neumann conditions can be written as
\begin{equation}
\partial_+ X^i =\partial_- X^i \;\;\; \mbox{at} \;\;\; \sigma^+
-\sigma^- =0.   
\end{equation}
The next step is to specify the boundary conditions at $\sigma
=\pi$. After having done this, one can perform the doubling trick,
i.e.\ define a boson $\partial X^\mu$ on the interval $0\leq \sigma <2\pi$
analogously to the definition of $\varepsilon$ in (\ref{doubled}) where
$\partial_{\pm} X$ take the role of $\epsilon_{\pm}$. As the reader
can easily verify, the outcome is that $\partial X^\mu$ is periodic
whenever we have chosen the same type of boundary conditions at the
two ends of 
the string in the $x^\mu$ direction. The corresponding open string
sectors are called DD (NN) according to the choice of Dirichlet
(Neumann) boundary conditions at the two ends. For an opposite choice
of boundary conditions (ND or DN strings) $\partial X^\mu$ will turn
out to be anti-periodic. In analogy to the closed string we call the
sector with periodic $\varepsilon$ R sector and the one with
anti-periodic $\varepsilon$ NS sector. For DD or NN strings this
implies that in the NS sector we take the boundary conditions at
$\sigma =\pi$ to be
\begin{equation}
\psi_+ ^i = -\psi_- ^i \;\;\; , \psi^a _+ = \psi^a _- \;\;\;
\mbox{at}\;\;\; \sigma =\pi .
\end{equation}
Defining a ``doubled'' worldsheet fermion $\Psi^\mu$ in analogy to
$\varepsilon$ (where the role of $\epsilon_\pm$ is taken over by the
$\psi_\pm ^\mu$) we find that for DD or NN strings $\Psi$ is
anti-periodic. In the R sector we take the boundary conditions
\begin{equation}
\psi_+ ^i = \psi_- ^i \;\;\; , \psi^a _+ = -\psi^a _- \;\;\;
\mbox{at}\;\;\; \sigma =\pi
\end{equation}
and obtain periodic boundary conditions. In the above we have used
that, for example, in the R-sector periodicity of $\varepsilon$ implies
that the boundary conditions of $\epsilon_\pm$ at $\sigma =\pi$ are
identical to the ones at $\sigma =0$. Plugging this back
into the supersymmetry transformations (\ref{suvarm}) and
(\ref{suvarp}) evaluated at $\sigma =\pi$ and taking into account the
boundary conditions for the bosons, we obtain the boundary conditions
of the worldsheet fermions at $\sigma =\pi$. This in turn determines
the periodicity of $\Psi^\mu$. Performing the same procedure for ND or DN
boundary conditions, one finds that $\Psi^\mu$ is periodic in the NS sector
and anti-periodic in the R sector whenever $x^\mu$ is a direction
with ND or DN boundary conditions. The ND or DN directions are
somewhat similar to the twisted sectors we met when discussing
${\mathbb Z}_2$ orbifolds.

At first, we will consider only the case of a single
D-brane. This means that we can have only DD or NN boundary conditions
depending on whether we are looking at a direction transverse or
longitudinal to the D-brane. Then $\partial X^\mu$ will always be
periodic and $\Psi^\mu$ (anti-)periodic in the (NS) R sector. 

\subsubsection{Quantization of the open string ending on a single
  D-brane}\label{openquant}

The quantization of the open string is very similar to the closed
superstring. In the following we will point out the differences.
At first, we solve the equations of motion again by taking a
superposition of left-moving and right-moving fields. For the bosons,
these are given in (\ref{leftsol}) and (\ref{rightsol}). The boundary
conditions relate this two solutions. 
(In addition, we need to replace $e^{-2in\sigma^\pm} \rightarrow
e^{-in\sigma^\pm}$.)\footnote{\label{length2}If we had chosen the open
  string half as long as 
the closed one we would not need this replacement.}
In Neumann directions, they
impose 
\begin{equation}
\alpha_n ^i = \tilde{\alpha}_n ^i .
\end{equation}
For Dirichlet directions, we obtain a similar relation and constraints
on the zero modes,
\begin{equation}
x^a = c^a \;\;\; ,\;\;\; p^a =0 \;\;\;,\;\;\; \alpha_n ^a =
-\tilde{\alpha}_n 
^a .
\end{equation}
The general solutions for the bosonic directions read 
\begin{eqnarray}
X^i & = & x^i + p^i\tau + i\sum_{n\not= 0}\frac{1}{n}\alpha_n ^i
e^{-in\tau} \cos 
n\sigma ,\label{neumannmode} \\   
X^a & = & c^a - \sum_{n\not= 0} \frac{1}{n} \alpha_n ^a e^{-in\tau}
\sin n\sigma .\label{dirichletmode}
\end{eqnarray}
The mode expansions for the NS sector fermions look as follows
\begin{eqnarray}
\psi_- ^i & = & \frac{1}{\sqrt{2}}\sum_{r \in {\mathbb Z} +\frac{1}{2}}
    b^i _r e^{-ir \sigma^-} ,\label{openfour}\\
\psi_+ ^i & = & \frac{1}{\sqrt{2}}\sum_{r \in {\mathbb Z}+\frac{1}{2}}
b^i _r e^{-ir\sigma^+} ,\\
\psi_- ^a & = & \frac{1}{\sqrt{2}}\sum_{r \in {\mathbb Z}+\frac{1}{2}}
b^a _r e^{-ir \sigma^-},\\
\psi_+ ^a & = & -\frac{1}{\sqrt{2}}\sum_{r \in {\mathbb
    Z}+\frac{1}{2}}
b^a _r e^{-r\sigma^+} .
\end{eqnarray}
For the R sector fermions, one obtains
\begin{eqnarray}
\psi_-^i & = & \sum_{n\in {\mathbb Z}}d_n ^i e^{-in\sigma^-},\\
\psi_+^i & = & \sum_{n\in {\mathbb Z}}d_n ^i e^{-in\sigma^+},\\
\psi_-^a & = & \sum_{n\in {\mathbb Z}}d_n ^a e^{-in\sigma^-},\\
\psi_+^a & = & -\sum_{n\in {\mathbb Z}}d_n ^a
e^{-in\sigma^+}.\label{openfourt} 
\end{eqnarray}
The next step is to eliminate two directions by performing the light
cone gauge. We take this to be the time-like (Neumann) direction and a
space-like Neumann direction, which we choose to be $x^1$.\footnote{We do not
consider an open string ending on a $D0$ brane, here. As in the case
of the closed
string, it is useful to take Lorentz invariance as a guiding principle
for a consistent quantization. For a $D0$ brane, the Lorentz group is
broken down to time reparameterizations which is too small for our
purposes. Later, we will see that we can obtain the $D0$ brane by
T-dualizing a higher dimensional D-brane.}
For the open string we have only an NS sector and an R sector. Since
the right movers are not independent from the left movers, the right
and left moving sectors do not decouple anymore. The constraints that
the expressions (\ref{enmopp}) -- (\ref{sucum}) vanish are not all
independent. The zero mode part of vanishing energy momentum tensor
again yields the mass shell condition (since the mode expansions
differ 
by factors of two, there is a difference of a factor of four as
compared to the closed string (\ref{massshelf})),
\begin{equation}\label{openmass}
M^2 = 2\left( N - a\right) ,
\end{equation}
where $a$ is the normal ordering constant and the number operator $N$ is
defined in (\ref{snumber}) for the NS sector and in
(\ref{rnumber}) for the R sector. In the NS sector the GSO projection
operator is as
in (\ref{gsoprojection}) with the second factor removed. The lowest
GSO invariant states in the NS sector are
\begin{equation}\label{openns}
b^i_{-\frac{1}{2}}\left| k\right> \;\;\; , \;\;\; b^a
_{-\frac{1}{2}}\left|k\right>, 
\end{equation}
where we have indicated again the momentum eigenvalue of the vacuum by
$k$. The first set of these states transforms in the vector
representation of $SO(p-1)$ -- the little group of the unbroken
Lorentz group. Hence, this state should be massless leading to the
consistency condition\footnote{In principle, we could combine the
  first states in (\ref{openns}) with one of the second states in
  order to form a massive vector representation as long as
  $p<d-1$. However, later we will see that the case $p=d-1$ is related
  by T-duality to the other cases. With this additional ingredient it
  follows that the states in (\ref{openns}) must be massless.}
\begin{equation}
a_{NS} =\frac{1}{2}.
\end{equation}
Like in the closed string, this translates into a condition on the number
of target-space dimensions
\begin{equation}
d=10.
\end{equation}
The first states in (\ref{openns}) (with label $i$) form a $U\left(
  1\right)$ gauge field. The states with label $a$ are scalars
  transforming in the adjoint of $U\left( 1\right)$ (here, this appears
  just as a pompous way of saying that they are neutral under $U(1)$,
  however, 
  later we will discuss non-abelian gauge groups where those fields
  are adjoints rather than singlets). Since the center of mass
  position of the open string is confined to be within the world
  volume of the D-brane, all the open string states correspond to
  target-space particles which are confined to live on the D-brane. 
The NS mass-spectrum is depicted in figure \ref{fig:openmass}.

%%%%%%%%%%%%%%%%%%%%%%%%%%%%%%%%%%%%
\begin{figure}
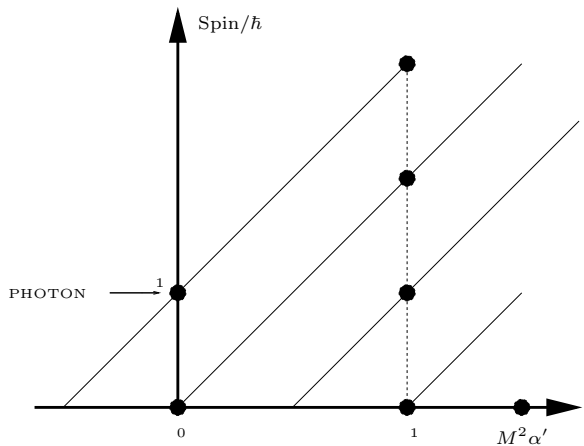
  
\begin{center}
\input openmass.pstex_t
\end{center}
\caption{NS mass spectrum of the open superstring}
\label{fig:openmass}
\end{figure}
%%%%%%%%%%%%%%%%%%%%%%%%%%%%%%%%%%%%

The construction of the R sector vacuum state goes along the same
lines as in the closed string. The ten dimensional Majorana spinor
decomposes into a couple of spinors with respect to the unbroken
Lorentz group $SO(p,1)$. We impose the GSO projection by
multiplying the states with one of the projection operators defined in
(\ref{rgso}). The sign is a matter of convention. The R-vacuum is
massless on its own. It leads to target space spinors providing all
fermionic degrees of freedom which are needed to obtain the maximal
rigid supersymmetry in $p+1$ dimensions.\footnote{The maximally
  possible amount of supersymmetry differs for rigid and local
  supersymmetry. In rigid supersymmetry, the highest occurring spin
  should not exceed one, whereas in locally supersymmetric field
  theories, spin two fields (the gravitons) are allowed. In $3+1$
  dimensions, the maximal supersymmetry is $N=4$ ($N=8$) for rigid
  (local) supersymmetry. From this one can deduce the maximally
  allowed amount of supersymmetry in higher dimensions by 
  viewing the $3+1$ dimensional theory as a toroidally compactified
  higher dimensional theory.}

In the following sections we will investigate systems with more than
one D-brane. This will lead to non-abelian field theories on a stack
of D-branes. But before doing so, we will briefly discuss the possible
D-brane setups which are in agreement with supersymmetry.

\subsubsection{Number of ND directions and GSO
  projection}\label{numberofnd} 

At first, consider the case that we have an open string with an odd
number of ND directions. Thus, we will have an odd number of
directions where the worldsheet fermions have zero-modes. For example
in the R sector, the zero-modes form a Clifford algebra in $p+1 =
odd$ dimensions. The representation of this algebra by the R ground
state will be irreducible (there is no notion of chirality in odd
dimensions). Therefore, we cannot perform the GSO projection on those
states. The theory will not lead to target-space
supersymmetry.

Let us now discuss the case of an even number of ND directions,
taken to be $8-2n$. Then the GSO projection operator on the R sector
ground state will be of the form
\begin{equation}\label{ndgso}
P_{GSO} =  1\pm 2^n\, d_0 ^2 \ldots d_0 ^{2n+1}.  
\end{equation}
Using some algebra this can be written as
\begin{equation}
P_{GSO} = 1\pm e^{i\pi\left(J_{23}+\ldots +J_{2n,2n+1}\right)},
\end{equation}
where the $J_{kl}$ are the generators of rotations in the $kl$ plane
\begin{equation}
J_{kl} = -\frac{i}{2} \left[ d_0 ^k ,d_0 ^l\right] .
\end{equation}
The eigenvalue of the ND Ramond groundstate under a $180^\circ$ rotation
in one plane is $\pm i$. Thus, the eigenvalues of the R groundstate
$\left| R\right>$ 
under $P_{GSO}$ will be
\begin{equation}
P_{GSO}\left| R\right> = \left( 1\pm i^n\right)\left| R  \right> .
\end{equation}
From this we deduce that the GSO projection is possible only if the
number of ND directions is an integer multiple of four. This means for
example that, if a lower dimensional D-brane lives inside the worldvolume
of a higher dimensional D-brane, the higher dimensional D-brane has to
extend in four or eight more directions. We could have deduced this
result faster by noting that (\ref{ndgso}) defines a projection
operator only if $n$ is a multiple of four since otherwise the second
term in (\ref{ndgso}) squares to $-1$.

\subsubsection{Multiple parallel D-branes -- Chan Paton factors}

In this section we will discuss sets of parallel
Dp-branes.\footnote{Recall that the worldvolume of a Dp-brane has p
  space like and one time like dimension.} First,
let us have a look at two parallel Dp-branes which are separated by a
vector $\delta c^a$ in the transverse space. (Later we will see that
the distance between parallel D-branes is a modulus, i.e.\ any value
is consistent.) From strings ending with both ends on the same
D-brane we obtain the same spectrum as discussed in the section
\ref{openquant}. In particular, we obtain a $U(1) \times U(1)$ gauge
symmetry where the corresponding gauge fields live on the first
brane for the first $U(1)$ and on the second brane for the second
$U(1)$ factor.

In addition, we have strings stretching between the two branes. There
are two such strings with opposite orientations. As compared to section
\ref{openquant}, only the mode expansion for the bosons in Dirichlet
directions is modified. The string starting on the brane at $c^a$ and
ending on the brane at $c^a + \delta c^a$ has the mode expansion
\begin{equation}
X^a  =  c^a +\frac{\delta c^a}{\pi}\sigma - \sum_{n\not= 0} \frac{1}{n}
\alpha_n ^a e^{-in\tau} 
\sin n\sigma .
\end{equation}
The string with the opposite orientation is obtained by replacing
$\sigma \rightarrow \pi -\sigma$. We rewrite the term with $\delta
c^a$ in a suggestive way
\begin{equation}
\delta c^a\sigma = \frac{1}{2}\delta c^a \left(\sigma^+
  -\sigma^-\right)
\end{equation}
and compare with the expressions (\ref{compmodl}) and
(\ref{compmoder}). The finite distance between the D-branes enters the
mode expansion in a very similar way as the winding number in the
toroidally compactified closed string does. This is also intuitively
expected -- the winding closed string is stretched around a compact
dimension. As a finite winding number also the finite distance
contributes to the mass of the stretched string state, it results in a
shift of
\begin{equation}\label{massshift-open}
\delta M^2 = \frac{\left(\delta c^a\right)^2}{\pi ^2} . 
\end{equation}
The strings stretching between the branes transform under $U(1)\times
U(1)$ with the charges $\left( 1,-1\right)$ and $\left( -1,1\right)$
depending on the orientation. We will see below that these charge
assignments are necessary for consistency. Pictorially, they are
obtained by the rule that a string starting at a brane has charge $+1$
with respect to the $U(1)$ living on that brane whereas it has charge
$-1$ if it ends on the brane. (The photon which starts and ends on the
same brane has net charge zero.) The $U(1)\times U(1)$ can be also
rearranged into a diagonal and a second $U(1)$ such that all states are
neutral under the diagonal $U(1)$.  

The lightest GSO-even states in the NS sector of the stretched
string are
\begin{eqnarray}
\psi^i _{-\frac{1}{2}} \left| 12\right> & , &
\psi^i _{-\frac{1}{2}}\left| 21\right> ,\label{strelo} \\ 
\psi^a _{-\frac{1}{2}}\left| 12\right> & , & \psi^a
_{-\frac{1}{2}}\left| 21\right> \label{stretra}
\end{eqnarray}
where $\left| 12\right>$ and $\left| 21\right>$ denote the
NS vacua for strings stretched between the two D-branes and we have
dropped the zero mode momentum eigenvalues $k$ in the notation. In
this 
sector, the lightest states form a vector and $d-(p+2)$ scalars. (Note
that, in the light cone gauge, we have to combine one of the
transverse excitations (\ref{stretra}) with the longitudinals
(\ref{strelo})  in order to obtain a massive vector.) The R sector
states provide the fermions needed to fill up supermultiplets. (The
amount of supersymmetry is the same as in the single brane case.)

Now, take the inter-brane distance to zero. We obtain two massless
vectors and $2\left( d -p -1\right)$ massless scalars. Together with the
massless  fields coming from strings ending on identical branes, the
vectors combine into a $U(2)$ gauge field, and the scalars combine
into $d- p -1$ scalars transforming in the adjoint of $U(2)$. (One can
split $U(2)$ into a diagonal $U(1)$ times an $SU(2)$. All 
fields are $SU(2)$ adjoints and neutral under $U(1)$.)
Moving the D-branes apart from each other can be viewed as a Higgs
mechanism from the target space perspective. The amount of
supersymmetry leads to flat directions for the scalars in the adjoint
of $U(2)$. This means that a scalar can have some non zero vev
breaking $U(2)$ to $U(1)\times U(1)$. With the given amount of
supersymmetry, all massless fields transform in the same representation
of the gauge group as the vector bosons ({\it viz.} in the adjoint). 
Therefore, 
the Higgs mechanism can only work for non-abelian gauge groups. Our
charge assignments of the open strings stretched between two different
D-branes thus lead to a consistent picture.

An economic way of studying systems with $N$ parallel D-branes is 
to replace all the different sectors corresponding to the
possibilities of strings stretching among the $N$ D-branes by one
matrix valued state
\begin{equation}
\left| \cdot \right> \rightarrow \left| \cdot ,ij\right>\lambda_{ji}
\end{equation}
where $\lambda $ is an $N\times N$ matrix. The component
$\lambda_{ji}$ corresponds to a string stretching between the $i$th
and the $j$th brane. 
The matrix $\lambda$ is called Chan-Paton factor.
Consider again the case where all the $N$
D-branes are separated in the transverse space. 
For the lightest NS sector states
the diagonal elements $\lambda_{ii}$ 
are $N$ $U(1)$ gauge fields and $d-p-2$ scalars. They are neutral under
$U(1)^N$, i.e.\ 
\begin{equation}
\lambda_{ii} = \lambda^\dagger _{ii}.
\end{equation}
The off-diagonal elements correspond to massive vectors and
scalars. Open string sectors with opposite orientation have opposite
charges under $U(1)^N$, i.e.\
\begin{equation}
\lambda_{ij} = \lambda^\dagger _{ji} .
\end{equation}
The Chan-Paton factor is a unitary $N\times N$ matrix. Maximal gauge
symmetry is obtained when all $N$ D-branes sit at the same point in
the transverse space. The diagonal and off-diagonal elements of
$\lambda$ combine and give rise to a $U(N)$ vector multiplet.

\subsection{D-brane interactions}\label{dbraneinteractions}
\setcounter{equation}{0}

Already at an intuitive level, one can deduce that D-branes
interact. This comes  about as follows. The two ends of an open string
ending on the same D-brane can join to form a closed string. The closed
string is no longer bound to live on the D-brane, it can escape into
the bulk of the target space. In particular, it may reach another
D-brane by which it is absorbed. The absorption process is inverse to
the emission process. The closed string hits the D-brane where it can
split into an open string which is constrained to live on the second
D-brane. D-branes talk to each other by exchanging closed strings. 
In figure \ref{treecyl} we have drawn such a process. In order to make
contact to conventions of the standard reviews on D-brane physics we
take the closed string twice as long as the open string (see also the
footnote \ref{length2}). This implies that in the closed string mode
expansions we replace $e^{-2in\sigma^\pm}\rightarrow
e^{-in\sigma^\pm}$.

%%%%%%%%%%%%%%%%%%%%%%%%%%%%%%%%%%%%
\begin{figure}
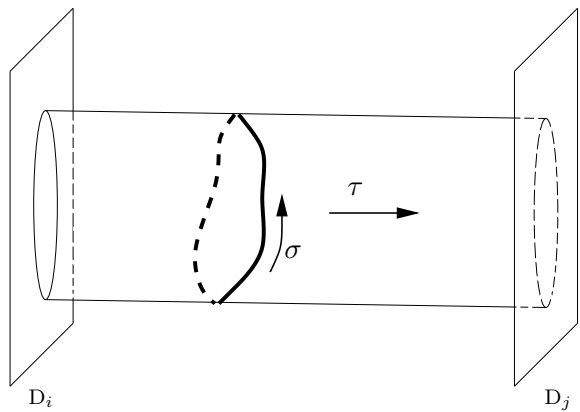
  
\begin{center}
\input treecyl.pstex_t
\end{center}
\caption{D-brane D$_i$ and D-brane D$_j$ talking to each other by
  exchanging a closed string.}
\label{treecyl}
\end{figure}
%%%%%%%%%%%%%%%%%%%%%%%%%%%%%%%%%%%%

We will compute
the process depicted in figure \ref{treecyl} in Euclidean worldsheet
signature. The range of the worldsheet coordinates is
\begin{equation}
0 \leq \sigma < 2\pi \;\;\; ,\;\;\; 0 \leq \tau < 2\pi l .
\end{equation}
The Euclidean worldtime $\tau $ is taken to be compactified on a
a circle of radius $2l$. The worldtime taken by a string to get from
one brane to the other one is $2\pi l$ -- this process can be
periodically continued such that one period lasts $4\pi l$. (The
factor of $2\pi $ is a matter of convention. It is introduced because
compact directions are usually specified by the radius of the
compactification circle rather than its circumference.)  
Note also, that $l$ has nothing
to do with the distance of the D-branes. The distance in target space
will appear later and will be denoted by $y$.
  
Since we have defined the D-branes in terms of open strings it will be 
useful to compute also the D-brane interactions in terms of open
strings. To this end, we perform a so called worldsheet duality
transformation, 
i.e.\ we interchange $\sigma$ with $\tau$. The resulting picture is an
open string one-loop vacuum amplitude. It is described by the annulus
diagram drawn in figure \ref{loopan}.  

%%%%%%%%%%%%%%%%%%%%%%%%%%%%%%%%%%%%
\begin{figure}
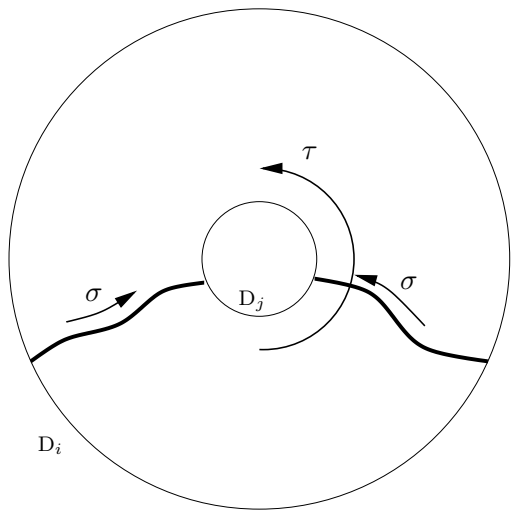
  
\begin{center}
\input loopan.pstex_t
\end{center}
\caption{D-brane D$_i$ and D-brane D$_j$ talking to each other by
  a pair of virtual open strings stretching between them.}
\label{loopan}
\end{figure}
%%%%%%%%%%%%%%%%%%%%%%%%%%%%%%%%%%%%

The parameter ranges for the open string are 
\begin{equation}\label{nonstarange}
0 \leq \tau < 2\pi \;\;\; ,\;\;\; 0 \leq \sigma < 2\pi l .
\end{equation}

The periodicity of closed string worldsheet fermions is related to the
behavior of open strings under shifts in $\tau$ by $2\pi$. The diagram
\ref{loopan} corresponds to a vacuum amplitude and thus to a trace in
the open string sector. This trace is actually a supertrace with
respect to worldsheet (and target space) supersymmetry. 
The additional sign in the trace over worldsheet fermions is imposed
by specifying the boundary condition 
\begin{equation}\label{peritra}
\psi_{\pm}^\mu\left( \tau + 2\pi ,\sigma\right) =\left( -\right)^F
\psi^\mu _{\pm}\left( \tau ,\sigma\right) .
\end{equation}
Thus, a $\left( -\right)^F$ insertion (canceling the $\left( -\right)^F$ in
(\ref{peritra})) corresponds to
closed string RR sector exchange whereas no $\left( -\right)^F$
insertion yields the closed string NSNS sector exchange.
From the open string perspective there is no exchange of
NSR or RNS sector closed strings between the D-branes. 
In the picture
\ref{treecyl}, the D-brane appears as a boundary state of the closed
string.\footnote{It will turn out that the closed strings which are
  exchanged in figure \ref{treecyl} are type II superstrings.} 
This boundary state is a superposition of an NSNS sector state
and an RR sector state. There are no NSR or RNS sector
contributions.  This can be explained by the
fact that the D-brane is a 
target space boson. It is specified by the target space vector $c^a$
in (\ref{dirichletcondition}) and hence transforms as a vector and not
as a spinor under rotations in the space transverse to the brane. We
will not present the details of the boundary state formalism here, and
recommend the review \cite{Gaberdiel:2000jr} instead. To be slightly
more 
specific let us just present the defining equation for a boundary
state in closed string theory (as usual we label the Neumann
directions by $i= 0, \ldots ,p$ and the Dirichlet directions by $a =
p+1, \ldots , 9$) 
\begin{equation}\label{boundarystate}
\partial_\tau X^i \left| \mbox{D-brane}\right> = \left( X^a
  -c^a\right)\left| \mbox{D-brane}\right> = 0.
\end{equation}
This relates the right moving and left moving bosonic excitations the
boundary state can carry. 
Applying the worldsheet supersymmetry transformations (\ref{suvarm})
and (\ref{suvarp}) on (\ref{boundarystate}) and requiring that there
is a combination of the two supersymmetries which annihilates the
boundary state tells us that the boundary state should have the same
number of right moving and left moving fermionic excitations. Also,
when $\epsilon_+$ is taken to be (anti)-periodic then $\epsilon_-$
should have the same periodicity. The boundary state cannot be
excited by NS fermions in, say, the right moving sector and R fermions
in the left moving sector. It has only an NSNS and an RR sector.

Instead of the non-standard range for the open string worldsheet
coordinates we would like to have the standard range
\begin{equation}
0 \leq \tau < 2\pi t \;\;\; , \;\;\; 0\leq \sigma < \pi .
\end{equation}
In order to achieve this we redefine $ \tau \rightarrow
\tau t$ and $\sigma \rightarrow \frac{\sigma}{2l}$.
Under this redefinition, the Hamiltonian (which is obtained by
integrating the $\tau\tau$ component of the energy momentum tensor
over $\sigma$)
transforms according to
\begin{equation}
H \rightarrow 2l t^2 H .
\end{equation}
Further, we want the time evolution operator when going once
around the annulus to transform as ($2\pi t$ should be identified with
the worldsheet time it takes the open string to travel around the
annulus once)
\begin{equation}
e^{-2\pi H} \rightarrow e^{-2\pi t H}.
\end{equation}
This yields the relation
\begin{equation}\label{dictionary-cyl}
tl = \frac{1}{2}.
\end{equation}
The annulus vacuum amplitude in figure \ref{loopan} yields the vacuum
energy of an open string starting on the D-brane D$_i$ and ending on
the D-brane D$_j$. This can be expressed as
\begin{eqnarray}
\lefteqn{-\frac{1}{2}\log \det H  =  -\frac{1}{2}\mbox{tr}\log H
  =  \frac{1}{2}\lim_{\epsilon\to 0} \mbox{tr}\frac{d
 H^{-\epsilon} }{d\epsilon}=} \nonumber\\ 
& & \mbox{tr}\left(\lim_{\epsilon \to
 0}\frac{d}{d\epsilon}\left( \epsilon \int_0 ^\infty \frac{dt}{2t} t^\epsilon
 e^{-2\pi t H}\right)+\frac{1}{2}\log 2\pi -\frac{1}{2}
  \Gamma^\prime\left( 1\right)\right) . 
\end{eqnarray}
At a formal level this expression is correct. However, the next step is
to take the limit of $\epsilon \to 0$ before performing the integral
over $t$. This would be allowed only if the integral were
converging. This is not the case in most of the applications (for
example, the integral diverges if $H$ is just a number and no trace is
taken). But the
error done is some unknown additive constant contribution which is not
of interest for us\footnote{In our case the trace is actually a
  supertrace which vanishes when taken over a constant. However, since
  the corresponding series does not converge absolutely the result
  depends on the ordering in which we take the trace.}. Together 
with this unknown constant we also drop 
the $\frac{1}{2}\left(\log 2\pi - \Gamma^\prime \left(
    1\right)\right)$ and obtain for the 
amplitude in figure \ref{loopan} (reinstalling $\alpha^\prime$)
\begin{equation}\label{anulusloop}
\int_0 ^\infty \frac{dt}{2t} \,\mbox{Str}\; e^{-2\pi \alpha^\prime tH} 
\end{equation}
(here we have replaced the trace by a supertrace. It refers to target
space supersymmetry, i.e.\ the 
trace receives an additional minus sign for target space spinors. The
integral over $t$ is usually regulated by a UV cutoff near $t=0$.)
The expression (\ref{anulusloop}) has also an intuitive
interpretation. The supertrace describes a process where a pair of
open strings is created from the vacuum, then propagates for a time
$2\pi t$ and annihilates. This corresponds to the diagram drawn in
figure \ref{loopan}. Further, we integrate over all possible moduli
$t$ of the annulus with the measure $\frac{dt}{2t}$. The Hamiltonian is
$p^2 + M^2$ which can be expressed by use of (\ref{openmass}) and
(\ref{massshift-open}) as follows
\begin{equation}
H = p^2 + \frac{y^2}{\pi^2} +2\left( N - a\right),
\end{equation} 
where $y$ is the distance between the two D-branes, and $a$ is the
normal ordering constant ($\frac{1}{24}$ per bosonic 
direction, $\frac{1}{48}$ per fermionic direction in the NS sector,
and $-\frac{1}{24}$ per fermionic direction in the R sector). 
Recalling that in this expression we have set $\alpha^\prime =
\frac{1}{2}$ gives (just multiply with appropriate powers of
$2\alpha^\prime$ to get the mass dimension right)
\begin{equation}
\alpha^\prime H = \alpha^\prime p^2 +\frac{y^2}{4\pi^2 \alpha^\prime}
+\left( N-a\right) .
\end{equation}
It is useful to split (\ref{anulusloop}) into several contributions
\begin{eqnarray}
\lefteqn{\int \frac{dt}{2t} \mbox{Str}\, e^{-2\pi\alpha^\prime tH}
  =}\nonumber 
\\ & &  \int
\,\frac{dt}{2t}\,\mbox{tr}_{\mbox{\tiny
      ZERO}\atop\mbox{\tiny  MODES}}\left( e^{-2\pi
    t \alpha^\prime H_0}\right)
\mbox{tr}_{\mbox{\tiny BOSONS}}\left(e^{-2\pi t
    H_B}\right)\nonumber \\ 
& & \left( \mbox{tr}^{GSO} _{\mbox{\tiny NS}\atop
      \mbox{\tiny FERMIONS}} \left( 
      e^{-2\pi t H_{NS}}\right) - \mbox{tr}^{GSO} _{\mbox{\tiny R}\atop
        \mbox{\tiny FERMIONS}} \left( e^{-2\pi tH_{R}}\right)\right).
\end{eqnarray}
We have split the Hamiltonian into 
\begin{equation}\label{openhama}
\alpha^\prime H =\alpha^\prime  H_0 + H_B + H_{NS/R},
\end{equation}
with
\begin{eqnarray}
H_0 & = & p^2 +\frac{y^2}{4\alpha^{\prime 2}\pi ^2},\label{openham0} \\ 
H_B & = & \sum_{i=1} ^8 \left( \sum_{n=1}^\infty \alpha_{-n}^i
  \alpha_{n}^i -\frac{1}{24}\right) ,\label{bosham}\\
H_{NS} & = & \sum_{i=1} ^8 \left( \sum_{r=\frac{1}{2}}^\infty r
  b_{-r}^i b^i _{r} -\frac{1}{48}\right) , \\
H_{R} & = & \sum_{i=1}^8\left(\sum_{n=1}^\infty n d_{-n} ^i d_n ^i
+\frac{1}{24}\right) \label{openhame}
.
\end{eqnarray}
An additional minus sign in the R sector contribution is due to the
fact that we take the supertrace with respect to space time
supersymmetry (R-sector states are space time fermions). The
superscript GSO indicates that the trace is taken over GSO even
states. We will clarify this point later.  

The trace over the zero modes is
\begin{equation}\label{openzerotr}
\mbox{tr}_{\mbox{\tiny
      ZERO}\atop\mbox{\tiny  MODES}} = 2 V_{p+1} \int \frac{d^{p+1}
      k}{\left( 2\pi\right)^{p+1}}\,
      e^{-2\pi t\alpha^\prime  k^2 -\frac{t y^2}{2\pi \alpha^\prime}
      } = 2V_{p+1}\,\left( 8\pi^2 \alpha^\prime
      t\right)^{-\frac{p+1}{2}}\; 
      e^{-\frac{t y^2}{2\pi\alpha^\prime}}  ,
\end{equation}
where the factor of two counts the possible orientations of the open
string traveling through the annulus. The factor $V_{p+1}$ denotes
formally the worldvolume of the parallel $D_p$ branes. It arises due
to the normalization of states with continuous momentum ($\langle
p|p\rangle = \delta^{(p+1)}\left( 0\right) =
V_{p+1}/\left(2\pi\right)^{p+1}$).
To express oscillator traces, it is useful to define the following set
of functions\footnote{These are related to the Jacobi theta functions,
  which are also often used in the literature.}
\begin{eqnarray}
f_1\left( q\right) = q^{\frac{1}{12}}\prod_{n=1} ^\infty
  \left( 1 - 
  q^{2n}\right) & , & f_2 \left(q\right) =
  q^{\frac{1}{12}}\sqrt{2}\prod_{n=1} ^\infty \left( 1 + q^{2n}\right),
  \nonumber \\
f_3\left( q\right) = q^{-\frac{1}{24}}\prod_{n=1} ^\infty \left(1 + q^{2n
  -1}\right) & , & f_4\left( q\right) = q^{-\frac{1}{24}}\prod_{n=1}
  ^\infty \left( 1 - q^{2n -1}\right) .\label{thetafunctions}
\end{eqnarray}
These functions satisfy the identity
\begin{equation}\label{abstruse}
f_3 ^8\left( q\right) = f_2 ^8\left( q\right) + f_4 ^8 \left( q
\right) .
\end{equation}
In order to translate the open string calculation back to the closed string
process (figure \ref{treecyl}) we will make use of the modular
transformation properties,
\begin{equation}\label{modulartra}
f_1\left( e^{-\frac{\pi}{s}}\right) = \sqrt{s}
    f_1\left( e^{-\pi 
    s}\right), \;\; f_3\left( e^{-\frac{\pi}{s}}\right) = f_3\left(
    e^{-\pi s}\right),\;\; f_2\left( e^{-\frac{\pi}{s}}\right) = f_4
    \left( e^{-\pi s}\right) .
\end{equation}
Next, we are going to compute the trace over the worldsheet
bosons. The sum over the coordinate label $i$ in (\ref{bosham}) can be
written in front of 
the exponential as a product. Nothing depends explicitly on the
direction $i$, therefore this gives a power of eight to the result for
a single bosonic direction. The second sum can be decomposed into a level
part and an occupation number part, giving the result
\begin{equation}\label{numberpart}
\mbox{tr}_{\mbox{\tiny BOSONS}}\; e^{-2\pi t H_B} = \left( e^{\frac{\pi
  t}{12}}
  \prod_{l=1}^\infty \sum_{k=0}^\infty e^{-2\pi t l k}\right)^8  .
\end{equation}
Here, $l$ denotes the level of a creator $\alpha_{-l}$ acting on the
ground state and $k$ is the occupation
number (the number of times this creation operator acts). 
The sum over $k$ is just a geometric series, and we obtain the result
\begin{equation}
 \mbox{tr}_{\mbox{\tiny BOSONS}}\; e^{-2\pi t H_B} = \frac{1}{f_1
 ^8\left( e^{-\pi t}\right)}.
\end{equation}

The calculation of the traces over the fermionic sectors is
similar. Let us just point out the differences. 
First of all, we have to take the trace only over GSO even
states. This is done 
by inserting the GSO projection operator into the trace and summing
over all states
\begin{equation}
\mbox{tr}^{GSO}\left( \cdots \right) = \frac{1}{2}\mbox{tr}\left(
  \cdots\right) 
+\frac{1}{2}\mbox{tr}\left( \left(-\right)^F \cdots\right).
\end{equation}
The second difference -- as compared to the bosonic calculation -- is
that for worldsheet fermions the occupation number can be only zero or
one (since the creators anticommute).
The NS trace without the $\left( -\right)^F$ insertion comes out to be
\begin{equation}
\frac{1}{2}\mbox{tr}_{\mbox{\tiny NS}}\left( e^{-2\pi t H_{NS}}\right)
= \frac{1}{2}f_3 ^8 \left( e^{-\pi t}\right) .
\end{equation}
Since the NS vacuum is GSO odd we assign an additional minus to states
with even and zero occupation if $\left( -\right) ^F$ is inserted
into the trace,
\begin{equation}
\frac{1}{2}\mbox{tr}_{\mbox{\tiny NS}}\left( \left( -\right)^F
  e^{-2\pi t H_{NS}}\right) 
= -\frac{1}{2}f_4 ^8 \left( e^{-\pi t}\right) .
\end{equation}
For the R-sector trace without the $\left( -\right)^F$ insertion one
obtains 
\begin{equation}
-\frac{1}{2}\mbox{tr}_{\mbox{\tiny R}} \left( e^{-2\pi t H_R}\right) =
-\frac{1}{2}f_2 ^8\left( e^{-\pi t}\right) ,
\end{equation}
where the 16-fold degeneracy of the R vacuum has been taken into
account by the factor of $\sqrt{2}$ in the definition of $f_2$
(\ref{thetafunctions}). The R sector trace with a $\left( -\right)^F$
insertion vanishes identically. Half of the R sector groundstates have
eigenvalue $+1$ whereas the other half has eigenvalue $-1$.
Adding up all the results and using the identity (\ref{abstruse}) we
find that the net result for the annulus amplitude vanishes.
This implies also that the closed string diagram \ref{treecyl}
vanishes, and a hasty interpretation of this would lead to the
conclusion that D-branes do not interact (at least not via the
exchange of closed strings). However, as we will argue now, this is
not the case. The situation rather is that repulsive and attractive
interactions average up to zero. In order to see this\footnote{Already
  in the annulus computation, there are signs for such a
  cancellation. For the result, the minus sign in front of the R sector
  contribution was essential. Since R sector states are target space
  fermions, this indicates that the result is due to target space
  supersymmetry.}, let us translate
the annulus result back to the tree channel. Further, we would like to
filter out the contributions of massless closed string excitations. To
this end, we replace $t$ in terms of $l$ using
(\ref{dictionary-cyl}), and afterwards apply (\ref{modulartra}). The
contribution of 
massless closed string 
excitations is obtained by focusing on the leading contribution in the
$l \rightarrow \infty$ limit. (Massless interactions have infinite
range whereas the interactions carried by massive bosons have finite
range.) 
We collect the result of this straightforward calculation in table
\ref{table:closedint}. We separate  closed RR contributions from NSNS
sector contributions. The former ones correspond to the $\left(
  -\right)^F$ insertion 
and the latter to the $1$ insertion in the annulus
amplitude. 
 
%%%%
\begin{table}
\begin{center}
\begin{tabular}[h]{|l|l|} 
\hline 
\rule[0mm]{0mm}{9mm}{Sector} & $\int_{l\to \infty}dl\times$ \\[0.5cm]
\hline \hline 
\rule[0mm]{0mm}{9mm}{RR} & -$\frac{1}{2}\left(4\alpha^\prime
  \pi^2\right)^{-\frac{p+1}{2}}\; V_{p+1}\;  
l^{\frac{p-9}{2} }\; e^{-\frac{y^2}{4\pi \alpha^\prime l}}  $ \\[0.5cm]
\hline
\rule[0mm]{0mm}{9mm}{NSNS} &
$\frac{1}{2}\left(4\alpha^\prime\pi^2\right)^{-\frac{p+1}{2}}\; V_{p+1}\;  
l^{\frac{p-9}{2} }\; e^{-\frac{y^2}{4\pi\alpha^\prime l}}  $
\\[0.5cm]\hline 
\end{tabular}
\end{center}
\caption{Contributions of massless RR and NSNS sector closed strings
  to figure \ref{treecyl}.}\label{table:closedint}
\end{table}
%%%%%%%%%%%%%%%%%%%%%%%%%%% 

From table \ref{table:closedint} we deduce that interactions carried
by closed strings in the RR sector cancel interactions mediated by
closed strings in the NSNS sector. One can take the diagram
\ref{treecyl} to the field theory limit. In that limit the `hose'
connecting the two D-branes becomes particle propagators (lines). In
the NSNS sector we find propagators for the metric (fluctuations),
the dilaton and the anti-symmetric tensor $B_{\mu\nu}$. The D-branes
appear as source terms for those fields. The NSNS contribution to
diagram \ref{treecyl} tells us the strength of this coupling. In
particular, 
it yields the strength of the gravitational coupling which is
given by the
tension $T_p$. A detailed analysis of the effective field theory and
comparison 
with table 
\ref{table:closedint} leads to the result \cite{polchinski-book}
\begin{equation}\label{tension}
T^2 _p = \frac{\pi}{\kappa^2}\left( 4\pi ^2
  \alpha^\prime\right)^{3-p}e^{-2\Phi_0}  ,
\end{equation}
where $\kappa$ is the gravitational coupling in the effective theory
(see section \ref{effectiveactions}) and $\Phi_0$ denotes the constant
vev of the dilaton. Even though we did not derive
this explicitly here, let us make a few comments to motivate the
expression qualitatively. In a field theory calculation the propagator
of an NSNS field is accompanied by a power of
$\kappa^2e^{2\Phi_0}$. With $T_p$ 
defined as in (\ref{tension}) the $\kappa$ and $\Phi_0$ dependence
drop out. Since 
$\kappa^2 \sim \left(\alpha^\prime\right)^4$, the mass dimension of
$T_p$ is correct. (The $\alpha^\prime$ dependence in the string
calculation yields agreement with (\ref{tension}) after substituting
for the integration variable $l$ such that the $\alpha^\prime$
dependence in the exponent vanishes.) Further, the
exchange of massless particles should lead to a Coulomb interaction in
field theory. For the interaction between the two D-branes this means
that the potential should be given by the distance to the power of two
minus the number of transverse dimensions
\begin{equation}\label{D-coulomb}
V \sim y^{p-7}.
\end{equation}
In the string result (table \ref{table:closedint}), we can extract the
$y$ dependence after rescaling the integration parameter $l$ such that
the $y$ dependence in the exponent disappears. The result agrees with
(\ref{D-coulomb}). In order to fix the numerical coefficient, one
needs to do a more detailed analysis of the field theory
calculation. More details on this can be found in Polchinski's
book\cite{polchinski-book}.

The second line in table \ref{table:closedint} tells us that and how
string RR fields couple to the brane. We find the same Coulomb
law as for the gravitational interaction. The RR field should be a
$p+1$ form. The $p= even$ branes interact via closed
type IIA strings and the $p=odd$ branes via closed type IIB strings.  
The value of the RR contribution is exactly minus the value of the
NSNS contribution. This provides us with the RR charge of the
D-brane\footnote{The factor of $2\kappa^2$ has been introduced in
  order to match the charge definition to be used later in equ.\
  (\ref{dchargedef}).}
\begin{equation}\label{rrcharge}
\mu_p ^2 = 2\kappa^2 T_p^2e^{2\Phi_0}, 
\end{equation}
where we have taken into account the dilaton dependent RR field
redefinition performed in section \ref{effectiveactions}.
The signs are undetermined at this level. 
 
\subsection{D-brane actions}
\setcounter{equation}{0}

In the following we will specify the
actions for the field theory on the D-brane. We will also argue that
the D-brane interaction with the bulk field is obtained by adding the
action for fields living on the D-brane (the D-brane action) to the
effective type II 
action of section \ref{effectiveactions}. 
The previous calculation fixes the coefficient in front of the D-brane
action.  

\subsubsection{Open strings in non-trivial
  backgrounds}\label{sec:openbeta} 

In this section we will modify the calculation presented in section
\ref{betafunctions} such that it accommodates open string
excitations. 
We perform a Wick rotation such that the
worldsheet is of Euclidean signature. Further, we map the parameter
space of the open string worldsheet from a strip to the upper half
plane via the conformal transformation
\begin{equation}\label{upperhco}
z=e^{\tau + i\sigma}= z^1 + i z^2 .
\end{equation}
The discussion is performed for bosonic
strings but later we will also give the result for superstrings. Since
the open string couples naturally to closed strings (via joining its
ends) we also switch on non-trivial closed string modes. These are the
metric $G_{\mu\nu}$ and the antisymmetric tensor $B_{\mu\nu}$. We will
take those background fields to be constant. 
For the open string we take Neumann boundary conditions in all
directions at first. (For the superstring this is not consistent with
RR conservation. At the moment we will ignore this problem and return
to it later.) Later, we will discuss that T-duality maps Neumann to
Dirichlet boundary conditions. 
Hence, the restriction to Neumann
boundary conditions will not result in a loss of generality. 
At first, we consider a single brane setup. 
The massless open string excitation is now a $U(1)$ gauge field
$A_\mu$. We restrict ourself to the special case that the $U(1)$ field
strength $F_{\mu\nu}$ is slowly varying, i.e.\ we neglect
contributions containing second or higher derivatives of
$F_{\mu\nu}$. 
Under these conditions we will be able to perform the calculation to
all orders in $\alpha^\prime$ (in difference to section \ref{betafunctions}).
The nonlinear sigma model reads 
\begin{eqnarray}
S  &=& \frac{1}{2\pi \alpha^\prime}\left[ \int d^2 z\frac{1}{2} \big(
  \partial_\alpha 
  X_\mu \partial ^\alpha X^\mu \right. +
   i\epsilon^{\alpha\beta} B_{\mu\nu} 
  \partial_\alpha X^\mu \partial_\beta X^\nu\big)\nonumber \\
 && +\left. i{\displaystyle \int_{z^2 =0}} dz^1 A_\mu \partial_1
  X^\mu\right] .  
\label{opensigma}
\end{eqnarray}
Here, $A$ has been rescaled such that $\alpha^\prime$ appears as an
overall factor in front of the action. The target space indices
$\mu$,$\nu$ are raised and lowered with the constant background metric
$G_{\mu\nu}$. The worldsheet metric is taken to be the identity in the
$z^\alpha$ coordinates (\ref{upperhco}).\footnote{In principle this
  choice introduces gauge fixing ghosts as stated in section
  \ref{betafunctions}. Since their effect is not altered by the
  presence of the boundary we will not discuss the ghosts here. 
We should, however, mention that there are technical subtleties when
taking into account the dilaton in worldsheets with boundaries, see
e.g.\ \cite{Behrndt:1992dg}. (Recall that the ghosts contribute to the
dilaton beta function.)} 
Using Stoke's theorem the
term with the constant $B$ 
field can be rewritten as a surface integral
\begin{equation}
\int d^2 z \epsilon^{\alpha\beta} B_{\mu\nu}\partial_\alpha X^\mu
\partial_\beta X^\nu = 2\int_{z^2 = 0} dz^1 B_{\mu\nu} X^\mu \partial_1
X^\nu . 
\end{equation}
The term with the B-field can be absorbed into a redefinition of the
gauge field $A_\mu$ and we can put it to zero without loss of
generality. 
(It can be recovered by replacing $F \rightarrow F-2B$.)

In order to proceed we specify a classical configuration around which
we are going to expand. We denote this again by $\bar{X}^\mu$. For
freely varying ends the equation of motion and boundary conditions
read 
\begin{eqnarray}
\partial_z\partial_{\bar{z}} \bar{X}^\mu & = & 0 ,\\
\left( \partial_2 \bar{X}^\mu + i {F_\nu}^\mu \partial_1 \bar{X}^\nu\right)
_{\left|z^2 =0\right.} 
&= & 0.
\end{eqnarray}
The presence of the $U(1)$ gauge field $A_\mu$ results in inhomogeneous
Neumann boundary conditions. Since we have restricted ourselves to the
case 
where the target space metric is constant, the background field
expansion simplifies in comparison to the computation of section
\ref{betafunctions}. The fields can simply be Taylor
expanded. Neglecting second and higher derivatives of $F_{\mu\nu}$, the
background field expansion terminates at the third order in the
fluctuations, 
\begin{eqnarray}
\lefteqn{S\left[ \bar{X} +\xi\right]  =  S\left[ \bar{X}\right] +\frac{1}{2\pi
  \alpha^\prime} \int d^2 z \frac{1}{2}\partial^\alpha \xi_\mu
  \partial_\alpha \xi^\mu} \nonumber\\
& & \!\!\! +\frac{i}{2\pi \alpha^\prime}\! \int_{z^2=0}\!\!\!\!\!\!\!\!\!
  dz ^1\!\! 
  \left(\frac{1}{2}\! F_{\mu\nu} \xi^\nu\partial_1 \xi^\mu +
  \frac{1}{2}\!\partial_\nu F_{\mu\lambda}\xi^\nu \xi^\lambda
  \partial_1 \bar{X}^\mu   +\frac{1}{3}\! \partial_\nu
  F_{\mu\lambda}\xi^\nu\xi^\lambda \partial_1 \xi^\mu\!\!\right)\!\!
  .
\end{eqnarray}
Since we have chosen the worldsheet metric to be the identity (and the
geodesic curvature of the boundary to vanish) a suitable technique to
integrate out the fluctuations $\xi^\mu$ is given by a Feynman
diagrammatic approach. This means that we split the action into a free
and an interacting piece. The free piece determines the propagator
whereas the interacting one leads to vertices. As the free part of the
action we take
\begin{equation}
S_{free} = \frac{1}{4\pi\alpha^\prime} \int d^2z \partial^\alpha
\xi_\mu \partial_\alpha \xi^\mu +\frac{i}{4\pi\alpha^\prime}\int_{z^2
  =0}dz^1 F_{\mu\nu} \xi^\nu\partial_1 \xi^\mu .
\end{equation}
Hence, the interacting part is given by the rest
\begin{equation}
S_{int} = \frac{i}{2\pi\alpha^\prime}\int_{z^2 =0}  dz^1\left(
  \frac{1}{2}\partial_\nu F_{\mu\lambda} \xi^\nu\xi^\lambda\partial_1
  \bar{X}^\mu +\frac{1}{3}\partial_\nu F_{\mu\lambda}
  \xi^\nu\xi^\lambda \partial_1 \xi^\mu\right) .
\end{equation}
In order to compute the propagator we have to invert the two
dimensional Laplacian and to satisfy the (inhomogeneous Neumann)
boundary conditions arising from the variation of $S_{free}$ with
respect to $\xi^\mu$ (with free varying ends of the $\xi^\mu$),
\begin{eqnarray}
\frac{1}{2\pi\alpha^\prime} \Box \Delta_{\mu\nu}\left( z,z^\prime\right) &=&
-\delta\left( z- z^\prime\right) G_{\mu\nu},\\
\left(\partial_2 \Delta_{\mu\nu}\left( z, z^\prime\right) + i
  {F_\mu}^\lambda \partial_1 \Delta_{\lambda\nu}\left( z,
    z^\prime\right)\right) _{\left|z^2 =0\right.} & = & 0.
\end{eqnarray}
This boundary value problem can be solved (for example by borrowing
the method of mirror charges from electro statics) with the result
\begin{eqnarray}
\lefteqn{ \Delta^{\mu\nu}\left( z, z^\prime\right) = }\nonumber \\ & &
  -\alpha^\prime 
  \left[ G^{\mu\nu}\log\frac{\left| z-z^\prime\right|}{\left|
        z-\bar{z}^\prime\right|} +\left( \hat{G}^{-1}\right) ^{\mu\nu}
      \log\left| z - \bar{z}^\prime\right|^2 +\theta^{\mu\nu}
      \log\frac{z- \bar{z}^\prime}{\bar{z} - z^\prime}\right] .
\label{nappiyostprop}
\end{eqnarray}
Here, we have introduced the following matrices (an index $S$($A$) stands
for (anti-)symmetrization and $G^{\mu\nu}$ are the components of the
inverted target space metric $G^{-1}$, as usual)
\begin{eqnarray}
\left( \hat{G}^{-1}\right)^{\mu\nu}& = &\left( \frac{1}{G+ F}\right)_S
^{\mu\nu} = \left( \frac{1}{G+F}G\frac{1}{G-F}\right)^{\mu\nu} , \\
\left( \hat{G}\right)_{\mu\nu} & = & G_{\mu\nu} -\left( F
  G^{-1}F\right)_{\mu\nu} , \\
\theta^{\mu\nu} & = & \left( \frac{1}{G+F}\right)^{\mu\nu} _A = \left(
  \frac{1}{G+F} F\frac{1}{G-F}\right)^{\mu\nu} .
\end{eqnarray}

The interaction piece $S_{int}$ gives rise to two vertices ---one with
two and one with three legs--- located at
the boundary. The Feynman rules are summarized in figure \ref{fig:rules}.

 %%%%%%%%%%%%%%%%%%%%%%%%%%%%%%%%%%%%
\begin{figure}
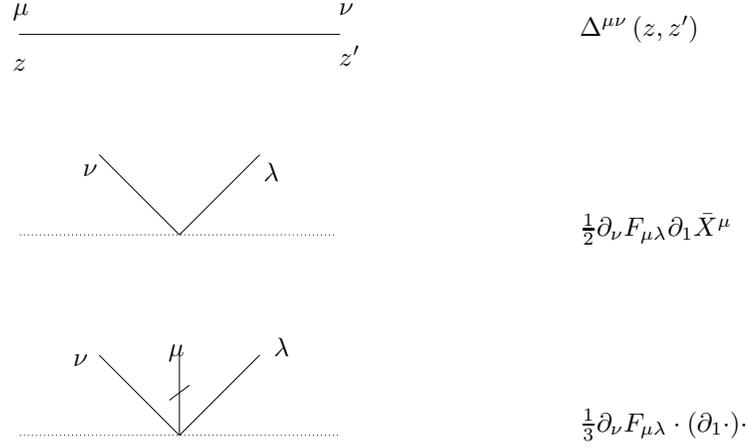
  
\begin{center}
\input rules.pstex_t
\end{center}
\caption{Feynman rules: The dotted line denotes the worldsheet
  boundary. The slash on the leg means that a derivative acts on the
  corresponding leg.}
\label{fig:rules}
\end{figure}
%%%%%%%%%%%%%%%%%%%%%%%%%%%%%%%%%%%

The propagator $\Delta^{\mu\nu}\left( z ,z^\prime\right)$ becomes
logarithmically divergent if the arguments coincide. Therefore, we
replace the logarithm of zero by its dimensionally regularized version
\begin{equation}
 \log \left| z -z^\prime\right|_{\left| z=z^\prime\right.} = -\int
 \frac{d^2 k}{2k^2} 
 {e^{k\left( z-z^\prime\right)}}_{\left|z=z^\prime\right.} =
 -\lim_{\epsilon \to 0} 
\mu^\epsilon \int \frac{d^{2-\epsilon}k}{2\left(k^2 +m^2\right)},  
\end{equation}
where the momentum integral extends over a two
dimensional plane. We introduced a mass scale $\mu$ which is needed in
order to keep the mass dimension fixed while changing the momentum
space dimension. In the last step we have introduced also an infrared
cutoff $m^2$.\footnote{A different (maybe more elegant) way to deal
  with infrared divergences is discussed e.g.\ in
  \cite{Chetyrkin:1982nn}, see also the appendix
  of\cite{Grisaru:1987wj}.} With our regularization prescription we
obtain
\begin{equation}
\log \left( z- z^\prime\right)_{\left| z=z^\prime\right.} =
\frac{1}{2}\pi^{\frac{2-\epsilon}{2}}\left(\frac{\mu}{m}\right)^\epsilon
\Gamma\left( 
\frac{\epsilon}{2}\right), 
\end{equation}  
which has a simple pole as $\epsilon$ goes to zero.

The bare background field (coupling) $A_\mu$ is infinite. 
By adding counterterms to the action
the bare field can be expressed in terms of a renormalized field which
is finite as $\epsilon$ goes to zero. The only counterterm arises from
the diagram in figure \ref{fig:counter}. The action is written in
terms of renormalized fields by adding 
\begin{equation}\label{aboprop}
\delta S =  -\frac{i}{2\pi\alpha^\prime} \int_{z^2 =0} dz^1
\frac{1}{2}\partial_\nu F_{\mu\lambda} \partial_1 \bar{X}^\mu
\Delta^{\nu\lambda}\left( z^1, z^{\prime 1}\right)_{| z^\prime \to z} .
\end{equation}
(and replacing the bare gauge field by the renormalized one. Hoping
that the renormalization program is sufficiently familiar we do not
introduce sub- or super-scripts indicating the difference between bare
and 
renormalized couplings).
The beta-function of $A_\mu$ is obtained by applying
$\mu\frac{d}{d\mu}$ on the renormalized couplings and using the fact
that the bare couplings are independent of the cutoff. This leads to
(now, $A_\mu$ denotes the renormalized coupling)
\begin{equation}
\beta_\rho ^A = \mu \frac{d}{d\mu} A_\rho = \partial_\nu
F_{\rho\lambda} \left( \hat{G}^{-1}\right)^{\lambda\nu} .
\end{equation}
In this case the vanishing of the beta function ensures conformal
invariance. (We do not encounter the subtleties which we met in
section \ref{betafunctions}. Partially, this is the case because we
have written the action always in a manifestly gauge invariant form,
i.e.\ in terms of the gauge field strength. By performing partial
integrations  differently we could have carried out the
calculation in 
a slightly more complicated way, with the same result.) The equation
of motion for the gauge field is 
\begin{equation}
\beta_\mu ^A =0 .
\end{equation}
This equation of motion can be lifted to the Dirac-Born-Infeld action
\begin{equation}\label{dbiaction}
S = \frac{\sqrt{\pi}}{\kappa}\left( 4\pi ^2
  \alpha^\prime\right)^{\frac{3-p}{2}} 
\int d^{p+1}x\, e^{-\Phi}\sqrt{\det\left( G+F\right)} ,
\end{equation}
where $p+1$ is the number of Neumann directions (i.e.\ for our
discussion $p+1 = 10 (26)$ for the super (bosonic) string\footnote{We
  view 
  the result of 
  our computation as a result for the bosonic modes of the
  superstring. Since we did not specify the effective action for the
  closed bosonic string the factor in front of (\ref{dbiaction}) is
  irrelevant for the bosonic string (see also the discussion of the
  Fischler Susskind mechanism below).}). The factor in
front of the integral in (\ref{dbiaction}) has not been fixed by our
current discussion. We will explain how to fix it below. 
The same applies to the dilaton dependence. (We discussed only the
case of a constant dilaton $\Phi$.)
Since we have
rescaled $A_\mu$ by powers of $\alpha^\prime$ such that the
$\alpha^\prime$ dependence appears as an overall factor in
(\ref{opensigma}), 
 %%%%%%%%%%%%%%%%%%%%%%%%%%%%%%%%%%%%
\begin{figure}
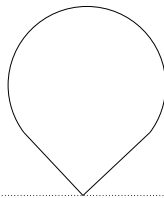
  
\begin{center}
\input counter.pstex_t
\end{center}
\caption{The logarithmically divergent Feynman diagram.}
\label{fig:counter}
\end{figure}
%%%%%%%%%%%%%%%%%%%%%%%%%%%%%%%%%%%
the $\alpha^\prime$ expansion of the action (\ref{dbiaction}) is not
obvious. Performing rescalings such that $A_\mu$ has mass-dimension one
(or zero) shows that the $\alpha^\prime$ expansion is a power
expansion in $F_{\mu\nu}$. Also, we note that the propagator
(\ref{aboprop}) contains higher orders in
$\alpha^\prime$. Alternatively, we could have chosen a propagator
satisfying homogenous Neumann boundary conditions at the price of
having an additional vertex operator. This additional vertex does not
lead to a one loop (leading order in $\alpha^\prime$) divergence since
it is antisymmetric in its legs. The leading order $\alpha^\prime$
equation is 
\begin{equation}
\partial_\mu F^{\mu\nu} = 0 .
\end{equation}
Lifting this to an action would give (in the small $\alpha^\prime$
approximation) 
\begin{equation}\label{alpha-prime}
S\sim \int d^{p+1}x e^{-\Phi}\sqrt{-G}F^2 ,
\end{equation}
where the $\Phi$ dependence has been taken such that the result
coincides with the small $\alpha^\prime$ expansion of (\ref{dbiaction}). 
Expanding  (\ref{dbiaction}) in powers of $F$ and
keeping only terms up to $F^2$ we find in addition to
(\ref{alpha-prime}) a contribution $\int d^{p+1}x
e^{-\Phi}\sqrt{-G}$. 
From a
field theory perspective this is a tree level vacuum energy. So far,
we did not properly couple the open string excitations to gravity. 
We included the effects of bulk fields on the equations of
motion for open string excitations, but we did not encounter a
back reaction, i.e.\ that the field $A_\mu$ enters the equations of
motion for the closed string excitations. The reason is that the
back reaction 
is an annulus effect. We will not present a detailed annulus
calculation but sketch the result. (In principle we have done the
necessary computations in the previous section.) Since the beta
functions depend on local features (short distance behaviors) one
would guess that for the beta function it may not matter whether the
worldsheet is an annulus or a disc. However, the annulus may
degenerate as depicted in figure \ref{fig:fishler}.

%%%%%%%%%%%%%%%%%%%%%%%%%%%%%%%%%%%%
\begin{figure}
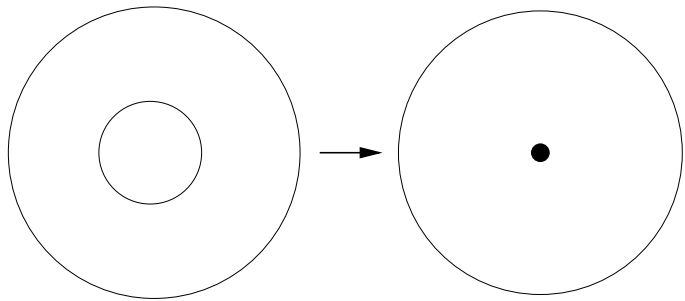
  
\begin{center}
\input Fishler.pstex_t
\end{center}
\caption{Fischler Susskind mechanism: The annulus degenerates into a
  punctured disc as the inner 
  circle shrinks to zero. This gives rise to a closed string
  counterterm depending on the open string excitations.}
\label{fig:fishler}
\end{figure}
%%%%%%%%%%%%%%%%%%%%%%%%%%%%%%%%%%%

This gives an additional short distance singularity. (The inner circle of
the annulus becomes short.) This singularity can be taken care of by
adding counterterms to the closed string action. The counterterms
depend on open string modes. This in turn leads to terms in the closed
string beta functions which depend on the open string modes. This
process is known as the Fischler Susskind mechanism. The net effect is
that we add the open string effective action to the closed string
effective action (\ref{dbiaction}) and obtain the equations of motion 
by varying the sum. This is the expected back reaction. (In particular
the Einstein equation now contains the energy momentum tensor of the
open string modes.) After taking the back reaction into account, the
coefficient in (\ref{dbiaction}) does matter. In the previous section
we have computed the tension of the D-brane (\ref{tension}). 
This fixes the coefficient and the dilaton dependence\footnote{Note
  also that this dilaton dependence agrees with our general discussion
  in section \ref{effectiveactions}. The Euler number of the disc
  differs by one from the Euler number of the sphere.} as given in 
(\ref{dbiaction}). According to our discussion in the previous
section, the presence of a D-brane should also back-react on the RR
background. We could not see this in the present consideration since
we did not take into account non trivial RR backgrounds. (In fact, it
is rather complicated to switch on non-trivial RR backgrounds in the
non-linear sigma model.) We will come back to the discussion of RR
contributions to the open string effective action below.

So far, we have studied the case of a single D-brane. How is this
discussion modified in the presence of multiple D-branes? We have
focused on the case where we have only Neumann boundary conditions. 
This means that multiple D-branes must sit on top of each other,
simply because there is no space dimension left in which they could be
separated. The effect of having more than a single brane is that the
gauge field $A_\mu$ is a $U(N)$ gauge field -- it is a matrix. 
Calling the expression (\ref{opensigma}) an action does not make much
sense 
anymore since we would have a matrix valued action. Therefore, one
takes just the bulk part of the action (the first line in
(\ref{opensigma})) and computes instead of the partition function the
Wilson loop along the string boundary\cite{Dorn:1986jf},
\begin{equation}
W = \left<tr\left( P e^{i\int_{\partial M} dt A_\mu
      \dot{X}^\mu}\right)\right> , 
\end{equation}
where we have denoted the boundary of the worldsheet by $\partial M$
and chosen some $t$ to parameterize this curve. The letter $P$ stands
for the path ordered product.   
The expectation value
is computed with respect to the bulk action only. Now, it is
problematic to get an expression containing all orders in
$\alpha^\prime$. The leading $\alpha^\prime$ contribution to the beta
function results in the Yang-Mills equation
\begin{equation}
\nabla _\mu F^{\mu\nu} = 0,
\end{equation}  
where $\nabla$ denotes a gauge covariant derivative. The effective
action in leading approximation can be obtained as follows. We expand
(\ref{dbiaction}) to first order in $F^2$. We replace $F^2$ by $tr
F^2$. In addition, we multiply the zeroth order term in $F$ by the
number $N$ of D-branes (the tension is $N$ times the tension of a
single D-brane). The generalization of the Dirac-Born-Infeld action
(\ref{dbiaction}) to non-abelian gauge fields is a subject of ongoing
research, see e.g.\ \cite{Tseytlin:1999dj,Bogaerts:2000rx}.

\subsubsection{Toroidal compactification and T-duality for open strings}

In the previous section we have discussed the case of having Neumann
boundary conditions in all directions. This means that the D-branes
have been space filling objects. In order to obtain results for
D-branes extending along less dimensions we will discuss
T-duality for open strings, now. 

At first, we focus on the case with trivial background fields. From
section \ref{t-duality} we recall that T-duality interchanges winding
with momentum modes. For the open string we have either winding or
momentum modes in compact directions. A string with DD boundary
conditions along the compact dimension can have non-trivial winding
modes. Since the ends of the string are tied to the D-brane it cannot 
unwrap. On the other hand, the DD string does not have quantized Kaluza
Klein momenta. The D-brane can absorb any momentum carried by the
string in the compact direction. 
For NN strings, opposite statements are true. If the string has Neumann
boundary conditions along the compact dimension, its ends can move
freely in that direction -- it can continuously wrap and unwrap the
compact dimension. On the other hand, the string cannot transfer
Kaluza-Klein 
momentum to the D-brane. The NN string has non-trivial momentum
modes. This consideration suggests that T-duality for open strings
interchanges Neumann with Dirichlet boundary conditions.

Let us substantiate these qualitative statements by studying the effect
of T-duality on the mode expansions. For the T-duality transformation
we use the ``recipe'' (\ref{recepy}). To be specific we choose the
ninth direction to be compact, i.e.\
\begin{equation}
x^9 \equiv x^9 + 2\pi R.
\end{equation}
For the string with NN boundary conditions in the ninth dimension
this implies that the center of mass momentum is quantized
\begin{equation}
p^9 = \frac{n}{R} ,
\end{equation}
with $n$ being an integer. There are no integer winding numbers in the
case of NN boundary conditions. We rewrite the mode expansion
(\ref{neumannmode}) in a suggestive way
\begin{eqnarray}
X^9 & = & X^9 _R + X^9 _L , \label{nnmot}\\
X^9 _R & = &\frac{x^9}{2} +\frac{n}{2R}\sigma^-  +\frac{i}{2}\sum_{n\not=
  0}\frac{1}{n}\alpha_n ^9 e^{-in\sigma^-} ,\\
X^9 _L & = & \frac{x^9}{2} +\frac{n}{2R}\sigma^+ +
\frac{i}{2}\sum_{n\not= 0}\frac{1}{n}\alpha_n ^9 e^{-in\sigma^+}.
\end{eqnarray}
Applying the recipe (\ref{recepy}), we obtain the mode expansion for
the T-dual coordinate
\begin{equation}\label{tduamod}
X^9 \;\;\;\stackrel{\mbox{\tiny T-DUALITY}}{\longrightarrow} 
\;\;\; \tilde{X}^9 =\frac{n}{R}\,\sigma +\sum_{n\not=
  0}\frac{1}{n}\alpha_n ^9 e^{-in\tau}\sin 
n\sigma 
.
\end{equation}
This mode expansion is zero at $\sigma =0$ and $2nR^\prime$ at $\sigma
= \pi$, where (see (\ref{tduality})) 
\begin{equation}
R^\prime = \frac{1}{2R} = \frac{\alpha^\prime}{R} .
\end{equation}
The interpretation is that the open string ends on a D-brane located at
$x^9 =0$\footnote{We could have obtained a different position by
  distributing the center of mass position of the NN string $x^9$
  asymmetrically among left and right movers.}. The open string
winds $n$ times around a circle of radius $R^\prime$. 
It is rather obvious that ---starting from a DD string with mode
expansion 
(\ref{tduamod})--- T-duality  will take us to an NN string with mode
expansion (\ref{nnmot}), (the center of mass position depends again on
the way we distribute a constant between left and right movers).
So, T-duality inverts the compactification radius and interchanges
Dirichlet with Neumann boundary conditions. We leave it to the reader
to verify that an investigation of the worldsheet fermions and of
ND directions is consistent with this picture.

In section \ref{dbraneinteractions} we have noticed that Dp-branes
with p even (odd) 
interact via the exchange of closed type IIA(B) strings. Our present
observation that T-duality along a compact direction interchanges
Dirichlet with Neumann boundary conditions implies that a Dp-brane
with even p is mapped onto a Dq-brane with odd q, and vice versa
($\mbox{q}=\mbox{p}\pm 1$). This goes along nicely with our earlier
statement (section \ref{t-super}) that T-duality along one circle
interchanges type IIA with type IIB strings.

Finally, let us discuss T-duality for open strings in the presence of
non-trivial background fields. For the closed string we have done this
in section \ref{tsigma}. Because the discussion of the closed string
background fields is not affected by the open string, we will focus on
the special case where only the open string gauge field is
non-trivial. For simplicity we also restrict to one D-brane only (for
multiple D-branes see e.g.\ \cite{Dorn:1996an})\footnote{We will
  comment briefly on the case of multiple D-branes at the end of this
  section.}. Let us first outline 
in words the procedure we are going to carry out. The compactification
has to be done in a Killing direction. (Shifts along the
compact direction are isometries.) We will take this dimension to be
the ninth. The next step is to gauge this isometry and to undo the
gauge by forcing the corresponding gauge field to be trivial. This
will be done again by adding a Lagrange multiplier times the field
strength of the isometry gauge field. The Lagrange multiplier will
become the T-dual coordinate in the end. In particular the Lagrange
multiplier lives on a circle whose radius is inverse to the original
compactification radius. This is derived from the requirement that the
radius of the isometry-gauge group ($U(1)$) agrees with the
compactification radius. We will not discuss the technical details of
this derivation (they are presented for example in the appendix of
\cite{Alvarez:1996up}). Instead, we will focus on a detailed discussion
of the boundary conditions. The boundary condition of the isometry
gauge fields is constrained by the boundary condition of the open
string. This will be implemented by a second Lagrange
multiplier which lives only at the boundary of the worldsheet. After
integrating out the isometry gauge fields the integration over this
second Lagrange multiplier will give the boundary condition for the
T-dual coordinate (the ``first'' Lagrange multiplier)\footnote{As we
  will see below integrating over the second Lagrange multiplier
  $\kappa$ boils down to setting an argument of a delta function to
  zero. This in turn implies boundary conditions on the Lagrange
  multiplier $\lambda$.}. 

After having described the strategy, we will now present the details of
the procedure. Setting $\alpha^\prime = \frac{1}{2}$ the open string
worldsheet action with a non-trivial $U(1)$ gauge field coupling to
the boundary reads (for convenience we use a rescaled $A_\mu$ as
compared to (\ref{opensigma}) and choose Minkowskian worldsheet
signature here)
\begin{equation}\label{startingt}
S = \frac{1}{2\pi}\left(\int_{\cal M} d^2z \,\partial_\alpha X^\mu
\partial^\alpha X^\mu +  \int_{\partial {\cal M}}dt\left(\, A_\mu
\partial_t X^\mu + V_{\mu} \partial_n X^\mu\right)\right) ,
\end{equation}
where ${\cal M}$ denotes the worldsheet and $\partial {\cal M}$ its
boundary (parameterized by $t$).  With $\partial_n $ we denote the
derivative into the direction normal to the boundary.
We specify the character of the boundary conditions in $X^9$
direction by the following
assignments\footnote{A fixed boundary condition on a variation means
  that this variation depends on the boundary values of variations of
  other fields (or is zero). In particular, if we do not vary the
  other directions we can replace ``fixed'' by ``zero'' in
  (\ref{bcass}). }
\begin{equation}
\begin{array}{| c|c|c|}
\hline
\mbox{Boundary Condition} & \delta X^9 & \partial_n \delta X^9 \\
\hline \hline
\mbox{Dirichlet} & \mbox{fixed} & \mbox{free} \\
\hline
\mbox{Neumann} & \mbox{free} & \mbox{fixed}\\
\hline
\end{array}.
\label{bcass}
\end{equation}
This implies that for Dirichlet boundary conditions we set $A_9 =0$
whereas 
for Neumann boundary conditions $V_9 =0$ is chosen. For the Neumann
boundary  
conditions (free varying ends) the variation of $S$ gives the boundary
condition (we denote the normal vector with $n^\alpha$ and the tangent
vector with $t^\alpha$)
\begin{equation}
n^\alpha\partial_\alpha X^9 = -\frac{1}{2} F^{9\nu}\partial_t x_\nu,
\end{equation}
where $F_{\mu\nu}$ is the field strength of the $U(1)$ gauge field
$A_\mu$. For Dirichlet conditions we obtain 
\begin{equation}\label{crypdir}
V_9 = 0.
\end{equation}
Since this should be in agreement with our assignment that the
variation of the end of the open string in the ninth direction is
fixed (possibly 
related to the variations in other directions), the function $V_\mu$
should be interpreted as a vector which is tangent to the
brane. Equation (\ref{crypdir}) then means that the D-brane is
localized in the ninth direction.   

Since we have chosen the simplified
case of trivial closed 
string backgrounds any direction (in cartesian target space
coordinates) is an isometry.  
Suppose that in addition
the $x^9$ derivative of the $U(1)$ gauge field is pure gauge, i.e.\
zero modulo gauge transformations. So, without loss of generality we
restrict ourselves to the case that the gauge background is $X^9$
independent. 
We also assume that the tangent vector
$V_\mu$ does not depend on $X^9$.
We specify the boundary condition on $X^9$ by the equation
\begin{equation}\label{genericbc}
b^\alpha \partial_\alpha X^9 _{\left|\partial {\cal M}\right.} =
  \mbox{independent 
  of}\; X^9,
\end{equation}
where $b^\alpha$ is a worldsheet vector with a given orientation to
the boundary. In case of Dirichlet boundary conditions, $b^\alpha $ is
parallel to the boundary ($b^\alpha = t^\alpha$). 
For free varying ends (Neumann boundary conditions) $b^\alpha$ is
normal to the boundary. 
In the
action (\ref{startingt}) $X^9$ does not mix with the other fields. We
focus on the $X^9$ dependent part
\begin{eqnarray}
S &=& \bar{S} + S^{(9)}, \\
S^{(9)} & = & \frac{1}{2\pi}\left(\int_{\cal M} d^2 z\, \partial_\alpha X^9
\partial^\alpha X^9 +  \int_{\partial {\cal M}}dt\left( A_9
\partial_t X^9 +V_9\partial_n X^9\right)\right) \! ,
\end{eqnarray}
where $\bar{S}$ stands for the $X^9$ independent part.
The action is invariant under constant shifts in $X^9$. We transform
this into a local symmetry by the replacement
\begin{equation}
\partial_\alpha X^9 \rightarrow D_\alpha X^9 = \partial_\alpha X^9 +
\Omega_\alpha ,
\end{equation}
where $\Omega_\alpha$ is the isometry gauge field. (We use this
terminology in order to avoid confusion with the open string excitation
mode $A_\mu$.) The isometry gauge field $\Omega_\alpha$ transforms
under local shifts in $X^9$ such that $D_\alpha X^9$ is invariant. We
introduce a bulk Lagrange multiplier $\lambda$ in order to constrain
the $\Omega$-field strength\footnote{In two dimensions we can hodge
  dualize the two form field strength to a scalar $f$.}
\begin{equation}
f = \epsilon^{\alpha\beta}\partial_\alpha \Omega_\beta
\end{equation}
to vanish. 
Further, we add a second boundary Lagrange multiplier $\kappa$ whose
task is to fix the boundary condition of $\Omega_\alpha$.
Taking into account the Lagrange multipliers, the gauged action
reads\footnote{As before we write the case of D and N boundary
  conditions into one formula. Recall that $A_9 =0$ and $b_\alpha =
  t_\alpha$ for D boundary conditions, and $V_9 =0$ and $b_\alpha =
  n_\alpha$ for N boundary conditions.}
\begin{eqnarray}
S^{(9)}_{gauged} & = & \frac{1}{2\pi}\int_{\cal M} d^2 z\, \left(
  \partial_\alpha 
  X^9\partial^\alpha X^9 +\Omega_\alpha\Omega^\alpha + 2\Omega_\alpha
  \partial^\alpha X^9 -
  2\epsilon^{\alpha\beta}\Omega_\beta\partial_\alpha\lambda 
  \right) \nonumber \\
& & \;+\frac{1}{2\pi}\int_{\partial {\cal M}} dt \, \left(
  A_9\partial_t X^9 + V_9\partial_n X^9\right)\nonumber\\
& & \; + \frac{1}{2\pi}\int_{\partial {\cal M}} dt\, \left( A_9 t_\alpha
+V_9n_\alpha  + \kappa b_\alpha 
  +2\lambda t_\alpha\right) \Omega^\alpha ,
\label{opengauged}
\end{eqnarray}
where for later convenience we have performed partial integrations
such that no derivative of $\Omega_\alpha$ appears in the action.
The worldsheet vector $t_\alpha$ denotes the tangent vector to the
boundary. 
The T-dual model will be obtained by integrating out
$\Omega_\alpha$. The T-dual coordinate will be $\lambda$. Its boundary
conditions are going to be fixed by the integration over
$\kappa$. Before going through the steps of this prescription, let us
verify that the gauged action is equivalent to the ungauged
one. Integration over $\lambda$ leads to 
\begin{equation}
\Omega_\alpha =\partial_\alpha \rho ,
\end{equation}
where $\rho$ is an arbitrary worldsheet scalar. Integrating out
$\kappa$ leads to the boundary condition
\begin{equation}
b^\alpha \partial_\alpha \rho = 0 .
\end{equation}
Because neither the background (nor $b^\alpha$) depend on $X^9$,
the 
scalar $\rho$ can be absorbed completely into a redefinition of $X^9$
without spoiling the boundary condition (\ref{genericbc}). (In
addition $\rho$ needs to live on a circle with radius equals the
compactification radius. This issue has been addressed in \cite{
  Alvarez:1996up}.
The discussion given there leads to the observation that $\lambda$
lives on a circle with inverted radius.) Hence, the
gauged and ungauged models are equivalent. 

In order to construct the T-dual model we first integrate out 
$\Omega_\alpha$. Because the action (\ref{opengauged}) does not
contain any derivatives of $\Omega_\alpha$ (it is ultra local with
respect to the isometry gauge field), the functional integral over
$\Omega_\alpha$ factorises into a bulk integral and a boundary
integral
\begin{equation}\label{ultralocal}
\int {\cal D} \Omega_{{\cal M}\cup \partial {\cal M}}\left(
  \ldots\right) = \int {\cal D}\Omega_{\cal M}\left(
  \ldots\right)\times \int {\cal D}\Omega_{\partial {\cal M}}\left(
  \ldots\right). 
\end{equation}
Integrating out $\Omega$ in the bulk leads to the ungauged bulk action
with $X^9$ replaced by $\lambda$. This is exactly as in the closed
string computation (up to a boundary term)
\begin{eqnarray}
\tilde{S}^{(9)}_{bulk} & = & \frac{1}{2\pi }\int_{\cal M} d^2z \,
  \left( 
  \partial_\alpha \lambda \partial^\alpha \lambda + 2
  \epsilon^{\alpha\beta} \partial_\alpha \lambda  \partial_\beta
  X^9\right)\\
& = & \frac{1}{2\pi} \int_{\cal M} d^2 z \, \partial_\alpha \lambda
  \partial^\alpha\lambda + \int_{\partial {\cal M}}dt \, 2 \lambda
  \partial_t X^9 ,\label{boundterm}
\end{eqnarray}
where in the second line we have used Stokes theorem.

The additional ingredient comes from the
second factor in (\ref{ultralocal}). This gives a two dimensional
delta function
\begin{equation}
\int {\cal D}\Omega_{\partial {\cal M}} e^{-S_{gauged, \partial {\cal
      M}}}\sim \delta^2\left( A_9 t_\alpha +V_9 n_\alpha + \kappa
      b_\alpha +2i\lambda 
      t_\alpha\right) .
\end{equation}
Let us evaluate this delta function for the two cases: $X^9$ has
Dirichlet boundary conditions ($b_\alpha = t_\alpha$)
or Neumann conditions ($b_\alpha =n_\alpha$). In
the first case, the evaluation of the delta function fixes $\kappa $ in
terms of $\lambda$ and sets $V_9=0$. This means that $\lambda$ has
free varying ends, i.e.\ Neumann boundary conditions. Taking into
account the boundary term in $(\ref{boundterm})$ we obtain that the
dual $U(1)$ gauge field is determined by the position of the original
D-brane, 
\begin{equation}\label{dualdir}
\tilde{A}_\lambda = -2{X^9}_{\left|\partial {\cal M}\right.} .
\end{equation}
Recall that the original Dirichlet boundary condition may depend on
the other directions, i.e.\ the rhs of (\ref{dualdir}) is some fixed
function. 

If $X^9$ satisfies Neumann conditions, the evaluation of the delta
function leads to $\kappa =0$ and the Dirichlet boundary condition
\begin{equation}\label{dualneu}
\lambda _{\left|\partial {\cal M}\right.} = -\frac{1}{2}A_9.
\end{equation}
In the T-dual string theory there is a D-brane located in $x^9$ along
the curve $A^9$ (note that $A_9$ may depend on the coordinates
different from $x^9$). Note also that plugging the boundary condition
(\ref{dualneu}) into (\ref{boundterm}) cancels the original $A_9$
coupling to the boundary. 

To summarize, we have seen that T-duality interchanges Dirichlet with
Neumann boundary conditions. The position of the D-brane is
interchanged  with the $U(1)$ gauge field component in the T-dualized
directions. Starting with Neumann boundary conditions it is easy to
see that gauge transformations do not change the sigma model for the
string, i.e.\ the field equations of the string excitations do not
depend on gauge transformations. Via T-duality this translates to
changes of the position of a D-brane, in particular constant shifts
are moduli of the theory.
From the above expressions it is also clear
that performing the T-duality twice will result in the original
theory. 

With these considerations we can go back to the effective action
(\ref{dbiaction}) and generalize it to non space filling branes. This
is done by simply replacing the $A_\mu$ components where $\mu$ labels
a Dirichlet direction by scalars. These scalars are the collective
coordinates of the lower dimensional D-brane. One can also
parameterize the worldvolume of the D-brane by an arbitrary set of
parameters. In this case one needs to replace bulk fields by the
induced quantities. The effective D-brane action for lower dimensional
D-branes can be also computed in the sigma model approach directly. This
has been done in \cite{Leigh:1989jq}.

Finally, let us comment briefly on the case of multiple branes. We
start with Neumann boundary conditions. The gauge field $A_9$ is now a
matrix. Suppose that this matrix is diagonal. In this case the above
discussion is valid if we just replace $A_9$ by a diagonal matrix
everywhere. In the T-dual theory, the position of the D-brane is a
diagonal matrix. The interpretation is that each entry corresponds to
the position of a single D-brane. The matrix describes a set of
D-branes. The more general case of non-diagonal gauge fields is rather
complicated. It is addressed e.g.\
in\cite{Dorn:1996an,Dorn:1997xk,Dorn:1998vz}. 

\subsubsection{RR fields}
So far, we have discussed D-brane effective actions only for trivial
RR backgrounds. The reason was mainly of technical origin. It is
rather complicated to describe non-trivial RR backgrounds in a sigma
model approach. Later in section \ref{adssection}, 
we will use such a description for a particular background. Now, we
will not discuss the RR background in a sigma model. Instead we will
use our computation of section \ref{dbraneinteractions} and field
theoretic arguments. 

In section \ref{dbraneinteractions} we have seen that the Dp-brane
carries RR charge with respect to a $p+1$ form RR gauge potential of
type 
II theories. In section \ref{sec:openbeta} we argued that the
interaction of D-branes via closed strings is obtained by adding the
effective D-brane action to the effective type II action (IIA for even
p, and IIB for odd p). Combining these two observations, we infer that
the effective D-brane action contains an additional piece
\begin{equation}
S_1 = S_{DBI} +\frac{\sqrt{\pi}}{\kappa}\left( 4\pi^2
  \alpha^\prime\right)^{\frac{3-p}{2}}\int d^{p+1}x A_{1,\ldots,p+1},
\end{equation}
where we assume that the D-brane worldvolume extends along the first
$p+1$ dimensions. (In general, the D-brane can be parameterized by a
set of $p+1$ parameters. In this case, the D-brane action is written
in terms of induced fields.) We have abbreviated the action
(\ref{dbiaction}) with $S_{DBI}$. Further, we used the result
(\ref{rrcharge}) to fix the coefficient in front of the RR coupling. 

The label in $S_1$ has been introduced because now we will argue that
there are 
further couplings to RR fields. These occur if another a D-brane lies
within the worldvolume of the considered D-brane, or a D-brane
intersects 
the considered D-brane. In such a case there will be
strings starting and ending on different D-branes. They give rise to
massless 
fields transforming in the fundamental representation of the gauge
group living on the considered D-brane. Under certain circumstances
there may be chiral fermions leading to potential gauge anomalies. Such
anomalies can be canceled by assigning anomalous gauge
transformations to certain bulk RR fields and adding an 
interaction term to the effective D-brane action. This procedure has
been carried out in detail in\cite{ Green:1997dd, Douglas:1995bn}.  
Here, we just briefly give the result. 

In cases that there is an anomaly, this anomaly can be canceled by
adding a Chern-Simons term to the D-brane action 
\begin{equation}
S = S_1 + S_{CS},
\end{equation} 
with (for $N$ coincident D-branes -- for $N>1$ also the DBI action
needs to 
be modified as discussed in the end of section \ref{sec:openbeta})
\begin{equation}
S_{CS} = \int_{{\cal B}_p} C\wedge \left(\mbox{tr} \,
  e^{\frac{iF}{2\pi}}\right)\sqrt{\hat{\cal A}\left( R\right)} . 
\end{equation}
The way of writing the Chern-Simons term needs explanation. The
integral is taken over the worldvolume of the Dp-brane which is
denoted by ${\cal B}_p$. The integral is a formal expression in
differential forms. It is understood that only $p+1$ forms out of this
expression are kept. 

The first form $C$ is an RR $q+1$ form where
$q$ is the spatial dimension of the surface in which the two D-branes
(or sets of D-branes) overlap. The last term contains the so called
A-roof genus. This is a polynomial in the curvature two-form
(for an explicit definition see e.g.\ \cite{Green:1997dd}). In
addition to adding $S_{CS}$ to the D-brane action the RR form $C$
receives a contribution under gauge transformations. This comes about
as follows. The definition of the RR field strength receives a
correction (the correction is related to a Chern-Simons form whose
explicit form is not needed here)
\begin{equation}
H = dC + \mbox{correction} 
\end{equation}
such that
\begin{equation}\label{anoM}
dH = 2\pi \delta\left( {\cal B}_p \rightarrow M_{10}\right)\mbox{Tr}_N
e^{\frac{iF}{2\pi}} \sqrt{\hat{\cal A}\left( R\right)},
\end{equation}
where the delta function means that this correction is supported on
the worldvolume of the D-brane, only. Even though the right hand side
of (\ref{anoM}) is gauge invariant, $C$ has to change under gauge
transformations in order to ensure that $H$ is invariant. The
construction is such that the change of $S_{CS}$ under gauge
transformations cancels the anomaly.   

\subsubsection{Noncommutative geometry}

It is interesting to observe that the D-brane action can be expressed
as a noncommutative gauge theory. Here, noncommutative must not be
confused with non Abelian. It does not refer to the gauge group but
to a property of space. Before sketching the connection to
string theory, we will briefly give some basic ingredients of
noncommutative field theory. In difference to commutative field theory
it 
is assumed that the coordinates of
${\mathbb R}^n$ do not commute (we indicate this by putting a hat on
the coordinate)  
\begin{equation}
\left[ \hat{x}^i , \hat{x}^j\right] = i\theta^{ij},
\end{equation}
where we restrict to the case that $\theta^{ij}$ are c-numbers.
Because of the non commuting coordinates one has to specify the
ordering in say complex functions. For our purpose the Weyl ordering
is appropriate. The Weyl ordering is constructed as follows. The
starting point is the pair of the function and its Fourier transform
in commutative space (first with commuting coordinates)
\begin{equation}\label{fourp}
\phi\left( x\right) = \frac{1}{\left(
    2\pi\right)^{\frac{n}{2}}} \int d^n k\, e^{ikx}\tilde{\phi}\left(
    k\right) . 
\end{equation}
The Weyl ordered functions are defined by replacing the commuting
coordinates $x^i$ with the non commuting ones $\hat{x}^i$
 in (\ref{fourp}) (but keeping $k$ as a commutative
integration variable), 
\begin{equation}\label{weylo}
\phi_W\left( \hat{x}\right) = \frac{1}{\left(
    2\pi\right)^{\frac{n}{2}}} \int d^n k\, e^{ik\hat{x}}\tilde{\phi}\left(
    k\right) . 
\end{equation}
A natural prescription to multiply two Weyl ordered 
functions is
\begin{equation}\label{moyawe}
\left( \phi_W \star \psi_W\right)\left( \hat{x}\right)\equiv
\phi\left(\hat{x}\right) \psi\left( \hat{x}\right) = \frac{1}{\left(
    2\pi\right)^n}
\int d^n k d^nq\, e^{-i(k+q)\hat{x}}e^{ik\hat{x}}\tilde{\phi}\left( q+
    k\right)\tilde{\psi}\left( -k\right) .
\end{equation}
Multiplying the two exponentials on the rhs of (\ref{moyawe}) using the
BCH formula and afterwards dropping the hat on the coordinates leads
to a natural way to deform the algebra of ordinary functions
($\phi$: commuting ${\mathbb R}_n\rightarrow {\mathbb C}$) by
replacing 
the ordinary product by the Moyal product
\begin{equation}\label{moyalpro}
\left(\phi \star \psi\right)\left( x\right)  
= e^{i\theta^{ij} \frac{\partial}{\partial
    x^i}\frac{\partial}{\partial y^j}}\phi\left( x\right) \psi\left(
  y\right)_{|x=y} .
\end{equation}
This deformed algebra is noncommutative but still associative. In the
limit $\theta^{ij} \rightarrow 0$ it becomes the familiar commuting
algebra (ordinary multiplication in ${\mathbb C}$). 

Noncommutative field theories -- as we will meet them on D-branes-- 
are roughly obtained as follows. One takes the ordinary action for the 
field theory and replaces products of fields by the Moyal product
(\ref{moyalpro}). (This is only a very rough prescription since for
example any ``zero'' can be expressed as the commutator with respect
to the ordinary product which becomes something non-trivial after the
deformation. An additional principle is for example given by the
requirement that the deformed action should posses the same (but
possibly deformed) symmetries as the commutative one.)  

Our starting point for connecting D-branes to noncommutative field
theory is a slightly rescaled version of the non linear sigma model
(\ref{opensigma})\footnote{The index $i$ instead of $\mu$ indicates
  here that we focus on space like target space dimensions. The
  rescalings of fields have been done mainly in order to achieve
  agreement with the literature\cite{Seiberg:1999vs}.}
\begin{equation}\label{opensigma-prime}
S = \frac{1}{4\pi\alpha^\prime}\int_{\cal M} d^2 z \left(
  G_{ij}\partial_\alpha X^i \partial^\alpha X^j - 2\pi i \alpha^\prime
  B_{ij} \epsilon^{\alpha\beta}\partial_\alpha X^i \partial_\beta
  X^j\right) .
\end{equation}
A possible $U(1)$ gauge background could be absorbed into
$B_{ij}$ by use of Stoke's theorem. However, we will restrict first
to the case that $B_{ij}$ is constant, and add a $U(1)$ gauge field
coupling to the boundary later.
Note also that $B_{ij}$ has now mass dimension two -- the
canonical dimension of a gauge field strength. We consider the case
that all coordinates $X^i$ have Neumann boundary conditions
(coordinates with Dirichlet boundary conditions do not play a role
here and may be added as spectators).
The propagator for the $X^i$ can be easily obtained from the expressions
(\ref{nappiyostprop}). 
The redefined quantities are
\begin{eqnarray}
\left(\hat{G}^{-1}\right)^{ij} &=& \left( \frac{1}{G+2\pi
    \alpha^\prime B}\right)_S ^{ij} = \left( \frac{1}{G +
    2\pi\alpha^\prime B} G\frac{1}{G-2\pi\alpha^\prime
    B}\right)^{ij} ,\\
\hat{G}_{ij} & = & G_{ij} -\left( 2\pi\alpha^\prime\right)^2 \left(
    BG^{-1}B\right)_{ij} ,\\
\theta^{ij} & = & 2\pi\alpha^\prime \left( \frac{1}{G+2\pi\alpha^\prime
    B}\right)^{ij} _A \nonumber \\ & = &  -\left(
    2\pi\alpha^\prime\right) ^2 \left( 
    \frac{1}{G + 2\pi\alpha^\prime B}B\frac{1}{G-2\pi\alpha^\prime
    B}\right)^{ij} .\label{thetaeijey}
\end{eqnarray}
In particular the open string ends propagate
according to (call $z^1 = \tau$)
\begin{equation}\label{seiwiprop}
\langle X^i\left( \tau\right)X^j\left(
  \tau^\prime\right)\rangle
= -\alpha^\prime \left(\hat{G}^{-1}\right)^{ij}\log \left( \tau
  -\tau^\prime\right)^2 + 
  \frac{i}{2}\theta^{ij}\, \epsilon\left( \tau -\tau^\prime\right),
\end{equation} 
where the epsilon function is equal to the sign of its argument, and
zero for vanishing argument. 

Let us pause for a moment and explain how the last term in
(\ref{seiwiprop}) arises. The propagator (\ref{nappiyostprop})
contains a term ( a factor of $\alpha^\prime$ appears now in the
definition of $\theta^{ij}$ (\ref{thetaeijey}))
\begin{equation}\label{lastinnapyo}
-\frac{\theta^{ij}}{2\pi}\log\frac{z -\bar{z}^\prime}{\bar{z} -
 z^\prime} .
\end{equation}
We take $z = \tau +i\sigma$ (hoping that this does not cause confusion
due to the fact that now $\tau$ and $\sigma$ parameterize the upper
half plane whereas they parameterized a strip earlier (and will so in
later sections)). Ordering with 
respect to real and imaginary part, one obtains for (\ref{lastinnapyo})
\begin{equation}
-\frac{\theta^{ij}}{2\pi}\log \left( \frac{ \left( \tau
 -\tau^\prime\right)^2 + 2i \left( \sigma +\sigma^\prime\right)\left(
 \tau -\tau^\prime\right)}{\left( \tau -\tau^\prime\right)^2 +\left(
 \sigma +\sigma^\prime\right)^2}\right) .
\end{equation}
Using the relation 
$$\log z = \log \left| z\right| + i\, arg\left( z\right) $$
and taking the limit $\sigma + \sigma^\prime \to + 0$ one obtains
\begin{equation}
-\frac{i}{2}\theta^{ij} \left( 1 - \epsilon\left( \tau -
 \tau^\prime\right)\right) .
\end{equation}
Dropping an irrelevant constant, this yields the last term in
(\ref{seiwiprop}). 

In the following we will be interested in the $\alpha^\prime \to 0$
limit (while keeping $\theta^{ij}$ fixed), where the propagator
(\ref{seiwiprop}) takes the form 
\begin{equation}
\langle X^i\left( \tau\right) X^j\left( 0\right)\rangle =
\frac{i}{2}\theta^{ij}\epsilon\left(\tau\right) .
\end{equation}
With this propagator one can compute the
following operator product
\begin{equation}
:e^{ip_ix^i\left( \tau\right)}::e^{iq_ix^i\left( 0\right)}:\, =
  e^{-\frac{i}{2}\theta^{ij} p_i q_j \epsilon\left(\tau\right)} :
  e^{ip_i X^i\left(\tau\right) + i q_i X^i\left( 0\right)}:,
\end{equation}
where the normal ordering means that self contractions within the
exponentials are subtracted. By use of Fourier transformation one
can deduce the operator product for generic functions
\begin{equation}
:\phi\left( X\left(\tau\right)\right)::\psi\left( X\left(
    0\right)\right):\, =\,
    :e^{\frac{i}{2}\theta^{ij}\frac{\partial^2}{\partial X^i\left(
    \tau\right)\partial X^j\left( 0\right)}}\phi\left(
    X\left(\tau\right)\right)\psi\left( X\left( 0\right)\right): .
\end{equation}
In the limit of coincident arguments the operator product can be
related to the Moyal product
\begin{equation}\label{conom}
\lim_{\tau \to +0} :\phi\left( X\left(\tau\right)\right)\psi\left(
  X\left( 0\right)\right) :\, =\, \left(\phi \star \psi\right)\left(
  X\left(0\right)\right) .
\end{equation}
This expression suggests that we are likely to obtain noncommutative
field theory if we use the limiting procedure on the lhs of
(\ref{conom}) as a way to regularize composite operators. This
regularization technique is known as point splitting. In composite
operators well defined (normal ordered) parts are taken at different
points, and then the limit to coinciding points is performed (after
adding counterterms if needed). 

In the following we are going to argue that we obtain an effective
noncommutative theory on the D-brane if we use the point splitting
regularization instead of dimensional
(or 
Pauli-Villars) regularization. For
a trivial worldsheet metric point splitting simply means that we cut
off short distances by keeping
\begin{equation}\label{pointsplit}
\left| \tau - \tau^\prime\right| > \delta ,
\end{equation}
and take $\delta$ to zero in the end.
First, we add the following interaction term to
(\ref{opensigma-prime})
\begin{equation}
S_{int} = -i\int d\tau A_i\left( X\right) \partial_\tau X^i .
\end{equation}
Classically this term is invariant under a gauge transformation
\begin{equation}
\delta A_i = \partial_i \lambda .
\end{equation}
Now, we are going to observe that whether or not the partition
function is invariant depends on the regularization prescription. To
this end, note that $\delta Z$ contains a term
\begin{equation}
\delta Z  = -\langle \int d\tau A_i\left( X\right)\partial_\tau X^i
\cdot \int d\tau^\prime \partial_{\tau^\prime}\lambda\rangle +\ldots .
\end{equation}
Schematically this integral has the form
\begin{equation}
\int d\tau \int d\tau^\prime \partial_{\tau^\prime} f\left( \langle
  X^i\left(\tau\right) X^j \left( \tau^\prime\right)\rangle\right) =
\left. \int d\tau f\left( \langle
  X^i\left(\tau\right) X^j \left(
  \tau^\prime\right)\rangle\right)\right|_{\tau^\prime = \tau
  +\delta}^{\tau^\prime =\tau-\delta}.
\end{equation} 
If we treat the divergence at $\tau =\tau^\prime$ 
with dimensional regularization (as we did in section \ref{sec:openbeta})
this expression vanishes since it does not matter from which side we
approach the singularity. (The epsilon function in the propagator is
zero at $\tau =\tau^\prime$ and the logarithms are replaced by the
regularized expressions.) 

If, however, we choose the point splitting method (\ref{pointsplit})
instead, we obtain 
\begin{eqnarray}
\delta Z &=& -\int d\tau : A_i\left(X\left(\tau\right)\right)\partial_\tau
X^i\left(\tau\right): :\lambda\left( X\left( \tau -0\right)\right)
  -\lambda\left( X\left( \tau +0\right)\right)\nonumber \\
& = & -\int d\tau :\left( A_i \star \lambda -\lambda \star A_i\right)
\partial_\tau X^i: + \ldots ,
\end{eqnarray}
where in the second step the connection between the operator product
and the Moyal product (\ref{conom}) has been used.
Hence, when using the point-splitting regularization
(\ref{pointsplit}), the string partition function is not invariant
under ordinary gauge transformations. However, the lack of invariance
can be cured by replacing the gauge field $A_i$ with a
``noncommutative'' gauge field $\hat{A}_i$ with the deformed gauge
transformation 
\begin{equation}
\hat{\delta}\hat{A}_i = \partial_i \lambda + i\lambda \star \hat{A}_i
- i\hat{A}_i\star \lambda .
\end{equation}
Such a transformation is a gauge symmetry in the noncommutative
version of ($U\left( 1\right)$) Yang-Mills theory. 
The gauge invariant field strength is
\begin{equation}
\hat{F}_{ij} = \partial_i \hat{A}_j -\partial_j\hat{A}_i -i
\hat{A}_i\star \hat{A}_j + i\hat{A}_j\star \hat{A}_i .
\end{equation}
Indeed, computing
the effective action of the open string with the point-splitting
method, one finds the noncommutative version of the Dirac-Born-Infeld
action (\ref{dbiaction}). We will not go through the details here, but
refer the interested reader to\cite{Seiberg:1999vs} and further
references to be given in the end of this review.

The effective D-brane action was obtained by setting open string beta
functions to zero. Now, we have seen that the outcome can depend on
the way we regularize singularities: commutative Dirac-Born-Infeld
e.g.\ for dimensional regularization and noncommutative
Dirac-Born-Infeld for point-splitting. From quantum field theory it is
known that beta functions which differ by the way of
renormalization should be identical up to redefinitions of the
couplings. In our example the couplings are $A_i$ in the commutative
case, and $\hat{A}_i$ in the noncommutative one. Therefore, there should
exist a field redefinition relating commutative gauge theory to
noncommutative one. Indeed, such a field redefinition has been
found in\cite{Seiberg:1999vs}, it is sometimes called the
Seiberg-Witten map. 

The connection between D-branes and noncommutative field theory has
many interesting aspects, which we will, however not further discuss
in this review.

%%% Local Variables: 
%%% mode: latex
%%% TeX-master: t
%%% End: 

\section{Orientifold fixed planes}\label{orientifolds}

In this section we will introduce an extended object which is called
orientifold fixed plane. This is nothing but the orbifold plane of
section \ref{sec:orbifold} whenever the corresponding discrete target
space mapping is combined with a worldsheet parity inversion. (Recall
that an orbifold fixed plane was defined as an object being invariant
under an element of a discrete group acting on the target space.) 

At first we will study unoriented closed (type II) strings. These are
closed strings which can be emitted or absorbed by an orientifold
fixed plane. Afterwards we will investigate how orientifold fixed
planes interact via closed strings. We will learn that orientifold
fixed planes carry tension and RR charges. In particular, RR charge
conservation implies that orientifold fixed planes cannot exist
whenever they possess compact transverse dimensions. However, by
adding D-branes one can construct models containing orientifold planes
with 
transverse compact dimensions. Such constructions are known as
orientifold compactifications. We will present the type I theory and
an orientifold analogon of the K3 orbifold discussed in section
\ref{k3orbi}. (Type I theory is actually not a compactification. Here,
the orientifold planes are space filling and do not have transverse
dimensions. However, the construction falls into the same category as
orientifold compactifications.)

\subsection{Unoriented closed strings}\label{unorclosed}
\setcounter{equation}{0}
Recall the mode expansions for type II strings (now with $0\leq\sigma
< 2\pi$). The general solution to the equation of motion for the bosons
is 
\begin{equation}
X^\mu = X^\mu _R\left(\sigma^-\right) + X^\mu _L\left( \sigma^+\right)
,
\end{equation}
with
\begin{eqnarray}
X^\mu _R &=& \frac{1}{2}x^\mu +\frac{1}{2} p^\mu \sigma^- +
\frac{i}{2}\sum_{n\not= 0}\frac{1}{n}\alpha_n ^\mu e^{-in\sigma^-} , \\
X^\mu _L &=& \frac{1}{2}x^\mu +\frac{1}{2}p^\mu\sigma^+
+\frac{i}{2}\sum_{n\not= 0} \frac{1}{n}\tilde{\alpha}_n ^\mu
e^{-in\sigma^+} .
\end{eqnarray}
The mode expansions for the worldsheet fermions are
\begin{eqnarray}
\psi_- ^\mu & = & \sum_{n\in{\mathbb Z}}d_n ^\mu e^{-in\sigma^-}, \\
\psi_+ ^\mu & = & \sum_{n\in{\mathbb Z}}\tilde{d}_n ^\mu
e^{-in\sigma^+},
\end{eqnarray}
in the R sectors, and
\begin{eqnarray}
\psi_- ^\mu & = & \sum_{r\in {\mathbb Z}+\frac{1}{2}}b_r ^\mu
e^{-ir\sigma^-},\\
\psi_+ ^\mu & = & \sum_{r\in{\mathbb Z}+\frac{1}{2}}\tilde{b}_r ^\mu
e^{-ir\sigma^+} 
\end{eqnarray}
in the NS sectors. 

We define an operator $\Omega$ which changes the orientation of the
worldsheet. For the closed string the action of $\Omega$ is
\begin{equation}
\Omega : \; \sigma \leftrightarrow -\sigma .
\end{equation}
For left handed fermionic modes, we introduce an additional
sign such that the product of a left with a right handed fermionic
mode is $\Omega$ invariant (recall that fermionic modes from the left
moving sector anti-commute with fermionic modes from the right moving
sector). In formul\ae , this means 
\begin{equation}
\begin{array}{c c c}
\Omega \alpha_n ^\mu \Omega^{-1} = \tilde{\alpha}_n ^\mu , & \Omega
b_r ^\mu\Omega^{-1} 
= \tilde{b}_r ^\mu , & \Omega \tilde{b}_r ^\mu \Omega^{-1} = -b_r ^\mu
, \\ 
& \Omega d_n ^\mu\Omega^{-1} =\tilde{d}_n ^\mu ,&\Omega \tilde{d}_n
\Omega^{-1} = -d^\mu _n .
\end{array}
\label{omega-trafo}
\end{equation}
From this we see that $\Omega$ is a symmetry in type IIB theory --
the only closed superstring which is left-right symmetric.
(Note that the GSO projection operator (\ref{rgso}) in the R sector
contains an 
even number of $d_0$'s. Hence, the sign in the transformation
(\ref{omega-trafo}) cancels out and e.g.\ $P_{GSO}^+$ is interchanged
with $\tilde{P}_{GSO}^+$.) 

Let us study the action of $\Omega$ on the massless sector of type IIB
excitations. We take the vacuum to be invariant under worldsheet
parity reversal. The massless NSNS sector states are (in light cone
gauge) 
\begin{equation}
b_{-\frac{1}{2}}^i \tilde{b}_{-\frac{1}{2}}^j\left| k\right>
\end{equation}
The action of $\Omega$ on this state interchanges the indices $i$ and
$j$. Thus the states surviving an $\Omega $ projection are symmetric
in $i,j$ -- these are the graviton $G_{ij}$ and the dilaton $\Phi$.
Since $\Omega$ relates the NSR with the RNS sector only invariant
superpositions are kept. Thus we obtain only one gravitino (56
components) and one dilatino (8 components). Half of the target space
supersymmetry is broken by the $\Omega$ projection. The massless
states in the RR sector are obtained from the tensor product of the
left with the right moving R vacuum. The R vacua are target space
spinor components and $\Omega$ interchanges the left with the right
moving vacuum. Because spinor components anti-commute the
antisymmetrized tensor product survives the $\Omega$ projection. This
is the 28 dimensional $SO(8)$ representation -- the antisymmetric
tensor $B^\prime _{ij}$. 
We obtain the field content of the heterotic string without the
internal fermions $\lambda_+ ^A$. As we stated before, a theory with
such a massless spectrum suffers from gravitational anomalies. In the
heterotic theory we actually needed 32 worldsheet fermions $\lambda_+
^A$ whose quantization provided exactly the gauge multiplets needed to
obtain an anomaly free massless spectrum. Later we will see that one
needs to add D-9-branes to the unoriented type IIB theory, for
consistency. Before going into that let us study for a while the
unoriented closed stringtheory -- even though it is not consistent
yet. 

The theory of unoriented type IIB strings contains
orientifold-nine-planes -- or short O-9-planes. An O-plane is a set of
target space points which is fixed under an element of a discrete
group which contains $\Omega$ (the element must contain
$\Omega$). Because $\Omega$ alone does not act on the target space
geometry the full target space is fixed under $\Omega$. The fixed set
of points is space filling -- it is an O-9-plane. 

We have seen that when we compactify the type IIB string on a circle
and perform a T-duality we obtain type IIA theory compactified on a
circle with inverted radius. Let us study what happens to the
O-9-planes in this process. Formally, we have the expression ($X^9$
stands for the bosonic string coordinate)
\begin{equation}
\Omega X^9 \Omega ^{-1} \stackrel{\mbox{\tiny T-DUALITY}}{\longrightarrow} 
T\Omega T^{-1}  T X^9 T^{-1} \left(T\Omega T^{-1}\right)^{-1} .
\end{equation}
We want to know the T-dual of $\Omega$ which is denoted by $T\Omega
T^{-1}$. This can be computed as follows. We first perform a
T-duality, then act with $\Omega$ on the T-dual coordinate, and
finally T-dualize back. These steps are collected in the following
diagram (we use (\ref{recepy}) for T-duality)
\begin{equation}\label{TomT}
X^9 _L + X^9 _R \stackrel{ T}{\longrightarrow} X^9_L -
X^9_R\stackrel{\Omega}{\longrightarrow} X^9 _R - X^9 _L
\stackrel{T^{-1}}{\longrightarrow} -X^9 _R - X^9 _L
\end{equation}
Thus we see that $T\Omega T^{-1}$ reflects the dimension in which T
acts, and also interchanges left with right movers (the second
statement can be easily verified by drawing the diagram (\ref{TomT})
for the left or right moving piece alone). Thus, for T-duality in
$X^9$ direction we can write
\begin{equation}
T\Omega T^{-1} = R_9 \Omega ,
\end{equation}
where $R_9$ is the ${\mathbb Z}_2$ element
\begin{equation}
R_9: X^9 \rightarrow -X^9 .
\end{equation}
The action on the worldsheet fermions can be studied likewise. Now we
go to the decompactification limit on the type IIA side. Instead of an
O-9-plane we have an O-8-plane, because now only points with $X^9 =0$
are fixed under the action of $\Omega R_9$. Repeating this
argumentations for more than one T-dualized circle we conclude that we
have O-p-planes with even (odd) p in type IIA (B) theory. For an
O-p-plane with even p, $\Omega$ comes combined with a ${\mathbb Z}_2$
operator reflecting an odd number of dimensions. In particular, this
combination interchanges e.g.\ $P_{GSO}^+$ with $\tilde{P}_{GSO}^-$,
i.e.\ it is indeed a symmetry of type IIA strings. The closed string
is unoriented only when it is 
located on an O-plane. A string off the O-plane is oriented. Its
counterpart with the opposite orientation is the $R_9$ image
of the string.   

\subsection{O-plane interactions}
\setcounter{equation}{0}

An O-plane is defined as an object where closed strings become
unoriented when they hit it. Topologically this can be depicted by a
crosscap as illustrated in figure \ref{fig:crosscap}.

%%%%%%%%%%%%%%%%%%%%%%%%%%%%%%%%%%%%
\begin{figure}
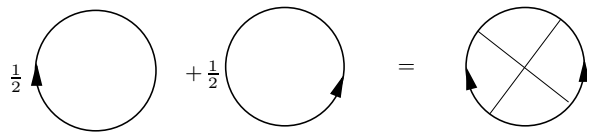
  
\begin{center}
\input crosscap.pstex_t
\end{center}
\caption{The superposition of two strings with opposite orientation can
  be viewed as a crosscap. The crosscap is a circle with diagonally
  opposite points being identified.}
\label{fig:crosscap}
\end{figure}
%%%%%%%%%%%%%%%%%%%%%%%%%%%%%%%%%%%%

The opposite process is a crosscap decaying into a pair of strings with
different orientations. Only one string out of this pair is physical
-- the other one is the $\Omega R$ image, where $R$ now stands for a
target space mapping leaving the O-plane fixed. Thus O-planes can emit
or absorb oriented strings. They possibly interact via the exchange of
closed oriented strings. This indicates that there is an interaction
among O-planes and between D-branes and O-planes. We are going to
study these interactions in the following two subsections. 

\subsubsection{O-plane/O-plane interaction, or the Klein bottle}
In figure
\ref{fig:o-charge} we have drawn a process in which two O-planes
interact via the exchange of closed strings. We restrict to the
special case that the orientifold group element $\Omega R$ squares to
one. (Combining orbifold compactifications with orientifolds, one can
have the more general situation that the orientifold group elements
square to a nontrivial orbifold group element. This has to be the
same for the two O-planes. Then a twisted sector closed string is
exchanged.) 

%%%%%%%%%%%%%%%%%%%%%%%%%%%%%%%%%%%%
\begin{figure}
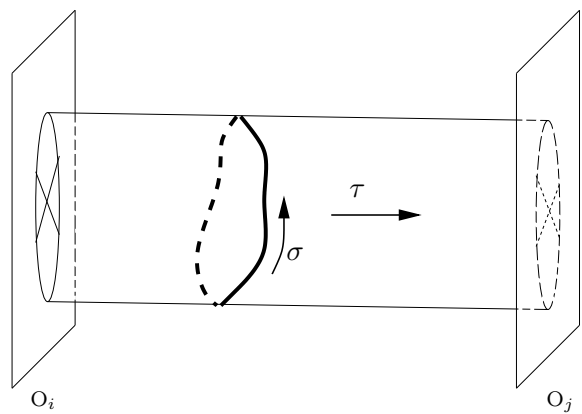
  
\begin{center}
\input o-charge.pstex_t
\end{center}
\caption{O-plane O$_i$ and O-plane O$_j$ talking to each other by
  exchanging a closed string.}
\label{fig:o-charge}
\end{figure}
%%%%%%%%%%%%%%%%%%%%%%%%%%%%%%%%%%%%

In the following we are going to compute this process. As in the
D-brane computation, we take the O-planes to be parallel. The range for 
the worldsheet coordinates is
\begin{equation}
0\leq \sigma < 2\pi \; , \; 0\leq \tau < 2\pi l .
\end{equation}
Like in section \ref{dbraneinteractions} we want to perform the
computation in the scheme where the role of $\tau$ and $\sigma$ are
interchanged -- i.e.\ in the worldsheet dual channel. 
In this dual channel, a virtual pair of closed strings pops out of the
vacuum -- one of the strings changes its orientation before they
rejoin. Therefore, this is called the loop channel.
Before performing the transformation to the loop channel, we need to
describe the tree channel process fig.\ \ref{fig:o-charge} such that
it is periodic in time. The method of doing so differs slightly from
the D-brane/D-brane interaction. It is best explained 
by looking at
the triangulated version of picture \ref{fig:o-charge} and its double
cover which is a torus. We draw this in figure \ref{fig:triangu}.   
In the left picture, the shaded region is the triangulated version of
figure \ref{fig:o-charge}. 
%%%%%%%%%%%%%%%%%%%%%%%%%%%%%%%%%%%%
\begin{figure}
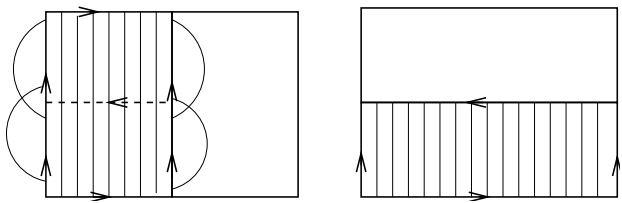
  
\begin{center}
\input triangu.pstex_t
\end{center}
\caption{The triangulated version of fig. \ref{fig:o-charge} on the
  left. By manipulations preserving the topology this can be mapped
  onto the triangulated version of a Klein bottle on the right.}
\label{fig:triangu}
\end{figure}
%%%%%%%%%%%%%%%%%%%%%%%%%%%%%%%%%%%%
The half-circles indicate the identifications of the crosscaps.
The white region shows our intention to obtain a description which is
periodic in $\tau$, with a period $4\pi l$. 
Now, one cuts the shaded part along the
dotted line (with the indicated orientation), and flips the upper
rectangle once around its right vertical edge and afterwards shifts
it down in the vertical direction.  We obtain a process which is indeed
periodic in $\tau$, and now $\tau \in \left[ 0, 4\pi l\right)$. 
(This periodicity appears due to the crosscap identifications indicated
by the left half circles. The crosscap identifications for the other
O-plane ensure that one can glue the upper rectangle to the lower one
after the flip and the shift.) 
It is
difficult to describe this process as a tree channel closed string
exchange. Instead we can interchange the roles of $\sigma$ and
$\tau$. Then the interpretation is that a pair of closed strings of
length $4\pi l$ pops out of the vacuum, one of the closed strings
changes its orientation before they annihilate after a worldsheet time
$\pi$. This is a vacuum loop amplitude which has the topology of the
Klein bottle. The parameter range is
\begin{equation}
0\leq \sigma < 4\pi l \; ,\; 0\leq \tau < \pi .
\end{equation}
As in the D-brane case (section \ref{dbraneinteractions}) we want to
rescale the dualized worldsheet coordinates such that their ranges are
the canonical ones, which (now for the closed string) are
\begin{equation}\label{canpark}
0\leq \sigma < 2\pi  \; ,0\leq \tau < 2\pi t .
\end{equation}
This can be achieved by the redefinitions
\begin{equation}
\tau \rightarrow \tau 2t \; , \sigma \rightarrow
\frac{\sigma}{2l} .
\end{equation}
For the Hamiltonian this induces a rescaling
\begin{equation}
H \rightarrow 2l \left( 2t\right)^2 H .
\end{equation}
Analogous to the annulus discussion in section
\ref{dbraneinteractions} we 
require that the action of the rescaling on the time evolution
operator is 
\begin{equation}
e^{-\pi H}\rightarrow e^{-2\pi t H} .
\end{equation}
This yields a relation between $l$ and $t$
\begin{equation}\label{ltklein}
lt =\frac{1}{4} .
\end{equation}
Periodic boundary conditions on fermions along the vertical axis of
the lhs of figure \ref{fig:triangu} correspond to a $\left(-\right)^F=
\left( -\right)^{\tilde{F}}$
insertion whenever the vertical axis is identified with the worldsheet
time on the rhs of figure \ref{fig:triangu}. Only closed strings for
which 
the rightmoving $\left( -\right)^F$ eigenvalue equals the leftmoving
one contribute to the Klein bottle amplitude. (In the R sector an
additional sign may occur depending on whether we are looking at type
IIA or IIB strings. This does not matter here since the 
R-sector contributions with a $\left( -\right)^F$ insertion vanish
anyway.) 

Since the connection between tree level periodicities and loop channel
insertions is a bit less obvious than in the D-brane/D-brane
interaction, let us explain it in some detail here. We take the
parameter range (\ref{canpark}). We are interested in the behavior of
worldsheet fermions
under shifts in $\tau$ by $4\pi t$. Periodic behavior corresponds to
tree level RR exchange whereas anti-periodicity translates to NSNS
exchange. Fig.\ \ref{fig:triangu} tells us how to continue in $\tau$
beyond $2\pi t$
\begin{equation}\label{contfour}
\psi_{\pm}^\mu\left( \tau + 4\pi t , \sigma\right) =\left(
  -\right)^{F+\tilde{F}} 
  \psi_{\pm}^\mu\left( \tau + 2\pi t, 
  2\pi -\sigma\right) ,
\end{equation}
where the $\left( -\right)^{F+\tilde{F}}$ reflects the boundary
condition on worldsheet fermions under $2\pi t$ shifts in $\tau$.
However, in the Klein bottle amplitude only states with $\left(
  -\right)^F =\left( -\right)^{\tilde{F}}$ contribute because of an
$\Omega$ insertion in the trace over states. Therefore, the additional
factor in (\ref{contfour}) is not relevant.
Now, the $2\pi t$ shift in $\tau$ can be replaced by acting with the
trace insertion $\left( -\right)^{F} \Omega$. The $\left( -\right)^F$
insertion just cancels a sign included in the definition of the trace
over fermions for the right movers. By the same token we have to
insert a $\left( -\right)^{\tilde{F}}$ for the left movers.
Thus we obtain
\begin{equation}
\psi_{\pm}^\mu \left( \tau + 4\pi t , \sigma\right) =\left(
  -\right)^{\tilde{F}} 
  \Omega \psi_{\pm}^\mu\left( \tau , 
  2\pi -\sigma\right) \Omega^{-1} = \psi_{\pm}^\mu \left( \tau ,
  \sigma\right) , 
\label{RRident}
\end{equation}
where in the last step we used our definition for $\Omega$
(\ref{omega-trafo}) and the $2\pi$ periodicity in $\sigma$. Thus the
$\left( -\right)^F$ insertion in the 
loop channel filters out the RR tree level exchange, indeed.
Strictly speaking the above consideration is correct only when the
fermions point in directions longitudinal to the O-plane (where the
${\mathbb Z}_2$ reflection $R$ acts as the identity). For the other 
directions there are two signs canceling each other and leading to the
same result. At first, there is an additional minus sign in
(\ref{contfour}) because the half-circles in fig.\ \ref{fig:triangu}
now contain also the (non-trivial) action of the ${\mathbb Z}_2$
reflection $R$. This sign is canceled when we replace $\Omega$ by
$\Omega R$ in (\ref{RRident}).

We want to filter out the contribution due to RR exchange in the tree
channel. 
Then, the loop channel vacuum amplitude is given by the following
expression
\begin{eqnarray}
\lefteqn{ \int \frac{dt}{2t}\;\mbox{Str}\left( \Omega R
    \frac{\left(-\right)^F}{2} 
    e^{-2\pi\alpha^\prime 
    t H}\right) =}\nonumber\\ 
&& \int \frac{dt}{2t} \mbox{tr}_{\mbox{\tiny
    ZERO}\atop \mbox{\tiny MODES}}\left( \Omega Re^{-2\pi\alpha^\prime
    tH_0}\right)
\mbox{tr}_{\mbox{\tiny BOSONS}}\left( \Omega Re^{-2\pi t\left( H_B
    +\tilde{H}_B\right)}\right)\nonumber\\ 
& & \mbox{tr}_{\mbox{\tiny NSNS}\atop \mbox{\tiny
    FERMIONS}}\left( \Omega R\frac{\left( -\right)^F}{2} e^{-2\pi t
    \left( H_{NS} + 
    \tilde{H}_{NS}\right)}\right) .
\end{eqnarray}
Here, we split the Hamiltonian into right and left moving parts $H$ +
$\tilde{H}$ and these in turn into
\begin{equation}
\alpha^\prime H =\alpha^\prime H_0 + H_B + H_{NS}
\end{equation}
with\footnote{From our treatment in section \ref{typeII} we
  would get a factor of $\alpha^\prime 8/2 =2$ instead of
  $\alpha^\prime 4/2 =1$ in the oscillator
  contributions. Recall, however, that we have changed the $\sigma$
  range from $\left[ 0,\pi\right)$ to $\left[ 0, 2\pi\right)$,
  meanwhile. We have distributed the zero mode contribution
  symmetrically on $H$ and $\tilde{H}$. Taking into account the effect
  of rescaling, this gives the factor of $1/4$.}   
\begin{eqnarray}
H_0 & = & \frac{p^2}{4} \\
H_B & = & \sum_{i=1}^8\left( \sum_{n=1}^\infty \alpha_{-n}^i\alpha_n
  ^i  -\frac{1}{24}\right) ,\\
H_{NS} & = & \sum_{i=1}^8\left( \sum_{r=\frac{1}{2}}^\infty r
  b_{-r}^i b_r ^i -\frac{1}{48}\right) ,
\end{eqnarray}
and the corresponding expressions for the left moving sector. 
The $\Omega R$ insertion projects out contributions of states with
zero mode momenta perpendicular to the O-planes, since those states
are mapped onto states with the negative momentum in the perpendicular
direction by the $\Omega R$ insertion. The result for the zero mode
contribution reads
\begin{equation}
\mbox{tr}_{\mbox{\tiny ZERO}\atop \mbox{\tiny MODES}} = 2V_{p+1} \int
\frac{d^{p+1}k}{\left( 2\pi\right)^{p+1}} \, e^{-2\pi \alpha^\prime t
  \frac{k^2}{2}} 
  = 2\cdot 2^{\frac{p+1}{2}}V_{p+1} 
\left( 8\alpha^\prime\pi ^2 
  t\right)^{-\frac{p+1}{2}} .
\end{equation}
Also here there is an additional factor of two, due to the possible
orientations of the closed string. (The trace is taken over oriented
strings with an $\Omega$ insertion. Another point of view would be
that one needs to add to the picture on the rhs of figure
\ref{fig:triangu} a picture with reversed orientations on the
horizontal edges.)  
For the traces over excited states we note that the insertion $\Omega
R$ in the trace means that only states contribute which are
eigenstates of $\Omega R$. This means that the left moving excitations
have to be identical to the right moving ones. Thus, it is
straightforward to modify the calculation presented in section
\ref{dbraneinteractions} by just changing the power of the arguments
in the functions \ref{thetafunctions} by two (since the identical left
and right moving contributions add).  
We obtain
\begin{equation}
\mbox{tr}_{\mbox{\tiny BOSONS}}\left( \Omega R e^{-2\pi t\left( H_B
        +\tilde{H}_B\right)} \right) = \frac{1}{f^8 _1\left( e^{-2\pi
        t}\right)} 
\end{equation}
for the trace over the bosonic excitations, and
\begin{equation}
\mbox{tr}_{\mbox{\tiny NSNS}\atop \mbox{\tiny FERMIONS}}\left( \Omega
  R \frac{\left( -\right)^F}{2}e^{-2\pi t\left(
  H_{NS}+\tilde{H}_{NS}\right)}\right)  =
  -\frac{1}{2}f_4 ^8\left( 
  e^{-2\pi t}\right).
\end{equation}
Thus we obtain
\begin{eqnarray}
\lefteqn{\int \frac{dt}{2t}\mbox{Str}\left( \Omega R \frac{\left(
 -\right)^F}{2} e^{-2\pi t H}\right) =}\nonumber\\
& &  -\frac{1}{2}2^{\frac{p+1}{2}} V_{p+1} \int \frac{dt}{2t} \left(
  8\alpha^\prime \pi^2 
  t\right)^{-\frac{p+1}{2}}\frac{f_4 ^8\left( e^{-2\pi t}\right)}{f_1
  ^8\left( e^{-2\pi t}\right)} .
\end{eqnarray}

Now we undo the worldsheet duality by expressing $t$ in terms of $l$
(\ref{ltklein}). We use the transformation properties
(\ref{modulartra}) and take the limit $l\to\infty$ in which the
contribution of massless closed string excitations dominates. This
leads to the expression 
\begin{equation}\label{KBresult}
-\frac{1}{2}\int_{l\to \infty} dl\, 2^{p+1}\, V_{p+1} \left(
  4\alpha^\prime  
  \pi^2\right)^{-\frac{p+1}{2}} l^{\frac{p-9}{2}}.
\end{equation}
We see that the result has almost the same structure as the one we
obtained for the D-brane/D-brane interaction in table
\ref{table:closedint}. (Recall that now, we 
separated out the RR sector exchange.) The differences are that we do
not have the exponential dependence on the distance and that we do
have an additional factor of $2^{p+1}$. The explanation for the
missing exponent is very simple. Since the orientifold planes are all
located at a fixed point of the ${\mathbb Z}_2$ action $R$, they
cannot be separated in target space. (However, we could for example
compactify the dimensions transverse to the brane. In that case
winding modes would play the role of the distance.) 

Before we can deduce the ratio of the O-plane RR to the D-brane RR
charge, we need to discuss a subtlety appearing because we have
modded out reflections in the transverse directions. This has the
effect that each transverse direction is ``half as long'' as in the
D-brane computation. The implications of this effect are best seen in
a field theory consideration. The field theory result gives a
``Coulomb potential'' which is of the structure
charge-squared times density. (The density appears as the inverse of a
second order differential operator.) The charge is obtained as an
integral over the transverse space (analogous to $Q=\int d^3 x j^0$ in
electro-dynamics). In the O-plane case this gives a factor of a half
per transverse direction as compared to the D-brane/D-brane
interaction. On the other hand the density is multiplied by a factor of
two per transverse direction. Hence, the overall net-effect of this
transformations is an additional factor of $2^{p-9}$ which we need to
put by hand into the O-plane/O-plane result, before we can compare it
with the D-brane/D-brane calculation.\footnote{If we performed a
  detailed field theory calculation we would find this factor due to
  the different target spaces (as argued in the text).
Later, we will
  compactify the transverse dimensions. Then this factor will appear
  ``automatically'' due to a Poisson resummation of the sum over the
  winding modes. This must be the case since in the compactified
  theory D-branes and O-Planes will have the same transverse
  space.\label{promise}}   
Taking this into account, the ratio of the D-p-brane RR charge $\mu_p$ to
the O-p-plane RR charge $\mu^\prime _p$
\begin{equation}\label{orr}
\mu^\prime _p = \mp 2^{p-4}\mu_p . 
\end{equation}
We cannot fix the sign by the present calculation since the charges
enter quadratically the expressions we derived so far.
Computing also the contributions without the $\left( -\right)^F$
insertions to the Klein bottle, one obtains the square of the O-plane
tension. Here, we infer the result by supersymmetry instead. Since the
$\Omega R$ projection leaves half of the supersymmetries unbroken, the
total one loop amplitude should vanish. This tells us that the ratio
of the D-brane tension $T_p$ to the O-plane tension $T^\prime _p$ is
\begin{equation}\label{ot}
T_p ^\prime = \mp 2^{p-4} T_p ,
\end{equation}
where at the present stage of the calculation the sign is not
known. In order to fix the signs in  (\ref{orr}) and (\ref{ot}) we
need to study the interaction between D-branes and O-planes. We will
do so in the next subsection.

\subsubsection{D-brane/O-plane interaction, or the M\"obius strip}

So far, we have seen that D-branes as well as O-planes interact via
the exchange of closed type II strings. This suggests that also
D-branes interact with O-planes. Such a process is depicted in figure
\ref{fig:treemoe}. We consider the case of parallel D-branes and
O-planes. This implies that the D-brane is located in directions where
the ${\mathbb Z}_2$ reflection acts with a sign and extended along the
other directions. 

%%%%%%%%%%%%%%%%%%%%%%%%%%%%%%%%%%%%
\begin{figure}
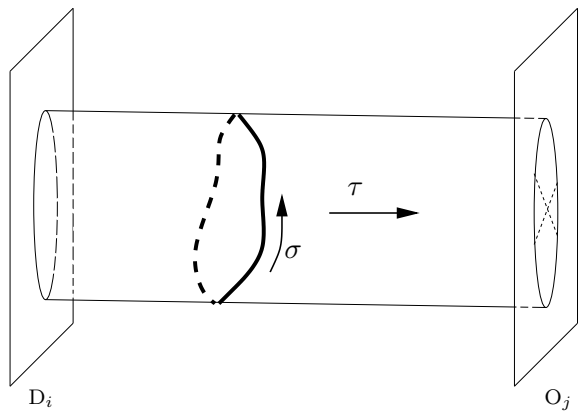
  
\begin{center}
\input treemoe.pstex_t
\end{center}
\caption{D-brane D$_i$ and O-plane O$_j$ talking to each other via the
  exchange of closed strings.}
\label{fig:treemoe}
\end{figure}
%%%%%%%%%%%%%%%%%%%%%%%%%%%%%%%%%%%%

Again, the range for the worldsheet coordinates is
\begin{equation}
0\leq \sigma < 2\pi \;\;\; ,\;\;\; 0\leq \tau < 2\pi l.
\end{equation}
In order to understand how to perform the worldsheet duality
transformation it is useful to study the triangulated version of the
diagram \ref{fig:treemoe}. The result of this investigation is drawn
in figure \ref{fig:triamoe}. The right picture is obtained by cutting
the left one along the dashed line flipping the upper rectangular
around its right edge and afterwards shifting it down. Looking at the
left picture with time passing along the vertical axis we see a
process in which a pair of open strings pops out of the vacuum. Both
ends of the strings are in the worldvolume of the brane D$_i$. As time
goes by one of the open strings changes its orientation before they
finally annihilate. The topology of this diagram is called M\"obius
strip (or M\"obius band).

%%%%%%%%%%%%%%%%%%%%%%%%%%%%%%%%%%%%
\begin{figure}
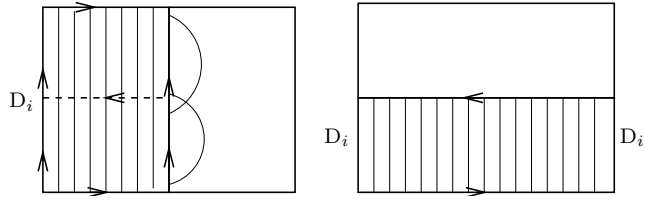
  
\begin{center}
\input triamoe.pstex_t
\end{center}
\caption{The triangulated version of fig.\ \ref{fig:treemoe} on the
  left. By manipulations preserving the topology this is mapped onto
  the triangulated version of a M\"obius strip on the right.}
\label{fig:triamoe}
\end{figure}
%%%%%%%%%%%%%%%%%%%%%%%%%%%%%%%%%%%%

The range for the worldsheet coordinates after interchanging the role
of time and space is
\begin{equation}
0\leq \sigma < 4\pi l\;\;\; ,\;\;\; 0\leq\tau < \pi ,
\end{equation}
whereas the canonical range for the open string parameters is
\begin{equation}
0\leq \sigma < \pi\;\;\; , \;\;\; 0\leq \tau < 2\pi t .
\end{equation}
Hence, we perform the rescaling
\begin{equation}
\tau \rightarrow \tau 2t \;\;\; ,\;\;\; \sigma \rightarrow
\frac{\sigma}{4l} .
\end{equation}
For the Hamiltonian this induces
\begin{equation}
H\rightarrow 4l\left( 2t\right)^2 H .
\end{equation}
Finally the time evolution operator should take its canonical form
\begin{equation}
e^{-\pi  H} \stackrel{!}{\rightarrow} e^{-2\pi t H}.
\end{equation}
This tells us how to relate $l$ and $t$
\begin{equation}\label{texpl}
lt = \frac{1}{8}.
\end{equation}

Now, we would like to identify which of the loop channel contributions
corresponds to an RR exchange in the tree channel. Periodicity under
$4\pi t$ shifts in $\tau$ translates to RR tree level exchange and
anti-periodicity to NSNS tree level exchange. We use fig.\
\ref{fig:treemoe} to identify
\begin{equation}\label{moebid}
\psi^\mu_{\pm}\left( \tau + 4\pi t, \sigma\right)  =
  \left( -\right)^F\psi^\mu_{\pm}\left( 
  \tau + 2\pi t,\pi -\sigma\right) ,
\end{equation}
where the factor of $\left( -\right)^F$ appears due to the
anti-periodic boundary condition of worldsheet fermions under shifts
of $2\pi t$ in $\tau$.
Let us study the case where we insert in the loop channel trace just
$\Omega$ (possibly combined with some target space ${\mathbb Z}_2$
action which we will discuss below). Taking into account the
sign when a trace is taken over worldsheet fermions we obtain 
\begin{eqnarray}
\psi^\mu_{\pm}\left( \tau + 4\pi t, \sigma\right) & = &
 \Omega \psi_{\pm}^\mu \left( \tau , \pi
  -\sigma\right) \Omega^{-1} \nonumber\\
&=&  \Omega \psi^\mu _\mp \left( \tau , \sigma
  -\pi\right)\Omega^{-1} , 
\label{dedutria}
\end{eqnarray}
where in the second step we have used the mode expansions
(\ref{openfour})--(\ref{openfourt}). In the open string sector we
define the action of $\Omega$ as taking $\sigma \rightarrow \pi
-\sigma$.
This finally results in
\begin{eqnarray}
\psi^\mu_{\pm}\left( \tau + 4\pi t, \sigma\right) & = & \psi_{\mp}^\mu
\left( \tau , 2\pi -\sigma\right) \nonumber \\
& = & \psi_{\pm}^\mu\left(\tau , \sigma - 2\pi\right) ,
\label{Rboundaryc}
\end{eqnarray}
where once again the mode expansion has been used. We deduce that open
string R sector contributions correspond to closed string RR
exchange. (This can also easily be seen in the mode expansions
(\ref{openfour})--(\ref{openfourt})). The above consideration is
correct only in NN directions (in directions in which the D-brane
extends). For DD directions there are a couple of signs which cancel
each other such that one gets the same result. Since the ${\mathbb
  Z}_2$ reflection $R$ acts with a sign in those directions, the first
line in (\ref{dedutria}) receives an additional minus sign. Looking at
the mode expansion (\ref{openfour})--(\ref{openfourt}) in DD
directions we observe that this sign is canceled when going to the
second line in (\ref{dedutria}).  Because now we need to replace
$\Omega$ by $\Omega R$, the first line in (\ref{Rboundaryc}) contains
an additional minus sign which again is canceled by using the DD
mode expansion when going to the second line in (\ref{Rboundaryc}). 

In the above consideration we have only specified how $\Omega$ acts on
the oscillators, and not how it acts on the vacuum. (In the closed
string we tacitly took the vacuum as being 
invariant under $\Omega$ leaving in the NSNS sector the metric
invariant and projecting out the $B$-field.) The computations of the
D-brane/D-brane and the O-plane/O-plane interactions provided the
absolute values of the corresponding RR charges. The result for the
D-brane/O-plane calculation will give the product of the O-plane times
the D-brane charge. This should be compatible with our previous
result. As we will see in a moment this leaves the two choices that
the $\Omega R$ eigenvalue of the open string R-vacuum
is $\pm 1$. We 
will choose the minus sign. This corresponds to a D-brane.
The action on the NS sector can be
inferred by supersymmetry, i.e.\ it should be such that the complete
one loop M\"obius strip amplitude vanishes. The result is that the
massless states have eigenvalue minus one. (This holds as well
for Neumann directions as for Dirichlet directions, since a sign due
to the different mode expansions cancels a sign due to the non-trivial
action of $R$ in Dirichlet directions.)  

Now, we have collected all the necessary information needed to write
down the loop channel amplitude which gives the tree channel RR
exchange (recall that a $\left( -\right)^F$ insertion leads to a
vanishing R sector trace)
\begin{eqnarray}
\lefteqn{-\int \frac{dt}{2t}\mbox{tr}_R\Omega R\frac{1}{2} 
e^{-2\pi \alpha^\prime
    H}= }\nonumber\\
& & -\int\frac{dt}{2t}\mbox{tr}_{\mbox{\tiny ZERO}\atop \mbox{\tiny
    MODES}} \left( \Omega R e^{-2\pi t \alpha^\prime
    H_0}\right)\mbox{tr}_{\mbox{\tiny BOSONS}}\left( \Omega R e^{-2\pi
    t H_B}\right)\nonumber\\
& & \mbox{tr}_{\mbox{\tiny R}\atop \mbox{\tiny FERMIONS}}\left( \Omega
    R\frac{1}{2} e^{-2\pi t H_R}\right) .
\end{eqnarray}
The expressions for the Hamiltonians can be directly taken from
(\ref{openhama})--(\ref{openhame}) with the difference that we put
$y=0$ in (\ref{openham0}) (because of the $\Omega R$ insertion in the
trace only D-branes at distance zero from the O-plane
contribute). With this difference the trace over the zero modes gives
(see (\ref{openzerotr})
\begin{equation}
\mbox{tr}_{\mbox{\tiny
      ZERO}\atop\mbox{\tiny  MODES}} = 2 V_{p+1} \int \frac{d^{p+1}
      k}{\left( 2\pi\right)^{p+1}}\,
      e^{-2\pi t\alpha^\prime  k^2 
      } = 2V_{p+1}\,\left( 8\pi^2 \alpha^\prime
      t\right)^{-\frac{p+1}{2}}  ,
\end{equation}
From the mode expansion (\ref{neumannmode}), (\ref{dirichletmode}) we
learn that
\begin{equation}
\Omega R\, \alpha_{-n}^\mu\, \left( \Omega R\right) ^{-1} = \left(
  -1\right)^n \alpha_{-n}^\mu .
\end{equation}
Modifying the expression (\ref{numberpart}) accordingly we obtain
\begin{equation}
\mbox{tr}_{\mbox{\tiny BOSONS}}\; \Omega R e^{-2\pi t H_B} = e^{-i\pi
  \frac{2}{3}}\frac{1}{f_1 
  ^8\left( e^{-\pi (t +\frac{i}{2})}\right)} .
\end{equation}
The next step is to split the product over integers in the definition
of $f_1$ (\ref{thetafunctions}) into a product over even times a
product over odd numbers.  This gives finally
\begin{equation}
\mbox{tr}_{\mbox{\tiny BOSONS}}\; \Omega R e^{-2\pi t H_B} =
    \frac{1}{f^8 _1\left( 
    e^{-2\pi t}\right) f^8 _3\left( e^{-2\pi t}\right)} .
\end{equation}
The mode expansion on the fermions 
(\ref{openfour})--(\ref{openfourt})
yields
\begin{equation}
\Omega R\, d_n ^\mu \,\left( \Omega
    R\right)^{-1} = e^{i\pi n} d_n ^\mu .
\end{equation}
Manipulations analogous to the bosonic trace give the result (recall
that we have chosen the $\Omega R$ eigenvalue of the R vacuum to be
minus one)
\begin{equation}
-\mbox{tr}_{\mbox{\tiny R}\atop \mbox{\tiny FERMIONS}} \left(\frac{
 \Omega R }{2}e^{-2\pi t 
 H_R}\right) = f_2 ^8\left( e^{-2\pi t}\right) f_4
 ^8\left( e^{-2\pi t}\right) ,
\end{equation}
where the 16-fold degeneracy of the R vacuum has been taken into
account. 
We are interested in the contributions due to tree channel RR exchange
and have computed now everything we need to obtain the
result. However, in order to specify the action of $\Omega R$ on the
NS vacuum one needs to compute the tree channel NSNS exchange. The
requirement that this cancels the RR interaction determines the action
of $\Omega R$ on the open string NS vacuum. We leave this as an
exercise. The result is that the massless vector is odd under $\Omega
R$. In the computation of the open string NS sector trace it is useful
to apply the identity (\ref{abstruse}) on the $f$ functions with the
shifted arguments and afterwards to proceed as we did above, i.e.\ to
split the product in the definitions of the $f$'s into a product over
even and over odd numbers.

So far, we obtained the result
\begin{equation}
-\int\frac{dt}{2l}\Omega R\mbox{tr}_R \frac{1}{2}e^{-2\pi\alpha^\prime
 H} =  
 V_{p+1}\int\frac{dt}{2t}\left(
 8\pi^2\alpha^\prime t\right)^{-\frac{p+1}{2}}
\frac{f_2 ^8\left( e^{-2pi t}\right)f_4 ^8\left( e^{-2\pi
 t}\right)}{f_1 ^8\left( e^{-2\pi t}\right)f_3 ^8\left( e^{-2\pi
 t}\right)} .
\end{equation}
Expressing $t$ in terms of $l$ via (\ref{texpl}) and using the
properties (\ref{modulartra}) yields finally the tree channel infrared
asymptotics 
\begin{equation}\label{MBresult}
2\frac{1}{2} V_{p+1}\int_{l\to \infty} dl\left(
 4\pi^2\alpha^\prime\right)^{-\frac{p+1}{2}} 2^{p-4} l^{\frac{p-9}{2}}
 .
\end{equation}
This expression has to be compared with the second line (RR
contribution) of table \ref{table:closedint} and (\ref{KBresult}),
where (\ref{KBresult}) has to be multiplied with $2^{p-9}$ as
discussed earlier. (For the M\"obius strip we do not need to put such
a factor since there is a cancellation between the O-plane charge and
the density.) In (\ref{MBresult}) we have pulled out a
factor of 
two. If we take the D brane distance $y$ to zero in
(\ref{table:closedint}) we can write down the cumulative infrared
asymptotics for a system consisting out of one D-brane and one O-plane
(situated at the origin in the transverse space)
\begin{equation}
-V_{p+1}\left( 4\pi^2\alpha^\prime\right)^{-\frac{p+1}{2}}\int_{l\to
 \infty} dl  l^{\frac{p-9}{2}}\left( 1 - 2^{p-4}\right)^2 .
\end{equation}
In field theory one obtains a result proportional to $\left( \mu_p
  +\mu^\prime _p\right)^2$, (recall that $\mu_p$ is the D-brane charge
  and $\mu^\prime _p$ the O-plane charge). Thus we obtain finally the
  ratio between D-brane and O-plane RR charges
\begin{equation}\label{mpmps}
\mu^\prime _p = -2^{p-4} \mu_p.
\end{equation}
The M\"obius strip computation fixed also the sign of this
ratio. However, if we assigned an $\Omega R$ eigenvalue of $+1$ to the
open string R vacuum we would obtain an additional minus sign in
(\ref{mpmps}). There is an ambiguity here. In the next section we will
use our results to construct consistent string theories containing
D-branes and O-planes. In this construction this ambiguity cancels
out. (In some sense it will turn out that our present choice is the
``natural'' one.)
The ratio of the D-brane tension to the O-plane tension can be inferred
by supersymmetry
\begin{equation}
T_p ^\prime = -2^{p-4}T_p .
\end{equation}

\subsection{Compactifying the transverse dimensions} 
\setcounter{equation}{0}
When we are trying to compactify the transverse directions of a D-brane
and/or an O-plane we immediately run into problems. These arise as
follows. The equation of motion for the RR field under which the
D-p-brane (or the O-p-plane) is charged reads (for otherwise trivial
background) 
\begin{equation}
d\star F_{p+2} = \star j_{p+1} ,
\end{equation}
where $j_p$ is the external $U\left( 1\right)$ current indicating the
presence of the D-brane (O-plane). Integrating this equation over a
compact transverse space gives zero for the left hand side and the
D-brane (O-plane) charge on the right hand side. Therefore, the RR
charge on the rhs has to vanish. To overcome
this problem one may want to add D-branes
and O-planes such that the net RR charge is zero.
Since one needs more than one D-brane in order to
achieve a vanishing net RR charge, one has to specify how $\Omega R$
acts on a set of multiple D-branes. For example it could (and actually
will) happen that $\Omega R$ (anti)symmetrises strings starting and 
ending on different D-branes. Technically, we define a (projective)
representation of the ${\mathbb Z}_2$ (generated by $\Omega R$) on the
Chan-Paton labels carried by open string in case of multiple
D-branes. The generating element of this representation is denoted by
$\gamma_{\Omega R}$. The $\Omega R$ action on an open string is
\begin{equation}\label{gammaomact}
\Omega R: \;\;\;\;\;\; \left| \psi,ij\right> \rightarrow \left(
  \gamma_{\Omega R}\right)_{i i^\prime} \left| \Omega
  R\left(\psi\right) , j^\prime 
  i^\prime\right> \left( \gamma_{\Omega R} ^{-1}\right)_{j^\prime j} .
\end{equation}
Here, $\psi$ denotes the oscillator content of the string on which
$\Omega R$ acts in the same way as discussed previously. In addition,
the order of the Chan-Paton labels is altered due to the orientation
reversal.  Acting twice with $\Omega R$ should leave the state
invariant. This leads to the condition
\begin{equation}\label{gamompos}
\gamma_{\Omega R} = \pm \gamma_{\Omega R} ^T,
\end{equation}
i.e.\ $\gamma_{\Omega R}$ is either symmetric or antisymmetric. By a
choice of basis this gives the two possibilities
\begin{equation}
\gamma_{\Omega R} =I \;\;\; \mbox{or}\;\;\; \gamma_{\Omega R} =
\left( \begin{array}{c c } 0 & iI \\ -iI & 0\end{array}\right) .
\end{equation}
Let $N$ be the number of D-branes ($\Omega R$ images are
counted). Then $I$ denotes an $N\times N$ identity matrix for
symmetric $\gamma_{\Omega R}$ and an $\frac{N}{2}\times \frac{N}{2}$
identity matrix for antisymmetric $\gamma_{\Omega R}$. 

The trace in the open string amplitudes (annulus and cylinder) includes
also a trace over the Chan-Paton labels. For the annulus this is
simply
\begin{equation}\label{antrcp}
\sum_{i,j =1}^N\left< i j\right| \left. ij\right> =\sum_{i,j =1}^N
\delta_{ii}\delta_{jj} = N^2 .
\end{equation}  
In the M\"obius strip amplitude the trace over the Chan-Paton labels
yields the additional factor
\begin{equation}
\sum_{i,j =1 }^N \left< ij\right| \Omega R \left| ij\right> =
\mbox{tr}\left( \gamma_{\Omega R} ^{-1}\gamma_{\Omega R}^T\right) =
\pm N,
\end{equation}
with the (lower) upper sign for (anti-) symmetric $\gamma_{\Omega
  R}$.

\subsubsection{Type I/type I$^\prime$ strings}

In the following we are going to investigate the case where the
compact space is a torus. 
The next issue we need to discuss are zero mode contributions due to
windings in the compact transverse dimensions. 
For open strings windings can appear in Dirichlet directions.
Since $\Omega R$
leaves the winding number of a state invariant these contribute to the
annulus, the Klein bottle and the M\"obius strip. Including the sum
over the winding numbers into the corresponding traces leads to
additional factors. The transverse space is a $9-p$--torus: 
\begin{equation}
T^{9-p} =\underbrace{S^1 \times \cdots\times S^1}_{9-p\; \mbox{\tiny
    factors}}.
\end{equation}
For simplicity we take the radii of these $S^1$s to be identical and
denote them by $r$. It is useful to introduce a dimensionless parameter
\begin{equation}
\rho = \frac{r^2}{\alpha^\prime}
\end{equation}
for the size of the compact space.

With this ingredients the sum over the winding modes gives the
following factors (under the $\frac{dt}{2t}$ integral):
\begin{eqnarray}
\left( \sum_{w=-\infty}^{\infty} e^{-2\pi t\rho w^2}\right)^{9-p}
& & \mbox{for the annulus},\\
\left( \sum_{w=-\infty}^{\infty} e^{-\pi t\rho w^2}\right)^{9-p}
& & \mbox{for the Klein bottle},\\
\left( \sum_{w=-\infty}^{\infty} e^{-2\pi t\rho w^2}\right)^{9-p}
& & \mbox{for the M\"obius strip} .
\end{eqnarray}
In the annulus we have restricted ourselves to the special case that
all D-branes are situated at the same point. This configuration
gives the correct leading infrared contribution to tree channel
amplitude. One can also include distances among the D-branes into the
computation. In that case the trace over the Chan-Paton labels cannot
be directly taken as in (\ref{antrcp}) because the zero mode
contribution depends on the Chan-Paton label. Taking the infrared
limit in the tree channel removes this dependence on the Chan-Paton
labels and gives the same result as our slightly simplified
computation.\footnote{In the limit $l\to\infty$ the distance dependent
  exponential function in table \ref{table:closedint} becomes one.}
Now, we express $t$ in terms of $l$ using
(\ref{dictionary-cyl}), (\ref{ltklein}) and (\ref{texpl}) and apply
the Poisson resummation formula
\begin{equation}\label{poisson}
\sum_{n=-\infty}^\infty e^{-\pi\frac{\left( n-b\right)^2}{a}} =
  \sqrt{a}\sum_{s=-\infty}^\infty e^{-\pi a s^2 +2\pi i sb}.
\end{equation}
We obtain
\begin{eqnarray}
l^{\frac{9-p}{2}}\rho^\frac{p-9}{2}\left( \sum_{w=-\infty}^{\infty} e^{-\frac{\pi l
    w^2}{\rho} 
    }\right)^{9-p}  
& & \mbox{for the annulus},\\
l^{\frac{9-p}{2}}\rho^\frac{p-9}{2}2^{9-p}\left( \sum_{w=-\infty}^{\infty}
    e^{-\frac{4\pi 
    lw^2}{\rho}}\right)^{9-p} 
& & \mbox{for the Klein bottle},\\
l^{\frac{9-p}{2}}\rho^\frac{p-9}{2}2^{9-p}
\left( \sum_{w=-\infty}^{\infty} e^{-\frac{4\pi l w^2}{\rho}}\right)^{9-p}
& & \mbox{for the M\"obius strip} .
\end{eqnarray}
In the IR limit $l\to\infty$ the sums become a factor of one. The
common $\rho$ dependent factor is a dimensionless quantity
representing the volume of the transverse space (sometimes denoted by
$v_{9-p}$). In the Klein bottle as well as in the M\"obius strip there
is an additional factor of $2^{9-p}$. In the M\"obius strip this is
simply the number of O-planes. (The number of R-fixed points is two
per $S^1$.) Since the Klein bottle amplitude is proportional to the
square of the total O-plane charge we would expect another factor of
$2^{9-p}$ here. However, this is ``canceled'' by the correction
factor we put earlier in by hand. As promised in footnote
\ref{promise} this factor appeared automatically after we compactified
the transverse dimensions. This is good because now we would miss the
argument for putting it in by hand. 

Together with our previous results table \ref{table:closedint},
(\ref{KBresult}) and (\ref{MBresult}) we obtain for the infrared limit
of the total (tree level RR channel) amplitude
\begin{equation}\label{totaltadp}
-\frac{1}{2}\left( 4\alpha^\prime \pi^2 \right)^{-\frac{p+1}{2}}
 V_{p+1}\rho^{\frac{p-9}{2}} \int_{l\to\infty} dl \, 
 \left( N^2 + 32^2 \mp 2N 32 \right) ,
\end{equation}
where the $\mp$ sign is correlated with the $\pm$ sign in
(\ref{gamompos}). We observe that the contributions of the
D-brane/D-brane, O-plane/O-plane and D-brane/O-plane interaction add
up to a complete square, proportional to
\begin{equation}\label{numberofd}
\left( 32 \mp N\right)^2.
\end{equation}
For consistency the total RR charge has to vanish. Thus, we are lead
to the conditions
\begin{equation}\label{gamomsym}
\gamma_{\Omega R} = \gamma_{\Omega R} ^T
\end{equation}
and
\begin{equation}
N=32.
\end{equation}
Note that the condition (\ref{gamomsym}) is related to our choice that
the $\Omega R$ eigenvalue of the R vacuum (in the open string sector)
is minus one. Since (\ref{gamomsym}) implies that we can choose a
basis such that
\begin{equation}
\gamma_{\Omega R} = I,
\end{equation}
our choice of the $\Omega R$ eigenvalue of the R vacuum seems
natural. The case $p=9$ can be obtained from our previous 
considerations in two ways. The simplest is to set $p=9$ in
(\ref{totaltadp}). Requiring this expression to vanish yields
equations (\ref{gamomsym}) and (\ref{numberofd}) independent of
$p$.\footnote{Recall that a $p$-dependence cancels in the product of
  O-plane charges with the number of O-planes.} An alternative way is
to perform T-dualities with respect to all compact directions and to
take the decompactification limit in the T-dual model. Both methods
lead to the same results. The massless closed string spectrum has been
discussed in section  \ref{unorclosed}. In the NSNS sector one finds
the metric $G_{ij}$ and the dilaton $\Phi$. The RR sector provides the
antisymmetric tensor $B^\prime _{ij}$.  The $\Omega$ invariant
combinations of the 
NSR with the RNS sector massless states yield the space time fermions
needed to fill $N=1$ supermultiplets.

It remains to study the open string sector. The massless NS sector
states are
\begin{equation}\label{nstypeI}
\psi_{-\frac{1}{2}}^i\left| k,mn\right> \lambda^{mn}.
\end{equation}
The $\Omega R$ image of these states is
\begin{equation}
-\psi_{-\frac{1}{2}}^i\left| k,mn\right> \left(\lambda^T\right)^{mn}.
\end{equation}
Hence, the Chan-Paton matrix $\lambda$ must be antisymmetric
\begin{equation}\label{chanpatypeI}
\lambda = -\lambda^T .
\end{equation}
The states (\ref{nstypeI}) are vectors and thus should be interpreted
as gauge fields of a certain gauge group. In order to identify the
gauge group, we require that the state (\ref{nstypeI}) transforms in
the adjoint and that gauge transformations preserve the condition
(\ref{chanpatypeI}). Thus the commutator of a generator of the gauge
group with a $32\times 32$ antisymmetric matrix ($\lambda$) should be a
$32\times 32$ antisymmetric matrix. This consideration determines the
gauge group to be $SO(32)$. The R sector provides fermions in the
adjoint of $SO(32)$. NS and R sector together yield an $N=1$ $SO(32)$
vector multiplet. The list of consistent closed string theories in
ten dimensions (table \ref{tab:tab2}) is supplemented by the ten
dimensional theory containing (unoriented) closed strings and open
strings in table \ref{tab:tabopen}.

\begin{table}
\begin{center}\begin{tabular}[h]{|l|l|l|l|l|}
\hline 
   & \# of $Q$'s& \# of $\psi_\mu$'s & \multicolumn{2}{c|}{massless
   bosonic spectrum}\\  
\hline
\hline
type I & 16 & 1 &NSNS  & $G_{\mu\nu}$,  $\Phi$\\ \cline{4-5}
$SO(32)$ & & &open string &$A_{\mu}^a$ in
adjoint of $SO(32)$ \\ \cline{4-5}
 & & &RR & $B^\prime _{\mu\nu}$
\\ \hline\end{tabular}\end{center} 
\caption{Consistent string theories in ten dimensions containing open
  strings.} \label{tab:tabopen}
\end{table} 

\subsubsection{Orbifold compactification}\label{sec:gpmodel}

So far, we have studied the consistency conditions implied by a torus
compactification of the transverse dimension. Since type I theory is a
ten dimensional $N=1$ supersymmetric theory, torus compactifications
will result in extended supersymmetries in lower dimensions (e.g.\
$N=4$ in four dimensions). For phenomenological reasons it is
desirable to obtain less supersymmetry. This can be achieved by taking
the transverse space to be an orbifold. In the following we will add
O-planes and D-branes to the orbifold compactification considered in
section \ref{k3orbi}. We supplement the ${\mathbb Z}_2$ action
(\ref{o-projection}) with an $\Omega R$ action, where $R$ acts on the
target space in the same way as given in (\ref{o-projection}). Hence, our
discrete group is generated by $R$ and $\Omega R$. The third
non-trivial element is the product of the two generators:
$\Omega$. Thus, the theory contains O-5-planes and O-9-planes. We
expect that we need to add D-5-branes and D-9-branes in order to
preserve RR charge conservation. Before, studying the open strings
induced by those D-branes let us discuss the untwisted and twisted
closed string sector states. We focus on the massless part of the
spectrum. All the information needed to find the untwisted massless
states is given in table \ref{untwisted2}. In addition to the
${\mathbb Z}_2$ symmetry we also need to respect the $\Omega$ and
$\Omega R$ symmetry. This is done by symmetrization in the NSNS sector
and antisymmetrization in the RR sector. Thus the untwisted NSNS
sector contains the metric $G_{ij}$, the dilaton $\Phi$ and ten
scalars. In the RR sector one finds  a selfdual and an anti-selfdual
two form and twelve scalars. The relevant twisted sector states are
listed in table \ref{twisted2}. Taking into account that there are 16
fixed points we obtain 48 massless scalars in the twisted NSNS sector,
and 16 massless scalars in the twisted RR sector. Adding the fermions
in the $\Omega$ and $\Omega R$ invariant combinations of NSR and RNS
sector states, one obtains finally a $d=6$, $N=1$ supergravity
multiplet, one tensor multiplet and 20 hypermultiplets. (We should
emphasize again that the present review is not self contained as far as
the supergravity representations are concerned. The reader may view
the arrangement of the massless states into multiplets as some
additional information which is not really employed in the forthcoming
discussions. In order to obtain a nice overview about supermultiplets
in various dimensions we recommend\cite{Salam:1989fm}.) 

As already mentioned, we need to add D-5- and D-9-branes in order to 
cancel the O-plane RR charges. Thus the Chan-Paton matrix is built out
of the following blocks: $\lambda^{(99)}$ corresponding to strings
starting and ending on D-9-branes, $\lambda^{(55)}$ corresponding to
strings starting and ending on D-5-branes, $\lambda^{(59)}$ and
$\lambda^{(95)}$ corresponding to  open strings with ND boundary
conditions in the compact dimensions. The action of $\Omega$ and
$\Omega R$ on the Chan-Paton labels is as described in
(\ref{gammaomact}). The $\gamma_\Omega$ and $\gamma_{\Omega R}$ posses
also a block structure distinguishing between the action on a string
end at a D-5- or D-9-brane, e.g.\ $\gamma_{\Omega }^{(9)}$ represents
the $\Omega$ action on a Chan-Paton label corresponding to an open
string end on a D-9-brane. Finally, we specify the representation of
the ${\mathbb Z}_2$ element $R$ (\ref{o-projection}) as follows 
\begin{equation}
R:\;\;\;\;\;\; \left|\psi , ij\right> \rightarrow
\left(\gamma_{R}\right)_{ii^\prime} \left| R\left( \psi\right) , i^\prime
    j^\prime\right> \left( \gamma_{R}^{-1}\right)_{j^\prime j} .
\end{equation}
Also, the $\gamma_R$ can be split into two blocks: $\gamma_R ^{(9)}$
and 
$\gamma_R ^{(5)}$. The requirement that performing twice the same 
${\mathbb Z}_2$ action should leave the state invariant leads to the
conditions that every gamma-block containing an $\Omega$ in the
subscript must be either symmetric or anti-symmetric, whereas the
gamma-blocks without an $\Omega$ in the subscript must square to the
identity\footnote{We fix a possible phase to one.}. At this point, we
need to discuss a subtlety of the five-nine 
sector, i.e.\ strings with ND boundary conditions along the compact
dimensions. In that sector the Fock space state (without the
Chan-Paton label) has an $\Omega^2$ and an $\left(\Omega R\right)^2$
eigenvalue of minus one. Unfortunately, we did not develop the
techniques needed to show this, in this review. An argument employing
an isomorphism between the algebra of vertex operators and Fock space
states can be found in\cite{Gimon:1996rq}. Since $\Omega^2$ and
$\left( \Omega R\right)^2$ should leave states invariant, this minus
sign needs to be canceled by an appropriate action on the Chan-Paton
labels. For example a symmetric $\gamma_{\Omega}^{(9)}$ implies an
anti-symmetric $\gamma_{\Omega}^{(5)}$, and a symmetric $\gamma_{\Omega
  R}^{(5)}$ implies an anti-symmetric $\gamma_{\Omega R}^{(9)}$.   

Let us now study the amplitudes in the loop channel. For strings
starting and ending on D-9-branes there is a tower of Kaluza-Klein 
momentum modes but no winding modes. The D-9-branes are space filling
and thus must lie on top of each other. Open strings with both ends on
a D-5-brane can have winding modes and can be separated in the compact
directions. (In our previous computation we did not consider this
separation, because it is not relevant in the tree channel infrared
limit. In order to see explicitly that the dependence on the distance
among D-5-branes drops out, we will take it into account, here. We call
the component of the position of the $i$th brane $c_i ^a$.)
Further, all amplitudes obtain an additional insertion
\begin{equation}
\frac{1+R}{2},
\end{equation}
ensuring that we trace only over states which are invariant under the
orbifold group. For terms containing an $R$ insertion the D-5-brane
must be located at $R$ fixed points, since otherwise the states in the
$(55)$ sector are not eigenstates under $R$. For the same reason the
winding or Kaluza-Klein momentum modes have to vanish in the presence
of an $R$ insertion.  
Further, there will be
additional signs for oscillators pointing into a compact dimension. This
modifies the oscillator contribution to the trace in a straightforward
way. (We leave the details as an exercise.) Taking into account all
these effects and the discussion in section \ref{dbraneinteractions}
one finds for the annulus amplitude ${\cal A}$
%\footnote{There
%  is a difference of a factor 1/2 between the formul\ae\ presented
%  here and the ones in\cite{Gimon:1996rq}. In\cite{Gimon:1996rq} there
%  is an $\Omega /2$ insertion instead of an $\Omega$ insertion,
%  here. This is because $\Omega /2$ is viewed as part of the
%  projection operator $\left( 1+\Omega\right)/2$. By the same token
%  also the annulus has a factor of $1/2$ corresponding to the other
%  term in the projection operator. Since in the end we are going to
%  require the infrared leading contribution to vanish, this overall
%  factor does not matter.} %%%% There is a facotr of 4 difference!!!
%
\begin{eqnarray}
{\cal A} &=& -\frac{V_6}{4}\int \frac{dt}{t^4}\left( 8\pi^2
  \alpha^\prime\right)^{-3} \left[ \frac{f_4 ^8\left( e^{-\pi
  t}\right)}{f_1 ^8\left( e^{-\pi t}\right)}\left\{ \left(
  \mbox{tr}\, \gamma_1 ^{(9)}\right)^2\left( \sum_{n=-\infty}^\infty
  e^{-\frac{2\pi t n^2}{\rho}}\right)^4\right.\right. \nonumber\\  
& & \; \left.\left. + \sum_{i,j \in 5} \left(
  \gamma_1 ^{(5)}\right)_{ii}\left( \gamma_1 ^{(5)}\right)_{jj}
  \prod_{a=6}^9 \sum_{w= -\infty}^{\infty}e^{-t\frac{\left( 2\pi w
  \sqrt{\rho\alpha^\prime} +c_i ^a - c_j
  ^a\right)^2}{2\pi\alpha^\prime}}\right\}\right. \nonumber \\
& & \;  -2 \frac{f_2^4\left(
  e^{-\pi t}\right) f_4 ^4\left( e^{-\pi t}\right)}{f_1 ^4\left(
  e^{-\pi t}\right)f_3 ^4\left( e^{-\pi t}\right)}\left(\sum_{I=1}^{16}
  \mbox{tr}\, 
  \gamma_{R,I} ^{(5)}\right)\left( \mbox{tr}\, \gamma_R ^{(9)} \right)
\nonumber \\  
& & \; \left. + 
4\frac{f_3 ^4\left( e^{-\pi t}\right) f_4 ^4\left( e^{-\pi
  t}\right)}{f_1 ^4\left( e^{-\pi t}\right) f_2 ^4\left( e^{-\pi
  t}\right)} \left\{ \left( \mbox{tr}\, \gamma_R ^{(9)}\right)^2 +
  \sum_{I=1}^{16} \left( \mbox{tr} \,\gamma_{R,I}
  ^{(5)}\right)^2\right\} \right] ,
\end{eqnarray}
where we have formally assigned a gamma with subscript $1$ to the
action of the identity element of the orbifold group on the Chan-Paton
labels. The sum over $i,j\in 5$ means that we sum over all Chan-Paton labels
belonging to an open string end on a D-5-brane. 
The index $I=1,\ldots , 16$ labels the fixed 5-planes, and a
corresponding subscript at a $\gamma^{(5)}$ indicates that the
D-5-brane is located on the $I$th fixed plane.

Next, we want to compute the Klein bottle amplitude ${\cal K}$. It
contains the insertions $\Omega$ and $\Omega R$. In principle, we have
to take the trace over untwisted and twisted sector states (with the
$\left( -\right)^F$ insertion). Because half of the RR sector states
have the opposite $\left( -\right)^F$ eigenvalue than the other half, RR
sector states do not contribute to the trace with a $\left(
  -\right)^F$ insertion. The same applies to RR and NSNS
twisted sector states.
Eigenstates of $\Omega$ have zero winding numbers whereas for eigenstates
of $\Omega R$ the Kaluza-Klein momenta are zero. With this ingredients
we find
\begin{eqnarray}
\lefteqn{{\cal K} =
 -8 \frac{V_6}{4}\int \frac{dt}{t^4}\left( 8\pi^2
  \alpha^\prime\right)^{-3}}\nonumber\\
& & \frac{f_4 ^8\left( e^{-2\pi t}\right)}{f_1
  ^8\left( e^{-2\pi t}\right)}\left\{ \left( \sum_{n=-\infty}^\infty
  e^{-\frac{\pi tn^2}{\rho}}\right)^4 +\left( \sum_{w=-\infty}^\infty
  e^{-\pi t \rho w^2}\right)^4\right\}.
\end{eqnarray}

Finally, for the M\"obius strip amplitude ${\cal M}$ we need to trace
over R sector states with an $\Omega +\Omega R$ insertion. Eigenstates
correspond to open strings starting and ending on the same
brane. According to our earlier assignments, the $\Omega R$ eigenvalue
of the R vacuum corresponding to a string ending on a D-5-brane is minus
one, and so is the $\Omega$ eigenvalue of the R vacuum corresponding
to a string ending on D-9-branes. To determine the remaining
eigenvalues one has to act with $R$ on the Ramond vacuum. $R$ can be
viewed as a $180^\circ$ rotation and the R vacua as target space
spinors. Hence, half of the Ramond vacua have $R$ eigenvalue minus one
and the other half plus one. For this reason, only D-9-branes
contribute to the term with the $\Omega$ insertion whereas only
D-5-branes give a non-vanishing result for the trace containing an
$\Omega R$ insertion. The result for the M\"obius strip is
\begin{eqnarray}
{\cal M}& = & \frac{V_6}{4}\int \frac{dt}{t^4}\left( 8\pi^2
  \alpha^\prime\right)^{-3}\frac{f_2 ^8 \left(e^{-2\pi t}\right) f_4
  ^8\left( e^{-2\pi t}\right)}{f_1 ^8\left( e^{-2\pi t}\right)f_3
  ^8\left( e^{-2\pi t}\right)} \nonumber \\
& & \; \left\{ \mbox{tr}\left( \left(
  \gamma_{\Omega}^{(9)}
  \right)^{-1}\left(\gamma_{\Omega}^{(9)}\right)^T\right) \left( \sum_{n=
  -\infty}^\infty e^{-\frac{2\pi tn^2}{\rho}}\right)^4
  \right. \nonumber\\
& & \; \left. +\mbox{tr} \left( \left( \gamma_{\Omega R}
  ^{(5)}\right)^{-1} \left(\gamma_{\Omega R}^{(5)}\right)^T
  \right)\left( \sum_{w= -\infty}^\infty e^{-2\pi t \rho
  w^2}\right)^4\right\} .
\end{eqnarray}

With the next steps necessary to compute the total RR charge of the system we
are familiar by now. We replace $t=\frac{1}{2l}$ in the annulus,
$t=\frac{1}{4l}$ in the Klein bottle and $t=\frac{1}{8l}$ in the M\"obius
strip. In order to be able to read off the infrared (large $l$)
asymptotics we use formul\ae\ (\ref{modulartra}) and
(\ref{poisson}). The final result is 
\begin{eqnarray}
{\cal A}+{\cal K}+{\cal M }& \longrightarrow &
-\frac{V_6}{4} \int_{l\to\infty} \!\!\! dl\, \left(
  4\pi^2\alpha^\prime\right)^3 \nonumber \\   
& & \, \left[ \rho^2 \left\{ \left(\mbox{tr} \gamma_1^{(9)}\right)^2 -32
    \mbox{tr} 
    \left( \left( \gamma_{\Omega}^{(9)}\right)^{-1}\left( \gamma_\Omega
      \right) ^T \right) + 32^2\right\}\right. \nonumber \\ 
& & \; +\frac{1}{\rho^2} \left\{ \left(\mbox{tr} \gamma_1^{(5)}\right)^2 -32 
  \mbox{tr} 
    \left( \left( \gamma_{\Omega R}^{(5)}\right)^{-1}\left( \gamma_\Omega
      \right) ^T \right) + 32^2\right\}\nonumber \\
& & \; \left. +\frac{1}{4}\sum_{I=1}^{16} \left(
    \mbox{tr} \gamma_R^{(9)} + 4
    \mbox{tr}\gamma_{R,I}^{(5)}\right)^2\right] . 
\label{gptadpole}
\end{eqnarray}  
The setup respects RR charge conservation if (\ref{gptadpole}) vanishes. Thus,
we need 32 D-9-branes and 32 D-5-branes. (A gamma representing the identity is
of course the identity matrix.) 
Further, we take 
\begin{equation}
\gamma_{\Omega R}^{(5)} =\left( \gamma_{\Omega R}^{(5)} \right)^T \,\,\, ,
\,\,\ \gamma_{\Omega }^{(9)} =\left( \gamma_{\Omega }^{(9)} \right)^T .
\end{equation}
Our previous discussion of the (59) sector implies
\begin{equation}
\gamma_{\Omega }^{(5)} =-\left( \gamma_{\Omega }^{(5)} \right)^T\,\,\, ,
\,\,\, \gamma_{\Omega R}^{(9)} =-\left( \gamma_{\Omega R}^{(9)} \right)^T .
\end{equation}
The remaining representation matrices can be found by imposing that
the gammas should form a projective\footnote{``Projective" means up to
  phase factors,  which drop out since the gamma acts in combination
  with its inverse on the Chan-Paton label.}  
representation of the orientifold group (${\mathbb
  Z}_2\times {\mathbb Z}_2$). We simply choose
\begin{eqnarray}
\gamma_R^{(5)} &=&\gamma_{\Omega R}^{(5)} \gamma_\Omega^{(5)}\\
\gamma_R^{(9)}& =&\gamma_{\Omega R}^{(9)} \gamma_\Omega^{(9)}.
\end{eqnarray}
By fixing a basis in the Chan-Paton labels we obtain
\begin{equation}
\gamma_{\Omega R}^{(5)} = \gamma_\Omega ^{(9)} = I ,
\end{equation}
where the rank of the identity matrix is 32. The antisymmetric form is
\begin{equation}\label{gamomfiv}
\gamma_\Omega^{(5)} = \gamma_{\Omega R}^{(9)} = \left(
\begin{array}{cc} 0 & iI\\
-iI & 0
\end{array}\right) ,
\end{equation}
with $I$ being a $16\times 16$ identity matrix, here. Note that our
choice is consistent with the requirement that $\gamma_R ^{(\cdot )}$
squares to the identity. So far, we did not take into account that the
last term in (\ref{gptadpole}) has to vanish. With $\gamma_R ^{(\cdot)}$
being traceless this is ensured.

We have now all the ingredients needed to determine the open string
spectrum. Let us first study strings starting and ending on the
D-9-branes, or in short the (99) sector. We keep states which are 
invariant under each element of the orientifold group. (The D-9-branes
are fixed under each element of the orientifold group.)
In the NS sector we find massless vectors with the Chan-Paton matrix 
\begin{equation}\label{99vector}
\lambda^{(99)}_{\mbox{\tiny vector}} = \left( \begin{array}{cc}
A & S \\ -S & A\end{array}\right) ,
\end{equation}
where $A$ denotes a real antisymmetric and otherwise arbitrary
$16\times 
16$ matrix and $S$ stands for a real $16\times 16$ symmetric
matrix. For the scalars in the NS sector one finds
\begin{equation}
\lambda^{(99)}_{\mbox{\tiny scalars}} = \left( \begin{array}{ c c} A_1
    & A_2 \\ A_2 & -A_1\end{array}\right) ,
\end{equation}
where the $A_i$ are $16\times 16$ antisymmetric matrices. Let us
ignore the D-5-branes for a moment and determine the gauge group and
its action on the scalars in the (99) sector. Since the vectors are
in the adjoint of the gauge group, the gauge group should be $16^2$
dimensional. $U\left( 16\right)$ is a good candidate. Further, we know
that the vector should transform in the adjoint under global gauge
transformations under which it should not change the form specified by  
(\ref{99vector}). Thus, we define an element of the gauge group as
\begin{equation}
g^{(9)} = \mbox{exp}\left(\begin{array}{c c} A_g &S_g\\ -S_g &
    A_g\end{array}\right) ,
\end{equation} 
where $S_g$ ($A_g$) are real anti-(symmetric) matrices with infinitesimal
entries. 
A gauge transformation acts on the Chan-Paton matrix as
\begin{equation}\label{99scalar}
\lambda^{(99)} \rightarrow g^{(9)} \lambda^{(99)}
    \left(g^{(9)}\right)^{-1} =\left[ 
    \left(\begin{array}{c c} A_g &S_g\\ -S_g & 
    A_g\end{array}\right) ,\lambda\right]  .
\end{equation}
We observe that the vectors transform in the adjoint and the form of the
Chan-Paton matrix is preserved. Note also that $g^{(9)}$ is unitary
and has $16^2$ parameters. It is a $U\left( 16\right)$ element.
The $U\left( 1\right)$ subgroup corresponds to $A_g =0$ and $S_g$
proportional to the identity.
From our assignment that the Chan-Paton
matrix transforms in the adjoint of $U\left( 16\right)$ it is also
clear that the Chan-Paton label $i$ and $j$ transform in the {\bf
  16} and {\bf $\overline{\mbox{\bf 16}}$} of $U\left( 16\right)$ Thus, the
scalars can be decomposed into the antisymmetric {\bf 120} + 
  $\overline{\mbox{\bf 120}}$. One may also explicitly check that the
  form of the Chan-Paton matrix for the scalars is not altered by a
  gauge transformation.  
We leave the discussion of the fermions in the R sector as an
exercise. The result is that that half of them carry the Chan-Paton
matrix (\ref{99vector}) and the other half the matrix
(\ref{99scalar}). Altogether the (99) sector provides a vector
multiplet in the adjoint of $U\left( 16\right)$ and a hypermultiplet
in the {\bf 120} + {\bf $\overline{\mbox{\bf 120}}$}.

Now, we include the D-5-branes. Here, we have to distinguish between
the case that a D-5-brane is situated at a fixed plane or not. In the 
first case the (55) strings have to respect the $\Omega R$ and $R$
symmetry, whereas in the second case these orientifold group elements
just fix the fields on the image brane. Suppose we have $2m_I$
D-5-branes at the $I$th fixed plane. (The number of the D-5-branes per
fixed plane must be even, since otherwise they cannot form a
representation of the orientifold group.) 
The NS sector leads again to a
massless vector and massless scalars with almost the same Chan-Paton
matrices as in the (99) sector ((\ref{99vector}) and (\ref{99scalar})).
The only difference is that the antisymmetric and symmetric matrices
are now $m_I\times m_I$ instead of $16\times 16$. Hence, we obtain a
vector multiplet in the adjoint of $U\left( m_I\right)$ and a
hypermultiplet in the  $ \frac{\mbox{\bf m}_I}{\mbox{\bf 2}}
\left( \mbox{\bf m}_I \mbox{\bf -1}\right)$ + $\overline{ \frac{\mbox{\bf
    m}_I}{\mbox{\bf 2}} 
\left( \mbox{\bf m}_I \mbox{\bf -1}\right)}$.

Let $2n_j$ D-5-branes be situated away from the fixed plane (but on
top of each other). For the (55) sector belonging to those D-5-branes
we impose invariance under $\Omega $, only. The solution for 
$\gamma_\Omega ^{(5)}$ is given in (\ref{gamomfiv}). Together with the
minus eigenvalue on the massless NS sector Fock space state, this
leads to the result that the vector in the (55) sector is an element
of the $USp\left( 2n_j\right)$ Lie algebra in the adjoint
representation. Taking into account (part of) the R sector this is
promoted to a 
$USp\left( 2n_j\right)$ vector multiplet. The scalars together with
the remaining R sector states form a hypermultiplet in the
antisymmetric $\mbox{\bf n}_j\left( \mbox{\bf 2n}_j \mbox{\bf
    -1}\right)$ representation.

It remains to study the (95) sector. (Here, one has to take into
account that along the compact directions NS sector fermions are
integer modded whereas R sector fermions are half integer modded.
This is quite similar to the twisted sector closed string.
In particular, the (95) NS sector ground state is already 
massless. Hence, the NS sector does not give rise to massless
vectors. 
We do not impose $\Omega$ or $\Omega R$ invariance on (95) strings
since they are mapped onto (59) strings by the worldsheet parity
inversion. If the considered D-5-branes are situated at one of the
fixed planes we impose $R$ invariance. 
In this case, one finds in the NS sector two scalars with the Chan-Paton
matrix
\begin{equation}
\lambda ^{(95)} = \left( \begin{array}{cc} X_1 & X_2\\
-X_2 & X_1\end{array}\right) ,
\end{equation}  
where the $X_i$ are general $m_i \times 16$ matrices. Together with
the R sector this leads to a
hypermultiplet in the $\left(\mbox{\bf 16 , m}_I\right)$ of $U\left(
  16\right)\times U\left( m_I\right)$, (the hypermultiplet is neutral
under the gauge group living on D-5-branes not situated at the $I$th
fixed plane). For D-5-branes which are not a fixed plane the (95)
sector provides a hypermultiplet in the $\left( \mbox{\bf 16,
    2n}_J\right)$ of $U\left( 16\right)\times USp\left( 2n_j\right)$. 

Altogether we find the gauge group is
\begin{equation}
U\left( 16\right)\times \prod_{I=1}^{16} U\left( m_I\right) \times
\prod_J USp\left( 2n_j\right) ,
\end{equation}
where $j$ labels the D-5-brane packs away from fixed planes. In
addition the total number of D-5-branes has to be 32 (images are
counted), i.e.\ 
\begin{equation}
\sum_{I=1}^{16} 2m_I + 2\sum_{j} 2n_j = 32 .
\end{equation}
There are hypermultiplets in the representation  
\begin{eqnarray}
& & 2\left( \mbox{\bf 120 , 1, 1}\right) +\sum_{I=1}^{16} \left\{ 2\left(
  \mbox{\bf 1}, \frac{\mbox{\bf 1}}{\mbox{\bf 2}} \mbox{\bf m}_I  \left(
  \mbox{\bf m}_I\mbox{\bf -1}\right) ,\mbox{\bf
  1}\right)_I+\left(\mbox{\bf 16}, \mbox{\bf
  m}_I, \mbox{\bf 1}\right)_I\right\}\nonumber \\ 
& & \; +\sum_j \left\{ \left( \mbox{\bf 1} ,\mbox{\bf 1}, \mbox{\bf 
  n}_j\left( \mbox{\bf 2n}_j\mbox{\bf -1}\right)\mbox{\bf -1}\right)_j 
  +\left( \mbox{\bf 1},\mbox{\bf 1},\mbox{\bf 1}\right)
  +\left(\mbox{\bf 16}, \mbox{\bf 1}, \mbox{\bf 2n}_j\right)_j
  \right\} ,
\end{eqnarray}
where we have split the anti-symmetric representation of $USp\left( 2
  n_j\right)$ 
into its irreducible parts and an index $I$, $j$ refers to the gauge
group on the D-5-brane pack at a fixed plane and off a fixed plane,
respectively. It can be checked that the effective six dimensional
theory is free of anomalies. The models belonging to different
distributions of the D-5-branes on and off fixed planes can be
continuously transformed into each other. In the field theory
description this corresponds to the Higgs mechanism.

We have seen that adding to the orbifold compactification of section
\ref{sec:orbifold} O-planes and D-branes gives a very interesting
picture. Apart from the closed string twisted sector states located
at the orbifold fixed planes we obtain various fields from open
strings ending on D-branes. These D-branes can be moved within the
compact directions while keeping the geometry fixed. The 
techniques described in this section can be also applied to
phenomenologically more interesting setups leading to four dimensional
theories. A description of such models is beyond the scope of the
present review.

%%% Local Variables: 
%%% mode: latex
%%% TeX-master: t
%%% End: 

\newpage
\chapter{Non-Perturbative description of branes}\label{chap:nonpert}
\section{Preliminaries}
\setcounter{equation}{0}

In the previous sections we gave a perturbative description of various
extended objects: the fundamental string, orbifold planes, D-branes
and Orientifold planes. The string plays an outstanding role in the
sense that field theories on the worldvolumes of the other extended
objects are effective string theories. 
The quantization of the fundamental string is
performed in a trivial target space (i.e.\ the target space metric is
the Minkowski metric and all other string excitations are constant or
zero). Further, the worldsheet topology is specified to the spherical
(for closed strings)
or disc (for open strings) topology (after Wick rotating to Euclidean
worldsheet signature). Our treatment leads to a perturbative expansion
in the genus of the worldsheet (see section
\ref{effectiveactions}). The perturbative expansion is governed by
the string coupling 
\begin{equation}
g_s = e^{\langle\Phi\rangle} ,
\end{equation}
which needs to be small. Perturbative closed string theory has an
effective 
field theory description which contains supergravity. 
How does one obtain insight into regions
where $g_s$ is large? Clearly, the perturbation theory breaks down in 
this case, and indeed this region is rather difficult to study. There
are, however, a few results one can obtain also for strong couplings.    
Let us recall how non perturbative effects in Yang-Mills theory can be
studied. Apart from the trivial vacuum, (Euclidean) Yang-Mills theory
contains several other stable vacua, {\it viz.} the
instantons. Studying fluctuations around an instanton vacuum, one  
finds an additional weight factor in the path integral which comes from
the background value of the action and is of the form
$$ e^{-\frac{n}{g^2}} , $$
where $n$ is the instanton number and $g$ is the Yang-Mills gauge
coupling. As long as $g$ is small, the fluctuations around an
instanton 
vacuum are heavily suppressed. However, as soon as $g$ becomes large, 
the suppression factor becomes large. Thus, knowing about the instanton
solutions in Yang-Mills theory gives a handle on non perturbative
effects. But how can one know, that one does not have to include
strong coupling effects into the theory before deriving the
instanton solutions? The answer is that instantons are stable, they
are characterized by a topological number which cannot be changed in a
continuous way when taking $g$ from small to large. Therefore,
instantons can give information about strongly coupled Yang-Mills
theory even
though they are found as solutions to the perturbative formulation of
Yang-Mills theories. States (vacua) with such a feature are called
BPS states.\footnote{There are also stable non BPS states (for
  reviews see e.g.\ \cite{Sen:2000vx, Lerda:1999um,
    Gaberdiel:2000jr}). We will not discuss these.}\footnote{BPS
  stands for the names Bogomolny, Prasad and Sommerfield, and refers
  to the papers\cite{Bogomolny:1976de,  Prasad:1975kr}.}

Therefore, our aim will be to find BPS states in string theory. In the
low energy limit, the various superstring theories are described by  
supergravities. Insights into non-perturbative effects in string
theory can be gained by finding the BPS states of perturbative string
theory. As a guiding principle, we will look for solutions to the
effective equations of motion that preserve part of the supersymmetry
(i.e.\ are invariant under a subset of the supersymmetry
transformations). Roughly speaking, it is then the number of preserved
supersymmetries which cannot be changed continuously when taking the
string coupling from weak to strong. We will see that such solutions
can be viewed as branes. The number of branes takes the role of the
instanton number in the Yang-Mills example discussed above. We will be
very brief in our analysis and essentially only summarize some of the
important results. The classical review on branes as supergravity
solutions is\cite{Duff:1995an} and we will give more references in the
end of this review.

\section{Universal Branes}
\setcounter{equation}{0}

From section \ref{effectiveactions} we recall that all the closed
superstring theory effective actions contain a piece
\begin{equation}\label{eq6:universal}
S_{univ} = \frac{1}{2\kappa^2}\int d^{10} x \sqrt{-G} e^{-2\Phi}\left(
  R  +4\left( \partial \Phi\right)^2 -\frac{1}{12} H^2\right) .
\end{equation}
In the present section we will truncate all closed string effective
field theories to (the supersymmetric extension of)
(\ref{eq6:universal}). This is consistent because we will restrict on
backgrounds where the discarded part of the action vanishes and the
corresponding equations of motion are satisfied trivially. By adding
appropriate terms including fermions (\ref{universal}) can be promoted
to an $N=1$ supersymmetric theory. (For type II theories this is a
sub-symmetry of the $N=2$ supersymmetry.) 
The supersymmetric extension is usually given in the Einstein
frame. The action (\ref{eq6:universal}) is written in the string frame
were the string tension is a constant and independent of the
dilaton. The Einstein frame is obtained by the metric redefinition
\begin{equation}\label{einsteinframe}
g_{\mu\nu} = e^{-\frac{\Phi}{2}} G_{\mu\nu} ,
\end{equation}
where $G_{\mu\nu}$ is the string frame metric and $g_{\mu\nu}$ is the
Einstein frame metric. The action (\ref{eq6:universal}) takes the form 
\begin{equation}\label{einsteinac}
S_{E,univ} = \frac{1}{2\kappa^2}\int d^{10} x \sqrt{-g}\left( R
  -\frac{1}{2}\left( \partial \Phi\right)^2 -\frac{1}{12} e^{-\Phi}
  H^2\right). 
\end{equation}
We observe that (\ref{einsteinac}) starts with the familiar Einstein
Hilbert term (therefore the name ``Einstein frame''). Further, the
kinetic term of the dilaton has the ``correct'' sign now, and the
coupling of the $B$ field is $\Phi$ dependent. In the supersymmetric
extension, a gravitino and a dilatino are added. We do not give the
supersymmetric action explicitly. For us, it will suffice to know the
supersymmetry transformations of the gravitino and the dilatino. These
are 
\begin{eqnarray}
\delta \psi_\mu &=& D_\mu \epsilon +\frac{1}{96}
e^{-\frac{\Phi}{2}}\left( {\Gamma_{\mu}}^{\nu \rho \kappa} -
  9\delta_\mu ^\nu \Gamma^{\rho\kappa}\right)
H_{\nu\rho\kappa}\epsilon  ,\label{gravitino}\\ 
\delta\lambda & = & -\frac{1}{2\sqrt{2}}\Gamma^\mu\partial_\mu \Phi
\epsilon +\frac{1}{24\sqrt{2}}e^{-\frac{\Phi}{2}}
\Gamma^{\mu\nu\rho}H_{\mu\nu\rho} \epsilon \label{dilatino},
\end{eqnarray}
where $\psi_\mu$ denotes the gravitino and $\lambda$ the dilatino. 
The Gamma matrices with curved indices are obtained from ordinary
Gamma matrices ($16\times 16$ matrices satisfying the usual Clifford
algebra in ten dimensional Minkowski space) by transforming the flat
index with a vielbein to a curved one. A Gamma with multiple indices
denotes the anti-symmetrized product of Gamma matrices. 
The spinor $\epsilon $ is the
supersymmetry transformation parameter. 

Sometimes it is useful to formulate the theory in a slightly
different way. To this end, one adds to the action (\ref{einsteinac})
a Lagrange multiplier term providing the constraint of a fulfilled
Bianchi identity. Calling the Lagrange multipliers $A_{\mu_1\ldots
  \mu_6}$, such a term looks like
\begin{equation}\label{hodge}
\int d^{10}x\, \epsilon^{\mu_1 \ldots \mu_{10}}A_{\mu_1\ldots
  \mu_6}\partial_{\mu_7} H_{\mu_8\mu_9\mu_{10}} .
\end{equation}
The $A_{\mu_1\ldots\mu_6}$ equation of motion yields the Bianchi
identity of the $B$ field strength $H_{\mu\nu\rho}$. However, one can
alternatively solve the $B$ field equation of motion with the result
\begin{equation}
H_{\mu\nu\rho} \sim e^\Phi \epsilon_{\mu\nu\rho \mu_1\ldots \mu_7}
K^{\mu_1\ldots\mu_7} ,
\end{equation}
with
\begin{equation}
K_{\mu_1 \ldots \mu_7} = \partial_{\left[\mu_1\right. } A_{\left. \mu_2
  \ldots \mu_7\right]} . 
\end{equation}
This means that we can trade the antisymmetric tensor $B$ for a six
form potential $A$. Choosing an appropriate normalization for the
Lagrange multiplier terms (\ref{hodge}), the effective action
(\ref{einsteinac}) in terms
of the six form potential $A$ reads
\begin{equation}
\tilde{S}_{E,univ} =\frac{1}{2\kappa^2}\int d^{10} x\sqrt{-g}\left( R -
    \frac{1}{2}\left( \partial\Phi\right)^2 - \frac{1}{2\cdot
    7!}e^\Phi K^2\right) .
\label{fivebraneaction}
\end{equation}
Also in this form the action can be supersymmetrized. In terms of the
six form potential $A$ the gravitino and dilatino supersymmetry
transformations read
\begin{eqnarray}
\delta \psi_\mu &= & D_\mu \epsilon + \frac{1}{2\cdot
  8!}e^{\frac{\Phi}{2}}\left( 3 {\Gamma_\mu}^{\nu_1 \ldots\nu_7} -7
  \delta_\mu ^{\nu_1} \Gamma^{\nu_2\ldots\nu_7}\right) K_{\nu_1 \ldots
  \nu_7} \epsilon ,\label{gravifuenf} \\
\delta \lambda & = & -\frac{1}{2\sqrt{2}}\Gamma^\mu \partial_\mu \Phi
  \epsilon - \frac{1}{2\cdot 2\sqrt{2} \cdot 7!}
  e^{\frac{\Phi}{2}}\Gamma^{\mu_1 \ldots \mu_7}K_{\mu_1 \ldots
  \mu_7}\epsilon . \label{dilifuenf}
\end{eqnarray}

In the following two subsections, we will present two solutions
preserving half of the supersymmetry.

\subsection{The fundamental string}
\setcounter{equation}{0}

The solutions we are going to discuss in the present and subsequent
sections are generalizations of extreme Reissner--Nordstr\"om black
holes. Reissner--Nordstr\"om black holes are solutions of
Einstein--Hilbert 
gravity coupled to an electro magnetic field. They carry mass and
electric or magnetic charge. Extreme Reissner--Nordstr\"om black holes
satisfy a certain relation between the charge and the mass. 
(In our case such a relation will be dictated by the requirement of
partially preserved supersymmetry.)
Replacing
the electro magnetic field strength $F$ by its dual $\star F$
interchanges electric with magnetic charge. (For a more detailed
discussion of Reissner--Nordstr\"om black holes see
e.g.\cite{Townsend:1997ku}.) 

The action (\ref{einsteinac}) bears some analogy to four dimensional
Einstein gravity 
coupled to an electro magnetic field. The difference is that the
theory is ten dimensional 
instead of four dimensional and the electro magnetic field strength is
replaced by the three form $H$. In addition, there is the scalar
$\Phi$. Since the gauge potential is now a two form which naturally
couples to the worldvolume of a string, we look for ``extreme
Reissner--Nordstr\"om black strings'' instead of black holes. 
The corresponding ansatz for the metric is
\begin{equation}
ds^2 = e^{2A}\eta_{ij} dx^i dx^j + e^{2B}\delta_{ab}dy^a dy^b ,
\end{equation}
with $i,j =0,1$ and $a,b = 2,\ldots ,9$. Further, $A$ and $B$ are
functions of
\begin{equation}
r =\sqrt{ \delta_{ab}y^a y^b},
\end{equation}
only. Here, we have taken the second step before the first one in the 
sense that first we should have thought about what kind of isometries
we would like to obtain and only afterwards we should have written
down a general ansatz respecting the isometries. Therefore, let us
perform the first step now and discuss the isometries of the
ansatz. Clearly, there is an $SO\left(1,1\right)$ isometry acting on
the $x^i$. This means, that up to Lorentz boosts, $x^i$ spans the
worldvolume of a straight static string. There is no further
dependence on $x^i$ in the ansatz because we do not wish to
distinguish a point on the worldvolume of the string. 
The other isometry acts as an
$SO\left( 8\right)$ on the $y^a$. This is the natural extension of 
the $SO\left( 3\right)$ isometry associated with non-rotating four
dimensional black holes. It is $SO\left( 8\right)$ now because the
space transverse to the string is eight dimensional (whereas in 4d
black holes the space transverse to the hole is three dimensional).
The $r$ dependence respects the $SO\left( 8\right)$
isometry. Distinguishing between different values of $r$ means
specifying the position of the string, i.e.\ $r$ measures the radial
distance from the string. 

In order not to spoil the above symmetries, we choose for the
remaining fields the ansatz
\begin{equation}
B_{01} = -e^C \,\,\,\,\,\, ,\,\,\,\,\, \Phi = \Phi\left( r\right) ,
\end{equation}
where $C$ is also a function of $r$ only. 
All other components of $B$ are zero.
Viewed as a two form, $B$ is proportional to the invariant volume form
of the string worldvolume. The factor $e^C$ may depend on $r$.

The ansatz for the target space spinors is that they all vanish. As
mentioned earlier we are interested in situations where the solution
preserves part of the supersymmetry because this ensures that we can
continuously take the string coupling from weak to strong. In
particular, the unbroken supersymmetry is parameterized by spinors
$\epsilon$ for which the gravitino and dilatino values of zero do not
change under supersymmetry transformations, i.e.\ those $\epsilon$ for
which the rhs of (\ref{gravitino}) and (\ref{dilatino}) vanish. In
order to
find such solutions for our ansatz it is convenient to represent the
ten dimensional Gamma matrices $A,B = 0,\ldots ,9$,
\begin{equation}
\left\{ \Gamma_A ,\Gamma_B\right\} = 2\eta_{AB}
\end{equation}
as a tensor product of $2\times 2$ matrices $\gamma_i$
in $1+1$ and $8\times 8$ matrices
$\Sigma_a$ in $8$
dimensions\footnote{The corresponding algebras are $\left\{ \gamma_i
  ,\gamma_j\right\} =2\eta_{ij}$ and  $\left\{ \Sigma_a
  ,\Sigma_b\right\} = 2\delta_{ab}$.}
\begin{equation}
\Gamma_A = \left( \gamma_i \otimes \mbox{\bf I} , \gamma_3\otimes
  \Sigma_a\right) ,
\end{equation}
where {\bf I} is the $8\times 8$ identity matrix and 
\begin{equation}
\gamma_3 = \gamma_0 \gamma_1 
\end{equation}
squares to the $2\times 2$ identity matrix. Further, we have to take
into account that the ten dimensional $N=1$ supersymmetry parameter
$\epsilon$ is subject to the constraint
\begin{equation}\label{tendchirality}
\Gamma_{11} \epsilon = \epsilon .
\end{equation}
Under certain conditions to be specified below the variations of the
gravitino and the dilatino vanish for
\begin{equation}
\epsilon = e^{\frac{3\Phi}{8}}\varepsilon_0 \otimes \eta_0 ,
\end{equation} 
where $\varepsilon_0$ and $\eta_0$ are $SO\left( 1,1\right)$ and
$SO\left( 8\right)$ constant spinors, respectively, which satisfy the
lower dimensional chirality conditions
\begin{equation}\label{lowerchir}
\left( 1-\gamma_3\right) \varepsilon_0 = 0 \,\,\, ,\,\,\, 
\left( 1 -\prod_{a=2}^9\Sigma_a\right) \eta_0 = 0.
\end{equation}
This breaks the supersymmetry to half the amount of the
perturbative (trivial) vacuum. (The condition (\ref{tendchirality})
could be also satisfied by choosing simultaneously the opposite
chiralities in the two equations (\ref{lowerchir}).) 

We already mentioned that only under certain conditions we can find
unbroken supersymmetries at all. Requiring that asymptotically ($r\to
\infty$) we obtain the perturbative vacuum, these conditions read
\begin{eqnarray}
A &= &\frac{3}{4}\left( \Phi -\Phi_0\right) ,\label{sugstra}\\
B & = & -\frac{1}{4}\left( \Phi -\Phi_0\right) ,\label{sugstrb}\\
C & = & 2\Phi - \frac{3}{2}\Phi_0\label{sugstrc}
\end{eqnarray}
where $\Phi_0$ is the asymptotic value of $\Phi$. Hence supersymmetry
leaves only one function out of our ansatz undetermined. This function
can be taken to be the dilaton whose equation of motion boils down to
\begin{equation}\label{sugdileom}
\delta^{ab}\partial_a \partial_b e^{-2\Phi\left(r\right)} = 0,
\end{equation}
i.e.\ the ``flat'' Laplacian of the transverse space (spanned by the
$y^a$) acting on $e^{-2\Phi}$ has to vanish. As in the case of four
dimensional black holes, we solve this equation everywhere but at the
origin $r=0$, where there are additional contributions due to a source 
string. (We do not add the source string explicitly here, but will
infer its properties (tension and charge) in an indirect way
below. For the explicit inclusion of the source term see
e.g.\cite{Duff:1995an}.) The solution to (\ref{sugdileom}) reads
\begin{equation}
e^{-2\Phi} = e^{-2\Phi_0}\left( 1 + \frac{k}{r^6}\right) ,
\end{equation}
where $k$ is an integration constant which will be related to the
string tension below.
Plugging this back into (\ref{sugstra}) -- (\ref{sugstrc}) and in
the ansatz gives the final solution.

Next, we would like to deduce the tension of the string source from
our solution. This is done by studying the Newtonian limit of general
relativity. In particular, by comparing the Einstein equation with the
geodesic equation of a point particle (which has constant mass in the
Einstein frame) one finds that the Newton potential
of the string-source is encoded in the subleading term in a large $r$
expansion of $g_{00}$.    
Therefore, we first observe that for large $r$
\begin{equation}\label{approxasy}
g_{00} = -1+\frac{3k}{4r^6} + \ldots . 
\end{equation} 
The relation between $g_{00}$ and the Newton potential of a string is
explicitly such that\footnote{Equation (\ref{newtonpot}) has the same
  form as the equation satisfied by the Newton potential. The
  numerical factor of $\frac{3}{2}= 
  \frac{7-p}{4}$ on the rhs of (\ref{newtonpot}) is a matter of
  convention, which fixes the relation between $\kappa $, the speed of
  light ($c=1$), and Newton's constant (see
  e.g.\cite{Stephani:1982ac}). We choose our convention in agreement
    with\cite{Duff:1995an}, where explicit source terms (containing
    the tension) are added.}
\begin{equation}\label{newtonpot}
\frac{1}{r^7}\partial_r\left( r^7 \partial_r g_{00}\right) =
-\frac{3}{2}\kappa^2 
T_E \frac{\delta\left( r\right)}{\Omega_7 r^7}
\end{equation}
holds, with the understanding that terms denoted by $\ldots$ in
(\ref{approxasy}) are neglected. The string tension is denoted by
$T_E$. Further, the unit volume of a seven--sphere $\Omega_7$ enters the
expression. Hence, we obtain
\begin{equation}
T_E = \frac{3 k}{\kappa^2}\Omega_7 .
\end{equation}
We put the index $E$ at the tension in order to indicate that it is
measured in the Einstein frame. What we are actually interested in, is 
the tension in the string frame. This is readily obtained by noticing
that transforming back to the string frame (asymptotically) implies
\begin{equation}
\kappa^2 \to e^{2\Phi_0} \kappa^2 \,\,\, , \,\,\, \Omega_7 \to
e^{\frac{7\Phi_0}{2}} \Omega_7 .
\end{equation}
Thus in the string frame the tension is\footnote{On dimensional
  grounds, one would expect a different scaling of $T$. In order to
  obtain this, 
  one has to take into account that $k$ is a dimensionful quantity. In
  (\ref{stringstringframetension}) 
   $k$ is given Einstein frame units.} 
\begin{equation}\label{stringstringframetension}
T = \frac{3k}{\kappa^2} e^{\frac{3\Phi_0}{2}}\Omega_7 .
\end{equation}
Recalling that our ``elementary particle'' is a string of tension
$\frac{1}{2\pi \alpha^\prime}$ and requiring that any string like object
must consist out of an integer number $N$ of elementary strings we
finally determine the integration constant $k$ to be\footnote{Here we
  use the fact that for a superposition of BPS states there is no
  binding 
  energy, i.e.\ the total tension is obtained by simply summing the
  tensions of the individual BPS states.}
\begin{equation}
k = N \frac{\kappa^2 }{6\pi \alpha^\prime \Omega_7}
e^{-\frac{3\Phi_0}{2}}. 
\end{equation}

It remains to compute the $U\left( 1\right)$ charge carried by the
vacuum. This is basically done by integrating the $B$ equation of
motion 
over the transverse space. The result is
\begin{equation}
\mu = \frac{1}{\sqrt{2}\kappa}\int_{S^7}
e^{-\Phi}\star H , 
\end{equation} 
where the integration is over an asymptotic seven-sphere enclosing the
string source. The $U\left( 1\right)$ charge is
denoted by $\mu$. Expressing the result in terms of the string tension
one obtains
\begin{equation}\label{BPSbound}
\mu = \sqrt{2}\kappa \frac{N}{2\pi \alpha^\prime} .
\end{equation}
This equality is related to partially unbroken
supersymmetry. If the configuration was not stable the tension of the
bound state would be larger than the sum of the elementary
tensions. Hence, the rhs of (\ref{BPSbound}) is larger for general
(non BPS) states. The BPS state saturates a general inequality. Since
the BPS state is stable, there can be no state with less tension and 
the same charge since otherwise the BPS state would decay into such a
state. The lower bound on the tension set by the BPS state is called
the Bogomolnyi bound. 

\subsection{The NS five brane}
\setcounter{equation}{0}
In this subsection we repeat the analysis of the previous section,
however, with the action (\ref{fivebraneaction}) instead of
(\ref{einsteinac}). Thus, we will obtain the magnetic dual of the
previously discussed string solution. This is called the NS five
brane. Its properties (tension, charge) will be fixed in terms of the
string properties via the Dirac quantization condition.
(For generalizations of the Dirac quantization condition to extended 
objects see\cite{Nepomechie:1985wu,Teitelboim:1986yc}.) As the
derivation of the NS five brane solution goes along the same lines as
the one given in the previous subsection, we will be even more sketchy 
here. Instead of the two form potential, we have now the six form
potential $A$. Since an object which extends along five spatial
dimensions naturally couples to a six form potential, we choose the
following ansatz for the metric
\begin{equation}
ds^2 = e^{2A}\eta_{ij} dx^idx^j + e^{2B}\delta_{ab}dy^a dy^b ,
\end{equation}
where now $i,j = 0,\ldots ,5$ and $a,b = 6,\ldots ,9$. The five brane
worldvolume extends along the $x^i$ directions and the functions $A$
and $B$ are allowed to depend on the radial distance from the five
brane $r$ with
\begin{equation}
r =\sqrt{\delta_{ab}y^a y^b} .
\end{equation}
The ansatz for the six form potential is
\begin{equation}\label{sixfoansatz}
A_{012345} = - e^C ,
\end{equation}
where $C$ is a function of $r$.  The components of $A$ which cannot be
obtained by permuting the indices in (\ref{sixfoansatz}) are zero. The
final input is that also the dilaton depends only on $r$,
\begin{equation}
\Phi = \Phi\left( r\right) .
\end{equation}
All fermionic fields are again set to zero. There is an unbroken
supersymmetry if we can find a spinor such that the gravitino and
dilatino transformations (\ref{gravifuenf}) and (\ref{dilifuenf}) 
vanish. It turns out that half of the supersymmetry is preserved if
the following relations hold
\begin{eqnarray}
A & = & -\frac{1}{4}\left( \Phi -\Phi_0\right) ,\\
B & = & \frac{3}{4}\left( \Phi -\Phi_0\right) ,\\
C & = & -2\Phi +\frac{3}{2} \Phi_0 ,
\end{eqnarray} 
where $\Phi_0$ denotes again the asymptotic $r\to \infty$ value of the
dilaton. (In addition to partially unbroken supersymmetry we have once
again imposed that for large $r$ the vacuum should approach the
perturbative vacuum.) 
Under these conditions the equations of motion boil down to 
\begin{equation}
\delta^{ab}\partial_a\partial_b e^{2\Phi} = 0.
\end{equation}
We solve this equation everywhere but at $r=0$ where we allow for
additional contributions due to source terms. One finds
\begin{equation}
e^{2\left(\Phi -\Phi_0\right)} = 1 + \frac{\tilde{k}}{r^2} .
\end{equation}
The integration constant $\tilde{k}$ can be fixed by exploiting the
Dirac quantization condition. To this end we compute the charge
carried by the vacuum
\begin{equation}
\tilde{\mu} =\frac{1}{\sqrt{2}\kappa} \int _{S^3} e^\Phi \star K =
\frac{\sqrt{2} \Omega_3 \tilde{k}}{\kappa} e^{\frac{\Phi_0}{2}} ,
\end{equation}
where the integral is over an asymptotic three--sphere surrounding the
five brane and $\Omega_3$ denotes the volume of a unit three--sphere.  
Now, the Dirac quantization condition reads
\begin{equation}
\tilde{\mu}\mu = 2\pi \tilde{N}N ,
\end{equation}
where $\mu$ is the charge of $N$ elementary strings
(\ref{BPSbound}). The number of five branes (number of magnetic
charges) is $\tilde{N}$.  
This fixes the integration constant, 
\begin{equation}
\tilde{k} =\frac{\pi
  \tilde{N}}{T_S\Omega_3}e^{-\frac{\Phi_0}{2}} , 
\end{equation}
where $T_S$ is the elementary string tension ($T_S =\frac{1}{2\pi
  \alpha^\prime}$ 
in the string frame).
By computing the gravitational potential in the Newtonian limit, one
finds the tension of the five brane in the Einstein frame (with $T_S$
and $\kappa$ also in Einstein frame units)
\begin{equation}
\tilde{T}_E = \frac{\pi
  \tilde{N}}{T_S\kappa^2}e^{-\frac{\Phi_0}{2}} .
\end{equation}
The mass dimension of $\tilde{T}_E$ is six. Hence, we obtain
\begin{equation}
\tilde{T} = e^{-2\Phi_0}\frac{2\pi \alpha^\prime \pi \tilde{N}}{\kappa^2} 
\end{equation}
in the string frame. We observe that the five brane tension behaves
as $1/g_s ^2$. In the perturbative region the NS five brane is very
heavy whereas it becomes lighter when the string coupling increases.

The NS five brane is an extended object for which we did not give a
perturbative description. Indeed, such a description is not known. One
could try to quantize strings in the NS five brane background. This is
possible only in certain spatial regions. Firstly, for large $r$ the
background becomes flat, and we know how to quantize strings
there. But also in the background at $r \to 0$ (the near horizon
limit) one can find a
quantized string theory. In that limit the string frame metric reads
\begin{equation}
ds_s ^2 = \eta_{ij} dx^i dx^j +\tilde{k} \left( d\log r\right)^2 +
\tilde{k} d\Omega_3 ^2 ,
\end{equation}
and the dilaton is linear in $\rho =\log r$. With $d\Omega_3 ^2$ we
denote the metric of a unit three--sphere. The NSNS field strength $H$
is a constant times the volume element $d\Omega_3$.  The geometry
factorises into a $5+1$ dimensional Minkowski space times the
direction on which the dilaton depends linearly times an $S^3$. Since
$S^3$ is an $SU\left( 2\right)$ group manifold, string theory can be
quantized in such a background. For more details
see\cite{Callan:1991dj} or the review\cite{Callan:1991at}.
\section{Type II branes}\label{dbranesnonp}
\setcounter{equation}{0}

Like in the previous sections, we are interested in setups where only a
truncated version of the effective actions (see section
\ref{effectiveactions} ) is relevant. The bosonic part of the
truncated type II action reads
\begin{equation}
S = \frac{1}{2\kappa^2} \int d^{10}x\left( e^{-2\Phi}\left( R+4\left(
  \partial \Phi\right)^2\right) -
  \frac{1}{2\left( p+2\right)!} F_{p+2}^2\right) ,
\end{equation}
where $F_{p+2}$ denotes the field strength of an RR $p+1$ form
potential. For type IIA $p$ is even whereas it is odd for type IIB
theory. For $p=3$ the action has to be supplemented by the constraint
that the field strength is selfdual.   
The ($p+2$ form) field strengths  are not all
independent but 
related by Hodge duality to the field strength corresponding to $6-p$
(an $8-p$ form field strength). We will restrict the discussion to the
cases $0\leq p < 7$. For $p=7$ the solution presented
in\cite{Greene:1990ya} is relevant. The 8-brane appears as a solution
of massive 
type IIA supergravity\cite{Bergshoeff:1996ui}. How this is related to
string theory (or rather M-theory) is discussed in the recent
paper\cite{Haack:2001iz} (see also references therein).  
We will consider only a single
relevant $p$ at a time. The field redefinition (\ref{einsteinframe})
takes us to the action in the Einstein frame
\begin{equation}
S_E = \frac{1}{2\kappa^2}\int d^{10}x \sqrt{-g}\left( R
  -\frac{1}{2}\left( \partial\Phi\right)^2-
  \frac{1}{2\left( p+2\right)!}
  e^{\frac{3-p}{2}\Phi}F_{p+2}^2\right) .
\end{equation} 

The $p$-brane ansatz reads
\begin{equation}
ds^2 = e^{2A}\eta_{ij}dx^i dx^j + e^{2B}\delta_{ab} dy^a dy^b ,
\label{P-branemetr}
\end{equation}
with $i,j = 0, \ldots ,p$ and $a,b = p+1,\ldots ,9$. $A$ and $B$ are
functions of
\begin{equation}
r=\sqrt{\delta_{ab}y^a y^b}
\end{equation}
The dilaton is also taken to be a function of $r$. Let us first
exclude the case $p=3$ from the discussion. For the $p+1$ form gauge
field we choose 
\begin{equation}\label{pforman}
A_{0,\ldots, p} = -e^C ,
\end{equation}
where $C$ is a function of $r$, and all other components of $A$ (which
cannot be obtained by permuting the indices in (\ref{pforman})) are
zero. All the other fields (NSNS $B$-field, the remaining RR forms,
and the fermions) are zero. The BPS condition leaves one out of the
four functions $A$, $B$, $C$ and $\Phi$ undetermined. Choosing for 
convenience $C$ to be the undetermined function, these conditions read 
\begin{eqnarray}
A & = & \frac{7-p}{16} \left( C - C_0\right) ,\label{bpscondition1} \\
B & = & -\frac{p+1}{16} \left( C - C_0\right) ,\label{bpscondition2}\\
\Phi & = & \frac{p-3}{4}\left( C - C_0\right) +\frac{4C_0}{p-3}
,\label{bpscondition3}  
\end{eqnarray}
where again the boundary condition that for $r \to \infty$ the
background should be trivial has been imposed. $C_0$ denotes the
asymptotic value of $C$ which is related to the asymptotic dilaton
value
\begin{equation}
C_0 = \frac{p-3}{4}\Phi_0 .
\end{equation}
The equations of motion reduce to
\begin{equation}
\delta^{ab}\partial_a \partial_b e^{-C} = 0.
\end{equation}
We solve this by
\begin{equation}
e^{-C} = e^{-C_0} + \frac{k_p}{r^{7-p}}.
\end{equation}
The RR charge of the vacuum is
\begin{equation}\label{dchargedef}
\mu_p = \frac{1}{\sqrt{2}\kappa}\int_{S^{8-p}}e^{\frac{3-p}{2}\Phi}
  \star F_{p+2} = \frac{7-p}{\sqrt{2}\kappa}\Omega_{8-p} k_p ,
\end{equation}
where the integration is over an asymptotic $\left(8-p\right)$--sphere
surrounding 
the $p$ brane, and $\Omega_{8-p}$ is the volume of the unit
sphere. Now we try whether we can identify the type II p-branes with
the D-branes discussed in section \ref{D-branes}. This trial is
motivated by the observation that the D-branes
considered in section \ref{D-branes} are also extended objects 
carrying RR charge. In section \ref{D-branes} we computed the charge
of a single D-brane to be (see (\ref{rrcharge}) and (\ref{tension}))
\begin{equation}\label{eq:dbratension}
\mu_p ^{\mbox{\tiny single brane}} = \sqrt{2\pi} \left( 4\pi^2
  \alpha^\prime\right)^{\frac{3-p}{2}} .
\end{equation}
Assuming that the vacuum considered in the present section is composed
out of an integer number $N_p$ of single D-branes, we identify ($T_S$ 
denotes the frame dependent tension of a single fundamental string)
\begin{equation}\label{integrationconstantfixed}
k_p = N_p \frac{2\kappa \left(
    4\pi^2/T_S\right)^{\frac{3-p}{2}}\sqrt{\pi}}{\left(7-p\right)
    \Omega_{8-p}}, 
\end{equation}
such that
\begin{equation}
\mu_p = N_p \mu_p ^{\mbox{\tiny single brane}} .
\end{equation}
A first consistency check is to observe that the Dirac quantization
condition
\begin{equation}
\mu_p \mu_{6-p} = 2\pi N_p N_{6-p}
\end{equation}
is satisfied. After we have fixed the integration constant $k_p$ the
tension of the brane solution is determined. Because the vacuum
considered here and the D-branes considered in section \ref{D-branes}
are BPS objects the tension is related to the charge and we expect
that our tension should be in agreement with (\ref{tension}). Let us
nevertheless 
compute it explicitly. To this end, we write down the
asymptotic expansion of $g_{00}$
\begin{equation}\label{D:assy}
g_{00} = -1 +\frac{7-p}{8} \frac{k_p}{r^{7-p}}e^{\frac{p-3}{4}\Phi_0}
+\ldots . 
\end{equation}
The tension is given via (see also (\ref{newtonpot}))
\begin{equation}
\frac{1}{r^{8-p}}\partial_r\left( r^{8-p}\partial_r g_{00}\right) =
-\frac{7-p}{4} \kappa^2 T_{p,E} \frac{\delta\left(
    r\right)}{\Omega_{8-p} r^{8-p}} ,
\end{equation}
where $T_{p,E}$ denotes the tension in Einstein frame units and it is
understood that terms denoted by dots in (\ref{D:assy}) are
neglected. This yields
\begin{equation}
T_{p,E} = \frac{7-p}{2\kappa^2} \Omega_{8-p} k_p e^{\frac{p-3}{4}\Phi_0} .
\end{equation}
Since $T_{p,E}$ has mass dimension $p+1$, it receives a factor of
$e^{-\frac{p+1}{4}\Phi} $ under the transformation to the string
frame. Taking this and (\ref{integrationconstantfixed}) into account,
we find for the string frame tension
\begin{equation}
T_p = N_p e^{-\Phi_0}\kappa\sqrt{\pi} \left( 4\pi^2
  \alpha^\prime\right) ^\frac{3-p}{2}\frac{1}{\kappa} ,
\end{equation}
in agreement with (\ref{tension}). Thus, we found that the $p$-brane 
vacuum can be viewed as consisting out of an integer number of
``elementary'' (or magnetic) D-branes considered in section
\ref{D-branes}. 

So far, we have derived this result only in the case $p\not= 3$. In
the case $p=3$, the condition (\ref{bpscondition3}) 
is changed into  
\begin{equation}
\Phi = \Phi_0 \;\;\; ,\;\;\; C_0 = 0 .
\end{equation}
The selfduality condition can be imposed by replacing $F_5$ from our
ansatz with $F_5 + \star F_5$, 
\begin{equation} \label{selfdualityreplacement}
F_5 \rightarrow F_5 + \star F_5 .
\end{equation}
The solution for $C$ is
\begin{equation}\label{d3background}
e^{-C} = 1 + \frac{k_3}{r^4} .
\end{equation}
The ``electric'' charge is
\begin{equation}
\mu_ 3 = \frac{1}{\sqrt{2}\kappa}\int_{S^5} \star d A_{0123} =
\frac{4}{\sqrt{2}\kappa }\Omega_5 k_3 .
\end{equation}
The replacement (\ref{selfdualityreplacement}) implies that the
solution carries also a magnetic charge $\tilde{\mu}_p =\mu_p$. Thus,
the Dirac quantization condition yields ($N_3$ is the number of
D3-branes) 
\begin{equation}\label{dreibraneka}
k_3 = N_3 \sqrt{\pi}\frac{\kappa}{2\Omega_5},
\end{equation}
such that 
\begin{equation}
\mu_3 = \sqrt{2\pi}N_3 .
\end{equation}
For the tension, one obtains in the string frame 
\begin{equation}
T_3 = N_3 e^{-\Phi_0}\frac{\sqrt{\pi}}{\kappa} .
\end{equation}
Thus, also for $p=3$ we can consistently assume that the vacuum
solution is made out of an integer number of the D3-branes introduced
in section \ref{D-branes}. We will return to the solution for $p=3$ in
section \ref{adssection}.

We stop our discussion on the appearance of branes as vacua of the
effective actions at this point. We should, however, mention that
there are 
many more configurations which can be constructed. For example, one
can find vacua where branes lie within the worldvolume of other
higher dimensional branes. Such studies confirm the result of
section \ref{numberofnd} that supersymmetry is completely broken
unless the number of ND directions is an integer multiple of four.

A final remark about the BPS vacua of type I theory is in order.
Although we called this section
``type II branes'', the discussion applies to type I theory as well.  
In the closed string sector of type I theory the NSNS $B$ field is
projected out, and thus there is neither a fundamental string nor an
NS five brane vacuum in the effective type I theory. On the other hand, the
RR two form potential survives the projection. Hence, type I theory
possesses the D1 and the D5 brane vacua.
\newpage
\chapter{Applications}\label{chap:appl}
In this chapter, we are going to present some applications of the
branes discussed so far. In the following, we will show that branes
are a useful tool in 
supporting duality conjectures involving an interchange between strong
and weak couplings. As a first example we consider dualities among 
different string theories. Thereafter, field theory dualities will be 
translated into manipulations within certain brane setups. Next,
we 
want to present the AdS/CFT correspondence -- a duality between
closed and open string theory, or in first approximation between
gravity and gauge theory. 
Finally, we argue that branes allow constructions in which the string
scale is about a TeV. Such setups have the prospect of being
discovered in the near future.
There are many more applications of branes
in theoretical physics. Some of them we will list in chapter
\ref{furtherreading} containing suggestions for further reading.

\section{String dualities}

There are many excellent reviews on string dualities and we do not plan
to provide an introduction into this subject here. We just want to
summarize how branes are mapped among each other under duality
transformations. We start by drawing the M-theory star in figure
\ref{fig:mstar}. 
%%%%%%%%%%%%%%%%%%%%%%%%%%%%%%%%%%%%
\begin{figure}
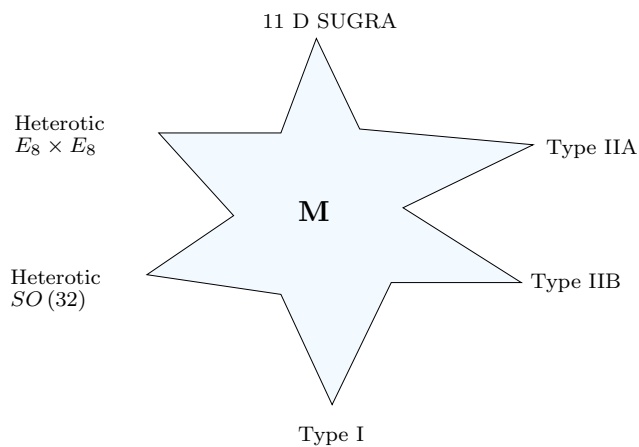
  
\begin{center}
\input mstar.pstex_t
\end{center}
\caption{M-theory star}
\label{fig:mstar}
\end{figure}
%%%%%%%%%%%%%%%%%%%%%%%%%%%%%%%%%%%%
The idea behind this picture is that the theories written at the tips
of the ``star'' are different descriptions of one underlying theory 
called M-theory. This underlying theory is not known. It is assumed to
possess a moduli space which looks like figure \ref{fig:mstar}.   
The picture is
supported by evidences for the conjecture that all the other theories
appearing in figure \ref{fig:mstar} are related by motions in
their moduli spaces. 

Let us briefly summarize how these theories are connected. We start at
the top of the star (11D SUGRA) and work our way down to the bottom
(type I), first counter clockwise. Compactifying eleven dimensional
supergravity on an interval ($S^1/{\mathbb Z}_2$) yields the effective
field theory of the heterotic $E_8\times E_8$ string. The dilaton is
related to the 
length of the interval such that the string coupling is small when the
interval is short. The $E_8\times E_8$ fields live as twisted sector
fields at the ends of the interval (the orbifold nine planes).  If we
take the string coupling of the $E_8\times E_8$ heterotic string
to be strong, 11D supergravity on an interval provides the more
suitable description. The connection between the $E_8\times E_8$ and
the $SO\left( 32\right)$ string was already discussed in section
\ref{t-super}. It does not relate strong with weak coupling but small
with large compactification radii in nine dimensions. The heterotic
$SO\left( 32\right)$ string is connected to type I strings by a
strong/weak coupling duality. Now, let us go back to the top of the
star and go down clockwise. Type IIA supergravity can be obtained by
compactifying 11 dimensional supergravity on a circle. The radius of
the circle determines the vev of the dilaton. For small string
coupling the circle is small, and for strong coupling it is large. The
connection between type IIA and type IIB strings is seen by
compactifying further down to nine dimensions and inverting the
radius, as argued in section \ref{t-super}.  Type I theory is obtained
by gauging worldsheet parity of type IIB strings and adding the
D-branes needed to ensure RR-charge conservation (in a sense these can
be viewed as twisted sector states).

Because the branes we have discussed are stable under deformations in
the moduli space, they should be mapped in a one-to-one way onto each
other by string dualities. Since eleven dimensional supergravity did
not appear until now in our discussion (it does not correspond to an
effective weakly coupled string theory), we have to list the relevant
BPS branes of 11 dimensional supergravity. 11 dimensional supergravity
contains a three form gauge potential which can be Hodge dualized to a
six form gauge potential. Analogously to the solutions found in the 
previous chapter, one finds thus a membrane (2 brane) and a five
brane.  

Let us now walk once around the star in figure \ref{fig:mstar} in a
clockwise direction and follow the branes along this journey. Upon
compactifying one of the eleven dimensions the momentum into this
direction becomes quantized. The off diagonal metric components
containing one 11 label become a Kaluza Klein gauge field -- a one
form potential, which can be Hodge dualized (with respect to the non
compact directions) to a seven form potential.   
The associated BPS states are zero and six branes. 
These become the D0 and the D6 branes in the type IIA picture.
For the branes which exist already in the uncompactified theory, there
are two options within the compactification. The compact dimension can
be transverse or longitudinal. Hence, the membrane will be either 
described by a fundamental string or by a D2 brane in weakly coupled
type IIA theory, and the five brane yields the D4 brane and the NS
five brane of type IIA theory.

Compactifying further down to nine dimensions and taking the
decompactification limit after a T-duality transformation, type IIA 
theory goes over into type IIB theory. The D-branes gain or lose one
spatial direction due to the T-duality, and hence we obtain all the
D-branes of type IIB theory. Type IIB theory possesses a symmetry
which is not depicted in figure \ref{fig:mstar}. This is an $SL\left( 2 ,
  \mathbb{ Z}\right)$ symmetry which we do not want to discuss in
detail. For later use we state that the $SL\left( 2,\mathbb{
    Z}\right)$ symmetry contains a transformation called S duality. S
duality interchanges strong with weak coupling, the D1 brane with a
fundamental string and the D5 brane with the NS five brane. The D3
brane stays a D3 brane under S-duality.\footnote{We do not include the
  D7 brane and its counterpart, the D instanton (related by Hodge
  duality), into the discussion.} 

Type I strings are obtained by projecting out worldsheet parity in type
IIB strings. This removes the fundamental string, the NS five brane,
and the D3 brane from the spectrum of BPS states. The remaining states
are the D1 and the D5 brane. Under the strong/weak coupling duality
mapping of type I theory to the $SO\left( 32\right)$ heterotic theory, 
these become the 
fundamental string and the NS five brane of the heterotic string. The
BPS spectrum is not affected when going over to the $E_8\times E_8$
heterotic string via T-duality. The $E_8\times E_8$ theory is supposed
to be dual to 11 dimensional supergravity on $S^1
/\mathbb{Z}_2$. Therefore, let us discuss which of the branes of 11
dimensional supergravity survive the ${\mathbb Z}_2$ projection. First
of all, the zero and the six branes are projected out since the
Kaluza-Klein gauge field is odd under changing the sign of the
eleventh coordinate. In order to deduce the ${\mathbb Z}_2$ action on
the three form potential $C$, we note that the action of 11
dimensional supergravity contains a Chern Simons term
$$ \int C \wedge dC\wedge dC .$$
This term is symmetric under the ${\mathbb Z}_2$ if $C$ receives an
additional sign, i.e.\ a $C$ component containing an 11 label is even
under the ${\mathbb Z}_2$. Conversely, the ${\mathbb Z}_2$
even components of the dual six form potential do not contain an 11
label. From the ten dimensional perspective, a one brane and a five
brane survive the ${\mathbb Z}_2$ projection. These are the
fundamental string and the NS five brane in the heterotic
description. 

Hence, we have seen that continuous changes of M theory moduli
preserve the spectrum of BPS branes. We have identified dual
descriptions of branes. Note also that not all tips of the star in
figure \ref{fig:mstar} are connected by continuous changes of
moduli. For example, 11 dimensional supergravity on $S^1$ is not
continuously
connected to 11 dimensional supergravity on $S^1/{\mathbb
  Z}_2$.   
Therefore, the BPS branes of the circle compactified 11 dimensional
supergravity have a one-to-one description in type IIA theory, but the
type IIA BPS branes cannot all be given a heterotic description, and
so on. 
\section{Dualities in Field Theory}

Another area where supersymmetry allows insight into strongly coupled
regions of perturbatively formulated theories are supersymmetric
field theories. In this section we will focus on four dimensional $N=1$ 
gauge theories with matter in the fundamental representation
(supersymmetric QCD). For the various other examples we refer to the 
literature (see chapter \ref{furtherreading}). In supersymmetric
theories, non-renormalization theorems allow to study the moduli 
space in strongly coupled regions. In $N=1$ theories, the
superpotential must be holomorphic in the fields. This often restricts
its form, and the moduli space is found by searching for flat
directions in the superpotential. A thorough analysis of $N=1$
$SU\!\left( N_c\right)$ gauge theory with $N_f$ chiral multiplets in
the fundamental representation led Seiberg to the conjecture that
perturbatively completely different looking theories are connected in
moduli space. Analyzing results on beta functions in such theories,
one finds that for $\frac{3}{2}N_c < N_f < 3N_c$  the
beta function becomes zero at a certain (strong) coupling. Hence,
such gauge theories flow to a conformal fixed point in the infrared
(they are asymptotically free). The amazing result of Seiberg's
analysis is that an $SU\!\left( N_f - N_c\right)$ theory with $N_f$
chiral 
  multiplets in the fundamental representation of $SU\!\left( N_f
    -N_c\right)$ and $N_f ^2$ gauge singlets flows to the same
  infrared fixed point as the above mentioned $SU\!\left( N_c\right)$
  theory. Thus, the moduli spaces of the two theories are connected in
  the strong coupling region. The field theory analysis involves first
  finding a duality map between conformal primaries at the infrared
  fixed point and to test whether the picture is consistent under
  continuous deformations. Another quite non trivial consistency check
  is that the `t Hooft anomaly matching conditions are satisfied.
In the present section we will sketch how
  the moduli spaces of the two theories mentioned above can be
  connected by simply playing around with branes. Throughout this
  section we will neglect the back reaction of branes on the target
  space geometry, i.e.\ we take a limit where gravity decouples. 

Our first task is to translate $N=1$ $SU\!\left( N_c\right)$
supersymmetric gauge theory with $N_f$ chiral multiplets in the
fundamental representation into a brane setup. A setup yielding the
desired theory is drawn in figure \ref{fig:braneqcd}.
%
%%%%%%%%%%%%%%%%%%%%%%%%%%%%%%%%%%%%
\begin{figure}
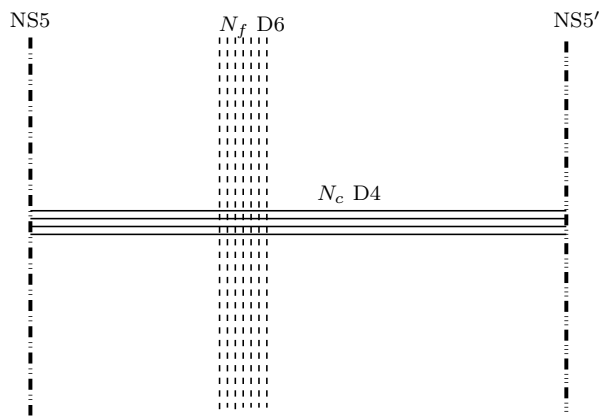
  
\begin{center}
\input seiberbrane-a.pstex_t
\end{center}
\caption{Brane setup for supersymmetric QCD. It has to be looked at in
  combination with table \ref{tab:bsetupa}.}
\label{fig:braneqcd}
\end{figure}
%%%%%%%%%%%%%%%%%%%%%%%%%%%%%%%%%%%%
%
Since it is difficult to draw
pictures in ten dimensions we supplement the figure by table
\ref{tab:bsetupa} where
hyphens stand for longitudinal and dots for transverse dimensions.
\begin{table}
\begin{center}\begin{tabular}[h]{| c || c|c|c|c|c|c|c|c|c|c|}
\hline
   & 0 & 1  & 2  & 3  & 4  & 5  & 6     & 7     & 8     & 9 \\
\hline\hline
NS5 & -- & -- & -- & -- & -- & -- & $\cdot$ & $\cdot$ & $\cdot$&
$\cdot$ \\ 
\hline
NS5$^\prime$ & -- & -- & -- & -- & $\cdot$ & $\cdot$ & $\cdot$ &
$\cdot$ & -- & -- \\
\hline
D4 & -- & -- & -- & -- & $\cdot$ & $\cdot$ & -- & $\cdot $ & $\cdot $
& $\cdot $ \\
\hline
D6 & -- & -- & -- & -- & $\cdot $ & $\cdot $ & $\cdot $ & -- & -- & --\\ 
\hline
\end{tabular}\end{center}
\caption{Brane setup for supersymmetric QCD. The numbers in the first
  line label the dimensions. Hyphens denote longitudinal and dots
  transverse dimensions.}\label{tab:bsetupa}
\end{table}
The D4-brane stretches in the sixth direction between the two NS5
branes. Hence, its extension along the sixth dimension is given by the
finite distance of the NS5 and the NS5$^\prime$ brane. If this
distance is shorter than the experimental resolution, the theory on
the D4-branes is effectively $3+1$ dimensional. The positions of the
NS5, NS5$^\prime$ and the D4 in the seventh dimension must coincide 
(simply for geometrical reasons). 
We take $N_c$ of such D4-branes in order to obtain $SU\left(
  N_c\right)$ gauge theory.  The 
position of the D4-branes in the transverse directions is fixed by the
condition that it stretches between the NS5 and NS5$^\prime$
brane. The scalar fields in the adjoint of the gauge group correspond
to collective coordinates for those positions. They are projected out
by the boundary condition. Therefore, the theory on the D4-branes
can admit at most $N=1$ supersymmetry (viewed from a $3+1$ dimensional
perspective). We will argue in a moment that there is partially
unbroken supersymmetry in the above setup. Before, let us comment on
the role of the D6-branes. A string starting on a D6 and ending on a
D4-brane transforms in the fundamental representation of $SU\!\left( 
  N_c\right)$. If we take $N_f$ D6-branes we obtain the $N_f$ desired
multiplets in the fundamental representation of the gauge group. The
$SU\left( N_f\right)$ gauge theory becomes the flavour symmetry of our
supersymmetric QCD. (Indeed, the effective four dimensional coupling
is obtained by integrating over the extra dimensions. It is inversely
proportional to the volume of the extra dimensions. Since the D6-brane
worldvolume contains non-compact extra dimensions, the $SU\left(
  N_f\right)$ dynamics decouples and we are left with a global
symmetry.) 

After we successfully constructed a brane setup for the gauge theory we
are interested in, we should check whether this brane setup is
consistent. One could, for example, couple it to gravity and look for
explicit solutions describing such a setup. This is, however, rather
complicated. What we will do instead, is to take a setup from which we
know that it is consistent and connect it to the above setup through a
chain of string dualities. A setup of which we know that it is
consistent is given in table \ref{tab:bsetupdual}.
\begin{table}
\begin{center}\begin{tabular}[h]{| c || c|c|c|c|c|c|c|c|c|c|}
\hline
   & 0 & 1  & 2  & 3  & 4  & 5  & 6     & 7     & 8     & 9 \\
\hline\hline
D5 & -- & -- & -- & -- & -- & -- & $\cdot$ & $\cdot$ & $\cdot$&
$\cdot$ \\ 
\hline
D5$^\prime$ & -- & -- & -- & -- & $\cdot$ & $\cdot$ & $\cdot$ &
$\cdot$ & -- & -- \\
\hline
F1 & -- & $\cdot $ & $\cdot$ & $\cdot $ & $\cdot$ & $\cdot$ & -- &
   $\cdot $ & $\cdot $ 
& $\cdot $ \\
\hline
D3 & -- & $\cdot$ & $\cdot$  & $\cdot $ & $\cdot $ & $\cdot $ & $\cdot
   $ & -- & -- & --\\  
\hline
\end{tabular}\end{center}
\caption{Dual brane setup for supersymmetric QCD. This can easily
   be checked to be consistent.}\label{tab:bsetupdual}
\end{table}
Here, a fundamental string (F1) stretches between two D5-branes (D5
and D5$^\prime$). This is consistent by the definition of a
D-brane. Above, we have argued that at most $N=1$ supersymmetry
survives on the (interval compactified) D4-brane. An argument that
this is exactly the case can be given by counting the ND directions of
the setup in table \ref{tab:bsetupdual} (see also section
\ref{numberofnd}). The D5 and the D3-brane 
provide eight ND directions (all but the zeroth and sixth
dimension). An open string starting on a D5$^\prime$-brane and ending
on a D3-brane has four ND directions (1237). Finally, the string
stretched between the D5 and D5$^\prime$-brane has mixed (ND)
boundary conditions along the four dimensions: 4589. Hence, the number
of ND directions is always an integer multiple of four, indicating
that the spectrum possesses some supersymmetry. (There are also more
precise methods to investigate the number of preserved
supersymmetries. One can study the conditions for vanishing gravitino
and dilatino variations in the rigid limit, see e.g.\
\cite{Elitzur:1997fh} for such an analysis within the present
context.) It remains to see that the setup in table
\ref{tab:bsetupdual} is connected to the one which we are interested
in (table \ref{tab:bsetupa}). Table \ref{tab:bsetupdual} contains
branes of type IIB theory. Therefore, we can apply an S-duality
(shortly described in the previous section) on this system. This takes
the D5 and D5$^\prime$-brane to the NS5 and NS5$^\prime$ brane of
table \ref{tab:bsetupa}. The fundamental string (F1) turns into a D1
string and the D3-brane remains invariant under S-duality. Performing
a T duality along the first, second and third dimension (replacing
type IIB by IIA) yields the configuration of table \ref{tab:bsetupa}. 

In the following we will describe a path in the moduli space of the
setup in figure \ref{fig:braneqcd} taking us to the dual theory found by
Seiberg. We will do so by essentially interchanging the position of
the NS5 with the NS5$^\prime$ brane. This involves however some
subtleties which we will mention but not elaborate on. For more
details we ask the interested reader to consult\cite{Elitzur:1997fh}
or literature to be given in chapter \ref{furtherreading}. 
Our first step is to move the D6-branes to the left of the NS5
brane. When the D6-branes cross the NS5 brane, $N_f$ additional
D4-branes  
stretching between the D6-branes and the NS5 brane are
created\cite{Hanany:1997ie}. After the D6-branes have been moved to
the left of the NS5 brane, there is a point in moduli space where
there are no D4-branes stretching between the NS5 and the NS5$^\prime$ 
brane. This can be achieved by connecting the $N_c$ D4-branes
stretching between 
NS5$^\prime$ and NS5 branes with $N_c$ out of the $N_f$ D4-branes
which stretch between D6-branes 
and NS5$^\prime$ branes. The result of performing this first step in
moduli space is drawn in figure \ref{fig:braneqcdd6l}.
%%%%%%%%%%%%%%%%%%%%%%%%%%%%%%%%%%%%
\begin{figure}
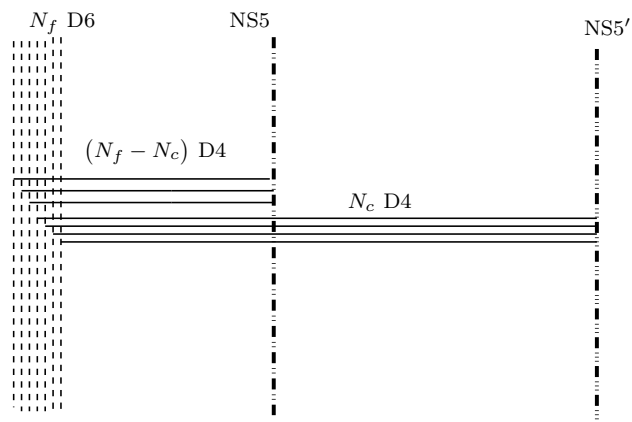
  
\begin{center}
\input seiberbrane-b.pstex_t
\end{center}
\caption{Brane setup of figure \ref{fig:braneqcd} after pushing the
  D6-branes past the NS5 brane.} 
\label{fig:braneqcdd6l}
\end{figure}
%%%%%%%%%%%%%%%%%%%%%%%%%%%%%%%%%%%%

Now, the boundary conditions are such that we can displace the NS5
brane in the seventh dimension. After doing so, it can be moved to the
right of the NS5$^\prime$ brane along the sixth
dimension.\footnote{Here, it is important that the NS branes do not
  meet when passing each other.} As soon as
the NS5 brane is situated to the right of the NS5$^\prime$ brane, we 
realign it in the seventh dimension with the positions of the
NS5$^\prime$ and the D4-branes. There are now $\left(N_f -N_c\right)$
D4-branes starting at the D6-branes passing through the NS5$^\prime$
branes and ending on the NS5 brane. These we break on the NS5$^\prime$
brane. The 
picture drawn in figure \ref{fig:braneqcdns5r} emerges.
%%%%%%%%%%%%%%%%%%%%%%%%%%%%%%%%%%%%
\begin{figure}
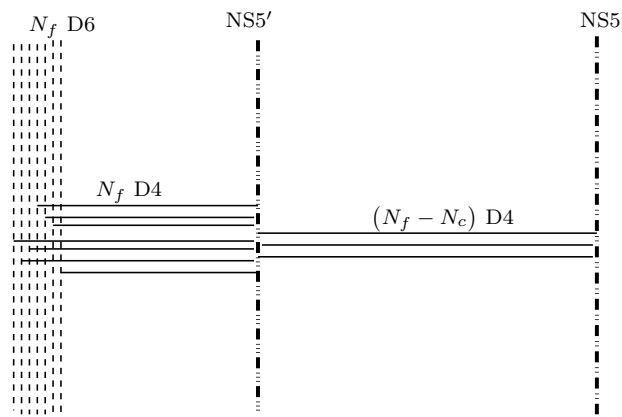
  
\begin{center}
\input seiberbrane-c.pstex_t
\end{center}
\caption{Brane setup for the dual gauge theory.}
\label{fig:braneqcdns5r}
\end{figure}
%%%%%%%%%%%%%%%%%%%%%%%%%%%%%%%%%%%%

Finally, we need to read off the perturbative formulation of the 
field theory corresponding to figure \ref{fig:braneqcdns5r}. The gauge
group of the theory living on the D4-branes stretching between the NS5
and the NS5$^\prime$ brane is $SU\left( N_f -N_c\right)$. There are
$N_f$ chiral multiplets in the fundamental of the gauge group coming
from strings stretching between the $N_f$ D4-branes on the left and
the $N_f -N_c$ D4-branes in the middle. The D4-branes to the left can
move (fluctuate) in the eighth and ninth dimension. This gives rise to
$N_f^2$ chiral multiplets which are singlets under the gauge group. 

In this section we have seen that branes can be useful tools in
deriving (or at least illustrating) quite nontrivial connections
between gauge field theories. Our purpose was to provide 
the rough ideas on how this works within an example. The reader who
found this 
interesting is strongly advised to check the literature (chapter
\ref{furtherreading}) for more details and subtleties. 
\section{AdS/CFT correspondence}\label{adssection}

In this section we will describe a duality between gravity and field
theory, or from a stringy perspective between closed string
excitations and open string excitations. We will focus on the most
prominent example where the field theory is ${\cal N}=4$
superconformal 
$SU\left(N\right)$ Yang-Mills theory\footnote{In
  order to distinguish between the number of branes and the number of
  supersymmetries we use ${\cal N}$ for the number of supersymmetries
  in the present section.} (the theory of open string
excitations ending on D3 branes) and gravity lives on an $AdS_5\times
S^5$ space (the near horizon geometry of D3 branes). In the next
subsection we will state the duality conjecture and mention the most
obvious consistency checks. Instead of elaborating on the various more
involved consistency checks which have been performed in the
literature, we will discuss an application of the duality. We will use
the gravity side of the conjecture (the theory of closed string
excitations) to compute a Wilson loop in field theory. This will be
done in a semiclassical approximation. We will also discuss next to
leading order corrections. In order to avoid disappointment, we should
mention here that we will not give a quantitative result for the next
to leading order corrections.

\subsection{The conjecture}
\setcounter{equation}{0}
From section \ref{dbranesnonp} we recall that in the case of the D3
branes the truncated action in the Einstein frame and in the string
frame look almost the same. We will work in the string frame and
absorb the constant dilaton into the definition of the gravitational
coupling $\kappa$. Choosing in addition a convenient\footnote{The
  precise relation is (in Einstein frame units) $\kappa^2 = 16 \pi^7
  {\alpha^\prime}^4$, where $\Omega_5 = \pi^3$ has been used (see
  (\ref{dreibraneka})). Plugging 
  this into (\ref{BPSbound}), one finds that for this choice the NSNS
  charge of the fundamental string equals the RR charge of the D1 string
  (\ref{eq:dbratension}). This implies that the numerics involved in
  the S duality of type IIB simplifies to $g_s \rightarrow
  \frac{1}{g_s}$.}    
numerical relation
between $\alpha^\prime$ and $\kappa$, we can write (see
(\ref{d3background})) 
\begin{equation}
e^{-C} = 1+\frac{4\pi g_s N {\alpha^\prime}^2}{r^4} ,
\end{equation}
where $g_s$ denotes the string coupling, and $N=N_3$ is the number of
D3 branes. Recall also that the metric 
is (\ref{P-branemetr})
(use (\ref{bpscondition1}) and (\ref{bpscondition2}) and $i,j = 
0,\ldots ,4$, and we parameterize the transverse space by polar
coordinates, i.e.\ $d\Omega_5 ^2$ is the metric of a unit five sphere) 
\begin{equation}\label{adscrosssmetric}
ds^2 = e^{\frac{C}{2}}\eta_{ij}dx^i dx^j
+e^{-\frac{C}{2}}\left( dr^2 + r^2 d\Omega_5 ^2\right) .
\end{equation}

Now we take the near horizon limit following the prescription
\begin{equation}\label{maldacenalimit}
\alpha^\prime \to 0 \;\;\; \mbox{and} \;\;\; U\equiv
\frac{r}{\alpha^\prime} \;\;\; \mbox{fixed}. 
\end{equation}
The first limit ensures that the field theory on the brane decouples
from gravity living in the bulk. The second limit implies that we zoom
in on the near horizon region. It is taken such that the mass of an
open string stretching between the $N$ D3 branes and some probe brane
at a finite distance is constant. Performing the limit
(\ref{maldacenalimit}) the metric (\ref{adscrosssmetric})
becomes\footnote{One can obtain this metric directly when dropping the
  requirement of an asymptotically trivial background in the search
  for BPS branes in section \ref{dbranesnonp}.}
\begin{equation}\label{ads5timess5}
ds^2 = \alpha^\prime \left[ \frac{U^2}{\sqrt{4\pi g_s
      N}}\eta_{ij}dx^idx^j + \sqrt{4\pi g_s N}\frac{dU^2}{U^2} +
      \sqrt{4\pi g_s N} d\Omega_5 ^2\right] .
\end{equation}
This describes an $AdS_5 \times S^5$ geometry. Before taking a short
detour on the description of $AdS_5$ spaces as hypersurfaces of a six
dimensional space let us check the validity region of
(\ref{ads5timess5}) by focusing on the $S^5$ part. The radius of the
$S^5$ is
\begin{equation}\label{radius}
R^2 = \alpha^\prime \sqrt{4\pi g_s N} .
\end{equation}
In order to avoid high curvature (where higher derivative corrections
become important, or even the effective gravity description may break
down) one should take this radius to be large, i.e.
\begin{equation}\label{thooftlarge}
g_s N \gg 1 .
\end{equation}

In addition we should keep the string coupling small which implies
that the number of D3 branes we look at has to be large.
Now, we recall that the field theory living on the D3 branes is ${\cal
  N} = 4$ supersymmetric $SU\left( N\right)$ gauge theory. The gauge
coupling is $g_{YM}^2 = 2\pi g_s$ (see 
(\ref{alpha-prime})). So, at first sight it seems that the gauge
coupling is small whenever the string coupling is small. However, we
should also impose the condition (\ref{thooftlarge}), in particular
the large $N$ limit. For large $N$ $SU\left(N\right)$ gauge theories
`t Hooft developed a perturbative expansion in the parameter (the `t
Hooft coupling) $g_{YM}^2 N$\cite{'tHooft:1974jz}. The condition
(\ref{thooftlarge}) 
implies that the `t Hooft coupling is large whenever the effective
gravity description is reliable. We will argue below that gravity (or
closed type IIB strings) in the space (\ref{ads5timess5}) is dual to
the gauge theory living on the D3 branes. Because of
(\ref{thooftlarge}) this is a strong/weak coupling duality.
One of the first things one should check before publishing a
conjecture on dual pairs is that the global symmetries of the dual
descriptions should match. (Global symmetries are observable.)
Therefore, let us take a short detour and describe the $AdS_5$ space
as a hypersurface in a six dimensional space. This will enable us to
see the isometries of $AdS_5$ in much the same way as one sees the
$SO\left( 6\right)$ isometry of $S^5$ when viewing it as a
hypersurface in six dimensional space.

The space in which we will find an $AdS_{p+2}$ space as a hypersurface
is a 
$2+p+1$ dimensional space with the metric
\begin{equation}\label{embeddingspace}
ds^2 = -dX_{-1}^2 - dX_0 ^2 +\sum_{\alpha =1}^{p+1} dX_\alpha ^2 .
\end{equation}
Analogously to a sphere, the $AdS_{p+2}$ space is defined as the set
of points satisfying the condition
\begin{equation}\label{adsdefinition}
- X_{-1}^2 - X_0 ^2 +\sum_{\alpha=1}^{p+1} X_\alpha ^2  = - R^2 ,
\end{equation}
where $R$ is called the radius of the $AdS_{p+2}$ space. We solve this
equation by
\begin{eqnarray}
X_{-1} + X_{p+1} & = & U \label{parameta},\\ 
\mbox{for $i = 0,\ldots ,p$:} \;\;\;\;\;\; X_i & = & x^i
\frac{U}{R} ,\label{parametb}\\
X_{-1} - X_{p+1} & = & \frac{x^2 U}{R^2} +\frac{R^2}{U}\label{parametc} ,
\end{eqnarray} 
where $x^2 = \eta_{ij} x^i x^j$ and
$U$ and $x^i$ parameterize the hypersurface
(\ref{adsdefinition}).  Plugging (\ref{parameta}) -- (\ref{parametc})
into (\ref{embeddingspace}), we obtain the $AdS_{p+2}$ metric
\begin{equation}
ds^2 =\frac{U^2}{R^2} \eta_{ij} dx^i dx^j + R^2 \frac{dU^2}{U^2} .
\label{adsmetricforlater}
\end{equation}
Comparison with (\ref{ads5timess5}) shows us that the limit
(\ref{maldacenalimit}) took us to an $AdS_5 \times S^5$ space where
the radii of the $AdS_5$ and the $S^5$ coincide and are given by
(\ref{radius}). 

After this detour we can easily read of the isometries of
(\ref{ads5timess5}). The isometry is $SO\left( 4, 2\right) \times
SO\left( 6\right)$. These isometries show up in the field theory on
the D3 branes as follows.  The $SO\left( 6\right)$  ($=SU\left(
  4\right)$) is the R symmetry of ${\cal N} =4$ supersymmetric
Yang-Mills theory. The beta function of the gauge theory vanishes
exactly, i.e.\ the gauge theory is a conformal field theory. The
$SO\left( 4,2\right)$ part of the isometry corresponds to the
conformal group which is a symmetry in the gauge theory. Taking into
account the preserved supersymmetries\footnote{These are enhanced in
  the near horizon limit.}, one observes that the isometry group
$SO\left( 4,2\right) \times SO\left( 6\right)$
can be extended to the superconformal group acting in the field
theory. Thus the global symmetries of the two descriptions
match. In the asymptotic region $U\to\infty$ the AdS part of the
metric (\ref{ads5timess5}) becomes (up to a conformal factor) the 3+1
dimensional Minkowski space. This is the boundary of the AdS space. 
The $SO\left( 4,2\right)$ isometry acts as
the group of conformal transformations on the Minkowski space. In
this sense, one can identify the boundary of the AdS space with the
location of the D3 branes, although one should not think of the two
descriptions simultaneously, because whenever the parameters are such
that the gravity description is reliable, the perturbative description
of the gauge theory breaks down and vice versa.  

Moreover, one can identify the $SL\left( 2, {\mathbb Z}\right)$
duality of the type IIB string with the Montonen Olive
duality\cite{Montonen:1977sn,Witten:1978mh,Osborn:1979tq,Witten:1979ey}
of ${\cal N} =4$ super Yang Mills theory.

Thus, we have seen some evidence that the AdS/CFT correspondence
conjecture holds. More checks have been performed, but we will not
discuss those here. In the following section we want to illustrate the
duality by computing Wilson loops in gauge theory using type IIB
superstrings. Before doing so, let us summarize the AdS/CFT
correspondence (duality) conjecture. 
\begin{itemize}
\item Type IIB superstrings living in an $AdS_5\times S^5$ background
  are dual to open superstrings ending on a stack of D3 branes.
\item The $AdS_5$ and the $ S^5$ have the same radius whose value (in
  units of $\alpha^\prime$) is
  related to the `t Hooft coupling of the gauge theory via equation
  (\ref{radius}), and $g_{YM}^2 = 2\pi g_s$. 
\item The type IIB string theory is in its perturbative regime if
  $g_s$ is small, and higher curvature effects are not dangerous as
  long as (\ref{thooftlarge}) holds. In this region, the gauge theory
  is in the large $N$ limit and strongly coupled.
\end{itemize}
In a somewhat weaker statement, one should replace ``type IIB
superstrings'' by ``type IIB supergravity'' and ``open strings ending on
D3 branes'' by ${\cal N}=4$\ $SU\left( N\right)$ gauge theory. We will
take the duality conjecture as stated in the items.

\subsection{Wilson loop computation}
\setcounter{equation}{0}
\subsubsection{Classical approximation}\label{wilsonclassiapo}
A Wilson loop is the (normalized) partition function of gauge theory
in the presence of an external quark anti-quark pair. A perturbative
description of this situation in a D3 brane setup for static quarks is
drawn in figure \ref{fig:wilsonpert}.
%%%%%%%%%%%%%%%%%%%%%%%%%%%%%%%%%%%%
\begin{figure}
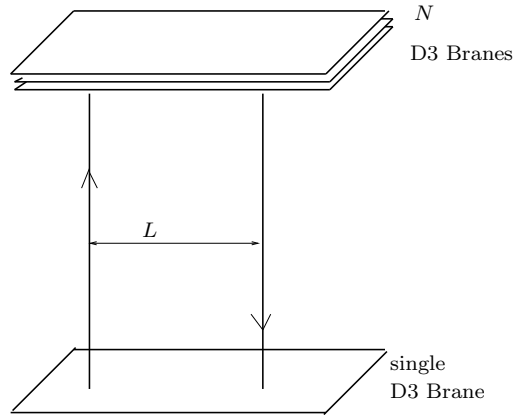
  
\begin{center}
\input wilsonpert.pstex_t
\end{center}
\caption{The perturbative Wilson loop setup. The quark anti-quark pair 
  corresponds to the ends of open strings on the $N$ D3 branes. The
  open strings have opposite orientation. The quark anti-quark pair is
  chosen to be static. The dynamics of the quarks decouples as long
  as the single D3 brane is very far away from the $N$ D3 branes. The
  distance between the quark and the anti-quark is $L$.}
\label{fig:wilsonpert}
\end{figure}
%%%%%%%%%%%%%%%%%%%%%%%%%%%%%%%%%%%%
In order to employ the AdS/CFT duality conjecture, we need to translate
figure \ref{fig:wilsonpert} to type IIB strings living on $AdS_5\times
S^5$. The prescription is that the open strings in figure
\ref{fig:wilsonpert} translate into a background string of type IIB
theory  on
$AdS_5\times S^5$. In the previous section we have argued that the
position of the $N$ D3 branes is translated to the boundary of the
$AdS_5$ space. (We should point out again that the emphasis is on
``translated'' since the gauge theory description breaks down whenever
the $AdS$ prescription is reliable.) Therefore, the background string 
should fulfill the boundary condition that its ends on the $AdS_5$
boundary are separated by a distance $L$. Classically, the background 
string is then uniquely determined by the requirement of minimal
worldsheet area. As we will see in a moment, the picture drawn in
figure \ref{fig:wilsonnon} arises.  
%%%%%%%%%%%%%%%%%%%%%%%%%%%%%%%%%%%%
\begin{figure}
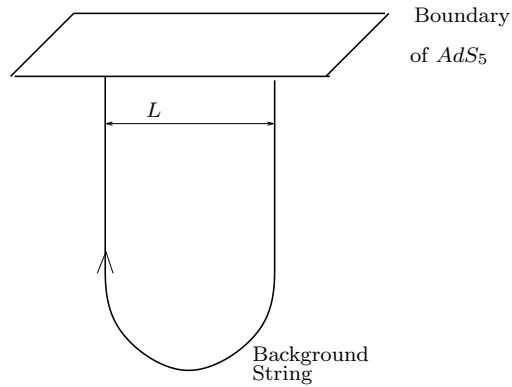
  
\begin{center}
\input wilsonnon.pstex_t
\end{center}
\caption{The non perturbative Wilson loop setup. The quark anti-quark pair 
  corresponds to a background string ending on the $AdS_5$ boundary. }
\label{fig:wilsonnon}
\end{figure}
%%%%%%%%%%%%%%%%%%%%%%%%%%%%%%%%%%%%
The fact that the string with minimal area goes down from the boundary
into the 
$AdS$ space and up again to satisfy the boundary condition is a result
of the non trivial metric. The corresponding calculation can be
carried out explicitly. For the sake of a minor simplification, we 
redefine the coordinate $U = \frac{R^2}{\alpha^\prime} u$ such that
the metric (\ref{ads5timess5}) reads
\begin{equation}\label{antikreuzschale}
ds^2 = R^2 \left[ u^2 \eta_{ij}dx^idx^j +\frac{du^2}{u^2} +d \Omega_5
  ^2\right] .
\end{equation}
The worldsheet area of the background string is
\begin{equation}\label{wil:nambugoto}
S = \frac{1}{2\pi\alpha^\prime} \int d\tau d\sigma \sqrt{-\det g}
\end{equation}
where $g_{\alpha\beta}$ is the induced metric (\ref{induced}). As an
ansatz for the background string we take
\begin{equation}\label{ansawilson}
X^0 = \tau \;\;\; , \;\;\; X^1 = \sigma \;\;\; ,\;\;\; X^4 = U\left( 
  \sigma\right) ,
\end{equation}
and the rest of the string positions is constant in $\sigma$ and
$\tau$. The indices are assigned in the order in which coordinates 
appear in (\ref{antikreuzschale}), and $X^4 = U$ (the capital $U$
denotes the string position in the space with the metric
(\ref{antikreuzschale}) and should not be confused with the capital
$U$ in (\ref{maldacenalimit})). 
The first two equations in (\ref{ansawilson}) represent the static
gauge and the sigma dependence of $U$ allows for the string to
describe the curve of figure \ref{fig:wilsonnon}. This is the simplest 
consistent ansatz for the given boundary conditions. 

The induced metric is
\begin{equation}\label{adsinduced}
\frac{ds^2_{ind}}{R^2} = -U^2 d\tau^2 + \left( U^2 +
  \frac{\left(\partial_\sigma U\right)^2}{U^2}\right) d\sigma^2 .
\end{equation}
and thus the Nambu-Goto action (\ref{wil:nambugoto}) reads
\begin{equation}\label{wil:action}
S = \frac{TR^2}{2\pi}\int d\sigma \sqrt{\left(\partial_\sigma
  U\right)^2 
  + U^4} ,
\end{equation}
where $T$ denotes the time interval we are considering and we have
set $\alpha^\prime$ to one ($R^2$ is then a dimensionless quantity
giving the $AdS$ radius in units of $\alpha^\prime$). The action
(\ref{wil:action}) (and also the Lagrange density ${\cal L}$ obtained
by dividing 
$S$ by $T$) does not depend explicitly on $\sigma$. This implies
\begin{eqnarray}
0 & = & \frac{\partial {\cal L}}{\partial\sigma}\nonumber \\
 & = & -\left(\partial_\sigma U\right) \frac{\partial {\cal L}}{\partial
 U} -\left( \partial^2 _\sigma U\right) \frac{\partial {\cal
 L}}{\partial \left(\partial_\sigma U\right)} + \frac{d{\cal
 L}}{d\sigma}\nonumber \\ 
 & = & -\frac{d}{d\sigma}\left( \left(\partial_\sigma U \right)
 \frac{\partial {\cal L}}{\partial \left( \partial_\sigma U\right)} -
 {\cal L}\right) \label{classicalmechanics},
\end{eqnarray}
where the Euler-Lagrange equations have been used in the last
step. For our system we obtain
\begin{equation}\label{wilsonclasssol}
\partial_\sigma U = \pm \frac{U^2}{U_0 ^2}\sqrt{U^4 - U_0^4} ,
\end{equation}
where $U_0$ is a constant related to the integration constant of the
last equation in (\ref{classicalmechanics}). $U_0$ is the lower bound
on the curve in figure (\ref{fig:wilsonnon}) at $\sigma =0$. Thus, we 
can solve for $\sigma$ as a function of $U$
\begin{equation}\label{sigmasolu}
\sigma = \pm \int_{U_0}^{U} d\tilde{U} \frac{U_0 ^2}{\tilde{U}^2
  \sqrt{\tilde{U}^4 - U_0 ^4}} ,
\end{equation}
where the plus-minus sign appears due to the two branches of the curve
in figure \ref{fig:wilsonnon} ($\sigma$ is a horizontal coordinate
and $U$ a vertical one in this figure). At the boundary ($U\to
\infty$) 
the difference between the two values of $\sigma$ should be $L$.
Some straightforward manipulations with the integral in
(\ref{sigmasolu}) yield 
\begin{equation}
\frac{L}{2} = \frac{1}{4U_0} B\left( \frac{3}{4} ,\frac{1}{2}\right) ,
\end{equation}
where 
$$ B\left( \alpha ,\beta\right) = \int_0 ^1 dx x^{\alpha -1}\left( 1 -
  x\right) ^{\beta-1} = \frac{\Gamma\left( \alpha\right) \Gamma\left(
    \beta\right)}{\Gamma\left( \alpha +\beta\right)}$$
denotes Euler's Beta function. Using the identities
$$ x\Gamma\left( x\right)=\Gamma\left( x+1\right) \;\;\; ,
\;\;\;\Gamma\left( x\right)\Gamma\left( 
  1- x\right) = \frac{\pi}{\sin \pi x}\;\;\; , \;\;\; \Gamma\left(
  \frac{1}{2}\right) = \sqrt{\pi}$$
one finds for the integration constant
\begin{equation}
U_0 = \frac{\left( 2\pi\right)^{\frac{3}{2}}}{\Gamma\left(
    \frac{1}{4}\right) ^2 L} .
\end{equation}
This shows our earlier statement that the background string is
uniquely determined by the boundary condition. The Wilson loop
$W\left[ C\right]$ is the partition function for the background
string. For the classical approximation we find
\begin{equation}
W\left[ C\right] = e^{-TE},
\end{equation}
with 
\begin{equation}
E = \frac{R^2}{2\pi}\int d\sigma \sqrt{\left(\partial_\sigma
    U\right)^2 +U^4}.
\end{equation}
Plugging in the classical solution (\ref{wilsonclasssol}) (and taking
into account a factor of two due to the two branches) yields
\begin{equation}
E = \frac{R^2}{\pi}\int_{U_0}^\infty dU \frac{U^2}{\sqrt{U^4 - U_0
    ^4}}.
\end{equation}
Now we split this integral into two pieces (the motivation for this
will become clear below)
\begin{equation}\label{splitintocoloumbandself}
E = E_c + E_s ,
\end{equation}
with 
\begin{equation}
E_s = \frac{R^2}{\pi} \int_{U_0}^\infty dU \frac{U^4 +
  U_0^4}{U^2\sqrt{U^4 -U_0 ^4}} ,
\end{equation}
and 
\begin{equation}
E_c = -\frac{R^2}{\pi} \int_{U_0}^\infty dU\frac{U_0 ^4}{U^2\sqrt{U^4 -
    U_0 ^4}}.
\end{equation}
Let us first discuss the integral $E_s$. This integral is divergent
due to the upper integration bound
and we regularize it by a cutoff $U_{max}$. The asymptotic expansion
for large $U_{max}$ is
\begin{eqnarray}
E_s & = & \frac{R^2 U_0}{\pi}\int_1 ^{\frac{U_{max}}{U_0}} dy\frac{y^4
    +1}{y^2\sqrt{y^4 -1}} \nonumber \\
& = & \frac{R^2 U_0}{\pi}\left[ \frac{\sqrt{y^4
    -1}}{y}\right]^{\frac{U_{max}}{U_0}} _{1} = \frac{R^2}{\pi}
    U_{max} +\ldots ,
\end{eqnarray}
where the dots stand for terms going to zero as $U_{max}$ is taken to
infinity. Thus, $E_s$ corresponds to the self energy of the two
strings in figure \ref{fig:wilsonpert}. It does not depend on the
distance $L$ and diverges as the length of the string is taken to
infinity. Here, we observe one interesting feature of the AdS/CFT
correspondence. From the AdS perspective $U_{max}$ is a large
distance, i.e.\ an IR cutoff. On the field theory side this appears as
a cutoff for high energies, i.e.\ a UV cutoff. This interchange
between infrared and ultraviolet cutoffs is a general characteristics
of the correspondence\cite{Susskind:1998dq}. Let us give a technical
remark in connection with the integral $E_s$. Plugging in our
classical solution (\ref{wilsonclasssol}) into the induced metric
(\ref{adsinduced}), we find the classical value of the induced 
metric (for later use we call this $R^2 h_{\alpha\beta}$)
\begin{equation}\label{classicalmetric}
ds_{class}^2 \equiv R^2 h_{\alpha\beta}d\sigma^\alpha d\sigma^\beta = R^2
\left( -U^2 d\tau^2 + \frac{U^6}{U_0^4}d\sigma^2 \right).
\end{equation}
The scalar curvature computed from $h_{\alpha\beta}$
reads 
\begin{equation}\label{correspondingcurvature}
R^{(2)} = 2\frac{U^4 + U_0 ^4}{U^4}.  
\end{equation}
With this information it is easy to verify that the structure of the
self energy integral is
\begin{equation}\label{selfenrgyincurv}
E_s = \frac{R^2}{4\pi T} \int d^2 \sigma \sqrt{-h} R^{(2)}.
\end{equation}
(This does not contradict the Gauss--Bonnet theorem because the
worldsheet of the background string is not compact.)

Now, let us come to the second contribution in
(\ref{splitintocoloumbandself}). This will turn out to be the more
interesting one. Its computation is quite similar to the computation
of $U_0$ in terms of $L$. Therefore, let us just give the result 
\begin{equation}\label{coulomblawads}
E_c = -\frac{4\pi^2 \sqrt{2g_{YM}^2 N}}{\Gamma\left(
    \frac{1}{4}\right)^4 L}, 
\end{equation}
where (\ref{radius}) and $g_{YM}^2 = 2\pi g_s$ has been used. This is
the part of the quark anti-quark potential which arises due to gluon
exchange among the two quarks. It is a Coulomb potential. Since $L$ is
the only scale appearing in the setup and ${\cal N}=4$ supersymmetric 
Yang--Mills theory has a conformal symmetry, there can be only a
Coulomb potential. Anything else would need another scale to produce
an energy, but this cannot appear due to conformal invariance. 

In that
respect models with less or none supersymmetry are more interesting
because one can observe confinement in those models. The corresponding 
literature is listed in chapter \ref{furtherreading}. The case we are
considering here, is the one where the AdS/CFT correspondence is
perhaps best understood. We will study a question which is interesting
from a more theoretical perspective, namely whether there are
corrections to the result (\ref{coulomblawads}).

\subsubsection{Stringy corrections}

Before discussing corrections to (\ref{coulomblawads}) we should
envisage the possibility that (\ref{coulomblawads}) is an exact
result. There are some results which may lead to this conclusion. By
analyzing the structure of possible corrections to the $AdS_5\times
S^5$ geometry,
physicists\cite{Banks:1998nr,Metsaev:1998it,Kallosh:1998zx}
found that this geometry is exact. Still, there is a very simple
argument destroying the hope that (\ref{coulomblawads}) might be
exact. Namely, the above Wilson loop computation can also be performed
in the perturbative regime, where the `t Hooft coupling is small. Then
one finds, of course, also a Coulomb law but the dependence on the `t
Hooft coupling is linear instead of a square root dependence (which
actually cannot be obtained in a perturbative calculation). This does
not contradict the result (\ref{coulomblawads}) but tells us that
taking the 
`t Hooft coupling smaller should result in corrections such that
finally for very small `t Hooft coupling the square root like
dependence goes over into a linear one. 

After we have excluded corrections to the $AdS_5\times S^5$ geometry, 
we will study fluctuations of the IIB string around the background
string in 
figure \ref{fig:wilsonnon}. That is, we consider the Wilson loop as
the quantum mechanical partition function
\begin{equation}\label{wilsonformal}
W\left[ C\right] = \int \left[ {\cal D} \delta X\right] \left[ {\cal
    D} \delta \theta\right] e^{-S_{IIB}\left( X +\delta X ,\delta
    \theta\right)} ,
\end{equation}
where $\delta X$ denote bosonic fluctuations and $\delta \theta$
fermionic ones (the fermionic background of the string is
trivial). Before going into the details of the computation, let us 
describe the expansion we are going to perform. From
(\ref{wil:nambugoto}) and (\ref{antikreuzschale}) we see that the
square root of the `t Hooft coupling appears as an overall constant in
front of the metric. (This is also true for terms containing
fermions.) Therefore, the expansion parameter $\hbar$ (or 
$\alpha^\prime$ in section \ref{betafunctions}) is identified with the
inverse square root of the `t Hooft coupling. The expression
(\ref{wilsonformal}) can be computed as a power series in this
parameter. In particular, the next to leading order correction to 
(\ref{coulomblawads}) will not
depend on the `t Hooft coupling. It is this correction we will discuss
in some detail in the following. In order to be able to use
(\ref{wilsonformal}) for explicit calculations we need to know the
type IIB string action in an $AdS_5\times S^5$
background. Fortunately, this has been constructed in the
literature\cite{Metsaev:1998it}. These authors gave a type IIB action
in the Green Schwarz formalism, which is appropriate in the presence
of non-trivial RR backgrounds. The construction is similar to 
the one discussed in section \ref{greenschwarzstring}. One
uses target space supersymmetry and kappa symmetry as a guide. The 
technicalities are rather involved and we will not discuss them
here. Because we will restrict ourselves to terms second order in
fluctuations we only need a truncated version of the action of type
IIB strings on $AdS_5\times S^5$. The complete action does not contain
terms with an odd number of target space fermions, in particular no
terms linear in target space fermions. Since the fermionic background
is trivial, contributions quadratic in fluctuations can either have
two fermionic fluctuations and no bosonic fluctuation or only bosonic
fluctuations. The part of the action which is quadratic in the
fluctuations consists of a sum of terms with only bosonic fluctuations
and terms with only fermionic fluctuations. 

Let us discuss the bosonic fluctuations first. The type IIB action is
a kappa symmetric extension of (\ref{wil:nambugoto}). For the bosonic
fluctuations only the contribution (\ref{wil:nambugoto}) is relevant
(for the lowest non trivial contribution). As in section
\ref{betafunctions} we parameterize the fluctuations by tangent
vectors to geodesics connecting the background with the actual value,
i.e.\ we perform a normal coordinate expansion. The quantum fields are
\begin{equation}
\xi^a = E_\mu ^a \xi^\mu ,
\end{equation}
where $E_\mu ^a$ are the vielbein components obtained by taking the
square root of the diagonal metric components. A number as a label on
a $\xi$ will stand for a flat index ($a$), unless stated explicitly
otherwise. The computation of the term second order in the $\xi^a$ is
a bit lengthy but straightforward with the information given in
section \ref{betafunctions}. (The only difference to section
\ref{betafunctions} is that we expand now a Nambu-Goto action instead
of a Polyakov action.) Before giving the result it is useful to
perform a local Lorentz rotation in the space spanned by the
$\xi^a$. The
rotation is\footnote{We employ the symmetry of the background string
  and take $0\leq \sigma \leq \frac{L}{2}$ and the plus sign in 
(\ref{wilsonclasssol}).}
\begin{equation}
\left( \begin{array}{c}
\xi^\parallel \\
\xi^\perp \end{array} \right) = \left( \begin{array}{cc} \cos \alpha &
\sin \alpha \\ -\sin \alpha & \cos \alpha \end{array}\right) \left(
\begin{array}{c} \xi^1 \\ \xi^4 \end{array}\right) ,
\label{rotation}
\end{equation}
with 
\begin{equation}\label{rotangle}
\cos \alpha = \frac{U_0 ^2}{U^2} \;\;\; , \;\;\; \sin \alpha
=\frac{\sqrt{U^4 -U_0 ^4}}{U^2} .
\end{equation}
Note that the determinant of the matrix appearing in (\ref{rotation})
is one.
The fields $\xi^\parallel$ and $\xi^\perp$ describe fluctuations
parallel and perpendicular to the worldsheet, respectively.
This is illustrated in figure \ref{fig:longperp}.
%%%%%%%%%%%%%%%%%%%%%%%%%%%%%%%%%%%%
\begin{figure}
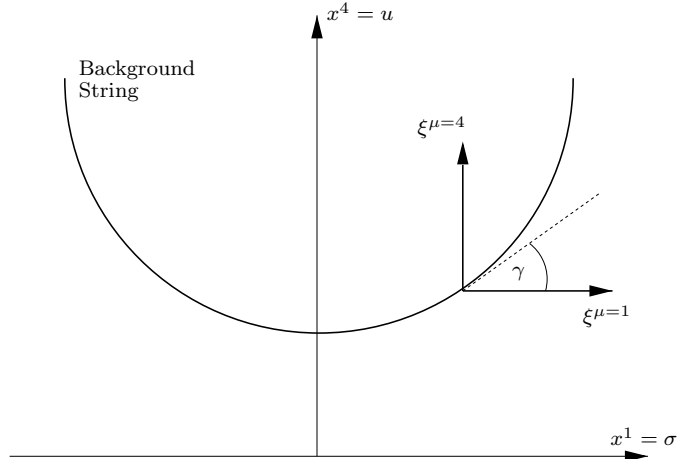
  
\begin{center}
\input longperp.pstex_t
\end{center}
\caption{Perpendicular and longitudinal fluctuations in the one--four
  plane. } 
\label{fig:longperp}
\end{figure}
%%%%%%%%%%%%%%%%%%%%%%%%%%%%%%%%%%%%
Fluctuations drawn into figure \ref{fig:longperp} carry Einstein
indices which we indicated explicitly. The angle $\gamma$ is given by the
slope of the background string
\begin{equation}
\tan \gamma = \partial_\sigma U .
\end{equation}
The combination
\begin{equation}\label{perpiflu}
\xi^{\mu =4} -\tan \gamma\, \xi^{\mu =1}
\end{equation}
vanishes for 
\begin{equation}
\tan \gamma = \frac{\xi^{\mu =4}}{\xi^{\mu =1}} ,
\end{equation}
i.e.\ if $\xi^{\mu =1} + \xi^{\mu =4}$ is tangent to the background
string. Thus, the combination in (\ref{perpiflu}) is normal to the
background string. Transforming the indices to flat ones (with the
vielbein) and ortho-normalization yields $\xi^\parallel$ and
$\xi^\perp$ with the given interpretation. When writing down the
Lagrangian second order in the fluctuations, we can set $R=1$ since we 
know 
the general $R$ dependence ({\it viz.} none) from the argument given
above. The Lagrangian for the bosonic fluctuations comes out to be
\begin{equation}
{\cal L}^{(2)}_{bosons} = {\cal L}^{(2)}_{AdS_5} + {\cal
  L}^{(2)}_{S^5} 
\end{equation}
with
\begin{eqnarray}
{\cal L}^{(2)}_{AdS_5} & = &\frac{1}{2}\sqrt{ -h} \left[
  \sum_{a=2,3,\perp} \xi^a \Delta \xi^a - 2\left( \xi^2\right)^2
  -2\left( \xi^3\right)^2 \right. \nonumber\\
& & +\left. \left( R^{(2)} -4\right) \left(
  \xi^\perp\right)^2\right] ,\label{bosofluc1} \\
{\cal L}^{(2)}_{S^5} & = & \frac{1}{2} \sqrt{-h} \sum_{a^\prime = 5}^9
  \xi^{a^\prime} \Delta \xi^{a^\prime} \label{bosofluc2},
\end{eqnarray}
where total derivative terms have been dropped (the fluctuations
should satisfy Dirichlet boundary conditions in order not to change
the classical boundary conditions). Further, $\Delta$ denotes the two
dimensional Laplacian with respect to the metric $h_{\alpha\beta}$
(\ref{classicalmetric}) and $R^{(2)}$ is the corresponding scalar
curvature (\ref{correspondingcurvature}). 
We observe that the longitudinal fluctuations $\xi^0$ and
$\xi^\parallel$ drop out of the action. Hence, we can fix the
worldsheet diffeomorphisms via
\begin{equation}
\xi^0 = \xi^\parallel = 0 .
\end{equation}
If the normalization of the functional integral in
(\ref{wilsonformal}) contains a division by the volume of the
worldsheet diffeomorphisms, we cancel the $\xi^0$ and $\xi^\parallel$
integration against this term in the normalization. This may be
problematic, and we will comment on this issue later.

It remains to study the fermionic fluctuations. Since the fermionic
background is trivial we just need to copy the Lagrangian
from\cite{Metsaev:1998it} (truncated to quadratic terms) and to plug
in our background. The result of the copying task is
\begin{eqnarray}
{\cal L}_F & = & -\frac{1}{2}\sqrt{-h} h^{\alpha\beta} \left( E_\alpha
  ^{\hat{a}} - i\bar{\theta}^I \hat{\gamma}^{\hat{a}}\left( D_\alpha
  \theta\right)^I\right) \left( E^{\hat{a}} _\beta -
  i\bar{\theta}^J \hat{\gamma}^{\hat{a}}\left( D_\beta
  \theta\right)^J\right) \nonumber \\
& & -i\epsilon^{\alpha\beta}E_\alpha ^{\hat{a}}\left( \bar{\theta}^1
  \hat{\gamma}^{\hat{a}}\left( D_\beta \theta\right)^1 -\bar{\theta}^2
  \hat{\gamma}^{\hat{a}} \left( D_\beta\right)^2\right)
  .\label{fermioniclagrangian} 
\end{eqnarray}
First, we need to explain some of the notation. The index $\hat{a} =
0, \ldots , 9$ labels the tangent space coordinates of $AdS_5 \times
S^5$. Later, we will use an index $a =0,\ldots, 4$ for the tangent
space of $AdS_5$ and $a^\prime = 5,\ldots ,9$ for the tangent space of
$S^5$. The vielbein with a worldsheet index is
\begin{equation}
E_\alpha ^{\hat{a}}=E_\mu ^{\hat{a}} \partial_\alpha X^\mu ,
\end{equation}
where $X^\mu$ is the position of the background string. The indices
$I,J =1,2$ label the two target space supersymmetries. The derivative
$D_\alpha$ is defined as
\begin{eqnarray}
\left( D_\alpha \theta\right)^I & = & \left[ \delta^{IJ} \left(
    \partial_\alpha  +\frac{1}{4}\left( \partial_\alpha
    X^\mu\right)\omega_{\mu}^{ab}\gamma^{ab}\right)
    -\frac{i}{2}\epsilon^{IJ}\left(\partial_\alpha X^\mu\right) E_\mu
    ^a\gamma^a\right]\theta^J \nonumber \\
& \equiv & \nabla_\alpha \theta^I -\frac{i}{2}\epsilon^{IJ} \left(
    \partial_\alpha X^\mu\right) E_\mu ^a\gamma^a \theta^J .
\label{fermionicpart1}
\end{eqnarray}
Here, $\omega_\mu^{ab}$ denotes the target space spin connection and
we have used the fact that our background is trivial in $S^5$
directions. 
The gamma matrices $\hat{\gamma}^{\hat{a}} =\left( \gamma^a ,
  i\gamma^{a^\prime}\right)$ satisfy $SO\left( 4,1\right)$ and $SO\left(
  5\right)$ Clifford algebras, respectively. 
The $\theta^I$ are
sixteen component spinors each. They are conveniently labeled by a
double spinor index $\theta^{\alpha\alpha^\prime}$ where $\alpha$ is a
spinor index in the tangent space of the $AdS_5$, and $\alpha^\prime$
a spinor index in the tangent space of $S^5$. The $\gamma^a$ and
$\gamma^{a^\prime}$ are four
times four matrices tensored with four times four identity
matrices. (In the following we will suppress target space spinor
indices in order to avoid confusion with worldsheet indices which are
also labeled by small Greeks.)

We do not intend to give a derivation of (\ref{fermioniclagrangian})
but let us have a brief look at its structure before
proceeding. Expression (\ref{fermionicpart1}) is a tensor
(density) with indices $\alpha \beta$ contracted either with
$h^{\alpha\beta}$ or $\epsilon^{\alpha\beta}$. The terms with
$h^{\alpha\beta}$ can be thought of as arising from the replacement
(\ref{Pi})\footnote{In general, the combination in (\ref{Pi})
contains also terms higher than quadratic order in the
fermions. This is because the superalgebra is altered in the
$AdS_5\times S^5$ case as compared to a flat target space. This
should be clear by noting that the isometries form a subgroup of the
supersymmetry transformations.} whereas the $\epsilon^{\alpha\beta}$
contracted terms 
come from the Wess Zumino term (\ref{kappaterm}) needed for kappa
symmetry. The details differ from the discussion in section
\ref{greenschwarzstring}  due to the different target space geometry
and the RR four form flux.

For our background, the Lagrangian (\ref{fermioniclagrangian}) can be
written in a compact way
\begin{equation}
{\cal L}_F = -\sqrt{-h}\left( \bar{\theta}^1,\, 
  \bar{\theta}^2\right)\left( \hspace{-0.2cm}\begin{array}{c c}
2i E_\mu ^a \left( \partial_\alpha X^\mu\right) \gamma^a {\cal
  P}_{-}^{\alpha\beta} \nabla_\beta &\hspace{-1cm} 1-{\cal
  B} \\
 -1-{\cal B} & \hspace{-1cm} 2iE_\mu ^a \left( \partial_\alpha
  X^\mu\right) \gamma^a 
  {\cal P}_{+}^{\alpha\beta}\nabla_\beta
\end{array} \hspace{-0.2cm}\right)
\left( \hspace{-0.2cm}\begin{array}{c}
\theta^1 \\ \theta^2\end{array}\hspace{-0.2cm}\right) ,
\end{equation}
where $X^\mu$ stands for the background position of the string and
\begin{eqnarray}
{\cal P}_{\pm}^{\alpha\beta} &=&\frac{1}{2}\left( h^{\alpha\beta} \pm
  \frac{\epsilon^{\alpha\beta}}{\sqrt{-h}} \right) ,\\
{\cal B} & = & \frac{1}{2\sqrt{-h}}\epsilon^{\alpha\beta} E_\mu ^a
  E_\nu ^b \left( \partial_\alpha X^\mu\right) \left( \partial_\beta
  X^\nu\right) \gamma^{ab} .
\end{eqnarray}
As usual, a gamma with a multiple index is the antisymmetrised product
of gamma matrices.

It is useful to perform the rotation (\ref{rotation}) also on the
spinors (with $\alpha$ as given in (\ref{rotangle})) 
\begin{equation}
\theta^I =\left( \cos\frac{\alpha}{2}
  -\sin\frac{\alpha}{2}\gamma^{14}\right) \psi^I .
\end{equation}
In order to compute the partition function, we should fix the kappa
symmetry. This is conveniently done in terms of the nilpotent matrices
\begin{equation}
\gamma^\pm =\frac{1}{2}\left( \gamma^0 \pm \gamma^1\right) .
\end{equation}
In analogy to section \ref{greenschwarzstring}, we choose the kappa
fixing conditions
\begin{equation}\label{adskappafix}
\gamma^- \psi^1 = 0\;\;\; , \;\;\; \gamma^+ \psi^2 = 0.
\end{equation}
We assume that the integration over spinors not satisfying
(\ref{adskappafix}) cancels the volume of kappa transformations
appearing as a normalization factor in the functional integral. This
may be problematic, and we will comment on this issue later. Spinors
satisfying the kappa fixing condition are then governed by the
Lagrangian
\begin{equation}
{\cal L}_F = -\sqrt{-h} \left( \bar{\psi}^1, \, \bar{\psi}^2\right) \left(
  \begin{array}{cc} i\gamma^+ \nabla_+ & 2 \\
-2 & i\gamma^- \nabla_-\end{array}\right) \left( \begin{array}{c}
 \psi^1 
 \\ \psi^2\end{array}\right) ,
\label{firstsimplify}
\end{equation}
where $\nabla_\pm$ are tangent space derivatives defined as follows
\begin{equation}\label{tangentspacederiv}
\nabla_\pm = e^\tau _0\nabla_\tau \pm e^\sigma _1\nabla_\sigma =
\frac{1}{U}\nabla_\tau \pm \frac{U_0 ^2}{U^3}\nabla_\sigma ,
\end{equation}
where $e_0 ^\tau$ and $e_1 ^\sigma$ are two dimensional (inverse)
vielbein components obtained from the square roots of the diagonal
elements of $h^{\alpha\beta}$. Note also that the covariant derivative
simplifies when acting on spinors satisfying
(\ref{adskappafix}). Defining partial tangent space derivatives
analogously to (\ref{tangentspacederiv}) one finds
\begin{equation}
\nabla_\pm \psi^I = \left( \partial_\pm \pm \frac{\omega
    _\pm}{2}\right) \psi^I ,
\end{equation}
where 
\begin{equation}
\omega_{\pm} = e_0^\tau\omega_\tau^{01} \pm
e_1^\sigma\omega_\sigma^{01} 
\end{equation}
are tangent space components of the two dimensional spin connection
$\omega^{01}_\alpha$ computed from the zweibeinen defined in
(\ref{tangentspacederiv}). Let us further define the matrices
\begin{equation}
\rho^+ =\left( \begin{array}{cc} 0 & 0 \\ \gamma^0 &
    0\end{array}\right) \;\;\; ,\;\;\; \rho^- =\left(\begin{array}{cc}
0 &-\gamma^0\\ 0 & 0\end{array} \right) .
\end{equation}
These are the same matrices as in (\ref{twodgammas}) with $i$ replaced
by $\gamma^0/2$. Finally, we rewrite (\ref{firstsimplify}) in a
suggestive way as follows
\begin{eqnarray}
{\cal L}_F & = & -\sqrt{-h}\left( \bar{\psi}_2 ,\, -\bar{\psi}_1\right)
\left( \begin{array}{cc} -2 & i\gamma^-\nabla_- \\
-i\gamma^+\nabla_+ & -2 \end{array}\right) \left( \begin{array}{c}
\psi_1 \\ \psi_2\end{array}\right) \nonumber \\
& =& -\sqrt{h}\left( \bar{\psi}_2,\, -\bar{\psi}_1\right) \left(
  \begin{array}{cc}  -2 & 2i\gamma^0\nabla_-\\ -2i\gamma^0\nabla_+
      & -2\end{array}\right) \left( \begin{array}{c}
\psi_1 \\ \psi_2\end{array}\right) \nonumber\\
& = & 2\sqrt{-h} \left( \bar{\psi}_2,\, -\bar{\psi}_1\right) \left(
  i\rho^m\nabla_m +1\right) \left( \begin{array}{c}
\psi_1 \\ \psi_2\end{array}\right) ,
\label{fermioflucfinal}
\end{eqnarray}
where in the second line (\ref{adskappafix}) has been used and a
repeated index $m$ stands for the sum over the labels $+$ and $-$. 
Comparison with the expressions in section \ref{worldshitsusy} shows
that the part of the action containing target space spinor
fluctuations `metamorphosed' into an action for worldsheet spinors
after imposing the kappa fixing condition (\ref{adskappafix}). The
difference is that the derivative contains the spin connection due to
the non trivial worldsheet metric  $h_{\alpha\beta}$
(\ref{classicalmetric}), and the mass terms appearing due to the
constant non vanishing curvature of the AdS space.

Now we have collected all the information needed to express the second
order fluctuation contribution to (\ref{wilsonformal}) in terms of
determinants of two dimensional differential operators. (For Dirac
operators one uses the formal identity $\det A = \sqrt{\det
  A^2}$.) 

Integration over the fluctuations leads to determinants of operators
which can be read off from (\ref{bosofluc1}), (\ref{bosofluc2}) and 
(\ref{fermioflucfinal}). The corrected expression for the Wilson loop
reads
\begin{equation}\label{correctedwilson}
W\left[ C\right] = e^{-TE_{class}}\frac{\det\left( -\Delta_F
    -\frac{1}{4}R^{(2)} +1\right)}{\det\left( -\Delta
      +2\right)\det^{\frac{1}{2}}\left( -\Delta +4 - R^{(2)}\right)
    \det ^{\frac{5}{2}}\left( -\Delta\right)} .
\end{equation}
The exponential is the classical contribution with $E_{class}$ given
by (\ref{splitintocoloumbandself}), (\ref{selfenrgyincurv}) and
(\ref{coulomblawads}). Note also that the operator appearing in the
numerator of (\ref{correctedwilson}) is a four times four matrix. The
Laplacian acting on worldsheet fermions $\Delta_F$ is
$\eta^{mn}\nabla_n\nabla_m$. Unfortunately it is not known how to
evaluate the determinants in (\ref{correctedwilson}) exactly. What is
known exactly are the divergent contributions. These are given in
(\ref{heatdiver}). They are of the form
\begin{equation}\label{divergentbad}
E_{div} \sim \int d^2\sigma \sqrt{-h} R^{(2)} .
\end{equation}
Comparing with (\ref{selfenrgyincurv}), we find that this divergence 
renormalizes the self-energy which is infinite anyway. A correction to
the Coulomb charge of the
quarks will be finite. Unfortunately, we cannot give it in a more
explicit way (as a number). 

In addition, there is also a conceptual
puzzle with the divergent contribution. Although for our problem it is
not 
relevant, it should not be there. The argument that
something might have gone wrong goes as follows. 
The string action is equivalent to a Polyakov type action, at least at
a classical level (see section (\ref{classbo})). The Polyakov action
is conformally invariant, and in a consistent string background the
conformal invariance should not be broken by quantum
effects. Therefore, divergences which introduce a cutoff (or
renormalization group scale) cannot occur. Indeed, it was argued
in\cite{Drukker:2000ep} that a treatment analogous to ours but with a
Polyakov instead of the Nambu-Goto action leads to a finite
result. This treatment is a bit more complicated since the worldsheet
metric appears as an independent field which also fluctuates. The
advantage is however that subtle contributions due to the occurring
integral measures are well understood. Such contributions are
typically of the structure
(\ref{divergentbad})\cite{Alvarez:1983zi,Friedan:1982is,Luckock:1989mv}.
(Note, however, that if the worldsheet metric is identified with the
induced metric, the term in (\ref{divergentbad}) is not really
distinguishable from $\left(\partial X\right)^2$ terms (see
e.g.\cite{Eisenhart})). In our derivation, we have mentioned already
two places where nontrivial measure contributions could arise. This
could happen when we cancel the integration over the longitudinal 
fluctuations against the volume of the worldsheet diffeomorphisms and
in the kappa fixing procedure. Unfortunately, the Nambu-Goto case is
less understood than the Polyakov formulation. (For a recent attempt
to fix the functional measures in the bosonic part
see\cite{Muck:2001nq}.) Fortunately, the result of the better
understood calculation in the Polyakov approach is identical to the
one given here (up to the irrelevant
divergence)\cite{Drukker:2000ep}. 

With these open questions we close our discussion on the AdS/CFT 
correspondence. The reader who wants to know more will find some
references in chapter \ref{furtherreading}. 
\section{Strings at a TeV}
\setcounter{equation}{0}

So far we have not determined the numerical value of the string scale
(set by $\alpha^\prime$) in terms of a number. We restricted our
discussions mostly to the massless excitations of the string. This was
motivated by the belief that the string scale (in energies) is large
compared to observed energy scales. Often it is comparable to the
Planck scale. This identification is motivated by studies of heterotic
weakly coupled strings which provided for a long time the most
promising starting point in constructing phenomenologically
interesting models. Such models are obtained by compactifying the ten
dimensional heterotic (mostly $E_8\times E_8$) string down to four
dimensions on a Calabi--Yau manifold. Let us give a rough estimate for
the resulting four dimensional couplings. The effective four
dimensional heterotic action is of the form
\begin{equation}\label{4dheterotic}
S_{het} = \int d^4 x \frac{V}{g_h ^2}\left( l_h^{-8}R^{(4)} - l_h^{-6}
  \mbox{tr} F^2 +\ldots \right) ,
\end{equation}
where we drop details which are not relevant for the present estimate
on scales. In (\ref{4dheterotic}) $l_h$ is the heterotic string scale
(set by $\alpha^\prime$), $g_h$ the heterotic string coupling (fixed
by the dilaton vev), and $V$ is the volume of the compact space. The
quantities in which four dimensional physics is usually described are
the four dimensional Planck mass $M_p$ and the gauge coupling
$g_{YM}$. These are related to the input data ($g_h$, $l_h$ and $V$)
as follows 
\begin{equation}
M_p ^2 = \frac{V}{g_h ^2 l_h^8} \;\;\; ,\;\;\; \frac{1}{ g_{YM}^2}
=\frac{V}{g_h^2 l_h^6} .
\end{equation}
Expressing $g_h$ in the first equation in terms of the second equation
and further defining the string mass scale as $M_h = 1/l_h$ the above
equation can be rewritten as
\begin{equation}\label{scalerelhet}
M_h =  g_{YM} M_p \;\;\; ,\;\;\; g_h = g_{YM}
\frac{\sqrt{V}}{l_h ^3} . 
\end{equation}
Now we assume that a  $ g_{YM} \sim 0.2$ is a realistic value. 
(This is the gauge coupling of the minimal supersymmetric standard
model at the GUT scale.) 
Plugging
$g_{YM} = 0.2$  into the first equation in (\ref{scalerelhet}) we find
that the 
heterotic string scale is
\begin{equation}
M_h \sim 10^{18} \mbox{GeV} ,
\end{equation}
i.e.\ of the order of the Planck scale. The second equation in
(\ref{scalerelhet}) implies that the compact space is also of the
Planck size if we want to stay within the region where the string
coupling is small.

Now let us investigate how the above estimates on scales are altered
in a theory containing branes. Phenomenologically interesting models
arise also as orientifold compactifications of type II theories. As we
have seen in section \ref{orientifolds}, these contain typically 
D-branes on which the gauge interactions are localized whereas the
gravitational sector corresponds to closed string excitations which
propagate in all dimensions. Assuming that the gauge sector (and
charged matter) is confined to live on D$p$-branes the effective action
for the orientifold compactification will be of the form
\begin{equation}
S_{ori} = \int d^{10}x \frac{1}{g_{II}^2 l_{II} ^8} R - \int d^{p+1}x
  \frac{1}{g_{II} l_{II}^{p-3}} \mbox{tr} F^2 ,
\end{equation}
where $l_{II}$ and $g_{II}$ are the string scale and coupling of the
underlying type II theory, respectively. Assuming further that our
orientifold construction is such that the compact space has dimensions
which are transverse to all relevant D-branes we denote by $V_\perp$
the volume of the compact space transverse to the branes and by
$V_\parallel$ the volume of the compact space longitudinal to the
branes (such that the overall compact volume is $V= V_\perp
V_\parallel$). With this notation the four dimensional action reads
\begin{equation}
S_{ori} = \int d^{4}x \frac{V_\parallel V_\perp}{g_{II}^2 l_{II} ^8} R
  - \int d^{4}x 
  \frac{V_\parallel}{g_{II} l_{II}^{p-3}} \mbox{tr} F^2 ,
\end{equation}
from which we obtain the four dimensional Planck length $l_p$ and gauge
coupling $g_{YM}$
\begin{equation}
\frac{1}{l_p ^2} = \frac{V_\parallel V_\perp}{g_{II}^2 l_{II}^8}
\;\;\; , \;\;\;
\frac{1}{ g_{YM}^2} = \frac{V_\parallel}{g_{II} l_{II}^{p-3}} . 
\end{equation}
Hence, the four dimensional Planck mass ($M_p = 1/l_p$) and the string
coupling $g_{II}$ are
\begin{equation}\label{typeonecouplings}
M_p ^2 = \frac{ v_\perp l_{II}^{-2}}{v_\parallel
  \left(  g_{YM}\right)^4} \;\;\; ,\;\;\; g_{II} =  g_{YM}^2
  v_{\parallel} ,
\end{equation}
where
\begin{equation}
v_\parallel = V_\parallel\, l_{II}^{3-p} \;\;\; ,\;\;\; v_\perp =
V_\perp\, l_{II}^{p-9}
\end{equation}
are dimensionless numbers describing the size of the compact space in
string scale units. The relations (\ref{typeonecouplings}) allow to
take the string length $l_{II}$ larger than the four dimensional Planck
length $l_p$. This can be achieved by taking $v_\perp$ large.  
The size of the parallel volume is taken to be of the ``string size'',
i.e.\ $v_\parallel \sim1$. If the parallel volume is smaller than the
string size, we T-dualize with respect to the smaller dimension. This
dimension 
will then contribute to the perpendicular volume since the string
changes boundary conditions. Hence, the $v_\parallel < 1$ case is T
dual to the considered case of large $v_\perp$. On the other hand if
$v_\parallel > 1$, the second equation in (\ref{typeonecouplings})
tells us that in this case the string coupling becomes strong, and our
description breaks down. (Moreover, it is problematic for gauge
interactions to be compactified on large volumes because the
corrections to the four dimensional gauge interactions are usually
ruled out by experimental accuracy.) 

Let us analyse in some detail what happens if we choose a TeV for
the string scale. This is about the lowest value which is just in
agreement with experiments. (For a lower value massive string
excitations should have shown up in collider experiments.)
With $M_p  \sim  10^{16} \mbox{TeV}$ we find
\begin{equation}
v_\perp \sim 10^{28} \;\rightarrow\; V_\perp = 10^{28}\frac{1}{\left(
    \mbox{TeV}\right)^{9-p}}. 
\end{equation}
The Planck length is about $10^{-33} \mbox{cm}$ and hence in our units
one 
TeV corresponds to $1/\left( 10^{-18}\mbox{m}\right)$. Thus we obtain 
\begin{equation}
V_\perp \sim 10^{28 - \left( 9-p\right) 18}\left(
  \mbox{m}\right)^{9-p} . 
\end{equation}
For the case $p=8$ (one extra large dimension) we obtain that the
perpendicular dimension is compactified on a circle of the size
\begin{equation}
p= 8 \; \rightarrow \; R_\perp \sim 10^{10} \mbox{km} .
\end{equation}
Such a value is certainly excluded by observations. (In the next
subsection we will compute corrections to Newton's law due to
Kaluza-Klein massive gravitons and see that the size of the compact
space should be less than a mm.) For $p=7$ we obtain (distributing the
perpendicular volume equally on the two (extra large) dimensions)
\begin{equation}
p=7 \rightarrow R_\perp \sim 0.1 \mbox{mm} .
\end{equation}
This value is just at the edge of being experimentally excluded. The
situation improves the more extra large dimensions there are. For
example in the case $p=3$ (and again a uniform distribution of the
perpendicular volume on the six dimensions ($V_\perp = R_\perp ^6$))
we obtain
\begin{equation}
p=3 \;\rightarrow\; R_\perp \sim 10^{-10} \mbox{m} ,
\end{equation}
which is in good agreement with the experimental value ($R_\perp
^{exp} =0... 0.1\mbox{mm}$). 

We have seen that D branes allow the construction of models where the
string scale is as low as a TeV. (Note also, that in the above
discussion we can perform T-dualities along the string sized parallel
dimensions. This changes $p$ but leaves the large extra dimensions
unchanged. Actually, it might be preferable to have $p=3$ in order to
avoid Kaluza--Klein gauge bosons of a TeV mass.) This gives the
exciting perspective that string theory 
might be at the horizon of experimental discovery. In near future
collider experiments, massive string modes would be visible. In
addition, the extra large dimensions could be also discovered
soon. This can happen either by the production of Kaluza-Klein
gravitons in particle collisions or by short distance Cavendish like
experiments. However, it might as well be the case that models with
less ``near future discovery potential'' are realized in nature. 

Apart from the prospect of being observed soon, strings at a TeV
scale are interesting for another reason. If the string scale is at a
TeV, we would call this a fundamental scale. Thus the hierarchy
problem would be rephrased. With the fundamental scale at a TeV we
should wonder why the (four dimensional) Planck scale is so much
higher, or why gravitational interactions are so much weaker than the
other known interactions. This hierarchy is now attributed to the size
of the extra large dimensions. Supersymmetry may not be necessary to
explain the hierarchy between the Planck scale and the weak scale.
Therefore, in the above models supersymmetry could be broken already
by the compactification. In such models the question of stability is
typically a problematic issue.

The above considerations are also interesting if one does not insist
on a direct connection to string theory. If one just starts `by hand'
with a higher dimensional setup containing branes, one would also
obtain the first equation in (\ref{typeonecouplings}). In this case,
one calls $l_{II}$ the higher dimensional Planck length, which in turn
can be chosen to be 1/TeV.  

\subsection{Corrections to Newton's law}\label{sec:newtonpotflat}

In the previous section we stated that observations provide
experimental bounds on the size of extra dimensions. In the brane
setup in which we found the possibility of large (as compared to the
Planck length) extra dimensions, these extra dimensions are typically
tested only by gravitational interactions. Therefore, let us describe
the influence of additional dimensions on the gravitational
interaction in some more detail. We will be interested in the
Newtonian limit of gravity. For simplicity, we assume that the space
is of the structure $M_4 \times T^n$, where $M_4$ is the $3+1$
dimensional compact space and $T^n$ is an $n$ dimensional torus of 
large volume. (There might be an additional compact space of Planck
size. This does not enter the computation carried out below.)

The analysis we will carry out here is similar to the discussion of
the massless scalar in section  \ref{KKSCALAR}, where the role of the
scalar is taken over by the Newton potential. Let us arrange the
spatial coordinates into a vector $\left( \mbox{\bf x}, \mbox{\bf
    y}\right)$, 
where {\bf x} corresponds to the $M^4$ and {\bf y} to the $T^n$. For
simplicity we assume that the torus is described by a quadratic lattice
and the uniform length of a cycle is $2\pi R$, i.e.\
\begin{equation}\label{new:per}
\mbox{\bf y} \equiv \mbox{\bf y} + 2\pi R .
\end{equation}  
The $n+4$ dimensional Newton potential $V_{n+4}$ of a point particle
with mass 
$\mu$ located at the origin is given by the equation
\begin{equation}\label{newtpothigh}
\Delta_{n+3} V_{n+4} =\left( n+1\right) \Omega_{n+2} G_{n+4}\,\mu\,
\delta^{\left(n+3\right)}\left( \mbox{\bf x}, \mbox{\bf y}\right) , 
\end{equation}
where $\Delta_{n+3}$ is the three dimensional flat Laplacian and
$\Omega_{n+2}$ is the volume of a unit $n+2$ sphere. 
Any solution to (\ref{newtpothigh}) should be periodic under
(\ref{new:per}). This can be ensured by expanding the potential in
terms of eigenfunctions $\psi_{\mbox{ \tiny\bf k}}\left( \mbox{\bf
    y}\right)$ of 
a Laplace 
operator. The eigenvalue equation is
\begin{equation}
\Delta_n \psi_{\mbox{\bf\tiny k}}\left( \mbox{\bf y}\right) = -m_k^2
\psi_{\mbox{\bf \tiny k}}\left(\mbox{\bf\small y}\right) .
\end{equation}
Thus an orthonormal set of eigenfunctions is
\begin{equation}
\psi_{\mbox{\bf\tiny k}} = \frac{1}{ \left( 2\pi
    R\right)^{\frac{n}{2}}} e^{i \frac{\mbox{\bf\tiny k}}{R}\mbox{\bf y}}, 
\end{equation}
where {\bf k} is an $n$ dimensional vector with integer entries. We
expand the higher dimensional Newton potential into a series of the
eigenfunctions with $r =\left|\mbox{\bf x}\right|$ dependent
coefficients 
\begin{equation}
V_{n+4} = \sum_{\mbox{\bf\small k}} \phi_{\mbox{\tiny \bf k}}\left( r\right)
\psi_{\mbox{\bf \tiny k}}\left( \mbox{\bf y}\right) .
\end{equation}
Plugging this ansatz into equation (\ref{newtpothigh}) determines the
Fourier coefficients\footnote{Here, one uses the completeness
  relation satisfied by the $\psi_{\mbox{\bf \tiny k}}$.}
\begin{equation}
\phi_{\mbox{\bf \tiny k}}\left( r\right) = -\frac{\Omega_n G_{n+4}\,
    \mu\, 
    \psi^\star _{\mbox{\bf\tiny k}}\left( 0\right) }{2} \frac{1}{r}
    e^{-\frac{\left|\mbox{\tiny\bf k}\right|}{R}} .
\end{equation}
Now, we consider the case that all particles with which we can test
the gravitational potential are localized at $\mbox{\bf y} =0$. (This
is natural from the brane picture since we can test gravity only with
matter which is confined to live on the brane. Recall that we
neglected the effects of the Planck sized longitudinal compact
dimensions.) 
We are interested in the Newton potential at $\mbox{\bf y} =0$. This
comes out to be
\begin{equation}
V_4 \equiv V_{n+4} = -\frac{G_4 \mu}{r} \sum_{\mbox{\bf\tiny k}} e^{-r
  \frac{\left|\mbox{\bf\tiny k}\right|}{R}} ,
\end{equation}
where the four dimensional and the higher dimensional Newton constant
are related via
\begin{equation}
G_4 = \frac{\Omega_n G_{n+4}}{2 \left( 2\pi R\right)^n} .
\end{equation}
For $\mbox{\bf k} = \mbox{\bf 0}$ we obtain the usual four dimensional
Newton potential. The other terms are additive Yukawa potentials. They
arise due to the exchange of massive Kaluza Klein gravitons.

Experimentalists usually parameterize deviations from Newton's law via
the expression \cite{Long:1998dk}
\begin{equation}
V_4\left( r\right) = -\frac{G_4 \mu}{r}\left( 1 +\alpha
  e^{-\frac{r}{\lambda}}\right). 
\end{equation}
In the paper\cite{Long:1998dk} the experimental values are
discussed. These maybe outdated by now but for us only the order of
magnitude is important (and the fact that so far no deviation from
Newton's law has been observed). Depending on the size of $\alpha$ an
upper bound on $\lambda$ varying from the $\mu$m range to the cm range
has been measured. This tells us that a scenario with two extra large
dimensions is almost excluded whereas setups with more than two extra
large dimensions are in agreement with the experimental tests of
Newton's law. 

\newpage
\chapter{Brane world setups}\label{chap:braneworlds}
In the last section of the previous chapter we have argued that branes
allow for scenarios with large extra dimensions transverse to the
brane. This is because those extra large dimensions can be tested only
via gravitational interactions which are (due to their weakness)
measured only at scales down to about 0.1 mm. We obtained such models
via 
investigations of string theory. One could, however, just postulate
the existence of branes (on which charged interactions are
located). In this last chapter we will take this latter point of view
and not worry whether the setups we are going to discuss have a
stringy origin. Because in the presence of branes we can attribute the 
hierarchy between the Planck and the weak scale to the size of the
transverse dimensions, we do not need supersymmetry in such
setups. Without supersymmetry, quantum effects usually create vacuum 
energies. A non vanishing vacuum energy on a brane will back react on
the geometry of the space in which the brane lives. Taking into
account such back reactions leads to so called warped
compactifications. This means that the higher dimensional geometry is
sensitive to the position of a brane. The most prominent example of
such warped compactifications are the Randall Sundrum models which we
will discuss next.   

\section{The Randall Sundrum models}
\subsection{The RS1 model with two branes}\label{sec:rs1}
\setcounter{equation}{0}
In the model we are going to describe in this section there is one
extra dimension which will be denoted by $\phi$. The five dimensional
space is a foliation with four dimensional Minkowski slices. The fifth
dimension is compactified on an orbifold $S^1 /{\mathbb Z}_2$. 3-branes are
located at the 
orbifold fixed planes (at $\phi =0$ and $\phi =\pi$) . Hence the
action is of the form 
\begin{equation}\label{RS1action}
S = S_{bulk} + S_{b1} + S_{b2},
\end{equation}
where $S_{b1}$ and $S_{b2}$ denote the actions on the branes. For the
bulk action we take five dimensional gravity with a bulk
cosmological constant,
\begin{equation}
S_{bulk} = \int d^4 x \int_{-\pi}^\pi d\phi \sqrt{-G}\left( 2M^3 R
  -\Lambda\right) , 
\end{equation}
where $M$ denotes the five dimensional Planck mass and $G_{MN}$ is
the five dimensional metric. The branes are located in $\phi$ and we
identify the brane coordinates with the remaining 5d coordinates
$x^\mu$, $\mu = 0,\ldots ,3$. Then the induced metrics on the branes
are simply
\begin{equation}
g_{\mu\nu}^{b1} = {G_{\mu\nu}}_{\left| \phi = 0\right.} \;\;\; , \;\;\;
g_{\mu\nu}^{b2} = {G_{\mu\nu}}_{\left| \phi =\pi\right.} .
\label{RSinduced}
\end{equation}
We assume that fields being localized on the brane are in the trivial
vacuum and take into account only nonzero vacuum energies on the
branes. Calling those vacuum energies $T_1$ and $T_2$, the brane 
actions read
\begin{equation}
S_{b1} + S_{b2} = -\int d^4 x\, \left( T_1 \sqrt{-g^{b1}} + T_2
  \sqrt{-g^{b2}}\right) ,
\end{equation}
where the first (second) term on the lhs is attributed to the first
(second) term on the rhs. Instead of working out the solutions to the
system on an interval $S^1/{\mathbb Z}_2$ it is technically easier to
construct a solution in a non compact space, such that the solution is
periodic in 
\begin{equation}\label{RSperiod}
\phi \equiv \phi + 2\pi ,
\end{equation}
and even under 
\begin{equation}
\phi \rightarrow -\phi .
\end{equation}
A vacuum with this property yields then automatically a compact
interval in $\phi$. (The equivalent\footnote{I acknowledge
  discussions with Rados\l aw Matyszkiewicz on this topic.} and more
complicated alternative 
is 
to define the theory on an interval from the very beginning and take
into account surface terms when deriving the equations of motion as
well as Gibbons Hawking\cite{Gibbons:1977ue} boundary terms (for a
discussion in the context of brane worlds see
also\cite{Dick:2000dt,Dick:2001sc}).)
With these remarks the Einstein equations of motion read (capital
indices run over all dimensions $M,N = 0 ,\ldots, 4$)
\begin{eqnarray}
\lefteqn{ \sqrt{-G} \left( R_{MN} -\frac{1}{2}G_{MN}\right) =} &
&\nonumber\\
& & -\frac{1}{4M^3}\left[ \Lambda \sqrt{-G} G_{MN} + \sum_{i=1}^2 T_i
  \sqrt{-g^{bi}} 
  g_{\mu\nu}^{bi} \delta^\mu _M \delta^\nu _N \delta\left( \phi
    -\phi_i\right) \right],\label{RSeinstein}
\end{eqnarray}
with $\phi_1 =0$ and $\phi_2 =\pi$. The delta functions appearing on
the rhs of (\ref{RSeinstein}) are defined on a real line. The most
general metric ansatz possessing a four dimensional Poincar\'e\
transformation as isometry is
\begin{equation}
ds^2 = e^{-2\sigma\left( \phi\right)} \eta_{\mu\nu}dx^\mu dx^\nu + r_c
^2 d\phi ^2 .
\label{RS1ansatz}
\end{equation}
We could rescale $\phi$ such that the $r_c$ dependence drops out, but
that would change the periodicity condition (\ref{RSperiod}). Plugging
this ansatz into the equations of motion (\ref{RSeinstein}) yields (a
prime denotes differentiation with respect to $\phi$)
\begin{eqnarray}
\frac{6{\sigma^\prime}^2}{r_c ^2} & = & -\frac{\Lambda}{4M^3}
,\label{RS1steom} \\ 
\frac{3\sigma^{\prime\prime}}{r_c ^2} & = & \frac{T_1}{4M^3
  r_c}\delta\left( \phi\right) + \frac{T_2}{4M^3 r_c}\delta\left( \phi
  -\pi\right) .\label{RS2ndeom}
\end{eqnarray}
The solution to (\ref{RS1steom}) is
\begin{equation}
\sigma = r_c\left| \phi\right| \sqrt{\frac{-\Lambda}{24 M^3}} ,
\end{equation}
where the modulus function is defined as usual in the interval $-\pi
<\phi < \pi$,
\begin{equation}\label{modulusf}
\left| \phi\right| = \left\{ \begin{array}{ll}
-\phi \; &, \; -\pi < \phi < 0\\
\phi \; &, \; 0 <\phi <\pi\end{array}\right. .
\end{equation}
This ensures that the solution is even under $\phi \to -\phi$. In
order to incorporate (\ref{RSperiod}), we define the modulus function 
on the real line by the periodic continuation of
(\ref{modulusf}). The resulting function is drawn in figure
\ref{fig:modulus}.
%%%%%%%%%%%%%%%%%%%%%%%%%%%%%%%%%%%%
\begin{figure}
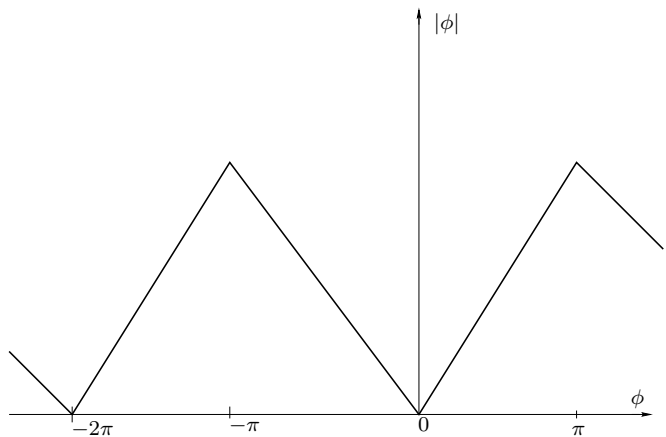
  
\begin{center}
\input modulus.pstex_t
\end{center}
\caption{The periodic modulus function.}
\label{fig:modulus}
\end{figure}
%%%%%%%%%%%%%%%%%%%%%%%%%%%%%%%%%%%%
Away from the points at $\phi =0$ and integer multiples of $\pi$, the 
second derivative of $\sigma$ vanishes and (\ref{RS2ndeom}) is
fulfilled in those regions. In order to take into account the delta
function sources in (\ref{RS2ndeom}), one integrates this equation
over an infinitesimal neighborhood
around the location of the brane sources. This gives rise to the
constraints
\begin{equation}\label{RSfine}
T_1 = -T_2 = 24M^3 k \;\;\;,\mbox{with}\;\;\; k^2 = -\frac{\Lambda}{24
  M^3} 
\end{equation}
on the parameters of the model. These constraints can be thought of as
fine tuning conditions for a vanishing effective cosmological constant
in four dimensions. We will come back to this point in section
\ref{sec:cosmo}. Our final solution is
\begin{equation}\label{RS1solu}
ds^2 = e^{-2kr_c \left|\phi\right|}\eta_{\mu\nu} +r_c ^2 d\phi ^2 ,
\end{equation}
where $k^2$ is defined in (\ref{RSfine}), and we take $k$ to be
positive (for a negative $k$ just redefine $\phi \to \pi -\phi$). 

We observe that by taking into account the back reaction of the branes 
onto the geometry, we obtain a metric which depends on the position in 
the compact direction. For the particular model we consider this
dependence is exponential. That opens up an interesting alternative
explanation for the large hierarchy between the Planck scale and the
weak scale. We take all the input scales ($M$, $\Lambda$ , $r_c$) to
be of the order of the Planck scale. First, we should check whether
this provides the correct four dimensional Planck mass. To this end,
we expand a general 4d metric around the classical solution
\begin{equation}\label{RSfluc}
ds^2 = e^{-2kr_c}\left( \eta_{\mu\nu} + h_{\mu\nu}\right) dx^\mu
dx^\nu + r_c ^2 d\phi^2 .
\end{equation}
In principle we should also allow the four-four component of the
metric $r_c^2$ to fluctuate. Since $r_c$ is an integration constant,
such fluctuations will be seen as massless scalars in the effective
four dimensional theory. This is a common problem known as moduli
stabilization problem. We will assume here that some unknown
mechanism gives a mass to the fluctuations of $G_{44}$ and take it to
be frozen at the classical value $r_c ^2$. The Kaluza-Klein gauge
fields $G_{\mu 4}$ are projected out by the ${\mathbb Z}_2$. 
Plugging (\ref{RSfluc})
into the action and integrating over $\phi$ yields the effective
action for four dimensional gravity
\begin{equation}
S_{eff} = M_p ^2 \int d^4 x \sqrt{-g} R^{(4)}\left( g\right) ,
\end{equation}
where $R^{(4)}\left( g\right)$ denotes the four dimensional scalar
curvature computed from $g_{\mu\nu} =\eta_{\mu\nu} +h_{\mu\nu}$ and
the four dimensional Planck 
mass $M_p$ is given by
\begin{equation}\label{RS1planckmass}
M_p ^2 = M^3 r_c \int_{-\pi}^\pi d\phi e^{-2kr_c \left| \phi\right|}
=\frac{M^3}{k}\left( 1 - e^{-2kr_c \pi}\right) .
\end{equation}
This tells us that choosing five dimensional scales of the order of
the Planck scale gives the correct order of magnitude for the four
dimensional Planck scale. 

Now, let us consider matter living on the branes. On the first brane
located at $\phi = 0$, the induced metric is just the Minkowski metric
and Lagrangians for matter living on that brane will just have their
usual form. On the other hand, matter living on the second brane
(located at $\phi = \pi$) feels the $\phi$ dependence of the bulk
metric. Let us focus on a Higgs field being located at the second
brane. Its action will be of the form
\begin{equation}\label{RShiggs}
S_{Higgs}^{b2} = \int d^4 x e^{-4kr_c\pi}\left\{
  e^{2kr_c\pi}\eta^{\mu\nu} 
  D_\mu H^\dagger D_\nu H - \lambda\left(  \left| H\right| ^2 -
  v_o^2\right)^2\right\} ,
\end{equation}
where the overall exponential factor originates from the determinant
of the induced metric. Rescaling the Higgs field $H$ such that
the kinetic term in (\ref{RShiggs}) takes its canonical form induces
the rescaling
\begin{equation}
v_0 \to v_{eff} = e^{-kr_c\pi}v_0 .
\end{equation}
This means that a symmetry breaking scale which is written as $v_0$
into the model effectively is multiplied by a factor of
$e^{-kr_c\pi}$. Repeating the above argument for any massive field, one
finds that any mass receives such a factor
\begin{equation}\label{RSmassre}
m_0 \to m_{eff} = e^{-kr_c\pi}m_0 ,
\end{equation}
when going to an effective description in which kinetic terms are
canonically normalized. 
Choosing $kr_c\approx 10$ (which is roughly a number of order one),
one can achieve that the exponential in (\ref{RSmassre}) takes Planck 
sized input masses to effective masses of the order of a TeV. Hence,
in the above model we can obtain the TeV scale from the Planck scale
without introducing large numbers, provided we live on the second
brane. 

\subsubsection{A proposal for radion stabilization}\label{joldberjer}

In the previous section, we have already mentioned that the internal 
metric 
component $G_{44}$ gives rise to a massless field in an effective
description. This means that its vev $r_c$ is very sensitive against
any perturbation and rather unstable. For the discussion of the
hierarchy problem it is important that the distance of the branes
$r_c$ is of the order of the Planck length. Therefore, it is
desirable to stabilize this distance, i.e.\ to give a mass to $G_{44}$
in the effective description. In the present section we briefly
present a proposal of Goldberger and Wise how a stabilization might be
achieved via an additional scalar living in the bulk. We will neglect
the back reaction of the scalar field on the geometry. This means that
we just consider a scalar field in the RS1 background constructed in
the previous section. The action consists out of three parts
\begin{equation}
S = S_{bulk}  + S_{b1} + S_{b2} ,
\end{equation}
where $S_{bulk}$ defines the five dimensional dynamics of the field
and $S_{b1}$ and $S_{b2}$ its coupling to the respective branes. We
choose
\begin{equation}
S_{bulk} = \frac{1}{2}\int d^4 x\int_{-\pi}^\pi d\phi \sqrt{-G}\left(
  G^{MN}\partial_M \Phi \partial_N\Phi - m^2 \Phi^2\right) ,
\end{equation}
where $\Phi$ is the scalar field and $G_{MN}$ is given in
(\ref{RS1solu}). The coupling to the branes is taken to be
\begin{eqnarray}
S_{b1} &=& -\int d^4 x \sqrt{-g^{b1}}\lambda_1 \left( \Phi^2 -
  v_1^2\right)^2 ,\\
S_{b2} & = & -\int d^4 x \sqrt{-g^{b2}}\lambda_2\left( \Phi^2 -
  v_2^2\right)^2 ,
\end{eqnarray}
where $v_i$ and $\lambda_i$ are dimensionfull parameters whose values
will be discussed below. With the ansatz that $\Phi$ does not depend
on the $x^\mu$ for $\mu = 0,\ldots, 3$ the equation of motion for the
scalar is 
\begin{eqnarray}
\lefteqn{e^{-4kr_c\left|\phi\right|}\left( -\frac{e^{4kr_c\left|
 \phi\right|}}{r_c ^2}\partial_\phi \left( e^{-4kr_c \left| 
 \phi\right|}\partial_\phi \Phi\right) + m^2 \Phi\right. } &&
 \nonumber\\
& & \left. + 4 \lambda_1 \Phi
 \left( \Phi^2 - v_1 ^2\right) \frac{\delta\left( \phi\right)}{r_c} +
 4\lambda_2 \Phi\left( \Phi^2 - v_2 ^2\right) \frac{\delta\left( \phi
 -\pi\right)}{r_c}\right) = 0 .
\label{GWeom}
\end{eqnarray}
With $\nu = \sqrt{4 +\frac{m^2}{k^2}}$ the solution inside the bulk $0
< \phi < \pi$ is written as
\begin{equation}
\Phi = e^{2kr_c \left| \phi\right|}\left( A e^{kr_c \nu\left|
      \phi\right|} + B e^{-kr_c \nu \left| \phi\right|}\right) ,
\end{equation}
where the integration constants $A$ and $B$ will be fixed
below. Plugging this solution back into the Lagrangian yields an $r_c$
dependent constant, i.e.\ a potential for the distance of the two
branes, 
\begin{eqnarray}
V\left( r_c\right) & = & k\left( \nu +2\right) A^2 \left( e^{2\nu
    kr_c\pi} -1\right) + k\left( \nu -2\right) B^2 \left( 1 -
    e^{-2\nu kr_c \pi}\right) \nonumber \\
& & \hspace{0.3cm} +  \lambda_1  \left( \Phi\left(
    0\right)^2 - v_1^2\right)^2 + \lambda_2e^{-4kr_c\pi} \left(
    \Phi\left( \pi\right) ^2
     - v_2 ^2 \right)^2 . 
\label{gwpotacu}
\end{eqnarray}
Because of the dependence of $\Phi$ on the modulus function (see
figure \ref{fig:modulus}) the second derivative in the first term in
(\ref{GWeom}) will lead to delta functions whose argument vanishes at
the position of the branes. Matching this with the delta
function source terms in (\ref{GWeom}) yields equations for the
integration constants $A$ and $B$. Instead of writing down and solving
those equations explicitly we suppose that $\lambda_1$ and $\lambda_2$
are large enough for the approximation
\begin{equation}
\Phi\left( 0\right) = v_1 \;\;\; , \;\;\; \Phi\left( \pi\right) =v_2 
\end{equation}
to be sufficiently accurate. In this approximation one obtains
\begin{eqnarray}
A &=& v_2 e^{-\left( 2+\nu\right) kr_c \pi} - v_1 e^{-2\nu kr_c\pi} ,\\
B & = & v_1 \left( 1 + e^{-2\nu kr_c\pi}\right) - v_2 e^{-\left(
    2+\nu\right) kr_c\pi} .
\end{eqnarray}
The next approximation lies in the assumption 
that
\begin{equation}
\epsilon = \frac{m^2}{4k}\ll 1 .
\end{equation}
In evaluating the potential $V\left( r_c\right)$ (\ref{gwpotacu}), we 
neglect terms of order $\epsilon^2$ but do not treat $\epsilon kr_c$
as a small number. This yields
\begin{eqnarray}
V\left( r_c\right) & = & k\epsilon v_1 ^2 + 4ke^{-4kr_c\pi}\left( v_2
  -v_1e^{-\epsilon kr_c \pi}\right)^2\left( 1
  +\frac{\epsilon}{4}\right)\nonumber \\
& & \hspace{0.3cm} - k \epsilon v_1 e^{-\left(
    4+\epsilon\right) 
  kr_c\pi}\left( 2v_2 - v_1 e^{-\epsilon kr_c\pi}\right) .
\end{eqnarray}
Up to orders of $\epsilon$, this potential has a minimum at 
\begin{equation}\label{gwminimum}
kr_c = \frac{ 4k^2}{\pi m^2}\log\left( \frac{v_1}{v_2}\right) .
\end{equation}
In figure
\ref{fig:gwpot}, we have drawn the potential in a neighborhood of the
minimum  (using Maple). (What is actually drawn is $V 
-k\epsilon v_1 ^2$.)
\begin{figure}[t]
\begin{center}
\begin{minipage}{8cm}
\centerline{}
\psfig{figure=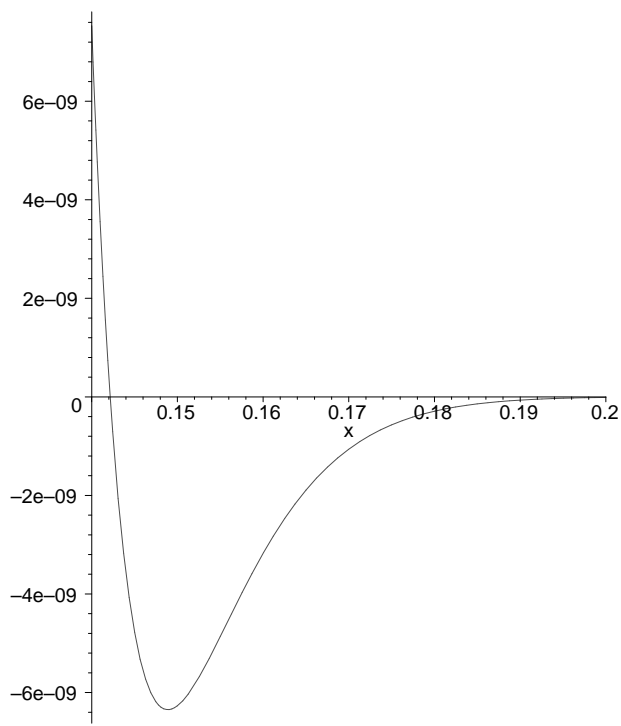,width=8cm}
\end{minipage}\\
\end{center}
%\vspace{-6mm}
\caption{
The Goldberger Wise potential for $k=10$, $m=9$, $v_2 =1$, $v_1 =3$.
The vertical axis shows $V- k\epsilon v_1 ^2$ whereas on the
horizontal axis $r_c$ is drawn.  
}\label{fig:gwpot}
\end{figure}
With the appropriate choice for the scales, the minimum of the
potential is clearly visible. One should note, however, the
exponentially 
suppressed height of the right wall of the potential. If we had chosen 
a larger scale for the drawing, figure \ref{fig:gwpot} would just show
a runaway potential which rapidly reaches its asymptotic value. This
might be a drawback of the stabilization mechanism.

The expression for the stable distance between the branes
(\ref{gwminimum}) shows that no extreme fine tuning is needed in
order to obtain the wanted value of about ten for $kr_c$. 
It remains to investigate whether the various approximations (including
the neglection of the back reaction) are sensible. This investigation
has been carried out in\cite{Goldberger:1999uk} by estimating the size
of next to leading order corrections. The result is that the
approximations are fine.

To close this section, we should mention that the described
stabilization method is often called ``Goldberger Wise mechanism'' in
the literature. We preferred to use the term ``proposal'' because we
are not certain that this mechanism is the commonly established method
for solving the problem of moduli stabilization. We decided to present
a brief description of the method because it is one of the most
prominent lines of thought in the context of the Randall Sundrum
model. In general, the problem of moduli stabilization is not very
well understood.

\subsection{The RS2 model with one brane}\label{sec:RS2}

In this section we are going to consider a variant of the model
presented in section \ref{sec:rs1} where the second brane is
removed. Since for the solution of the hierarchy it was essential that
the observers live on this second brane, we now give up the
goal of solving the hierarchy problem (at least temporarily).
The construction of the single brane solution is very simple. The
extra dimension is not compact anymore and therefore we use the
coordinate $y$ instead of $\phi$. We do not impose the periodicity
condition (\ref{RSperiod}) but still require a ${\mathbb Z}_2$
symmetry under
\begin{equation}
y \to -y .
\end{equation}
Further, we remove $S_{b2}$ from the action (\ref{RS1action}). Since
the extra dimension is not compact, we can perform rescalings of $y$ in
order to remove the $r_c$ dependence of the ansatz
(\ref{RS1ansatz}). Without loss of generality we take $r_c =1$. Thus,
in the single brane case, the solution for the metric is
\begin{equation}\label{RS2metric}
ds^2 = e^{-2k\left| y\right|}\eta_{\mu\nu}dx^\mu dx^\nu + dy^2.
\end{equation}
With a non compact extra dimension, one may worry that gravity is five 
dimensional now. However, taking the $r_c\to\infty$ limit of
(\ref{RS1planckmass}), one finds that the effective four dimensional 
Planck mass is finite. This means that the graviton zero mode is
normalizable and yields a four dimensional Newton law. Apart from the
zero modes, there will be also massive gravitons who lead to 
corrections of Newton's law. In the following subsection we will
investigate these corrections.

\subsubsection{Corrections to Newton's law}\label{sec:newtonpotcirc}

The Newton potential is obtained by studying fluctuations around the
background (\ref{RS2metric}), for example 
\begin{equation}
G_{00} = -e^{-2k\left| y\right|} -V\left( x,y\right) ,
\end{equation}
where $V$ denotes a fluctuation.
In the presence of a point particle with mass $\mu$ at the origin, the 
non relativistic limit of the linearized equation for $V$ reads
\begin{equation}\label{RS2newpotequ}
\left[ \Delta_3 + e^{-2k\left| y\right|}\left( \partial_y ^2 +
    4k\delta\left( y\right) - 4k^2\right)\right] V\left( x,y\right) =
    G\mu\, \delta^{(3)}\left( x\right) \delta\left( y\right) ,
\end{equation}
where $G$ is the five dimensional Newton constant. The fact that $V$
is indeed 
the Newton potential can be confirmed by studying the geodesic
equation of a point particle probe and comparing it with the Newton
equation of motion. The equation (\ref{RS2newpotequ}) is the warped
geometry analogon of equation (\ref{newtpothigh}). (The normalization
of the higher dimensional Newton constant is not really important
here.) 
It is useful to redefine the coordinate $y$ according to
\begin{equation}
z\equiv \frac{sgn\left( y\right)}{k}\left( e^{k\left| y\right|}
  -1\right) .  
\label{RSzcoo}
\end{equation}
With
\begin{equation}
\bar{V} = V\left( x,y\right) e^{\frac{k\left| y\right|}{2}}
\end{equation}
equation (\ref{RS2newpotequ}) takes the form
\begin{equation}
\left[ \Delta_3  +\partial_z ^2 - \frac{15k^2}{4\left(
      k\left| z\right| +1\right)^2 } +3k\delta\left(
      z\right)\right] \bar{V} = G\mu\, \delta^{(3)}\left( x\right)\delta\left(
z\right) 
\label{RSnewtonpoti}
\end{equation}
Analogous to section \ref{sec:newtonpotflat} we plan to expand the
solution $V$ into a series of eigenfunctions, i.e.\ in the case at
hand we are looking for solutions of the differential equation
\begin{equation}\label{RSeigen}
\left[ \partial_z ^2 - \frac{15k^2}{4\left(
      k\left| z\right| +1\right)^2 } +3k\delta\left(
      z\right)\right] \psi\left( m ,z\right) = -m^2 \psi\left(
      m,z\right) ,
\end{equation}
where we expect a continuous eigenvalue $m$ now, since the ``internal
space'' is not compact. 
Let us discuss first the zero mode, i.e.\ the solution to
(\ref{RSeigen}) with $m^2 =0$. The zero mode is found to
be\footnote{In forthcoming expressions we will always imply that
  $m>0$ when writing $\psi\left( m ,z\right)$.  The zero mode will
  be denoted by $\psi_0\left( z\right)$ from now on.} 
\begin{equation}
\psi_0\left( z\right)\equiv \psi\left( 0,z\right) =
    \frac{N_0}{\left( k\left| z\right| 
    +1\right)^\frac{3}{2}} ,
\end{equation}
where $N_0$ is an integration constant to be fixed later.
Note that
\begin{equation}
\partial _z \left| z\right| = sgn\left( z\right) \;\;\; ,\;\;\;
\partial_z sgn\left( z\right) = 2\delta\left( z\right). 
\end{equation}
Now, we take $m > 0$.
For $z>0$ the general solution to the above
equation can be written as a superposition of Bessel functions
\begin{equation}\label{besselsupo}
\psi\left( m,z\right) =\sqrt{\left| z\right| +\frac{1}{k}} \left(c_1
    J_2 \left( m\left( \left| z\right| 
    +\frac{1}{k}\right)\right) + c_2 Y_2 \left( m \left( \left|
    z\right| + \frac{1}{k}\right)\right)\right) ,
\end{equation}
where $J_\nu$ denotes the Bessel functions of the first kind whereas
$Y_\nu$ stands for the Bessel functions of the second kind and
$c_{1,2}$ are constants to be fixed below. Because the solution 
(\ref{besselsupo}) is written as a function of $\left| z\right|$, the 
second derivative with respect to $z$ in (\ref{RSeigen}) will yield a
term containing a $\delta\left( z\right)$ (and other terms). One can
fix the ratio $c_1/c_2$ by matching the factor in front of this delta
function with the factor in front of the delta function in
(\ref{besselsupo}). We will do this in an approximate way. The most
severe corrections to Newton's law are to be expected from gravitons
with small $m$ (because they carry interactions over longer
distances). In matching the coefficients of the delta functions, only
a neighborhood  around $z=0$ matters. Therefore, we replace the Bessel 
functions by their asymptotics for small arguments, which are
\begin{eqnarray} 
J_2\left( m\left( \left| z\right| +\frac{1}{k}\right)\right) & \sim &
\frac{m^2\left( \left| z\right| +\frac{1}{k}\right)^2}{8}
,\label{RSasy1}\\ 
Y_2\left( m\left(\left| z\right| +\frac{1}{k}\right)\right) &\sim & 
-\frac{4}{\pi m^2 \left( \left| z\right| +\frac{1}{k}\right)^2}
-\frac{1}{\pi} .\label{RSasy2}
\end{eqnarray}
Plugging the asymptotic approximation into (\ref{besselsupo}) and then
into (\ref{RSeigen}) one finds that the overall coefficient in front
of the delta function vanishes if
\begin{equation}
\frac{c_1}{c_2} = \frac{4k^2}{\pi m^2} .
\end{equation}
Hence, our general solution (\ref{besselsupo}) reads
\begin{equation}
  \psi\left( m ,z\right) = N_m \sqrt{\left| z\right|
  +\frac{1}{k}}\left[ Y_2 \left( m\left| z\right| +\frac{1}{k}\right)
  +\frac{4k^2}{\pi m^2} J_2\left( m\left( \left| z\right|
  +\frac{1}{k}\right)\right)\right] ,
\end{equation}
where we replaced $c_2 = N_m$ because this remaining integration
constant will turn out to depend on the eigenvalue $m$. 

Recall that the extra dimension $y$ (or $z$) is not compact. Thus the
eigenvalue $m$ is continuous. Therefore, we normalize
\begin{equation}\label{RSortho}
\int dz\, \psi\left( m,z\right) \psi\left( m^\prime ,z\right) =
\delta\left( m -m^\prime\right) ,
\end{equation}
for $m, m^\prime >0$. For $m\geq 0$ we impose the normalization
condition 
\begin{equation}\label{RSorthozero}
\int dz\, \psi_0\left( z\right) \psi\left( m, z\right) =\delta_{m,0},
\end{equation}
such that the completeness relation reads
\begin{equation}
\psi_0\left(z\right)\psi_0\left( z^\prime\right) + \int_0 ^\infty dm\, 
\psi\left( m , z\right) 
\psi\left( m ,z^\prime\right) = 
\delta\left( z- z^\prime\right) .
\label{RScomplete}
\end{equation}
The orthonormalization condition (\ref{RSortho}) fixes $N_m$. It turns
out that the computation simplifies essentially in the approximation
where the arguments of the Bessel functions are large, since the
corresponding asymptotics yields plane waves. Explicitly, for large
$m z$ the Bessel functions are approximated by
\begin{eqnarray}
\sqrt{z}J_2\left( mz\right) & \sim & \sqrt{\frac{2}{\pi m}}\cos \left(
  mz -\frac{5\pi}{4}\right) ,\\
\sqrt{z}Y_2\left( mz\right) & \sim & \sqrt{\frac{2}{\pi m}}\sin\left(
  mz -\frac{5\pi}{4}\right) .
\end{eqnarray}
Because we are mainly concerned about large ($>\mu m\gg 1/M_p$) 
distance modifications of Newton's law we focus on the contribution of
the ``light''  
modes ($\frac{m^2}{k^2}\ll 1$). (Recall that $k$ is of the order of
the Planck mass.) 
Then (\ref{RSortho}) yields for the
normalization constant (for $m>0$)
\begin{equation}
N_m = \frac{\pi m^{\frac{5}{2}}}{\left( 4k^2\right)} .
\end{equation}
The condition (\ref{RSorthozero}) is satisfied for $m>0$ to a good
approximation. Evaluating (\ref{RSorthozero}) for $m=0$ fixes 
\begin{equation}
N_0 = \sqrt{k}.
\end{equation}

Now, we expand  $\bar{V}\left( x, z\right)$   
into eigenfunctions $\psi_0\left(z\right)$ and $\psi\left( m,z\right)$
with $x$ dependent coefficients $\varphi_m\left( x\right)$
\begin{equation}
\bar{V}\left( x, z\right) = \varphi_0\left( x\right)\psi_0\left(
  z\right) + 
\int_0 ^\infty dm \,
\varphi_m\left( x\right) \psi\left( m,z\right) .
\label{RSpotexpanded}
\end{equation}
By plugging 
the ansatz (\ref{RSpotexpanded}) into (\ref{RSnewtonpoti}), we find 
that 
for $m\geq 0$ and $r = \left| x\right|$
\begin{equation}
\varphi_m\left( x\right) = -\frac{G\mu}{r}e^{-mr} a_m ,
\end{equation}
with the constants $a_m$ taken such that
\begin{equation}
a_0 \psi_0\left( z\right) +\int dm \, a_m \psi\left( m ,z\right) =
\delta\left( z\right) .
\end{equation}
Comparison with (\ref{RScomplete}) yields
\begin{equation}
a_0  =  \psi_0\left( 0\right) \;\;\; ,\;\;\; a_m = \psi\left( m
  ,0\right) .
\end{equation}
In the current setup we are interested in corrections to Newton's law
as an observer on the brane at the origin would measure them. Defining
the four dimensional Newton constant $G_4$ as
\begin{equation}
G_4 = Gk, 
\end{equation}
we find from (\ref{RSpotexpanded})
\begin{equation}
\bar{V}\left( x ,0\right) = V\left( x, 0\right) =
  -\frac{G_4\mu}{r}\left( 
  1 + \int_0 ^\infty dm \, \frac{ m}{k^2}e^{-mr}\right) ,
\label{RSalmostfinal}
\end{equation}
where once again we took into account only modes with $m/k \ll 1$ such
that we could use the asymptotics (\ref{RSasy1}) and (\ref{RSasy2}) in
order to evaluate $\psi\left( m,0\right)$. Finally, performing the
integral in (\ref{RSalmostfinal}) leads to
\begin{equation}
V\left( x,0\right) = -G_4 \frac{\mu}{r}\left( 1 +\frac{1}{r^2
    k^2}\right) .
\label{RSnewtonpotf}
\end{equation}
For $k$ being of the order of the Planck mass (\ref{RSnewtonpotf}) is
in very good agreement with the experimental values. This may look a
bit surprising. Even though the extra dimension is not compact, we
obtain a four dimensional Newton potential for observers who live on
the brane at $y=0$. This non trivial result finds its explanation in
the exponentially warped geometry. It is this geometry which is
responsible for the fact that the amplitude of the zero mode has its
maximum at the brane and vanishes rapidly for finite $z$. On the other
hand, the massive modes reach their maximal amplitudes asymptotically
far away from the brane. Therefore, they have very little influence on 
the gravitational interactions on the brane, although the masses of
the extra gravitons can be arbitrarily small. 

In the following subsection we are going to rederive
(\ref{RSnewtonpotf}) in a different way.

\subsubsection{... and the holographic principle}

In section \ref{adssection} we have described a duality between a
field theory living on the boundary of an $AdS_5$ space and a theory
living in the bulk of an $AdS_5$ space. This correspondence is
sometimes called the holographic principle since it allows to
reproduce bulk data from boundary data (and vice versa). Now we are
going to apply this principle to the RS2 setup. Before doing so, we
will establish that the RS2 setup has something to do with an $AdS_5$
space (namely it is a slice of an $AdS_5$ space). To this end, we
first write down the RS2 metric (\ref{RS2metric}) in terms of the
coordinate $z$ defined in (\ref{RSzcoo}). This results in
\begin{equation}
ds^2_{RS2} =\frac{1}{\left(k\left| z\right|
    +1\right)^2}\eta_{\mu\nu} dx^\mu dx^\nu + dz^2 .
\label{RSslice}
\end{equation}
For symmetry reasons the coordinate $z$ can be restricted to the half
interval between zero and infinity. The singularity at $z=0$ is caused
by the brane. 

Now, let us recall from section \ref{adssection} that the $AdS_5$
metric is (see (\ref{adsmetricforlater}))
\begin{equation}
ds^2_{AdS} = \frac{U^2}{R^2} \eta_{\mu\nu}dx^\mu dx^\nu + R^2
\frac{dU^2}{U^2} .
\label{adsmetricrecall}
\end{equation}
Changing the coordinates according to ($-R < z < \infty$)
\begin{equation}
U = \frac{R^2}{z +R} 
\label{Uzrelation}
\end{equation}
yields an $AdS$ metric of the form
\begin{equation}
ds_{Ads}^2 = \frac{1}{\left(
    1+\frac{z}{R}\right)^2}\eta_{\mu\nu}dx^\mu dx^\nu + dz^2 .
\label{RSads}
\end{equation}
Comparing (\ref{RSads}) with (\ref{RSslice}), we observe that the RS2 
geometry describes a slice of an $AdS_5$ space. The radius of the
$AdS_5$ space is $1/k$, and the space is cut off at $z=0$. Since the
boundary of the $AdS_5$ space is situated at $U\to \infty$, the cutoff 
at $z=0$ means that we lost the region between $U=R$ and the
boundary. Hence, the position of the brane in the RS2 setup can be
viewed as an infrared cutoff for gravity on an $AdS_5$
space. This suggests that we may apply the AdS/CFT conjecture on the
RS2 scenario. (Note however, that we do not have any supersymmetry
now. Without supersymmetry the AdS/CFT conjecture has passed less
consistency checks. Nevertheless, let us assume that the conjecture is
correct also without supersymmetry.) The field theory dual of the RS2
setup is thus a conformal field theory with a UV
cutoff\footnote{Recall from section  \ref{wilsonclassiapo} that an IR
  cutoff in the bulk theory corresponds to a UV cutoff in the dual
  field theory.} given by
$k$. (The cutoff actually breaks the conformal invariance. The
conformal anomaly induces a coupling of the field theory to gravity.)
In particular, we plan to employ the AdS/CFT duality conjecture for
the computation of corrections to Newton's law. As a preparation let
us sketch how Newton's potential is related to the gravity propagator
in four dimensions. If we did not have an extra dimension, the gravity  
propagator in momentum space is (up to a polarization tensor)
$1/\left( M_p ^2 p^2\right)$. The Newton potential can be obtained
from this 
propagator by 
formally setting the $p_0$ component to zero and Fourier
transforming\footnote{We use the following prescription for performing
  the Fourier transformation. Transforming the equation $\Delta_3
  f\left( x\right) = \delta^{(3)}\left( x\right)$ one finds that
  $1/p^2$ transforms into $1/r$. Later we will also have to compute
  Fourier transforms with additional powers of $p$ in the numerator or
  denominator. An additional power of $p$ in the numerator transforms
  into $\partial_r$ whereas powers of $p$ in the denominator can be
  generated by $\partial_p$ which in turn transforms into $r$.}
with respect to the spatial momentum components. The result in
position space is then $1/(M_p^2 r)$. Therefore, we will use the
AdS/CFT 
duality conjecture to compute the corrected graviton propagator and
deduce the corrected Newton potential via the above description.

The dual picture for the RS2 setup is that we have 
four dimensional gravity plus the CFT dual of gravity on
$AdS_5$ with a UV cutoff $k$. Corrections to four dimensional gravity
are caused by the interaction of gravity with the CFT. The 
effective graviton propagator is obtained by integrating over the CFT
degrees of freedom. 
The one loop corrected graviton propagator will be schematically of
the form 
\begin{equation}
\frac{1}{M_p ^2p^2}\left( 1 + \langle T_{CFT}\left( p\right)
  T_{CFT}\left( 
    -p\right)\rangle \frac{1}{M_p ^2 p^2}\right) ,
\label{holograpro}
\end{equation}
where $T_{CFT}$ stands for the energy momentum tensor of the CFT
dual. (The coupling of gravity to the CFT fields is given by the
energy momentum tensor.)
For any four dimensional CFT, the two point function of 
the energy momentum tensor is fixed to be of the form
\begin{equation}
\langle T_{CFT}\left( p\right) T_{CFT}\left( -p\right)\rangle =
cp^4 ,
\label{twopointepfunc}
\end{equation}
where we imposed that the UV cutoff is $k$. We will not derive this
result here, but just give two comments. First, notice that
(\ref{twopointepfunc}) is the four dimensional analogon of
(\ref{EPOPE}). The number $c$ quantifies the conformal anomaly.  The
second remark is, that the reader may get some impression on how the
expression (\ref{twopointepfunc}) arises by computing it explicitly
for pure 
gauge theory. We are interested in the order of magnitude of
$c$. 
This has been computed in\cite{Henningson:1998gx} to be
\begin{equation}
c \approx \frac{M_5 ^3}{k^3} = \frac{M_p ^2}{k^2} ,
\label{henning}
\end{equation}
where $M_5$ denotes the five dimensional Planck mass and the radius
of the $AdS$ space dual to the CFT is $1/k$. In the second equality
of (\ref{henning}) we used the relation between the five and four
dimensional Planck mass ($M_5$ and $M_p$) which, in the RS2 setup, is
obtained by taking $r_c\to \infty$ in (\ref{RS1planckmass}).

The corrected Newton potential is obtained by setting formally $p_0$
to zero in (\ref{holograpro}) and performing a three dimensional
Fourier transformation to the position space. Thus, the effect of 
integrating over the CFT fields results in the following
replacement of the Coulomb (Newton) potential 
\begin{equation}
-\frac{1}{r} \rightarrow -\frac{1}{r}\left( 1 + \frac{1}{k^2
 r^2}\right) .
\end{equation}
This result agrees with the expression (\ref{RSnewtonpotf}) computed
in the previous section. Thus, we have learned that integrating over
the CFT fields yields the same corrections to four dimensional gravity
as taking into account the massive ``Kaluza-Klein'' gravitons. Employing
the AdS/CFT correspondence, the computational effort decreases
substantially. We will make use of this fact when we combine the RS1
with the RS2 scenario in the next subsection.

\subsubsection{The RS2 model with two branes}  

In the previous two subsections we have seen that the RS2 setup has
the exciting feature of giving rise to effectively four dimensional
gravitational interactions even though the extra dimension is not
compact. On the other hand, we observed before that the RS1 model is
capable to explain the hierarchy between the Planck scale and the weak
scale without introducing large numbers. How can we combine these two
models? We should introduce a brane with the observers at $y=\pi r_c$
into the RS2 setup. However, this brane should not cause a change of
the RS2 metric (\ref{RS2metric}). The observers on the additional
brane (at $y =\pi r_c$) can achieve this by performing a fine tuning
such that the vacuum energy on their brane vanishes. In the following
we will call the brane at $y=\pi r_c$ the SM (Standard Model)
brane. The SM brane can be viewed as a probe in the RS2
background. The hierarchy can now be explained in the same way as it
is explained in the RS1 setup. What we should worry about are the
gravitational interactions as viewed by an observer on the SM
brane. In principle, these can be computed along the lines of section
\ref{sec:newtonpotcirc}. The situation is, however, slightly more
complicated since the approximation has to be refined. In particular, 
replacing the Bessel functions by their plane wave asymptotics in the
computation of $N_m$ is too rough an estimate. Now, this would imply
that the observer on the SM brane sees the Bessel functions as plane
waves. As argued in\cite{Lykken:1999nb} this is not the case, in
particular for the light continuum modes. The authors
of\cite{Lykken:1999nb} refined the approximation and obtained the
result
\begin{equation}
V\left( r, y=\pi r_c \right) = -\frac{G_4\mu}{r} \left( 1 + \frac{1}{k^2
    r^2}\right) - \frac{\mu}{M_w ^8 r^7}
\label{RLcorrection}
\end{equation}
for the Newton potential observed on the SM brane. Here, $M_w$ is of
the order of a TeV if we take $r_c$ such that the hierarchy problem
is solved. Instead of going through the tedious refinement of the
approximations performed in section \ref{sec:newtonpotcirc}, we employ 
the AdS/CFT correspondence to motivate (\ref{RLcorrection}). The
introduction of the SM brane modifies the RS2 dual such that it
consists out of four dimensional gravity, the CFT dual of the RS2
$AdS_5$ slice and the Standard Model of the probe brane. Note that
$y_c = r_c \pi$ is translated to $U_0- U_c = TeV$ in the course of the
coordinate transformations (\ref{RSzcoo}) and (\ref{Uzrelation}),
where $U_0$ denotes the position of the brane at the origin and $U_c$
the position of the SM brane.  This
means that SM fields and CFT fields interact via fields with masses of
the order of a $TeV$.\footnote{One may view the field theory dual as a
  stack of D-branes on which the CFT lives and the SM probe brane
  separated by a distance $1/TeV$ from the CFT branes. The interaction
  between the CFT and the SM can be thought of as arising due to open
  strings stretching between the corresponding branes.} Integrating
out those fields yields effective 
coupling terms between SM fields and CFT fields. (This is analogous to
generating the Fermi interaction via integrating out the $W$ and
$Z$ bosons.) The structure of the possible interaction terms is
restricted by symmetries to\cite{Arkani-Hamed:2000ds}
\begin{equation}
\frac{1}{M_w ^4} T^{\mu\nu}_{SM} T_{\mu\nu CFT} .
\label{SMCFTinter}
\end{equation}
Note the similarity to the coupling of the SM fields to
gravitons. Apart from charged interactions, the
SM fields interact via gravitons and via CFT fields. This suggests
that for an observer on the SM brane the effective graviton propagator
is
\begin{equation}
\frac{1}{M_p^2 p^2}\left( 1 + \langle T_{CFT}\left( p\right) T_{CFT}\left(
    -p\right)\rangle \frac{1}{M_p ^2p^2}\right) + \frac{1}{M_w^8}
\langle 
T_{CFT}\left( p\right) T_{CFT} \left( -p\right)\rangle , 
\end{equation}
where the first two terms are the same as in (\ref{holograpro}), and
the last term means that the observer will interpret the interaction
(\ref{SMCFTinter}) as gravitational interaction. In computing the
contribution due to the last
term we use (\ref{twopointepfunc}). Applying the recipe of the
previous section, we obtain out of 
the propagator the modified Newton potential
(\ref{RLcorrection}). This potential is still in agreement with the
observational bounds on deviations from Newton's law. Hence, adding a
probe brane at $y=\pi r_c$ in the RS2 setup one obtains a model which
explains the hierarchy and possesses effectively four dimensional
gravitational interactions, even though there is a non compact extra
dimension. However, we should remark that we discussed the setup only
classically and showing its stability against quantum corrections
may be a problematic issue. This corresponds to the technical hierarchy
problem which can be solved by supersymmetry in conventional four
dimensional models. Supersymmetric versions of the RS model appear in
the literature listed in chapter \ref{furtherreading}.

\section{Inclusion of a bulk scalar}
\setcounter{equation}{0}

In this section, we are going to modify the Randall Sundrum models of 
the previous section by introducing a bulk scalar $\Phi$ which couples
also to the branes. Actually, we have considered this modification
already in section \ref{joldberjer}, where we neglected the
back reaction of the scalar on the geometry. In the current section we
are going to take this back reaction into account. We will not return
to the stabilization mechanism of section \ref{joldberjer},
though. (The inclusion of back reaction into the Golberger Wise
mechanism is discussed in\cite{DeWolfe:1999cp}, with the result that
the mechanism works also when the back reaction is included.) 
Instead of addressing the question of how a scalar helps to stabilize 
the inter brane distance, we want to consider another question. As we
will see the cosmological constant problem is reformulated in a brane
world setup. We will investigate whether a scalar can help to find a
solution to the cosmological constant problem. Before doing so, we
briefly present a solution generating technique and consistency
conditions on the solutions.

\subsection{A solution generating technique}\label{solugener}
\setcounter{equation}{0}

Introducing a bulk scalar $\Phi$ modifies the action (\ref{RS1action})
to\footnote{For simplicity we set the five dimensional Planck mass to
  one. It can be introduced if needed by a simple analysis of the mass
  dimensions.}
\begin{equation}
S= \int \hspace{-1mm}d^4x \int\hspace{-1mm} dy
\sqrt{-G}\left( R 
    -\frac{4}{3}\left( 
    \partial \Phi\right)^2 - V\left( \Phi\right)\right) -
    \sum_{i}\int_{b_i}d^4 x \sqrt{-g^{bi}}f_i\left(\Phi\right) ,
\label{BULKSCALARIN}
\end{equation}
where $y$ is the coordinate labeling the extra dimension, and the sum
over $i$ stands for a sum over the branes. The index $b_i$ at the
integral means that $y$ is fixed to the position ($y_i$) of the brane 
$b_i$. The function $V\!\left( \Phi\right)$ is a bulk potential for the
scalar and $f_i\left( \Phi\right)$ is the coupling function of the
scalar to the brane $b_i$. 

For later use let us also generalize the ansatz (\ref{RS1ansatz}) to
\begin{equation}
ds^2 = e^{2A\left( y\right)}\bar{g}_{\mu\nu}dx^\mu dx^\nu + dy^2 ,
\label{Karchansatz}
\end{equation}
where $\bar{g}_{\mu\nu}$ denotes the metric of a four dimensional
maximally symmetric space, i.e.\
\begin{equation}
\bar{g}_{\mu\nu} = \left\{ \begin{array}{l l l}
diag\left( -1,1,1,1\right) & \mbox{for} &  M_4 \\
diag\left( -1, e^{2\sqrt{\bar{\Lambda}}t} , e^{2\sqrt{\bar{\Lambda}}t},
e^{2\sqrt{\bar{\Lambda}}t}\right) & \mbox{for} & dS_4 \\
diag\left(-e^{2\sqrt{-\bar{\Lambda}}x^3} ,
  e^{2\sqrt{-\bar{\Lambda}}x^3} , 
e^{2\sqrt{-\bar{\Lambda}}x^3} ,1\right) & \mbox{for} & AdS_4
\end{array} \right. .
\label{Cosmoansatz}
\end{equation}
The constant $\bar{\Lambda}$ is related to the constant curvature of
the de Sitter ($dS_4$) and the anti de Sitter ($AdS_4$) slices.

Let us first discuss the simplest case with $\bar{\Lambda} =0$. As
usual we consider fields which depend only on $y$ and denote a
derivative with respect to $y$ by a prime. The equations of motion for
$\bar{\Lambda} =0$ are
\begin{eqnarray}
\frac{8}{3} \Phi^{\prime\prime} +\frac{32}{3} A^\prime \Phi^\prime
-\frac{\partial V}{\partial \Phi}-\sum_i \frac{\partial f_i}{\partial
  \Phi}\, \delta\!\left( y-y_i\right) & 
= & 0 ,\label{Emunu} \\
6\left( A^\prime\right)^2 -\frac{2}{3}\left( \Phi^\prime\right)^2 +
\frac{V}{2} & = & 0 , \label{E55}\\
3A^{\prime\prime} +\frac{4}{3}\left( \Phi^\prime\right)^2
+\frac{1}{2}\sum_i f_i\, \delta\!\left( y-y_i\right) & = & 0
\label{PHI}. 
\end{eqnarray}
First, we analyze this system of equations in absence of the
branes. We start with the ansatz
\begin{equation}
A^\prime =  W\!\left( \Phi\right) .
\label{AWRELA}
\end{equation}
Equation (\ref{PHI}) fixes then
\begin{equation}
\Phi^\prime = -\frac{9}{4} \frac{\partial W}{\partial
  \Phi} .
\label{PHIW}
\end{equation}
the second equation (\ref{E55}) yields
\begin{equation}
V = \frac{27}{4} \left( \frac{\partial W}{\partial
    \Phi}\right)^2 - 12  W^2 .
\label{Ksuper}
\end{equation}
Finally, the first equation (\ref{Emunu}) is satisfied automatically.

With view on (\ref{Ksuper}), we could formally call $W$ a 
superpotential because such a relation is known from five dimensional
gauged supergravity\cite{Freedman:1999gp}. A solution in the absence
of branes can now be constructed as follows. Equation (\ref{Ksuper})
determines $W$ up to an integration constant. With a given $W$, one
can solve (\ref{PHIW}) for $\Phi$ up to another integration constant.   
Equation (\ref{AWRELA}) fixes $A$ up to an integration constant. So
altogether, there are three integration constants in the general
solution. 

Now, we take into account the source terms caused by the presence of
the branes. We are looking for solutions in which the fields are
continuous. Therefore, the first derivatives of the fields $A$ and
$\Phi$ are finite arbitrarily close to the position of the
branes. However, the first derivatives must jump when $y$ passes a
$y_i$. Integrating (\ref{PHI}) and (\ref{Emunu}) over $y = y_i -
\epsilon \ldots y_i + 
\epsilon$ and taking the limit $\epsilon \to 0$, one finds the jump 
conditions 
\begin{eqnarray}
3\left( A^\prime \left( y_i +0\right) - A^\prime\left( y_i -0\right)
\right) & = & -\frac{1}{2} f_i ,\label{DeJumpA}\\
\frac{8}{3}\left( \Phi^\prime\left( y_i +0\right) - \Phi^\prime\left(
    y_i - 0\right)\right) & =&  \frac{\partial f_i}{\partial \Phi} .
\label{DeJumpP}
\end{eqnarray}
For the ``superpotential'' $W$, this implies
\begin{eqnarray}
3\left( W_{\left| y= y_i +0\right.} - W_{\left| y = y_i
  -0\right.}\right) & = & -\frac{1}{2} f_i ,\label{Superjump} \\
\frac{3}{2}\left( {\frac{\partial W}{\partial \Phi}}_{\left| y = y_i
      +0\right. }
  -{\frac{\partial W}{\partial \Phi}}_{\left| y = y_i -0
    \right. }\right) & = & 
  -\frac{\partial f_i}{\partial \Phi} . \label{Superderjump}
\end{eqnarray}
This means that there are two additional conditions per brane. If we
safely want to obtain four dimensional gravity in the effective
theory, we should compactify the extra dimension. For an interval 
compactification we need at least two branes. The length of the
interval (the inter brane distance) enters the ansatz as a further
integration constant (e.g.\ $r_c$ in  (\ref{RS1ansatz}) now appears in
(\ref{RSperiod})). 
Therefore, four
integration constants are to be fixed by four conditions. However, we
should take into account that one of the integration constants
corresponds to constant shifts in $A$ which can be absorbed into a
rescaling of $x$. $A$ enters the equation of motions and the jump
conditions only with its derivatives. Therefore, one of the
integration constants is not fixed by the jump conditions. This means
that in a two (or more) brane setup at least one fine tuning of the
model parameters (appearing in $V\left( \Phi\right)$ and $f_i\left(
  \Phi\right)$) is necessary for the existence of a solution with
$\bar{\Lambda} =0$. 

For example in the RS1 model, we obtained two fine tuning conditions
(\ref{RSfine}). The fact that there is one more fine tuning condition
than expected by naive counting is related to the fact that the inter
brane distance $r_c$ is a modulus of the solution. This feature is
closely connected with the observation that we can remove the second
brane and still obtain four dimensional effective gravity. Even after
removing one brane the Randall Sundrum model requires one fine
tuning. We will come back to this point in section \ref{sec:cosmo}.

The fact that our solution requires fine tuning of parameters has its
origin in the $\bar{\Lambda}=0$ condition of the ansatz we have
considered so far. We can view $\bar{\Lambda}$ as an additional
integration constant in the ansatz (\ref{Karchansatz}). In general, 
constant shifts in $A$ can be absorbed in a rescaling of $x^\mu$ in
combination with a redefinition of $\bar{\Lambda}$. This suggests that
a mismatch in the fine tuning conditions results in a nonzero
$\bar{\Lambda}$. In order to see this more explicitly we write down
the equations of motion for $\bar{\Lambda}\not= 0$,
\begin{eqnarray}
\frac{8}{3}\Phi^{\prime\prime} +\frac{32}{3}A^\prime \Phi^\prime
-\frac{\partial V}{\partial \Phi}
-\sum_i \frac{\partial f_i}{\partial \Phi}\,\delta\!\left( y
  -y_i\right) 
& = & 0 ,\\
6\left( A^\prime\right)^2 -\frac{2}{3}\left( \Phi^\prime\right)^2
+\frac{V}{2} - 6\bar{\Lambda} e^{-2A} & = & 0 ,\label{Lambdanotzerob}\\
3A^{\prime\prime} + \frac{4}{3} \left( \Phi^\prime\right)^2 +
3\bar{\Lambda} e^{-2A} + \frac{1}{2}\sum_i f_i\, \delta\!\left( y 
  -y_i\right) & = & 0.\label{Lambdanotzeroc}
\end{eqnarray}
The jump conditions (\ref{DeJumpA}) and (\ref{DeJumpP}) are still of
the same form. We observe that a constant shift in $A$ enters the
equations of motion. Hence, there is no fine tuning to be expected if
we do not fix the value of $\bar{\Lambda}$ in the ansatz. 
For completeness, we note that the equations of motion can be reduced
to a set of first order equations like in the $\bar{\Lambda}=0$
case. The corresponding first order equations are
\begin{eqnarray}
V & = & \frac{27}{4} \frac{1}{\gamma\!\left( r\right)^2} \left(
  \frac{\partial W\!\left(\Phi\right)}{\partial \Phi}\right)^2 -12
  W\!\left( \Phi\right)^2 ,\label{Kcouple} \\
A^\prime & = & \gamma\!\left( r\right) W\!\left( \Phi\right) , \\
\Phi^\prime & = & -\frac{9}{4}\frac{1}{\gamma\!\left(
  r\right)}\frac{\partial W\!\left(\Phi\right)}{\partial\Phi} , \\
\gamma\!\left( r\right) & = & \sqrt{ 1 +\frac{\bar{\Lambda}}{W\!\left(
  \Phi\right)^2} e^{-2A}} .
\end{eqnarray}
To find a solution to this system of first order equations looks more
complicated than in the $\bar{\Lambda} = 0$ case. The equation
(\ref{Kcouple}) now couples to the rest of the equations due to the
 $\gamma$ dependent factor. 
 
\subsection{Consistency conditions}\label{Sec:Cons}
\setcounter{equation}{0}

In this subsection we are going to discuss consistency conditions
which any solution to the setup of the previous subsection has to 
satisfy. In 
principle, these consistency conditions constitute nothing but a check
whether there has been a computational error. They are, however,
useful in cases where the envisaged solution possesses singularities.  
Further, consistency conditions give sometimes informations about the
system without the need of constructing an explicit solution. 
The condition we are going do derive next is most simply expressed in
words. It states that the four dimensional effective cosmological
constant is compatible with the constant curvature of the four
dimensional slices. (This curvature is fixed by $\bar{\Lambda}$ in
(\ref{Cosmoansatz}).) Now let us translate this verbal statement into
fromul\ae .

In order to obtain the four dimensional effective cosmological
constant, we need to construct an effective action for four
dimensional gravity. We start with the five dimensional metric 
\begin{equation}
ds^2 = e^{2A\left( y\right)} \tilde{g}_{\mu\nu} dx^\mu dx^\nu +dy^2 , 
\end{equation}
where 
\begin{equation}
\tilde{g}_{\mu\nu} = \bar{g}_{\mu\nu } + h_{\mu\nu}
\end{equation}
is the metric on the four dimensional slices. It is taken to be
independent of $y$, and the background metric $\bar{g}_{\mu\nu}$ is
defined in (\ref{Cosmoansatz}). If we do not consider other
fluctuations than $h_{\mu\nu}$, the action for four
dimensional gravity will be of the general form
\begin{equation}
S_4 = M_p ^2\int d^4 x\sqrt{-\tilde{g}}\left( \tilde{R}^{(4)}
  -\lambda\right) ,
\label{genefour}
\end{equation}
where $\tilde{R}^{(4)}$ is the four dimensional scalar curvature computed
from $\tilde{g}_{\mu\nu}$. The cosmological constant $\lambda$ is
fixed by the condition that $\tilde{g}_{\mu\nu}=\bar{g}_{\mu\nu}$
should be a stationary point of (\ref{genefour}). This yields
\begin{equation}
\lambda = 6\bar{\Lambda} .
\label{barlambda-lambdak}
\end{equation}
We should also recall that the effective four dimensional Planck mass
is given by
\begin{equation}
M_p ^2 = \int dy\, e^{2A\left( y\right)} ,
\label{scalarinplanck}
\end{equation}
where $A$ takes its classical value. The vacuum value of the Lagrange
density in (\ref{genefour}) can be easily computed to be
\begin{equation}
\bar{\cal L}_4 = M_p ^2\left( \bar{R}^{(4)}-\lambda\right) =
6\bar{\Lambda}M_p ^2 ,
\label{4dlabackgr}
\end{equation}
where $\bar{R}^{(4)}$ is the scalar curvature computed from
$\bar{g}_{\mu\nu}$.

For consistency, $\bar{\cal L}_4$ should coincide with a result
obtained in the 
following way. We plug the solution of the equations of motion into
the five dimensional action and integrate over $y$. (This is exactly
the prescription of obtaining the classical value of the four
dimensional Lagrangian.) In order to do so, it is useful to write down
part of the equations of motion in a less explicit form than
before. The equations obtained from five dimensional metric variations
are the five dimensional Einstein equations
\begin{equation}
R_{MN} -\frac{1}{2} G_{MN} R = \frac{1}{2}T_{MN} .
\label{Einsteinwsca}
\end{equation}
For the model defined in (\ref{BULKSCALARIN}), the energy momentum
tensor $T_{MN}$ is
\begin{equation}
T_{MN} = \frac{8}{3} \partial_M\Phi\partial_N\Phi
-\frac{4}{3}\left(\partial \Phi\right)^2 G_{MN} -V\!\left(\Phi\right)
G_{MN} -\sum_i f_i\,\delta\!\left( y -y_i\right) g_{\mu\nu} \delta_M
^\mu \delta_N ^\nu ,
\label{EPtensorwbulksca}
\end{equation}
where $g_{\mu\nu}$ is the metric induced on the brane (see
(\ref{RSinduced})). The classical value of $R$ can be easily computed
by taking the trace of (\ref{Einsteinwsca}) with the result
\begin{equation}
 R  = \frac{4}{3}\left( \partial \Phi\right)^2 +\frac{5}{3}
 V\!\left(\Phi\right) +\frac{4}{3}\sum_i f_i\, \delta\!\left( y
 -y_i\right) .
\end{equation}
Plugging this into (\ref{BULKSCALARIN}), we obtain the classical value 
for the four dimensional Lagrangian
\begin{equation}
\bar{\cal L}_4 =\int dy\, e^{4A}\left( \frac{2}{3} V\!\left( \Phi\right)
  +\frac{1}{3}\sum_i f_i\, \delta\!\left( y -y_i\right) \right) ,
\label{vacuenedens}
\end{equation}
where it is understood that $A$ and $\Phi$ satisfy the equations of
motion. Comparing with (\ref{4dlabackgr}) and using
(\ref{EPtensorwbulksca}), we obtain finally the consistency condition 
\begin{equation}
-\frac{1}{3} \int dy\, e^{4A}\left( T_0 ^0 + T_5 ^5\right) =
 6\bar{\Lambda} M_p ^2 .
\label{Lavicons}
\end{equation}
We should emphasize again that (\ref{Lavicons}) is just a consequence
of the equations of motion. For $\bar{\Lambda} =0$, (\ref{Lavicons})
implies that the vacuum energy density of the solution has to vanish. 

Before closing this subsection we want to describe an alternative way
to obtain the same (or equivalent) consistency conditions.
First, we note that
\begin{equation}
\left( A^\prime e^{nA}\right)^\prime = e^{nA} \left(
  \frac{n-4}{9} 
  \left( \Phi^\prime \right)^2\hspace{-0.1cm} -\frac{nV}{12} +\left(
  n-1\right) 
  \bar{\Lambda} e^{-2A}\hspace{-0.1cm} -\frac{1}{6}\sum_i f_i\,
  \delta\!\left( y 
  -y_i\right) \hspace{-0.1cm}\right) .
\label{totalderA}
\end{equation}
This can be easily checked with the equations (\ref{Lambdanotzerob}) and
(\ref{Lambdanotzeroc}). With the expression (\ref{EPtensorwbulksca})
we rewrite (\ref{totalderA}) in the following way
\begin{equation}
\left( A^\prime e^{nA}\right)^\prime = e^{nA}\left( \frac{1}{6} T_0 ^0
  +\left( \frac{n}{12} -\frac{1}{6}\right) T_5 ^5 +\left( n-1\right)
  \bar{\Lambda} e^{-2A}\right) . 
\label{totalderEP}
\end{equation}
Assuming that for a consistent solution the integral over the total
derivative on the lhs of (\ref{totalderEP}) vanishes we find
\begin{equation}
-\frac{1}{3} \int dy\,  e^{nA}\left( T_0 ^0 +\left( \frac{n}{2} -1\right)
 T_5 ^5\right) = 2\left( n-1\right) \bar{\Lambda} \int dy\, e^{\left(
 n-2\right) A} .
\label{sumrules}
\end{equation} 
We observe that for $n=4$ this condition is identical to the
previously derived condition (\ref{Lavicons}). 
In the next subsection we will discuss solutions with
singularities. For those solutions, one could argue that the
preposition of condition (\ref{sumrules}) is not necessarily
satisfied. If there are singularities, an integral over a total
derivative may differ from zero, and one may not worry about
(\ref{sumrules}) in such a case. For $n=4$, we have shown that
(\ref{sumrules}) encodes the statement that the effective four
dimensional cosmological constant is compatible with the curvature of
the four 
dimensional slices. This should be the case also in the presence of
singularities. We leave it as an exercise to verify that the Randall
Sundrum models satisfy all the consistency conditions.  

\subsection{The cosmological constant problem}\label{sec:cosmo}
\setcounter{equation}{0}

In this section, we are going to discuss whether it is possible to
solve the cosmological constant problem within a brane world scenario 
containing a bulk scalar. Let us first state the problem as it arises
in conventional quantum field theory. The observational bound on the
value of the cosmological constant (as measured from the curvature of
the universe) is
\begin{equation}
\lambda M_p^2 \leq 10^{-120} \left( M_p\right)^4 .
\label{observation}
\end{equation}
Taking into account the leading order contribution of quantum field
theory, one obtains 
\begin{equation}
\lambda M_p ^2 = \lambda_0 M_p ^2 +\left( \mbox{UV-cutoff}\right)^4
Str\left( \mbox{\bf 1}\right) ,
\end{equation}
where $\lambda_0$ corresponds to a tree level contribution which can
be viewed as an input parameter of the model. The size of the
UV-cutoff is set by the scale up to which the effective field theory
at hand is valid. The supertrace is taken over degrees of freedom
which are light compared to the UV-cutoff. If for example we assume
that the standard model of particle physics is a valid effective
description of physics up to the Planck scale, we need to fine tune
120 digits of
the input parameter $\lambda_0 M_p ^2$ in order to obtain agreement
with (\ref{observation}). The situation slightly improves if we assume
that the standard model is a good effective description only up to a
supersymmetry breaking scale at (at least) about a TeV. In this case
we should take the UV-cutoff to be roughly a TeV. We still have to
fine tune 60 digits in $\lambda_0M_p ^2$ in order to match the
observation (\ref{observation}). 
To summarize, the cosmological constant problem is that a huge amount
of fine tuning of input parameters is implied by the observational
bound on the cosmological constant.

How could the situation improve in a brane world setup? Here, it may 
happen that the field theory produces a huge amount of vacuum energy
which however results only in a curvature along the invisible extra
dimension. In section \ref{solugener} we have seen that in a two (or
more) brane setup we need to fine tune input parameters such that
$\bar{\Lambda} =0$\footnote{Recent observations seem to hint at a
  small but non zero constant. For the discussion of the fine tuning
  problem this value is too small to be relevant.} is a solution of
the model. (See equation 
(\ref{barlambda-lambdak}) for the relation between $\bar{\Lambda}$ and
$\lambda$.) Actually, the amount of fine tuning needed in a two brane
setup is of the order of magnitude by which the vacuum energy on a
brane deviates from the observed value (\ref{observation}) because this
quantity enters the jump conditions.
One may hope to find a single brane model for which a solution without
fine tuning exists. This possibility is not excluded by our
investigations in section \ref{solugener}. However, we will prove
later that a single brane model with effectively four dimensional
gravity requires a fine tuning (as the RS2 model of section
\ref{sec:RS2} does). Before presenting the general (negative) result,
we would like to demonstrate the problems at an illustrative example.  

\subsubsection{An example}\label{KSSexample}

The model we are going to discuss is a special case of
(\ref{BULKSCALARIN}) with a single brane at $y=0$ as well as
$V\!\left( \Phi\right) \equiv 0$ and $f_0\!\left( \Phi\right) = T 
e^{b\Phi}$. Hence, the action reads
\begin{equation}
S =\int d^5 x \sqrt{-G}\left( R -\frac{4}{3}\left(\partial
    \Phi\right)^2\right) -\int d^4 x\sqrt{-g} T{e^{b\Phi}}_{\left|
    y=0\right.} ,
\label{KSSaction}
\end{equation}
where $b$ and $T$ are constants.
In what follows we will focus on the case $b\not= \pm
\frac{4}{3}$. The case $b =\pm \frac{4}{3}$ is similar and discussed
in\cite{Kachru:2000hf,Kachru:2000xs}, \cite{Arkani-Hamed:2000eg},
\cite{Forste:2000ps,Forste:2000ft}. We take the ansatz
(\ref{Karchansatz}) with $\bar{\Lambda} =0$.
From equation (\ref{E55}) one
finds that 
\begin{equation}
A^\prime = \pm \frac{1}{3} \Phi^\prime .
\end{equation}
We choose
\begin{equation}
A^\prime = \left\{ \begin{array}{ll}
\frac{1}{3}\Phi^\prime , & y < 0 \\
-\frac{1}{3}\Phi^\prime , & y >0 
\end{array} \right. .
\label{Aprimephiprime}
\end{equation}
The reader may verify that taking the same sign on both sides of the
brane does not lead to a consistent solution. The only other choice is
to interchange the signs in (\ref{Aprimephiprime}). This can be undone
by redefining $y \to -y$ and hence the ansatz (\ref{Aprimephiprime})
is general (for $b\not= \pm \frac{4}{3}$).
The rest of the equations of motion is easily solved with the result
\begin{equation}
\Phi\!\left( y\right) = \left\{ \begin{array}{l l}
\frac{3}{4}\log \left| \frac{4}{3} y + c_1\right| +d_1, & y <0 \\
-\frac{3}{4}\log \left| \frac{4}{3} y + c_2\right| + d_2, &  y>0 
\end{array}\right. ,
\end{equation}
where $c_i$ and $d_i$ are integration constants. The condition
that $\Phi$ should be continuous at $y=0$ fixes $d_2$ in terms of the
other integration constants. The jump conditions (\ref{DeJumpA}) and
(\ref{DeJumpP}) determine $c_1$ and $c_2$ in terms of  $d_1$
according to
\begin{eqnarray}
\frac{2}{c_2} &=&\left( -\frac{3b}{8} -\frac{1}{2}\right) T e^{bd_1}
\left| c_1\right|^{\frac{3b}{4}}, \\
\frac{2}{c_1} &=& \left( -\frac{3b}{8} +\frac{1}{2}\right) T
e^{bd_1}\left| c_1\right|^{\frac{3b}{4}} .
\end{eqnarray}
Together with possible constant shifts in $A$, two integration 
constants are not fixed by the equations of motion.

The next step is to ensure that an observer will experience four
dimensional gravitational interactions (plus possible small
corrections). This is the case only if the four dimensional Planck
mass is finite. 
The expression for the four dimensional Planck mass is given in
(\ref{scalarinplanck}). If the parameters ($T$ and $b$) of the model
are such that there is no singularity at some $y>0$ ($y <0$) the
integration region in (\ref{scalarinplanck}) extends to (minus)
infinity. In one or both of these cases the four dimensional Planck
mass diverges, and an effective four dimensional theory decouples from
gravity. This is not what we are interested in since with decoupled
gravity the problem of the cosmological constant does not
occur. Therefore, we have to choose our parameters such that there are
singularities at which we can cut off the integration over
$y$. Explicitly this imposes the conditions
\begin{eqnarray}
T\left( \frac{1}{2} -\frac{3b}{8}\right) & > & 0,\nonumber \\
T\left( -\frac{1}{2} -\frac{3b}{8}\right) & < & 0.\label{KSSparam}
\end{eqnarray}
These conditions are easy to satisfy without fine tuning of the
parameters. So far, it looks as if we have achieved to find a solution
with vanishing four dimensional curvature without the necessity of a
severe fine tuning of input parameters.

It remains to check whether the consistency condition (\ref{Lavicons})
is satisfied. Since we have taken the ansatz with $\bar{\Lambda}=0$, 
the condition states that the vacuum energy density of our solution
should vanish. The vacuum energy density is most easily computed from 
(\ref{vacuenedens}). To be specific, we fix the integration constant
in $A$ via $A =\frac{1}{3}\Phi$ for $y<0$. Taking further into account
that our background is static, we find for the vacuum energy density
\begin{equation}
{\cal E} = -\frac{1}{3}T {e^{4A + b\Phi}}_{\left|y=0\right.} =
-\frac{2}{3}\frac{8}{4-3b} e^{\frac{4}{3} d_1}\not= 0 .
\label{notconsistent}
\end{equation}
We see that the consistency condition is not satisfied. Since the
condition of vanishing vacuum energy density ${\cal E}=0$ is derived
from the equation of motion, (\ref{notconsistent}) implies that the
  equations of motion are not solved. Indeed, with the parameter
  choice (\ref{KSSparam}), the second derivatives of $\Phi$ and 
  $A$ contain delta functions which are not canceled by source terms 
  in the equations of motion. We have to cure this inconsistency by
  adding additional source terms to the setup, i.e.\ to extend the
  single brane scenario to a three brane scenario. From our
  considerations in section \ref{solugener}, we know already that this
  will lead to fine tuning conditions on the input parameters. For
  illustrative purposes, let us demonstrate the appearance of the fine
  tuning explicitly. We modify our action (\ref{KSSaction}) by two
  additional source terms, i.e.\ 
\begin{equation}
S \to S + S_+ + S_- ,
\end{equation}
with
\begin{equation}
S_\pm = -\int d^4 x T_\pm {e^{b_\pm \Phi}}_{\left| y = y_\pm\right. } .
\end{equation}
The quantities $b_\pm$ and $T_\pm$ are now input parameters of the
model. The value of $y_\pm$ gives the locations of the singularities,
\begin{equation}
y_- = -\frac{3}{4} c_1 \;\;\; ,\;\;\; y_+ = -\frac{3}{4} c_2 .
\end{equation}
The additional source terms give rise to four more jump conditions to
be satisfied by the solution. These jump conditions are
\begin{eqnarray}
\frac{8}{3}\left( \Phi^\prime\!\left( y_\pm + 0\right) -\Phi^\prime\!
  \left( y_\pm -0\right)\right) & = & b_\pm T_\pm e^{b_\pm \Phi\!\left(
  y_\pm\right)}, \label{extrajumpa}\\
\mp \left( \Phi^\prime\!\left( y_\pm +0\right) -\Phi^\prime\!\left(
  y_\pm -0\right)\right) & =& -\frac{1}{2}T_\pm e^{b_\pm\Phi\!\left(
  y\right)} .\label{extrajumpb}
\end{eqnarray}
Before solving these additional jump conditions we need to give a
prescription how to continue our solution beyond the
singularities. There are several possible descriptions. For example,
one may continue in such a way that the setup becomes periodic in
$y$. The simplest choice is to effectively cut off the space at the
singularities (at $y=y_\pm$) by freezing the fields to the singularity
values for $y\not\in\left[ y_- ,y_+\right]$ such that the first
derivatives vanish beyond the singularities. (The final conclusion is
not affected by the particular way of continuing the solution beyond
the singularities.)  With our prescription the conditions
(\ref{extrajumpa}) and (\ref{extrajumpb}) are solved by
\begin{equation}
b_\pm = \pm \frac{4}{3}
\end{equation}
and
\begin{equation}
T_- e^{-\frac{4}{3} d_1} = T_+ e^{\frac{4}{3}d_2} = -2 .
\end{equation}
One should recall that $d_2$ is already fixed by the jump conditions
at $y =0$. We observe that the input parameters need to be fine
tuned. 

The contribution of the branes at $y= y_\pm$ to the vacuum energy
density is
\begin{eqnarray}
{\cal E}_+ + {\cal E}_-  &=& -\frac{1}{3}\left( T_+ {e^{4A + b_+
      \Phi}}_{\left|y = y_+\right. } + T_- {e^{4A +
      b_-\Phi}}_{\left|y=y_-\right.}\right) 
      \nonumber \\
& = & \frac{2}{3} e^{\frac{4}{3}d_1}\frac{8}{4 -3 b} ,
\end{eqnarray}
where we have employed the jump conditions and fixed an integration
constant in $A$ by the choice $A = \frac{1}{3}\Phi$ for $y <0$. 
Hence, the contribution (\ref{notconsistent}) is exactly canceled by
the additional branes and the model is consistent now. However, we
failed to construct a brane setup yielding a vanishing effective four
dimensional cosmological constant without fine tuning of the
parameters. If the fine tuning is not satisfied, there exist
$\bar{\Lambda} \not= 0$ solutions\cite{Kachru:2000xs}. (The situation
is slightly different in the $b=\pm\frac{4}{3}$ case where the
possible value of $\bar{\Lambda}$ is fixed by the bulk potential $V$,
which needs to be fine tuned to zero for $\bar{\Lambda}=0$ to be a
solution\cite{Forste:2000ft}. In addition there is a fine tuning due
  to the necessity of additional branes for $b=\pm\frac{4}{3}$, too.) 

In the next subsection we will show that our failure to find a
$\bar{\Lambda}=0$ solution without fine tuning is not caused by an 
unfortunate choice of the model we started with but rather a generic
feature of brane models with a bulk scalar.

\subsubsection{A no go theorem}\label{sec:nogo}

The prepositions for the no go theorem for a ``brany'' solution to the
cosmological constant problem are:
\begin{itemize}
\item The model contains a single brane and $\bar{\Lambda} =0$. 
\item The four dimensional Planck mass is finite.
\item The model does not contain singularities apart from the one
  corresponding to the single brane source.
\item The bulk potential $V$ can be expressed in terms of the
  ``superpotential'' $W$ according to (\ref{Ksuper}).
\end{itemize}

In a first step, we are going to show that these prepositions imply
that the five dimensional space must be asymptotically (for large
$\left| y\right|$) an AdS space.
Suppose the warp factor asymptotically shows a
power like behavior,
\begin{equation}
e^A \sim \left| y\right| ^{-\alpha} .
\end{equation}
The four dimensional Planck mass is computed in
(\ref{scalarinplanck}). With a single brane and no further
singularities the integration is taken over $y \in \left( -\infty
  ,\infty\right)$. A necessary condition for
\begin{equation}
\int_{-\infty} ^{\infty} dy \, e^{2A} < \infty
\end{equation}
is
\begin{equation}
\alpha > \frac{1}{2} .
\end{equation}
On the other hand, equation (\ref{PHI}) tells us that in the bulk (in
particular asymptotically)
\begin{equation}
A^{\prime\prime} < 0 \Longrightarrow \alpha < 0.
\end{equation}
We conclude that $e^A$ cannot fall off with a power of $\left| y\right|$
as $\left| y\right| \to \infty$. 

Therefore, we assume an exponential fall off, i.e.\ for large $\left|
  y\right|$ 
\begin{equation}
A = -k\left| y\right| ,
\label{grojasy}
\end{equation}
with $k$ being a positive constant. 
In the following we will show that in this case there is a fine tuning
similar to the fine tuning of the RS2 model. Before going into the
details, let us sketch the outline of the proof. The asymptotic
behavior (\ref{grojasy}) suffices to reproduce the superpotential for
all $y$. Plugging this into the matching conditions (\ref{Superjump})
and (\ref{Superderjump}) will show that the input parameters of the
model need to be fine tuned. Let us now present the details of the
slightly tedious construction of $W$ from its asymptotics.

From equation (\ref{PHI})
we learn 
that $\Phi$ must be asymptotically constant. We denote the asymptotic
values of $\Phi$ by $\Phi_c ^\pm$ corresponding to the limits $y\to
\pm\infty$. Equation (\ref{PHIW}) implies that
\begin{equation}
\frac{\partial W}{\partial \Phi}_{\left|\Phi = \Phi_c ^\pm\right. } =
0. 
\label{firstsuperder}
\end{equation}
Plugging (\ref{grojasy}) into (\ref{AWRELA}) yields
\begin{equation}
W\!\left( \Phi_c ^+\right) < 0 \;\;\; \mbox{and}\;\;\; W\!\left(
  \Phi_c ^-\right) >0.
\label{asyWsign}
\end{equation}
Let us look again at equation (\ref{PHIW})
\begin{equation}
\Phi^\prime = -\frac{9}{4}\frac{\partial W}{\partial \Phi} ,
\label{PHIren}
\end{equation}
and view $\Phi^\prime$ as a function of $\Phi$. $\Phi$ should reach
its asymptotic values in a dynamical way which means that
$\Phi^\prime$ should be monotonically decreasing (increasing) as
$\Phi$ approaches $\Phi_c ^+$ ($\Phi_c ^-$). We obtain the conditions
\begin{equation}
\frac{\partial^2 W}{\partial \Phi ^2}_{\left|\Phi = \Phi_c ^+\right. }
> 0 \;\;\; ,\;\;\; \frac{\partial ^2 W}{\partial \Phi^2}_{\left| \Phi =
    \Phi_c ^-\right. } < 0.
\label{secondDerSign}
\end{equation}
(Equation (\ref{PHIren}) can be viewed as a renormalization group
equation, where the renormalization group scale is related to
$\Phi$. W is proportional to the
running coupling, and $\Phi^\prime$ (viewed as a function of $\Phi$)
is the beta function. The conditions (\ref{secondDerSign}) mean that
$\Phi = \Phi_c^+$ ($\Phi = \Phi_c^-$) correspond to stable UV (IR)
fixed points.)   
Equations (\ref{Ksuper}) and (\ref{firstsuperder}) fix the
asymptotic values of the superpotential according to
\begin{equation}
V\!\left( \Phi_c ^-\right) = -12 W\!\left( \Phi_c ^-\right)^2 \;\;\;
,\;\;\;
V\!\left( \Phi_c ^+\right) = -12W\!\left( \Phi_c ^+\right)^2 .
\label{WasymptValu}
\end{equation}
This implies that the asymptotic values of $V$ must be
negative. Further note that the asymptotic values of $W$ are fixed in
a unique way with the additional conditions (\ref{asyWsign}). 
So far, we know the asymptotic value of $W$ in terms of the input
parameters and the asymptotics of the first derivative of $W$
(\ref{firstsuperder}). 

In order to compute the higher derivatives of $W$, it is useful to 
express the $n$th derivative of $V$ in terms of $W$ via
(\ref{Ksuper}). The corresponding expression is
\begin{equation}
\frac{\partial^n V}{\partial \Phi^n} =\sum_{k=1}^n 2 {n-1 \choose
  k-1} \frac{\partial^k W}{\partial\Phi^k}\left(
  \frac{27}{4}\frac{\partial^{n-k+2} W}{\partial\Phi^{n-k+2}} - 12
  \frac{\partial ^{n-k}W}{\partial \Phi^{n-k}}\right) .
\label{nthderofV}
\end{equation}
This formula is most easily proven in the following way. First, apply
the Leibniz rule ($F$ and $G$ are arbitrary functions of $\Phi$)
\begin{equation}
\frac{\partial^n \left(FG\right)}{\partial \Phi^n} = \sum_{k=0}^n {n
  \choose k} \frac{\partial^k F}{\partial \Phi^k}\frac{\partial^{n-k} 
  G}{\partial \Phi^{n-k}}
\end{equation}
on $\frac{\partial W^2}{\partial \Phi^2} = 2 W\frac{\partial
  W}{\partial \Phi} $ in order to show that
\begin{equation}
\frac{\partial^n W^2}{\partial \Phi^n} = \sum_{k=1} ^n 2 {n-1 \choose
  k-1} \frac{\partial^{n-k} W}{\partial\Phi^{n-k}}\frac{\partial ^k
  W}{\partial \Phi^k} .
\label{nthderofsqua}
\end{equation} 
In a second step use (\ref{nthderofsqua}) with $W$ replaced by its
first derivative and redefine the summation index $k \to n+1 -k$. 

In the following we will employ (\ref{nthderofV}) to compute the
asymptotics of all derivatives of $W$. Since there are no
singularities between the brane and the asymptotic region, this will
enable us to expand $W$ in a Taylor series  yielding its value
arbitrarily close to the brane. 

The second derivative of $W$ needs some separate discussion. With the
result (\ref{firstsuperder}) we obtain the relation
\begin{equation}
\left[\frac{27}{2} \left(\frac{\partial^2 W}{\partial\Phi^2}\right)^2
  -24 W 
  \frac{\partial^2 W}{\partial \Phi^2} -\frac{\partial^2 V}{\partial
  \Phi^2}\right]_{\left| \Phi = \Phi_c ^\pm\right.} = 0 .
\label{EquForW2}
\end{equation}
This equation has real solutions for the asymptotics of the second
derivative of $W$ provided that
\begin{equation}
\frac{\partial^2 V}{\partial \Phi^2}_{\left| \Phi = \Phi_c
    ^\pm\right. } > \frac{8}{9} V\!\left( \Phi_c ^\pm\right) .
\end{equation}
Taking into account that the asymptotic value of $W$ is fixed uniquely
by (\ref{WasymptValu}) and (\ref{asyWsign}), and that the sign of the 
asymptotic value of the second derivative of $W$ is determined by
(\ref{secondDerSign}), one finds that (\ref{EquForW2}) can be solved
in a unique way. Note, that (\ref{asyWsign}) and (\ref{secondDerSign})
imply 
\begin{equation}
\frac{\partial^2 V}{\partial \Phi^2}_{\left|\Phi = \Phi_c ^\pm\right. } > 
0. 
\label{sign2ndDerOfV}
\end{equation}

The computation of the higher derivatives of $W$ in the large $\left|
  y\right|$ region is somewhat simpler. First, we notice that
  asymptotically on the rhs of (\ref{nthderofV})  the $n$th derivative
  of $W$ is the highest occurring derivative (see
  (\ref{firstsuperder})). Terms containing the $n$th derivative
  correspond to $k=2,n$. 
The expression (\ref{nthderofV}) evaluated
  at $\Phi_c ^\pm$ takes the form ($n>2$)
\begin{equation}
\frac{\partial^n V}{\partial \Phi^n}_{\left| \Phi = \Phi_c
    ^\pm\right. } =\left[  \frac{\partial^n W}{\partial\Phi^n}
     \left(\frac{27}{2} n \frac{\partial^2 W}{\partial
    \Phi^2} -24 W\right) + \ldots\right]_{\left| \Phi = \Phi_c
    ^\pm\right. } ,
\label{nthDerOfW}
\end{equation}
where the dots stand for terms containing lower derivatives of $W$. 
The relation (\ref{nthDerOfW}) allows to determine all derivatives of
$W$ 
provided that the coefficient at the $n$th derivative of $W$ differs
from zero. This is ensured by equations (\ref{EquForW2}) and 
(\ref{sign2ndDerOfV}). Indeed, requiring the coefficient in front of
the $n$th derivative of $W$ to vanish yields
\begin{equation}
\left[ \frac{\partial^2 V}{V\partial \Phi^2}\right]_{\left| \Phi =
    \Phi_c ^\pm\right. } = -\frac{32}{9n}\left( 1 -\frac{1}{n} 
\right) ,
\end{equation}
which is not compatible with (\ref{sign2ndDerOfV}) and
(\ref{WasymptValu}). 
We conclude that in the Taylor expansion
\begin{equation}
W\!\left(\Phi\right) = \left\{ \begin{array}{ll}\sum_{n=0}^\infty \frac{1}{n!}
  \frac{\partial^n W}{\partial \Phi^n}_{\left| \Phi = \Phi_c
  ^-\right.} \left( \Phi -\Phi_c ^-\right)^n , & y<0 \\
\sum_{n=0}^\infty \frac{1}{n!}
  \frac{\partial^n W}{\partial \Phi^n}_{\left| \Phi = \Phi_c
  ^+\right.} \left( \Phi -\Phi_c ^+\right)^n , & y>0 
\end{array}\right. 
\end{equation}
all the coefficients are fixed uniquely by the model parameters. Then
the jump condition (\ref{Superjump}) will fix the value of $\Phi$ at
$y=0$, whereas (\ref{Superderjump}) imposes generically a fine tuning
of the model parameters.

It may look somewhat disappointing to close a review with the proof of
a no go theorem. However, often no go theorems help to find a way
leading to the desired aim. This way should then start with a model
not satisfying the prepositions of the no go theorem. Indeed, there
have been proposals for not fine tuned solutions with $\bar{\Lambda}
=0$. These proposals are based on the idea of introducing more
integration constants without increasing the number of jump
conditions. We provide the corresponding references in section
\ref{furtherreading}. Here, we should remark that (so far) there is no
commonly accepted solution to the cosmological constant problem within
brane world setups. The explanation of the observed value of the
cosmological constant remains a great challenge. Whether branes will be
helpful in a solution of this problem has to be seen in the future. 
\newpage
\chapter{Bibliography and further reading}\label{furtherreading}
Throughout the text I have already given some references. However,
this I did only when I felt that a direct hint on results obtained in
the literature would be useful for the reader at that particular 
point. Of course, these notes are based on many more publications than 
already given in the text. In the present chapter, I will provide all
the references I used and give suggestions for further
reading. However, there are many more contributions to this field.  
I apologize to all those authors whose work could 
have been listed but is not.  

\section{Chapter \ref{chap:pert}}

\subsection{Books}\label{books:pert}

In
\cite{Green:1987sp,Green:1987mn}, \cite{Brink:1988nh},
\cite{Lust:1989tj}, \cite{polchinski-book}, \cite{Kaku:1988ap,
  Kaku:1999yd} 
I list the textbooks on string theory of which I am aware. In section
\ref{section1} I used mainly\cite{Green:1987sp} but
also\cite{Lust:1989tj}. For the discussion of orbifold planes in section
\ref{sec:orbifold} I borrowed some results presented
in\cite{polchinski-book}. D-branes and orientifolds are also covered
in\cite{polchinski-book}. Since string theory is a conformal field
theory the book\cite{DiFrancesco:1997nk} may be also a useful reference. 
The subject of Calabi-Yau compactifications entered the text rather as
a side remark. Apart from the discussions presented in the above
mentioned textbooks on strings, the book\cite{huebsch} is perhaps a useful
reference 
for people who are interested in Calabi-Yau spaces. 
Let me also mention three books on supersymmetry. Often the
conventions of the standard reference\cite{Bagger:1990qh} are used in
the literature. Ref.\cite{West:1990tg} contains (at least in its
second edition) a 
discussion of supersymmetry in two dimensions. Finally,
\cite{Salam:1989fm} is not really a textbook but a collection of
papers dealing with supergravity in various dimensions. For each
dimension there is a summary of the possible supermultiplets. 

\subsection{Review articles}

Four recent review articles on perturbative string theory are
\cite{Alvarez-Gaume:1992re}, 
\cite{Polchinski:1994mb}, \cite{Ooguri:1996ik} and
\cite{Kiritsis:1997hj}. In the context of perturbative string theory, 
the CFT lectures\cite{Ginsparg:1988ui} may be also useful.
The computation of beta functions in nonlinear
sigma models is reviewed in\cite{Tseytlin:1989md}. 
Various aspects of T-duality are presented
in\cite{Giveon:1994fu,Alvarez:1995dn}. 
Orbifold compactifications are covered in most of the textbooks
mentioned in \ref{books:pert}. Two review articles on orbifolds are
listed in\cite{Nilles:1987uy}, \cite{Dixon:1987bg}.
K3 and other Calabi-Yau compactifications are e.g\ discussed
in\cite{Hosono:1994av}, \cite{Greene:1996cy}, \cite{Aspinwall:1996mn},
\cite{Nahm:1999ps}. 
There are various reviews on D-branes:
\cite{Polchinski:1996fm,Polchinski:1996na},
\cite{Bachas:1996sc,Bachas:1998rg},
\cite{Taylor:1997dy},
\cite{Thorlacius:1998qa},
\cite{Johnson:1998pc,Johnson:2000ch},
\cite{DiVecchia:1999rh,DiVecchia:1999fx}. Readers who are interested
in 
D-branes on Calabi-Yau spaces should consult\cite{Douglas:2000be} (and
references therein). In\cite{Sagnotti:1997qj}, \cite{ Dabholkar:1997zd}
two lecture notes on orientifolds are listed. 
Phenomenological aspects of string theory are reviewed in
\cite{Louis:1992c,Louis:1998vu},
\cite{Quevedo:1996hx,Quevedo:1998uy}, \cite{Dienes:1997du}. There are
quite a few reviews on supersymmetry, e.g.\ \cite{Nilles:1984ge},
\cite{Sohnius:1985qm}, \cite{Lykken:1996xt}. Ref.\ \cite{deWit:1997sz}
presents 
supergravities in various dimensions.

\subsection{Research papers}

For early papers on string theory I refer to the excellent commented
bibliography given in \cite{Green:1987sp,Green:1987mn}. Although
there is still some overlap with the references
in\cite{Green:1987sp}, 
I want to start with section \ref{betafunctions}. Here, we
presented details which at some points differ from the discussion
in\cite{Green:1987sp}. A list of references about beta functions in
string sigma models (some of them about the
open string (section \ref{sec:openbeta})) is
\cite{Alvarez-Gaume:1981hn}, \cite{Braaten:1985is},
\cite{Lovelace:1984yv,Lovelace:1986kr},
\cite{Fradkin:1985qd,Fradkin:1985ys}, \cite{Callan:1985ia},
\cite{Sen:1985qt}, \cite{Candelas:1985en}, \cite{Abouelsaood:1987gd},
\cite{Banks:1987qs}, 
\cite{Tseytlin:1989iy,Tseytlin:1987ws}, \cite{Leigh:1989jq},
\cite{Curci:1987hi}, \cite{Dorn:1986jf}, \cite{Behrndt:1992dg} and
many others. The normal coordinate expansion technique in section
\ref{betafunctions} is taken from\cite{Alvarez-Gaume:1981hn}. In a
slightly different version it can be found in\cite{Braaten:1985is}. 
The Fischler Susskind mechanism is developed in \cite{Fischler:1986ci,
  Fischler:1986tb} and also discussed in e.g.\cite{Lovelace:1986kr},
\cite{Behrndt:1991tf}. Like the present text, most of the articles
do not include the discussion of non trivial backgrounds for massive
string modes. The corresponding sigma model is not
renormalizable. Some papers on beta functions for massive string modes
are \cite{Labastida:1989wi}, \cite{Brustein:1988qw},
\cite{Ellwanger:1989xf}, \cite{Lee:1990rd}, \cite{Foerste:1992uu},
\cite{ Buchbinder:1993ku}, \cite{ Buchbinder:1995rk}.

Concerning section \ref{effectiveactions} I give some references
related to the construction of the supergravity theories. The
existence of ten dimensional type II supergravities (and also 11
dimensional supergravity) was suggested in\cite{Nahm:1978tg}.  The
explicit construction has been carried out in\cite{Green:1983tk} (see
also\cite{Schwarz:1983wa,Schwarz:1983qr,Howe:1984sr}). Anomalies are
discussed in\cite{Alvarez-Gaume:1984ig}. That $N=1$ ten dimensional
supergravity coupled to $E_8\times E_8$ or $SO\left( 32\right)$ gauge
theory is anomaly free was demonstrated in\cite{Green:1984sg}.

T-duality for the circle compactified bosonic string is discussed
in\cite{Kikkawa:1984cp,Sakai:1986cs}. For compactifications on higher
dimensional tori
see\cite{Narain:1986jj,Shapere:1989zv,Giveon:1989tt}. 
The presentation in section \ref{tsigma} follows
closely\cite{Rocek:1992ps}. T-duality in non trivial backgrounds with
abelian isometries was originally studied
in\cite{Buscher:1985kb,Buscher:1987sk}. Some related papers are:
\cite{Meissner:1991ge}, 
\cite{Giveon:1992jj}, \cite{Giveon:1994ph}, \cite{Kiritsis:1994pb},
\cite{Alvarez:1994wj,Alvarez:1994qi}, 
\cite{Tseytlin:1991wr}, \cite{Giveon:1991sy}, \cite{Kiritsis:1991zt}.  
T-duality has been also discussed for backgrounds with non Abelian
isometries e.g.\ in \cite{delaOssa:1993vc},
\cite{Alvarez:1994qi,Alvarez:1994zr}, \cite{Giveon:1994ai},
\cite{Kiritsis:1994hg}, \cite{Klimcik:1995ux},
\cite{Elitzur:1995ri},
\cite{Curtright:1994be,Curtright:1995yr,Curtright:1996ig},
\cite{Alvarez:1996uc}, \cite{Tyurin:1995ey}, 
\cite{Alekseev:1996ym},
\cite{Mohammedi:1996mf},
\cite{Hewson:1997xj}. 
The T-duality relation between type IIA and type IIB strings can be
found in \cite{Dine:1989vu}, \cite{Dai:1989ua},
\cite{Bergshoeff:1995as}, \cite{Kehagias:1996ji}.  The
connection between compactified $E_8\times E_8$ and $SO\left(
  32\right)$ strings is presented in \cite{Ginsparg:1987bx}.

The techniques for orbifold compactifications of string theory have
been developed in \cite{Dixon:1985jw,Dixon:1986jc}.
More papers on orbifolds are (including explicit constructions of
phenomenological interest): \cite{Vafa:1986wx},
\cite{Hamidi:1987vh}, \cite{Bagger:1986wa}, \cite{Kawai:1987ah},
\cite{Ibanez:1987tp}, 
\cite{Ibanez:1987xa}, \cite{Ibanez:1987sn}, \cite{Font:1988mm},
    \cite{Font:1988tp}, \cite{Ibanez:1988pj}, \cite{Lerche:1989np},
    \cite{Font:1990aj}. 
T-duality for orbifold compactifications is for example discussed in
\cite{Lauer:1989ax,Lauer:1991tm}. 

The importance of D-branes was realized in\cite{Polchinski:1995mt},
where the connection to BPS solutions of supergravity was
discovered. In 
the text I have given conditions imposed by the requirement unbroken
supersymmetry on the number of ND directions. More generally, D-branes
can intersect at certain angles\cite{Berkooz:1996km}. The computation
of D-brane interactions is presented in \cite{Polchinski:1988tu},
\cite{Burgess:1987ah,Burgess:1987wt}. References concerning the beta
function approach are given together with the other references for the
beta function approach to effective field theories, above. D-brane
actions are also e.g.\ discussed
in\cite{Schmidhuber:1996fy}. 
The interchange of Dirichlet with Neumann boundary conditions via
T-duality has been pointed out in \cite{Dai:1989ua},
\cite{Horava:1989ga}, \cite{Green:1991et}.
For general backgrounds, T-duality for open strings with respect 
to abelian isometries is presented in\cite{Alvarez:1996up},
\cite{Dorn:1996an}. T-duality with respect to non-Abelian isometries
has been performed in \cite{Forste:1996hy}, \cite{Borlaf:1996na}. (The
boundary Lagrange multiplier has been introduced in
\cite{Forste:1996hy,Forste:1997ai}.) A different method of performing
T-duality transformations in general backgrounds has been proposed in
\cite{Klimcik:1996kw}. 
The Wess-Zumino term in the D-brane action has been derived (in steps)
in \cite{Douglas:1995bn}, \cite{Li:1996ed},
\cite{Bershadsky:1996qy,Bershadsky:1996sp}, \cite{Green:1997dd}. 
Our discussion of open strings and non commutative geometry follows
closely (the introductory section of) \cite{Seiberg:1999vs}. Constant
B-fields and non commutative geometry have been connected earlier in
e.g.\ \cite{Douglas:1998fm}, \cite{Connes:1998cr},
\cite{Schomerus:1999ug}. The connection between non commutativity and
the renormalization scheme is further investigated e.g.\ in 
\cite{Andreev:1999pv,Andreev:2000rm}.
A more abstract conformal field theory approach to D-branes can for
example be found in \cite{Fuchs:1998fu},
\cite{Recknagel:1998sb}. There are many more aspects of D-branes for
which I would like to ask the reader to consult one of the given
reviews and the references therein.

Orientifolds were introduced in\cite{Sagnotti:1987tw}. For early
papers on orientifold constructions see also \cite{Pradisi:1989xd},
\cite{Govaerts:1989md}, \cite{Horava:1989vt},
\cite{Bianchi:1990yu,Bianchi:1991tb}. 
The cancellation of divergences in string diagrams of type I
$SO\left( 32\right)$ strings is observed in\cite{Green:1985ed}. 
The model of section
\ref{sec:gpmodel} has been first constructed in
\cite{Bianchi:1990yu,Bianchi:1991tb}. The presentation in the text
follows \cite{Gimon:1996rq}. Indeed, it has been the paper
\cite{Gimon:1996rq} which triggered an enormous amount of research
devoted to orientifolds. This research resulted in a lot of papers out
if which I list only ``a few'':
\cite{Dabholkar:1996zi,Dabholkar:1996pc}, \cite{Gimon:1996ay},
\cite{Berkooz:1997dw}, \cite{Blum:1996hs}, \cite{Kakushadze:1997ku},
  \cite{Zwart:1998aj}, \cite{Forste:1997ur}, \cite{Forste:1998bd},
    \cite{O'Driscoll:1998mk}, \cite{Aldazabal:1998mr},
    \cite{Blumenhagen:1999md,Blumenhagen:1999ev},
    \cite{Cvetic:1999hb}, \cite{Pradisi:1999ii}, \cite{Forste:2000hx},
    \cite{Ibanez:2001nd}, \cite{Antoniadis:1998ki},
    \cite{Angelantonj:1999jh}, \cite{Aldazabal:1999jr},
    \cite{Angelantonj:1999ms}, \cite{Angelantonj:2000xf},
    \cite{Angelantonj:2000hi}, \cite{Rabadan:2000ma},
    \cite{Klein:2000hf}, 
    \cite{Klein:2000qw}, \cite{Klein:2000tf}, 
    \cite{Blumenhagen:2000vk}, \cite{Blumenhagen:2000ea},
    \cite{Forste:2001gb}, 
\cite{Rabadan:2001mt},
\cite{Blumenhagen:2001te},
    \cite{Aldazabal:2000cn,Aldazabal:2000dg},
    \cite{Cvetic:2001tj,Cvetic:2001nr}. 

\section{Chapter \ref{chap:nonpert}}

As far as I know there are no books devoted to solutions of ten
dimensional supergravity. 

\subsection{Review articles}\label{solitrew}

There are quite a few review articles to be mentioned in the context
of brane solutions to supergravity. In the text I used mainly results
presented in\cite{Duff:1995an}. BPS solutions to ten dimensional
supergravity are also derived in \cite{Callan:1991dj},
\cite{Stelle:1996tz}. The theories on the worldvolumes of the branes
are discussed e.g.\ in \cite{Townsend:1999hi}. Intersecting brane 
solutions are e.g.\ reviewed in\cite{Gauntlett:1997cv}.
In the text I did not discuss the relevance of the brane solutions to
black hole physics. A nice introductory review to black holes
is\cite{Townsend:1997ku}. Branes in the context of black hole physics
are reviewed e.g.\ in\cite{Maldacena:1996ky,Maldacena:1998tm},
\cite{Behrndt:1997jn}, 
\cite{Youm:1997hw}, \cite{Peet:1998es}, \cite{Skenderis:1999bs},
\cite{Duff:1999rk}, 
\cite{Mohaupt:2000gc,Mohaupt:2000mj}, \cite{Myers:2001ut}. 

\subsection{Research Papers}

The elementary string solution was found
in\cite{Dabholkar:1990yf}. The five brane solution has been considered e.g.
in \cite{Strominger:1990et}, \cite{Duff:1991wv},
\cite{Callan:1991dj}. The general p-brane solutions are presented in
\cite{Horowitz:1991cd}. For more references on the topic of brane
solutions to supergravity I would like to ask the reader to consult
the review articles mentioned in section \ref{solitrew}.

\section{Chapter \ref{chap:appl}}

The presented applications of branes are not a subject of a book. A
discussion of string dualities can be found in\cite{polchinski-book}. 

\subsection{Review articles}

There are many reviews devoted to the subject of string dualities:
\cite{Townsend:1996xj,Townsend:1997wg}, \cite{Aspinwall:1996mn},
\cite{Schwarz:1997bh,Schwarz:1997yk}, \cite{Vafa:1997pm},
\cite{Forste:1998yd}, \cite{deWit:1997sz}, \cite{Julia:1997cy},
\cite{Sen:1998kr}, \cite{Obers:1998fb}, \cite{Haack:1998gc}. 

A comprehensive review on the
relation between brane setups and field theory dualities is listed
in\cite{Giveon:1998sr}. (Another (shorter) review is
\cite{Karch:1998uy}.) 
In the text I mentioned only duality relations in $N=1$ supersymmetric
field theories. Such dualities are summarized
in\cite{Seiberg:1995ac,Intriligator:1996au},
\cite{Giveon:1996qs}, \cite{Peskin:1997qi},
\cite{Shifman:1997ua}. $N=2$ supersymmetric field theories are 
considered in\cite{Bilal:1995hc}, \cite{DiVecchia:1996sc},
\cite{Lerche:1997xu}, \cite{Alvarez-Gaume:1997mv},
\cite{Argyres:1998tq}. The duality of $N=4$ super Yang-Mills theory is
presented in \cite{Olive:1996sw}, \cite{Harvey:1996ur}.

The standard review article on the AdS/CFT
correspondence is\cite{Aharony:1999ti}. Two more introductory notes
are listed in \cite{Petersen:1999zh}, \cite{Klebanov:2000me}. Lecture
notes dealing with Wilson loops in the context of the AdS/CFT
correspondence are e.g.\cite{Sonnenschein:1999if}.

Settings where the string scale is the TeV scale are reviewed
in\cite{Antoniadis:1999ae}, \cite{Antoniadis:2001bh}.

\subsection{Research papers}

Early proposals of strong/weak coupling duality appear within the
context of the compactified heterotic string\cite{Font:1990gx},
\cite{Rey:1991xj}. This conjecture was supported by observations
reported in\cite{Sen:1993fr,Sen:1994fa,Sen:1994yi},
\cite{Schwarz:1994vs,Schwarz:1993mg}. 
The existence of 11 dimensional supergravity was suggested
in\cite{Nahm:1978tg}. The explicit construction was carried out
in\cite{Cremmer:1978km}. 
The M-theory picture was developed in the
papers\cite{Hull:1995ys}, \cite{Townsend:1995kk},
\cite{Witten:1995ex}. The duality between $SO\left( 32\right)$ type I
and heterotic strings was proposed in\cite{Polchinski:1996df}. The
$SL\left( 2, {\mathbb Z}\right)$ duality of type IIB strings is
discussed in\cite{Schwarz:1995dk}. The relation between the $E_8\times
E_8$ heterotic string and eleven dimensional supergravity is worked
out in\cite{Horava:1996qa,Horava:1996ma}.

Dualities in field theories were conjectured in\cite{Montonen:1977sn},
and shown to be exact in $N=4$ supersymmetric Yang Mills theory
in\cite{Witten:1978mh}, \cite{Osborn:1979tq}, \cite{Witten:1979ey}. 
Strong coupling results in $N=2$ gauge theories are presented in
\cite{Seiberg:1994rs}, \cite{Seiberg:1994aj}, for $SU\left(
  2\right)$. Extensions to other gauge groups are discussed in
e.g.\cite{Klemm:1995qs}, \cite{Klemm:1996wp}, \cite{Argyres:1995jj},
\cite{Ewen:1997uq}, \cite{Ewen:1997yj}.
The $N=1$ field theory dualities have been conjectured
in\cite{Seiberg:1994bz,Seiberg:1995pq}. Some out of many subsequent
papers are\cite{Leigh:1995ep}, \cite{Intriligator:1995ax},
\cite{Intriligator:1995id}, \cite{Intriligator:1995ne},
\cite{Elitzur:1995ap,Elitzur:1996xp}, \cite{Kutasov:1995np},
\cite{Kutasov:1996ss}, \cite{Brodie:1996vx}.
Studying field theories via manipulations in brane setups was
initiated in\cite{Hanany:1997ie}. The discussion in the text
follows\cite{Elitzur:1997fh}. There are many related works. Some
examples
are: \cite{deBoer:1997ck}, \cite{Elitzur:1997hc}, \cite{Evans:1997hk},
\cite{Brandhuber:1997ta},
\cite{Aharony:1997ju}, \cite{Tatar:1998xf}, \cite{Karch:1998yv}. The
connection between $N=2$ supersymmetric gauge theories and M-theory
branes is considered in\cite{Witten:1997sc}. There is also a larger
list of literature dealing with brane setups for $N=2$ theories,
for which, however, I would like to ask the reader to consult one of 
the reviews since this would lead to far away from the subjects discussed
in the text. 

The AdS/CFT correspondence is conjectured in\cite{Maldacena:1998re},
and further elaborated in\cite{Gubser:1998bc},
\cite{Witten:1998qj}. The computation of Wilson loops
within the conjecture is described in\cite{Maldacena:1998im},
\cite{Rey:1998ik}. Differently shaped Wilson loops are discussed
in\cite{Berenstein:1998ij}, \cite{Drukker:1999zq}. Breaking
supersymmetry by a finite temperature one can observe the confinement
of quarks\cite{Witten:1998zw}. Related papers
are\cite{Brandhuber:1998bs}, \cite{Rey:1998bq},
\cite{Brandhuber:1998er}, \cite{Gross:1998gk}, \cite{Gubser:1998nz},
\cite{Dorn:1998ee}, \cite{Pawelczyk:1998pb} and many others. 
The string action on $AdS_5\times S^5$ is constructed
in\cite{Metsaev:1998it}. This action is discussed further
in\cite{Kallosh:1998zx}, \cite{Kallosh:1998nx},
\cite{Kallosh:1998qv}, \cite{Pesando:1998fv}, \cite{Kallosh:1998ji},
\cite{Rajaraman:1999rc}. The construction of\cite{Metsaev:1998it}
leads also to the result that the $AdS_5\times S^5$ background is
exact. Different arguments for this statement are given
in\cite{Banks:1998nr}. The discussion of the stringy corrections to
the 
Wilson loop follows\cite{Forste:1999qn,Forste:1999cp}. A similar
approach (in the conformal gauge) and more examples are discussed
in\cite{Drukker:2000ep}. This paper also addresses the problem of the
divergence and gives a numerical estimate of the correction. String
fluctuations as a source for corrections to the Wilson loop are
also discussed in\cite{Greensite:1999jw}, \cite{Naik:1999bs},
\cite{Zarembo:1999bu}, \cite{Kinar:1999xu}, \cite{Janik:2000pp},
\cite{Loewy:2001pq}. Corrections to the field theory calculation are
derived in\cite{Erickson:1999qv}, \cite{Zarembo:2001jp},
\cite{Erickson:2000af},\cite{Plefka:2001bu}. An attempt to apply the
techniques for 
computing corrections to the Wilson loop on the M5 brane case is
reported in\cite{Forste:1999yj}.

That branes allow constructions with the string scale at a TeV has
been pointed out in\cite{Antoniadis:1998ig}. (Relating the hierarchy
problem to the size of extra dimensions has been proposed before in a
field theory context\cite{Arkani-Hamed:1998rs}.) The argument that in
compactifications of the perturbative heterotic string the size of the
compact space is of the order of the Planck size is given
in\cite{Kaplunovsky:1985yy}. Our discussion of corrections to Newton's
law follows\cite{Kehagias:1999my}. 

\section{Chapter \ref{chap:braneworlds}}

Since there are no books on the subject of brane world setups I start
directly with a list of review articles.

\subsection{Review articles}

The review articles on brane world setups with warped transverse
dimensions I am aware of are\cite{Kachru:2000tg},
\cite{Rubakov:2001kp}, 
\cite{Dick:2001sc}, \cite{Maartens:2001jx}. An overview on the
cosmological constant problem 
is presented in\cite{Weinberg:1989cp}, \cite{Witten:2000zk},
\cite{Binetruy:2000mh}. 

\subsection{Research papers}

Brane world models have been proposed already sometime back
in\cite{Rubakov:1983bb}, \cite{Akama:1982jy}.
The model discussed in section \ref{sec:rs1} is presented
in\cite{Randall:1999ee}. The stabilization mechanism is proposed
in\cite{Goldberger:1999uk}. The model of section \ref{sec:RS2} is
taken from\cite{Randall:1999vf}. An early paper on connecting the
Randall Sundrum model with the holographic principle
is\cite{Verlinde:1999fy}. The computation of the Newton potential via
the holographic principle has been pointed out by Witten in the
discussion session in a Santa Barbara Conference in 1999. (I have not
been there.) The presentation in the text is taken
from\cite{Gubser:1999vj} (see also\cite{Duff:2000mt}). The inclusion
of the second brane into the 
RS2 scenario is performed in\cite{Lykken:1999nb}. The computation of
the Newton potential via the AdS/CFT correspondence is taken
from\cite{Arkani-Hamed:2000ds} (see also \cite{Giddings:2000ay}).  
More discussions of the RS models from a holographic perspective can
be found e.g.\ in\cite{Anchordoqui:2000du},
\cite{Boschi-Filho:2000nh},
\cite{Rattazzi:2000hs}, \cite{Perez-Victoria:2001pa} ,
\cite{Karch:2000ct,Karch:2000fr}, \cite{Shiromizu:2001ve},
  \cite{Gherghetta:2001iv}, 
  \cite{Creminelli:2001th}. 
Supersymmetry within the context of the Randall Sundrum model is
discussed in\cite{Kallosh:2000tj}, \cite{Behrndt:2000tr},
\cite{Altendorfer:2000rr}, \cite{Gherghetta:2000qt},
\cite{Falkowski:2000er,Falkowski:2000yq},    
\cite{Bergshoeff:2000zn}, \cite{Zucker:2000ks}.

Section \ref{solugener} is closely related
to\cite{DeWolfe:1999cp}. 
The consistency condition that the effective
cosmological constant should be compatible with the metric on the
brane is also mentioned in\cite{DeWolfe:1999cp}. The derivation and
form of the consistency condition in section \ref{Sec:Cons} is presented
in\cite{Forste:2000ft}. The alternative method of integrating a total
derivative is developed in\cite{Ellwanger:1999pq}. The connection
between the two conditions has been pointed out
in\cite{Forste:2000ft}. The complete set of consistency conditions (as
it appears in the text) is given in\cite{Gibbons:2000tf}. (Different
consistency conditions are discussed in\cite{Kanti:1999sz}.)

That the cosmological constant problem is rephrased within a brane
world setup is discussed in\cite{Rubakov:1983bz}. The example of
section \ref{KSSexample} (and a closely related example) appear
in\cite{Kachru:2000hf}, \cite{Arkani-Hamed:2000eg}. That the effective
cosmological constant does not vanish in this models is observed
(simultaneously) in
\cite{Youm:2000dc}, \cite{Forste:2000ps}. To reach consistency by
adding branes and consequently fine tuning input parameters is proposed
in \cite{Forste:2000ps}. (Problems with singularities in warped
compactifications are considered e.g.\ also in\cite{Gubser:2000nd},
\cite{Kim:2000yq}.) 
The proof of the no go theorem is taken
from\cite{Csaki:2000wz}. 

There are too many papers on warped brane world scenarios to be 
listed. Therefore, the following list is restricted to papers dealing
with the cosmological constant problem (and most likely this list is
also incomplete): \cite{Chen:2000at}, \cite{deAlwis:2000qc},
\cite{Kanti:2000rd}, 
\cite{deAlwis:2000pr}, \cite{Horowitz:2000ds},
\cite{Grinstein:2000fn}, \cite{Barger:2000wj},
\cite{Krause:2000gp,Krause:2000uj}, \cite{Collins:2000ed},
\cite{Binetruy:2000wn}, \cite{Li:2000kx},
\cite{Kakushadze:2000ix}, 
\cite{KalyanaRama:2000dz}, \cite{Cline:2000ky},
\cite{Berezhiani:2001bn}, \cite{Brax:2001fh}, \cite{Carroll:2001zy},
\cite{Cline:2001nq}, \cite{Diemand:2001pb}, \cite{Brax:2001cx},
\cite{Hebecker:2001mz}. 
Papers containing proposals on avoiding the fine tuning problem of the
cosmological constant by going beyond the prepositions of the no go
theorem (section \ref{sec:nogo}) are \cite{Kehagias:2000dg},
\cite{Kim:2000mc,Kim:2001ez}, \cite{Csaki:2000dm,Grojean:2001pv}.
A different proposal for addressing the cosmological constant problem
in brane world scenarios is put forward in\cite{Verlinde:1999xm}, 
\cite{Schmidhuber:1999rc,Schmidhuber:2000cm,Schmidhuber:2001kw}.

Warped compactifications in the context of string theory are e.g.\
discussed in\cite{Lukas:1998yy,Lukas:1998tt}, \cite{Behrndt:2000zh},
\cite{Dudas:2000ff}, \cite{Blumenhagen:2000dc}, \cite{Brax:2001ds}.

Observational bounds on extra dimension scenarios are e.g.\ presented
in\cite{Long:1998dk}, \cite{Milton:2001np}, \cite{Milton:2000ti}.

\section*{Acknowledgments}

The topics of these notes represent the experience with string theory
that the author gained over the past years. During this period
I collaborated with and received invaluable help from various
people. First, I thank Hans-J\"org Otto and Harald Dorn 
who patiently taught me the basics of string theory, the beta function
approach to the effective field theory, and non critical strings. I
enjoyed very much collaborating with Klaus Behrndt
on\cite{Behrndt:1994zz,Behrndt:1994ev,Behrndt:1997na}. These
collaborations have been very efficient even at times when we
communicated only via email. I would like to thank the Jerusalem
group: Shmuel Elitzur, Amit Giveon and Eliezer Rabinovici for many
insights into the subjects of T-duality, marginal deformations of
conformal field theories, and strongly coupled supersymmetric gauge
theories. I also acknowledge many enjoyable discussions with Gautam
Sengupta. In Munich, I learned quite a lot about string dualities from
Jan Louis and Stefan Theisen. I would like to thank Stefan Schwager
and Alexandros Kehagias for collaborations on T-duality in open string
models. I had many interesting discussions with Kristin F\"orger,
Debashis Ghoshal, Jacek Pawelczyk and Emanuel Scheidegger. I thank
Debashis Ghoshal and Sudhakar Panda for collaborating on orientifold
constructions. I also enjoyed very much the collaboration with
Debashis Ghoshal and Stefan Theisen on stringy corrections to the
Wilson loop in the context of the AdS/CFT correspondence. 
Finally, it is my pleassure to express my gratitude to the high energy
physics group in Bonn. I acknowledge the fruitful collaboration with
Zygmunt Lalak, St\'{e}phane Lavignac and Hans-Peter Nilles on the
cosmological constant problem. Many thanks to Gabriele Honecker and
Ralph Schreyer for collaborating with me on orientifolds. While
writing these notes I had many helpful discussions with Jan Conrad,
Athanasios Dedes, Dumitru Ghilencea,
Stefan Groot Nibbelink, Gabriele Honecker, Mark
Hillenbach, Hanno Klemm,  
Marco Peloso, Ralph Schreyer and Martin Walter.
Special thanks to Gabriele Honecker and Ralph Schreyer for frequent
proof reading of the manuscript. Part of these notes I presented in a
lecture during the summer term 2001. I thank the students for the
stimulating atmosphere during these lectures and Martin Walter for
preparing and delivering accompanying exercises. Last but not least, I
would like 
to express my gratitude to Hans-Peter Nilles for his steady support
and encouragement.

My current work is supported in part by the European Commission RTN
programs \mbox{HPRN-CT-2000-00131}, 00148 and 00152. 
%\chapter{Applications}
\bibliography{literature}

\end{document}